\documentclass[10pt,letterpaper]{article}
\usepackage[top=0.85in,left=2.75in,footskip=0.75in]{geometry}

\usepackage{changepage}

\usepackage[utf8]{inputenc}

\usepackage{textcomp,marvosym}

\usepackage{fixltx2e}

\usepackage{amsmath,amssymb}

\usepackage{cite}

\usepackage{nameref,hyperref}

\usepackage[right]{lineno}

\usepackage{microtype}
\DisableLigatures[f]{encoding = *, family = * }

\usepackage{rotating}

\raggedright
\usepackage[aboveskip=1pt,labelfont=bf,labelsep=period,justification=raggedright,singlelinecheck=off]{caption}

\makeatletter
\renewcommand{\@biblabel}[1]{\quad#1.}
\makeatother

\date{}

\newcommand{\RR}{\mathbb R} 
\newcommand{\CC}{\mathbb C}

\newcommand{\LL}{\mathcal L}
\newcommand{\OO}{\mathcal O}
\newcommand{\dispdot}[2][-14mu]{\dot{#2\mkern#1}\mkern-#1}
\newcommand{\dispddot}[2][-14mu]{\ddot{#2\mkern#1}\mkern-#1}

\newcommand{\githuburl}{https://github.com/neuromethods/fokker-planck-based-spike-rate-models}
\newcommand{\secmethods}{Methods}
\newcommand{\secmeanfield}{Fokker-Planck system}
\newcommand{\secspecprops}{Remarks on the spectrum}
\newcommand{\supplement}{S1 Text}

\usepackage{mathtools}
\usepackage{braket}

\usepackage{xcolor} % text colors

\usepackage[normalem]{ulem} % for strike-through with \sout{...}

\usepackage{graphicx}

\usepackage{bbold}

\hypersetup{pdfborder={0 0 0}} % links wont be visualized % note that hyperref is in template too

\begin{document}
\vspace*{0.35in}

% Please capitalize all terms in the title except conjunctions, prepositions, and articles.
\begin{flushleft}
{\Large
\textbf\newline{Low-dimensional spike rate models derived from networks of adaptive %\textcolor{gray}{nonlinear} 
integrate-and-fire neurons: Comparison and %\textcolor{gray}{numerical} 
implementation}
% short title: Derived low-dimensional spike rate models
%alternative: Spike rate models derived from aEIF networks.
% title alternatives:  
% * Deriving low-dimensional spike rate models from adaptive integrate-and-fire neurons: implementation and comparison of two approaches based on the Fokker-Planck equation
% * Deriving low-dimensional models for the spike rate dynamics of adaptive integrate-and-fire neurons
% * A comparison of low-dimensional spike rate models derived from adaptive integrate-and-fire neurons
% * Benchmarking/Comparing low-dimensional/simple population activity/spike rate models derived from adaptive spiking/integrate-and-fire neurons
}
\\
Moritz Augustin\textsuperscript{1,2,\Yinyang,*}, %#COMMENTED FOR K.O. version
Josef Ladenbauer\textsuperscript{1,2,3,\Yinyang,*}, %#COMMENTED FOR K.O. version
Fabian Baumann\textsuperscript{1,2},
Klaus Obermayer\textsuperscript{1,2}
\\
\bigskip
\bf{1} Department of Software Engineering and Theoretical Computer Science, Technische Universit\"at Berlin, Germany
\\
\bf{2} Bernstein Center for Computational Neuroscience Berlin, Germany
\\
\bf{3} Group for Neural Theory, Laboratoire de Neurosciences Cognitives, École Normale Supérieure, Paris, France
\\
\bigskip

% Insert additional author notes using the symbols described below. Insert symbol callouts after author names as necessary.

% Primary Equal Contribution Note
\Yinyang These authors contributed equally to this work.

% Current address notes: COMMENTED FOR K.O. version
% \textcurrency a Moritz Augustin and Josef Ladenbauer, Technische Universität Berlin, Neural Information Processing Group, Marchstr. 23, MAR 5-6, 10587 Berlin, Germany
% \textcurrency b Insert current address of second author with an address update
% \textcurrency c Insert current address of third author with an address update

% Use the asterisk to denote corresponding authorship and provide email address in note below.
* augustin@ni.tu-berlin.de and josef.ladenbauer@tu-berlin.de

\end{flushleft}

% flatex input: [section/abstract.tex]
\section*{Abstract}

% \todofinal{Please keep the abstract below 300 words (from PLOS template)}

% \todofinal{TODOs for revision: proofread everything, optionally include diagram for LN cascade}
%\todo{computational complexity}
%\todo{more precise: which model works better, maybe leave as it is...}
%\todo{benchmark... scenarios...}
% the following text is based on the CNS 2015 (Prague) abstract (without refs)
% and should maybe partially be truncated and then parts moved to introduction
The spiking activity of single neurons can be well described by a nonlinear
% \textcolor{gray}{two-dimensional} 
integrate-and-fire model that includes 
somatic %\textcolor{gray}{/cell-intrinsic/neuronal} 
adaptation. 
%\textcolor{gray}{caused by slowly 
%decaying potassium currents}. 
When exposed to fluctuating inputs sparsely coupled populations of these model neurons exhibit stochastic collective dynamics that can be effectively 
characterized using the Fokker-Planck equation.
%To effectively characterize the stochastic population dynamics of these model neurons subject to  
%fluctuating inputs and sparsely coupled the Fokker-Planck equation provides a useful too
%For fluctuating inputs sparsely coupled spiking 
%model neurons exhibit stochastic population dynamics which can be effectively 
%characterized using the Fokker-Planck equation. 
This approach, however, leads to a 
model with an infinite-dimensional state space and non-standard boundary 
conditions.
Here we derive from that description four simple models for the spike 
rate dynamics in terms of low-dimensional ordinary differential equations 
using two different reduction %\textcolor{gray}{/approximation} 
techniques: one uses the spectral 
decomposition of the Fokker-Planck operator, the other is based on a cascade 
of two linear filters and a nonlinearity, which are
%fit to response properties of the Fokker-Planck equation.
determined %/extracted 
from the Fokker-Planck equation and semi-analytically approximated. 
%\textcolor{gray}{Although these approximation techniques are interrelated it is not clear which reduced model best reproduces the spike rate of the 
%original spiking network, depending on the statistics of the input.}
% Here we ﬁrst extend each of these reduction methods to account for neuronal 
% adaptation and then 
We evaluate the reduced models for a wide range of biologically plausible input statistics
and find that both approximation approaches lead to spike rate models 
that accurately reproduce the spiking behavior of the underlying adaptive
integrate-and-fire population. Particularly the 
cascade-based models are overall
most accurate and robust, especially in the sensitive 
region of rapidly changing 
%\textcolor{gray}{but weak}\footnote{\textcolor{gray}{Not only weak: cf. results for large background noise, or faster/larger input variations (small $ \tau_{ou}$ or $ \sigma_t $ or large $ \vartheta_\mu $)}} 
input. For the mean-driven regime, 
when input fluctuations are not too strong and fast,
%\footnote{\textcolor{gray}{multiple sense: background noise, as well as intensity and speed of input mean variations (spec2 performance also decreases more rapidly for fast sigma-variations)}}, 
however, the best performing model is based on the spectral decomposition. 
The low-dimensional models also well reproduce stable oscillatory spike rate dynamics that 
are generated either by %\textcolor{gray}{a loop of} 
recurrent synaptic excitation and neuronal adaptation 
or through delayed inhibitory synaptic feedback. 
The computational demands of the reduced
%low-dimensional 
models are very low but the implementation complexity differs 
between the different model variants. Therefore we have made available 
implementations %\footnote{\textcolor{gray}{Do not forget to add a link to 
%the (final) github repository before submission.}} 
that allow to numerically integrate 
the low-dimensional spike 
rate models as well as the %\textcolor{gray}{full/complete/original} 
Fokker-Planck partial differential equation in efficient ways for arbitrary model parametrizations
%\textcolor{gray}{together with example Python scripts} 
as open source software.
% and examples source software.
The derived spike rate descriptions %\textcolor{gray}{are computationally efficient and} 
retain a direct link to the properties of single neurons,
allow for convenient mathematical analyses of  
%\textcolor{gray}{asynchronous and rhythmic} 
network states, 
and are 
%and are thus 
well suited for %(i) 
application in 
neural mass/mean-field  
based brain network models. %having a link to single neuron properties retained and being computationally efficient, . %\footnote{Polish last sentence later.}
% flatex input end: [section/abstract.tex]

% flatex input: [section/authorsummary.tex]
\section*{Author summary}

% \todofinal{Please keep the Author Summary between 150 and 200 words. 
% Use first person.}

%\todo{improve: is a bit too long \& maybe too close to abstract (at least the conclusion)}

% Biophysically described neurons that are recurrently 
% coupled through chemical synapses show a wide range of interesting 
% dynamical states such as asynchrony, oscillations, chaos and bistability 
% thereof. 
% brain networks?

Characterizing the 
%network dynamics of biophysically modeled and 
%synaptically coupled neurons that are driven by in-vivo like fluctuating input 
dynamics of biophysically modeled, large neuronal networks  
usually involves extensive numerical 
simulations. 
% These can be very expensive, especially for significant parameter 
% space explorations, and additionally 
% are not susceptible to mathematical analysis beyond post-simulation statistics. 
As an alternative to this expensive procedure 
we propose efficient models that describe the 
network activity in terms of a few ordinary differential equations. % -- that is, systems, for which 
%linear stability analyses and powerful related methods 
%are readily applicable. 
These systems are simple to solve and allow for convenient investigations of asynchronous, oscillatory or chaotic network states because linear stability analyses
and powerful related methods are readily applicable.
We build upon two research lines 
% of computational biology
on which substantial efforts have been exerted in the last two decades: 
(i) the development of single neuron models of reduced complexity %have been developed 
that can accurately reproduce a large repertoire of observed neuronal 
behavior, %\textcolor{gray}{due to a second (slower) timescale in addition to the (faster) 
%membrane voltage dynamics}, 
and
% including spike-frequency adaptation 
% and have
% biophysically meaningful parameters that can be accurately fit to experimental data using 
% standard protocols.
% (adaptive exponential integrate-and-fire model).
(ii) different approaches %have been proposed 
to approximate the Fokker-Planck equation that represents the collective dynamics 
of large neuronal networks. %of these model neurons. 
We combine these advances and extend recent approximation methods of the latter kind to obtain 
spike rate models that surprisingly well reproduce
the macroscopic dynamics 
of the underlying neuronal network. At the same time the 
microscopic properties are retained through the single neuron model parameters.
%These models are therefore well suited to be employed in
%large-scale neuronal network models as well as for systematic 
%investigations of asynchronous, oscillatory or chaotic network states. 
To enable a fast adoption we have released an efficient Python 
implementation as open source software under a free license. 

% 
% 
% % We have developed a 
% \begin{itemize}
% 
%  \item link between macroscopic network activity and microscopic neuron properties via model parameters
%  \item benefits of low-dim. ODE form and available Python implementation under free 
%   license
% \end{itemize}

% flatex input end: [section/authorsummary.tex]

% \linenumbers

% flatex input: [section/introduction.tex]
\section*{Introduction}
% \todofinal{use for code the so-called data availability statement from plos}
% \textbf{Agenda}:
% \begin{itemize}
%  \item activity at population level important [cf. also gerstner literature intros] (neural mass signals \& coding 
%     properties [refs Harris2015NatNeuroscience Okun2015Nature] groups of neurons with similar properties produce low-dimensional overall spiking activity patterns): 
%  \item 2 modeling techniques available: first networks of spiking neurons are method of choice
%  \item particularly integrate-and-fire-neurons
%  \item nonlinear IF models + adaptation superior (cf. izhikevich and adex+properties)
%  \item simulation costly especially for large systems despite IF (HH would be even more problematic)
%  \item second modeling technique:  population measures such as firing rate and 
%   avg. membrane potential
%  \item motivate noisy setting
%  \item concrete setting we use
%  \item evaluation not only vs. ref but also vs. other models!
% \end{itemize}

There is prominent evidence that %the brain uses %(most/much) 
information in the brain, about a particular stimulus for example, is contained in the collective neuronal spiking activity 
averaged over populations of neurons with similar properties
%is used by the brain to transmit %/process/encode 
%information 
(population spike rate code) \cite{Shadlen1998,Gerstner2014}.
%In the neuroscience there is evidence for a spike rate code in the sense of 
%population activity that is averaged across a group of neurons 
%with similar properties [Shadlen1994,Shadlen1998,Gerstner2002,Gerstner2014]. 
Although these populations can comprise a large number of neurons \cite{Harris2015}, they often
%effectively  
exhibit low-dimensional collective spiking dynamics 
\cite{Okun2015} that can be measured using 
neural mass signals such as the local field potential or electroencephalography. 

The behavior of cortical networks at that level
is often studied computationally by employing 
%\textcolor{gray}{extensive} 
simulations of multiple (realistically large or subsampled) populations of synaptically coupled individual spiking model neurons. % \todo{some prominent refs. here?}. 
A popular choice of single cell description 
for this purpose are two-variable integrate-and-fire models \cite{Brette2005,Izhikevich2003} which describe the evolution of the fast (somatic) membrane voltage and 
an adaptation variable that represents a slowly-decaying potassium current. 
These models are computationally efficient and %their parameters 
can be 
successfully calibrated using 
electrophysiological recordings of real cortical neurons and standard stimulation protocols 
\cite{Brette2005,Badel2008JN,Naud2011,Harrison2015,Rauch2003} to accurately reproduce their subthreshold and spiking activity. 
%or using higher-dimensional Hodgkin-Huxley type models instead [Brette2005]. 
The choice of such (simple) neuron models, however, does not imply reasonable (short enough)
simulation durations for a recurrent network, especially when large numbers of neurons and synaptic connections between them are considered. 

%A faster alternative to simulations of large networks are population activity models  

A fast and mathematically tractable alternative to simulations of large networks are
population activity models in terms of low-dimensional ordinary 
differential equations (i.e., which consist of only a few variables) that 
typically describe the evolution of the spike rate. 
These reduced models can be rapidly solved and allow for convenient analyses
of the dynamical network states using well-known 
% \textcolor{gray}{(textbook)} 
methods that are simple to implement.
%Among those are the popular 
A popular example are the Wilson-Cowan equations \cite{Wilson1972}, which were also extended 
to account for (slow) neuronal adaptation \cite{Latham2000} 
and short-term synaptic depression \cite{Gigante2015}. 
Models of this type have been successfully applied to qualitatively 
characterize %the network behavior for a multitude of 
the possible dynamical states of coupled neuronal populations 
using phase space analyses 
\cite{Latham2000,Gigante2015,Wilson1972},
% spike rate oscillations, bistability of fixed points, 
% chaotic states etc [refs].
yet a direct link to more biophysically described %/original spiking neuron network 
networks of (calibrated) spiking neurons in terms of model parameters 
%such as conductances and time constants 
is missing. 
%
%\todo{put integral eq as first approach and then come to advantages of ODE...}

Recently, derived population activity models have been proposed that 
% \textcolor{gray}{mathematically/
% quantitatively 
bridge the gap between single neuron properties and mesoscopic 
% \textcolor{gray}{/macroscopic} 
network
% \textcolor{gray}{/collective} 
dynamics. 
These models are 
% \textcolor{gray}{expressed in/}
described by integral 
% \textcolor{gray}{/integro-differential} 
equations \cite{Naud2012,Schwalger2017} or partial differential equations 
\cite{Augustin2013,Nicola2015}
% \cite{Iyer2013,Augustin2013,Nicola2015}.
%For example, a mesoscopic model in terms of integral \textcolor{gray}{/integro-differential} equations ... or partial differential equations for the mean population activity 
%of spiking neurons has been developed \cite{Naud2012}.

%To bridge the gap between single neuron dynamics 
%and mesoscopic population activity  
%we take a bottom-up approach. 
Here we derive 
simple models in terms of 
low-dimensional ordinary differential equations (ODEs)
% (such as the Wilson-Cowan equations) 
for the spike rate 
dynamics of sparsely %-- sparse later?
coupled adaptive nonlinear integrate-and-fire neurons 
that are exposed to noisy synaptic input.
%These population models can be rapidly solved and allow for convenient analyses
%of the dynamical network states using well-known \textcolor{gray}{(textbook)} methods.
The derivations are based on %approximating 
a Fokker-Planck equation 
that describes the neuronal population activity 
%\textcolor{gray}{for sufficiently slow adaptation dynamics}
in the mean-field limit of large networks. 
%Using and extending recent methodological advances %works 
We develop reduced models using recent methodological advances on %for 
two different approaches: the first is based on 
a spectral decomposition of the Fokker-Planck operator
under two different slowness assumptions \cite{Mattia2002,Schaffer2013,Mattia2016Arxiv}. 
% \todofinal{replace Mattia2016 by arXiv version in mendeley and bib}
% \todofinal{remove urls and month from final bib file by mass replace}
In the second approach we consider a cascade of linear temporal filters and a nonlinear function 
which are determined from the Fokker-Planck equation and semi-analytically approximated, building upon \cite{Ostojic2011PLOS}. 
Both approaches are extended for an adaptation current, 
a nonlinear spike generating current 
% membrane voltage nonlinearity 
and recurrent coupling with distributed synaptic delays. 
%\todo{move summarized extensions from discussion here/delete in discussion or just use: sparsely coupled with distributed synaptic delays, adaptive nonlinear integrate-and-fire neurons}
%\footnote{\textcolor{gray}{More precise description, that filters are semi-analyt. fit using exponentials and damped oscillator functions too long here and explained later.}}  
%
%way we fit either exponential or damped oscillatory linear input filters
%and a nonlinear spike rate function to response properties of the 
%Fokker-Planck equation [Ostojic2011PLOS,Richardson?]. 
%\todo{should other reduction approaches (there are many -- but none arrives at low-dim. ODE) be mentioned already in the intro or only discussion?}
 
We 
evaluate the developed low-dimensional spike rate models quantitatively 
in terms of reproduction accuracy 
in a systematic manner over a wide range of biologically plausible parameter values. %(beyond the \textcolor{gray}{usual} point-wise tests/validations).
In addition, we provide numerical implementations for the different 
reduction methods as well as the Fokker-Planck equation under a free license as open source project.

For the derived models in this contribution we use the adaptive exponential integrate-and-fire (aEIF) model \cite{Brette2005} to describe individual neurons, which is similar to the model proposed by Izhikevich \cite{Izhikevich2003} but includes %incorporates 
biophysically meaningful parameters and a refined description of spike initiation. 
%\todo{remove oppsed... and turn ``improved'' description better.}
%improved spike initiation through exponential instead of quadratic nonlinearity
However, the presented derivations are equally applicable when using the Izhikevich model instead (requiring only a small number of simple substitutions in the code). % \textcolor{gray}{(for sufficiently slow adaptation)}. \todo{pick this up in the discussion perhaps}

Through their parameters the derived models retain a direct, quantitative link to the underlying spiking model neurons, and they are described in a well-established, convenient form (ODEs) that can be rapidly solved and analyzed. Therefore, these models are well suited (i) for mathematical analyses of dynamical states at the population level, e.g., linear stability analyses of attractors, and (ii) for application in multi-population brain network models.
Apart from a specific network setting, the derived models are also appropriate as a spike rate description of individual neurons under noisy input conditions. 

%\todo{add a paragraph based on the last sentence from abstract (brain network models: extremely slow simulation, and linear stability analysis of attractors)}

The structure of this article contains mildly redundant model specifications allowing the readers who are not interested in the methodological foundation to directly read the self-contained Sect.~Results.

% flatex input end: [section/introduction.tex]

% flatex input: [section/results.tex]
\section*{Results}

% flatex input: [section/results_models.tex]
\subsection*{Model reduction} %\textcolor{gray}{/Scenario/Models}
%\todo{consistently use membrane voltage vs. potential etc.}
% \todo{we are pretty sloppy with definitions, using $ =, :=, \equiv $ arbitrarily throughout the paper}
% \textbf{To Do}
% \begin{itemize}
%  \item instead of \textit{neglecting} recurrency put summarize diffusion approx. \& sparse coupling with delayed 
%  PSP pulses into the beginning of this passage \footnote{$\leadsto$ in the spectral2 model then replace $\mu_\mathrm{syn}$ by $\mu_\mathrm{ext}$ and $\sigma$ similar... and add complete version with delay etc.}
%  \item this yields the motivation for considering only open loop (either also in start of previous 
%   section (models) or at the beginning of the parameter exploration section
% \end{itemize}

The quantity of our interest is the population-averaged number of spikes 
emitted 
by a large homogeneous network of $N$ %adaptive exponential integrate-and-fire (aEIF) 
sparsely coupled aEIF model neurons 
%\textcolor{gray}{in vanishingly small time intervals /}
per small time interval, 
i.e., the spike rate $r_N(t)$. %\todo{Should we say that we here consider one, homog. pop. (identical model params) except for exc., inh. coupling perhaps?} 
% [NEW:]
%The dynamics of cell $i$ 
% in the presence of synaptic 
% bombardment due to recurrent and network-external
%is described by the piecewise continuous time evolution of membrane 
%potential $V_i$ and adaptation current $w_i$ depending on 
The state of neuron $ i $ at time $ t $ is described by the membrane voltage
$V_i(t)$ and adaptation current $w_i(t)$, which evolve 
% in a piecewise continuous manner 
piecewise continuously
in response to 
overall synaptic current 
$I_{\mathrm{syn},i} = I_{\mathrm{ext},i}(t) + I_{\mathrm{rec},i}(t)$.
This input current consists of fluctuating network-external drive %\textcolor{gray}{/activity}
$I_{\mathrm{ext},i} = C [ \mu_{\mathrm{ext}}(t) + \sigma_{\mathrm{ext}}(t) \xi_{\mathrm{ext},i}(t)]$ 
with membrane capacitance $C$, 
time-varying 
moments
$\mu_{\mathrm{ext}}$, $\sigma_{\mathrm{ext}}^2$ and 
% \textcolor{gray}{$\delta$-correlated/unit} 
unit Gaussian white noise process $\xi_{\mathrm{ext},i}$
%\textcolor{gray}{which fluctuates independently from that of other neurons 
%($\xi_j$ for $i\ne j$)}
as well as recurrent input $I_\mathrm{rec,i}$. The latter causes delayed %\textcolor{gray}{, rapidly rising and slowly decaying} 
postsynaptic potentials
(i.e., deflections of $ V_i $) of small amplitude $ J $ triggered by the spikes of $ K $ presynaptic neurons
%/ where every presynaptic spike triggers a jump of the 
%postsynaptic membrane voltage $V_i$ after a delay
%has passed 
(see Sect.~\secmethods{}{} for 
details).
Here we present two approaches
%\footnote{\textcolor{gray}{Should we put literature references also in this 
%results section or is it sufficient to have a detailled discussion of the method 
%including original papers in the methods part? JL: the latter seems better in my mind.}} 
of how the spike rate dynamics
of the 
large, %delay-coupled/
stochastic delay-differential equation system for the $2N$ states $(V_i,w_i)$
% , $i=1,\dots,N$ 
can be described 
by simple models in terms of low-dimensional ODEs. 
Both approaches (i) take into account adaptation current dynamics that are sufficiently slow, allowing 
to replace the individual adaptation current $w_i$ by 
its population-average $\langle w \rangle$, governed by 
\begin{equation}
\frac{d{\langle w \rangle}}{dt} = \frac{a 
\left(\langle V \rangle_\infty -E_w \vphantom{^2} \right) - 
  \langle w \rangle}{\tau_{w}} + b\, r(t), \label{eq_results_w_mean}
\end{equation}
where 
$a$, $E_w$, $b$, $\tau_w$ are the adaptation current model parameters 
(subthreshold conductance, reversal 
potential, spike-triggered increment, time constant, respectively), $\langle V \rangle_\infty$ is the steady-state 
membrane voltage averaged across the population (which can vary over time, see below), and $ r $ is the spike rate of the respective low-dimensional model. 
%and which depends on the spike rate $r$. -- is confusing at this point
Furthermore, both approaches (ii) are based on 
the observation that the collective %\textcolor{gray}{/population-averaged} 
dynamics of a large,
sparsely coupled (and noise driven) network of integrate-and-fire type neurons 
can be well described by a Fokker-Planck equation. 
%\textcolor{gray}{subject to boundary conditions accounting for 
%the discontinuous membrane voltage reset at spike times}. 
%
%in the mean-field limit \textcolor{gray}{($N \to \infty$ neurons)}
% distribution 
%of a \textcolor{gray}{large}
% (large) 
%sparsely coupled
%population
%the time evolution of the membrane potential 
%histogram $\{V_i\}$  can be described by a Fokker-Planck equation 
%\textcolor{gray}{subject to boundary conditions accounting for 
%the discontinuous reset-on-spike of the integrate-and-fire model, 
%see Fig.~\ref{fig1_intro}Z} 
%\textcolor{gray}{-- but in different ways}.
In this \textit{intermediate} Fokker-Planck (FP) model 
%\textcolor{gray}{(between spiking neuron network 
%and the different derived spike rate models/partial 
%differential eq.)} 
the overall synaptic input %\textcolor{gray}{, that is the sum 
%of external and recurrent input,} 
is 
approximated by a mean part with additive white %\todo{add unity of $\xi_i$?} 
Gaussian fluctuations,
$I_\mathrm{syn,i}/C \approx \mu_\mathrm{syn}(t,r_d) + \sigma_\mathrm{syn}(t,r_d) \xi_i(t)$, 
that are uncorrelated between neurons.  %\footnote{\textcolor{gray}{although 
%this $\xi_i$ is not exactly the one from above we could still use this symbol for simplicity. JL: prefer to distinguish via ext}}
The moments %$\mu_\mathrm{syn}$ and $\sigma_\mathrm{syn}^2$ 
of the overall synaptic input,
\begin{equation} \label{eq_res_musyn_sigmasyn}
\mu_\mathrm{syn} = \mu_\mathrm{ext}(t) + J K r_d(t), \qquad 
\sigma^2_\mathrm{syn} = \sigma_\mathrm{ext}^2(t) +  J^2 K r_d(t),
\end{equation} 
% depend on time (since we had time-varying... ext statistics) and 
depend on time via 
the moments of the external input and, due to 
recurrent coupling, on the delayed spike rate $r_d$. The latter is governed by
\begin{equation}
 \label{eq_results_delay}
 \frac{d r_d}{dt} = \frac{r - r_d}{\tau_d},
\end{equation}
%textcolor{gray}{i.e., the spike rate $r(t)$ is %\textcolor{gray}{exponentially} 
%filtered using an exponential kernel with decay constant $\tau_d$,} 
which corresponds to 
individual propagation delays drawn from an exponentially 
distributed random variable with mean $\tau_d$. 
%(other choices for the delays are also possible, for details see \secmethods).
% ... corresponds to exponentially distributed \textcolor{gray}{synaptic} propagation delays with mean $\tau_d$
The FP model involves solving a partial differential equation (PDE) to obtain the time-varying membrane voltage distribution $p(V,t)$ and the spike rate $r(t)$. 
%\textcolor{gray}{instantaneous 
%and delayed} spike rate \textcolor{gray}{$r(t)$ and $r_d(t)$, respectively}, and population-averaged adaptation current \textcolor{gray}{$\langle w \rangle(t)$}.
% \footnote{\textcolor{blue}{May be good to name these variables here, since they occur in Fig.1.}} 

% +  \mu_\mathrm{rec}$ 

%The first reduced spike rate description 
The first reduction approach is based on the spectral decomposition of the Fokker-Planck 
operator $\mathcal L$ and leads to the following two low-dimensional models: 
the ``basic'' model variant (spec$_1$) is given by a complex-valued differential equation describing 
the spike rate evolution in its real part,
\begin{equation} \label{eq_results_spectral1}
\frac{d \tilde r}{dt} = \lambda_1 (\tilde r - r_\infty), \qquad r(t) = \mathrm{Re}\{\tilde r\},
\end{equation}
where $\lambda_1(\mu_\mathrm{tot}, \sigma_\mathrm{tot})$ 
% \textcolor{gray}{(having always negative real part)} 
is the dominant eigenvalue of 
$\mathcal L$ 
and $r_\infty(\mu_\mathrm{tot}, \sigma_\mathrm{tot})$ is the steady-state spike rate. 
%\textcolor{gray}{as in the cascade based models, 
%cf. Eq.~\eqref{eq_results_ln_rate}}. 
Its parameters $ \lambda_1 $, $ r_\infty $, and $ \langle V \rangle_\infty $ (cf. Eq.~\eqref{eq_results_w_mean})
depend on the \textit{total} %\textit{overall} 
input moments given by 
% In this (and the following) spectrum based model 
% the parameters\textcolor{gray}{/quantities} 
% \textcolor{gray}{(here $\lambda_1$ and $r_\infty$)} 
% depend on the \textit{total} %\textit{overall} 
% input moments given by 
$ \mu_\mathrm{tot}(t) = \mu_\mathrm{syn} - \langle w \rangle / C$ and 
$\sigma_\mathrm{tot}^2(t) = \sigma_\mathrm{syn}^2$ 
which closes the model (Eqs.~\eqref{eq_results_w_mean}--\eqref{eq_results_spectral1}). 
%\textcolor{gray}{in contrast to the \textit{effective} (filtered) ones
%in the case of the linear-nonlinear models above}.
% \footnote{\textcolor{gray}{see footnote on next p.}}
The other, ``advanced'' spectral model variant (spec$_2$) is given by a 
real-valued second order differential equation for the spike rate,
\begin{equation} \label{eq_results_spectral2}
\beta_2 \, \ddot r + \beta_1 \, \dot r + \beta_0 \, r 
= r_\infty - r - \beta_c, 
% \left, 
% ( \langle w \rangle, r_d, \, \dot \mu_\mathrm{ext}, \, \ddot \mu_\mathrm{ext}, \, 
% \dispdot{\sigma_\mathrm{ext}^2}, \, \dispddot {\sigma_\mathrm{ext}^2} \right ),
\end{equation}
% \footnote{\textcolor{gray}{Suggest to remove bracket on r.h.s. and mention dependences in text for consistency reasons. Furthermore, would use $ \dot \mu_\mathrm{syn} $ etc. here and explain resulting dependency on delayed rate only in Methods (as in LN description below)}} 
where the dots denote time derivatives.
Its parameters $\beta_2$, $\beta_1$, $\beta_0$, $\beta_c$,  $ r_\infty $ and $ \langle V \rangle_\infty $
depend on the total input moments $(\mu_\mathrm{tot}, \sigma_\mathrm{tot}^2)$ as follows: the latter two parameters 
explicitly as in the basic model above, the former four indirectly via 
the first two dominant 
eigenvalues $\lambda_1$, $\lambda_2$ 
% $\lambda_1\textcolor{gray}{(\mu_\mathrm{tot},\sigma_\mathrm{tot})}$, $\lambda_2\textcolor{gray}{(\mu_\mathrm{tot},\sigma_\mathrm{tot})}$
% \textcolor{gray}{of $\mathcal L^{(\mu_\mathrm{tot},\sigma_\mathrm{tot})}$}, the 
%the steady state spike rate $r_\infty$, 
% $r_\infty\textcolor{gray}{(\mu_\mathrm{tot},\sigma_\mathrm{tot})}$, 
and via additional quantities obtained
from the (stationary and the first two nonstationary) 
eigenfunctions of $\mathcal L$ and its 
adjoint $\mathcal L^*$.  
Furthermore, the parameter $\beta_c$ depends explicitly on the population-averaged 
adaptation current $ \langle w \rangle$, the 
delayed spike rate $r_d$, and on 
the first and second order time derivatives of the external moments 
$\mu_\mathrm{ext}$ 
and $\sigma^2_\mathrm{ext}$.
% \textcolor{gray}{Furthermore, $\beta_1$, $\beta_0$, $\beta_c$ depend 
%on the adaptation current parameters $a$, $b$ and $\tau_w$, i.e., the 
%population-averaged adaptation current $\langle w \rangle$ is considered in the second 
%spectral model beyond subtracting from the input mean 
%(see also the explicit dependence on $\langle w \rangle$ in 
%% the coefficient of $\beta_c$ in 
%Eq.~\eqref{eq_results_spectral2}).}
%
% In the derivation of first spectral model, cf. 
% Eq.~\eqref{eq_results_spectral1}, temporal variations of the overall input moments 
% are completely neglected while the second description, Eq.~\eqref{eq_results_spectral2} 
% incorporates (slow) changes of the input moments through linear 
% terms 
% % (proportional to $\dot \mu_\mathrm{tot}$ or $\dispdot{\sigma^2_\mathrm{tot}}$) 
% and neglects \textcolor{gray}{(faster)} quadratic ones, 
% i.e., $\dot \mu^2_\mathrm{tot} \approx 0$, \
% $(\dispdot{\sigma^2_\mathrm{tot}})^2 \!\! \approx 0$, \ $\dot \mu_\mathrm{tot} \dispdot{\sigma^2_\mathrm{tot}} \approx 0$.
% % (omitting subscripts: $_\mathrm{syn}$).

The second approach is based on 
a Linear-Nonlinear (LN) cascade, in which the population spike rate is generated by applying to the time-varying mean and standard deviation of the 
overall synaptic
input, %\todo{@JL: $\sigma$ is not a moment}, 
$\mu_{\mathrm{syn}}$ and 
$\sigma_{\mathrm{syn}}$, separately %successively 
a linear temporal filter, %$D_\mu$ and $D_\sigma$, %respectively, 
followed by a common nonlinear function. % $F$
These three %model 
components -- two linear filters and a nonlinearity -- are extracted from the Fokker-Planck equation.  
%\textcolor{gray}{-- in particular, from %its response properties 
%the first order rate response to modulations of $\mu_{\mathrm{syn}}$ and 
%$\sigma_{\mathrm{syn}}$, and from the steady-state solution of the FP model.}
%a cascade of a linear input filter and a nonlinearity 
%which are extracted from response properties
%\textcolor{gray}{first order/linear rate response to input modulations and steady-state solution} 
%of the Fokker-Planck eq.
%-- yields the following 
%low-dim. model variants: the mean synaptic input $\mu_{\mathrm{syn}}(t)$ is either filtered by an
%exponential,
%
Approximating the linear filters using exponentials and damped oscillating functions yields two model variants:
In the basic ``exponential'' (LN$_{\mathrm{exp}}$) model the filtered mean $\mu_\mathrm{f} $ and standard deviation $\sigma_\mathrm{f} $ of the overall synaptic input are given by
\begin{equation} \label{eq_results_ln_exp}
\frac{d \mu_{\mathrm{f}}}{dt} = 
\frac{\mu_{\mathrm{syn}} - \mu_{\mathrm{f}}}{\tau_{\mu}}, \qquad
\frac{d \sigma_{\mathrm{f}}}{dt} = 
\frac{\sigma_{\mathrm{syn}} - \sigma_{\mathrm{f}}}{\tau_{\sigma}},
\end{equation}
where the time constants $\tau_\mu(\mu_{\mathrm{eff}},\sigma_{\mathrm{eff}})$, $\tau_\sigma(\mu_{\mathrm{eff}},\sigma_{\mathrm{eff}})$ depend on the \textit{effective} 
(filtered) input mean $\mu_{\mathrm{eff}}(t) = \mu_{\mathrm{f}} - \langle w \rangle/C$ and standard deviation
$\sigma_{\mathrm{eff}}(t) = \sigma_\mathrm{f}$. 
The ``damped oscillator'' (LN$_{\mathrm{dos}}$) model variant, on the other hand, describes the filtered input moments by
\begin{align} \label{eq_results_ln_dosc}
\ddot \mu_{\mathrm{f}} + \frac{2}{\tau} \dot \mu_{\mathrm{f}} + \left( \frac{2}{\tau^2} + \omega^2 \right) \mu_{\mathrm{f}}
&= \frac{1 + \tau^2 \omega^2}{\tau} \left( \frac{\mu_{\mathrm{syn}}}{\tau} + \dot \mu_{\mathrm{syn}} \right), \\
\frac{d \sigma_{\mathrm{f}}}{dt} &= 
\frac{\sigma_{\mathrm{syn}} - \sigma_{\mathrm{f}}}{\tau_{\sigma}},
\end{align}
where the time constants $\tau(\mu_{\mathrm{tot}},\sigma_{\mathrm{tot}})$, $\tau_\sigma(\mu_{\mathrm{tot}},\sigma_{\mathrm{tot}})$ and the angular frequency $\omega(\mu_{\mathrm{tot}},\sigma_{\mathrm{tot}})$ depend on the \textit{total} 
input moments defined above.
In both LN model variants the spike rate is obtained by the nonlinear transformation of the %\textcolor{gray}{respective} 
\textit{effective} 
input moments through the steady-state spike rate,
%\footnote{\textcolor{gray}{the 
%input-depencence of $\langle w \rangle_\infty$ is currently explicitly stated only in the 
%methods which is ok imho (M)}},
\begin{equation}
r(t) = r_\infty \! \left(\mu_{\mathrm{eff}}, \, \sigma_{\mathrm{eff}}\right), \label{eq_results_ln_rate}
\end{equation}
and the steady-state mean membrane voltage $ \langle V \rangle_\infty $ (cf. Eq.~\eqref{eq_results_w_mean})
is also evaluated at $ (\mu_{\mathrm{eff}},\sigma_{\mathrm{eff}}) $.

These four models (spec$_1$, spec$_2$, LN$_{\mathrm{exp}}$, LN$_{\mathrm{dos}}$) from both reduction approaches involve a number of parameters that depend on the strengths of synaptic input and adaptation current only via the \textit{total} or \textit{effective} input moments. We refer to these parameters as \textit{quantities} below to distinguish them from fixed (independent) parameters. %$r_\infty$, $\langle V \rangle_\infty$, $\tau_{\!\mu}$, $\tilde{f}$, $\tilde{\tau}$, $\tau_{\!\sigma}$ 
%[INCLUDE NO. OF QUANTS TO BE PRECOMPUTED PER MODEL FOR LOOKUP (to give a hint on complexities)]
%the functions $\langle V \rangle_\infty$, $\tau_{\!\mu}$, $\tilde{f}$, $\tilde{\tau}$, and $\tau_{\!\sigma}$ depend on the strengths of synaptic input and adaptation current only via the effective input mean $\mu_{\mathrm{f}} - \langle w \rangle/C$ and its standard deviation $\sigma_{\mathrm{f}}$. 
The computational complexity 
when numerically solving the models forward in time (for different parametrizations) can be greatly reduced by  precomputing those quantities for a range of values for the \textit{total}/\textit{effective} input moments %(see Fig.~\ref{LC_fig5}) 
and using look-up tables during time integration. Changing any parameter value of the external input, the recurrent coupling 
or the adaptation current does not require renewed precomputations, 
enabling rapid explorations of parameter space 
and efficient (linear) stability analyses of network states.
% asynchronous or oscillatory spike rate dynamics\textcolor{gray}{/network states}.

The full specification of the 
``ground truth'' system (network of aEIF neurons), the derivations of the 
intermediate description (FP model) and 
the low-dimensional spike rate models 
%\textcolor{gray}{, Eqs. \eqref{eq_results_ln_exp}, 
%\eqref{eq_results_ln_dosc}, \eqref{eq_results_spectral1} or \eqref{eq_results_spectral2},}
complemented by concrete numerical implementations
are provided in Sect.~\secmethods{} (that is complemented by the supporting material \supplement). 
In Fig.~\ref{fig1_example} we visualize 
the outputs of the different models using an example excitatory aEIF network exposed to external input with varying mean $\mu_\mathrm{ext}(t)$ and 
standard deviation $\sigma_\mathrm{ext}(t)$. 

% \todo{check consistency below: quantity vs. parameter}
% flatex input end: [section/results_models.tex]

% flatex input: [section/results_exploration.tex]
\subsection*{Performance for variations of the mean input} %\textcolor{gray}{/Approximation goodness}

% \todo{all figs with overlapping lines put note in caption}

% \todo{Maybe improve the beginning of this Sect. by inspiration of the model-common Methods Sect.}
% \todo{adjust $ r_N(t) $ in results / figures}

Here, and in the subsequent two sections, we assess the accuracy of the four low-dimensional
models to reproduce the spike rate dynamics of the underlying aEIF population. The \textit{intermediate} FP model is included for reference. 
The derived models %depend on the (combined) 
generate population activity in response to overall synaptic input moments 
$\mu_\mathrm{syn}$ and $\sigma_\mathrm{syn}^2$. These depend on 
time via the external moments 
$\mu_\mathrm{ext}(t)$ and $\sigma_\mathrm{ext}^2(t)$, and the delayed spike 
rate $r_d(t)$. Therefore, it is instrumental to first consider an uncoupled 
population %i.e., $K=0$, 
and 
suitable variations of external input moments 
that effectively mimic a range of biologically plausible 
presynaptic spike rate dynamics. 
This allows us to systematically compare the reproduction performance of 
the different %\textcolor{gray}{low-dimensional} 
models over a manageable parameter space (without $ K $, $ J $, $ \tau_d $),
% for a wide range of reasonable parameters. 
yet it provides useful information on the accuracy for recurrent networks.

For many network settings %instances\textcolor{gray}{/setups} 
the dominant effect of synaptic
coupling %\textcolor{gray}{through the delayed spike rate} 
is on the mean input %as compared to the input standard deviation, due to 
%\todo{mu rec, sigma rec not defined before}$\mu_\mathrm{rec}(t) = K J r_d(t) \textcolor{gray}{\,\propto r_d(t)}$ and   
%$\sigma_\mathrm{rec}(t) = J \sqrt{K r_d(t)}$ 
(cf. Eq.~\eqref{eq_res_musyn_sigmasyn}). 
Therefore, we consider first in detail time-varying mean but constant variance of the input.  
Specifically, to account for a wide 
range of oscillation frequencies for presynaptic spike rates, $\mu_\mathrm{ext}$ is described by an Ornstein-Uhlenbeck (OU) process 
%\textcolor{gray}{in stationary condition/with zero-initialization, i.e.,
%$\mu_\mathrm{ext}(0) = 0$}, \todo{discuss: change all $t_0$ here and in methods to $0$? JL: agree}
%\todo{discuss alternative: generic OU, e.g., $\dot x=\dots$ for $x=\mu_\mathrm{ext},\sigma_\mathrm{ext}^2$ (MA: I prefer the example version. JL: agree)}
\begin{equation} \label{eq_results_ou_mu}
\dot \mu_\mathrm{ext} = \frac{\bar\mu - \mu_\mathrm{ext}}{\tau^\mu_\mathrm{ou}} + \sqrt{\frac{2}{\tau^\mu_\mathrm{ou}}} \vartheta_\mu \xi(t), 
\end{equation}
where $\tau^\mu_\mathrm{ou}$ denotes the correlation time, 
$\bar\mu$ and $\vartheta_\mu$ are the mean and standard deviation 
of the stationary normal distribution, 
i.e., 
$\lim_{t\to\infty}\mu_\mathrm{ext}(t)\sim\mathcal N(\bar \mu,\vartheta_\mu^2)$, and 
$\xi$ is a unit Gaussian white noise process. 
%
%Particularly for a single simulation 
%we generate a sample path 
Sample time series generated from the OU process %, Eq.~\eqref{eq_results_ou_mu},
are filtered using a Gaussian kernel with a small standard deviation 
$\sigma_t$ to obtain %\textcolor{gray}{infinitely often} 
sufficiently differentiable time series 
$\tilde \mu_\mathrm{ext}$
(due to the requirements of the spec$_2$ model %\textcolor{gray}{, which depends on the first and second time-derivative of both input moments,} 
and the LN$_\mathrm{dos}$ model). %\textcolor{gray}{, which depends on the time-derivative of the mean input}). 
%spec$_2$ \textcolor{gray}{on $\dot \mu_\mathrm{ext}, \, \ddot \mu_\mathrm{ext}, \, 
%\dispdot{\sigma_\mathrm{ext}^2}, \, \dispddot {\sigma_\mathrm{ext}^2}$} 
%and LN$_\mathrm{dos}$ \textcolor{gray}{on $\dot\mu_\mathrm{ext}$}). 
The filtered 
realization $\tilde \mu_\mathrm{ext}(t)$ is then used for all models 
%\textcolor{gray}{-- the 
%aEIF network, the intermediate Fokker-Planck description and the low-dimensional 
%representations --} 
to allow for a quantitative comparison of the different spike rate responses to 
the same input.
The value of $\sigma_t$ we use 
in this study %we use  $\sigma_t = 1$~ms, which 
effectively removes very large oscillation frequencies which are rarely observed, 
while lower frequencies \cite{Brunel2003JN} are passed.
%larger than those 
%typically observed in 
%oscillatory activity of \textcolor{gray}{real cortical networks/}more detailed network models \cite{BrunelWang2003} while lower frequencies are passed\textcolor{gray}{/unaffected}.

The parameter space we explore covers large and small correlation times $\tau^\mu_\mathrm{ou}$, 
strong and weak input mean $\bar\mu$ and standard deviation $\sigma_\mathrm{ext}$, and for each of these combinations  
we consider an interval from small to large 
variation magnitudes $\vartheta_\mu$. 
The values of $\tau^\mu_\mathrm{ou}$ and $\vartheta_\mu$ determine how rapid and intense $\mu_\mathrm{ext}(t)$ fluctuates. 
%via 
%large and small correlation time $\tau^\mu_\mathrm{ou}$, 
%weak and strong mean input $\bar\mu$ 
%and variance  $\sigma^2_\mathrm{ext}$ and for each of these 
%combinations we consider an interval from small to very large 
%variation amplitudes $\vartheta_\mu$. 

We apply two performance measures, as in \cite{Ostojic2011PLOS}.
One is given by Pearson's correlation coefficient, 
\begin{equation}
\rho(r_N, r) \coloneqq \frac{\sum_{k=1}^M (r_N(t_k) - \bar r_N) 
  \left(r(t_k)-\bar r\right)}
  {\sqrt{ \sum_{k=1}^M \left( r_N(t_k) - \bar r_N \right)^2 \sum_{k=1}^M  ( r(t_k) - \bar r )^2 }}, 
\end{equation}
between the  %, $t_k \in [0,T]$) 
(discretely given) 
spike rates of the aEIF population and each derived model 
with 
time averages $\bar r_N = 1/M \sum_{k=1}^M r_N(t_k)$ 
and $\bar r = 1/M \sum_{k=1}^M r(t_k)$ over a time interval of length $t_M-t_1$. 
For comparison, we also include the correlation coefficient between the aEIF population spike rate and the time-varying mean input, $\rho(r_N,\mu_\mathrm{ext})$.
In addition, to assess absolute differences we calculate the 
root mean square (RMS) distance, 
\begin{equation}
\mathrm d_\mathrm{RMS}(r_N, r) \coloneqq \sqrt{ \frac{1}{M} \sum_{k=1}^M 
\Big(r_N(t_k) - r(t_k)\Big)^2 }\, ,
%d_\mathrm{RMS}(r_N, r) \coloneqq \sqrt{ \frac{1}{T} \sum_{k} 
%\Big(r_N(t_k) - r(t_k)\Big)^2 }.
\end{equation}
%\textcolor{gray}{which is related to the correlation coefficient by 
%$\rho = 1 - \frac{d^2_\mathrm{RMS}}{2 \mathrm{Var}(r)}$ if the 
%means and variances of both time series are identical} 
%following \cite{Ostojic2011PLOS}. %\todo{JL stopped here}
where $M$ denotes the number of elements of
the respective time series ($r_N, r$).
% \todo{@JL/MA: improve description of regimes (e.g. use the grey shaded 
% spike rate mean+-std and examples in Figs. 1\&2 for reference)}
% \todo{@JL/MA: improve interpretation/explanation with methodological model details}
% \textcolor{gray}{\textbf{I) Overall Results\\}}

We find that three of the four low-dimensional spike rate models 
(spec$_2$, LN$_\mathrm{exp}$, LN$_\mathrm{dos}$) 
very well reproduce
the spike rate $r_N$ of the aEIF neurons: for the LN$_\mathrm{exp}$ model $ \rho > 0.95 $ and for the spec$_2$ and LN$_\mathrm{dos}$ models $ \rho \gtrsim 0.8 $ (each) over the explored parameter space, see Fig.~\ref{fig2_mu_exploration}. 
%except for very strong variations of the mean input, $\mu_\mathrm{ext}(t)$, 
%i.e., when $\tau^\mu_\mathrm{ou}$ is small and $\vartheta_\mu$ large 
%(cf. Fig.~\ref{fig2_mu_exploration}). 
Only the basic spectral model (spec$_1$) is 
substantially less accurate. %\textcolor{gray}{than the other derived models} 
%on the whole parameter space.  
Among the best models, the simplest 
(LN$_\mathrm{exp}$) 
% is the most robust one and 
overall outperforms spec$_2$ and 
LN$_\mathrm{dos}$, in particular 
for fast and strong mean input variations. 
However, in the strongly mean-driven regime the best
performing model is spec$_2$.

%In particular,
We observe that 
the performance of any of the spike rate models decreases %monotonically
(with model-specific slope) with (i) increasing %\textcolor{gray}{mean input} 
variation strength $\vartheta_\mu$  
larger than a certain (small) value,
and with (ii) smaller $\tau^\mu_\mathrm{ou}$, i.e., faster changes of 
$\mu_\mathrm{ext}$. For small values of $\vartheta_\mu$ fluctuations of $r_N$, which are caused by the finite aEIF population size $ N $ and do not depend on the fluctuations of $\mu_\mathrm{ext}$, deteriorate the performance measured by $ \rho $ (see also \cite{Ostojic2011PLOS}, p.13 right). This explains why $ \rho $ does not increase as $\vartheta_\mu$ decreases (towards zero) for any of the models.
Naturally, the FP model is the by far 
most accurate spike rate description in terms of 
both measures, correlation coefficient $\rho$ and 
RMS distance. This is not 
surprising because the four low-dimensional models are derived from 
that %\textcolor{gray}{partial differential} 
(infinite-dimensional) representation. Thus, the performance of the FP system defines 
an upper bound on the correlation coefficient $\rho$ 
and a lower bound on the RMS distance for the low-dimensional models. 

In detail: for moderately fast changing mean input (large $\tau^\mu_\mathrm{ou}$) the three 
models spec$_2$, LN$_\mathrm{exp}$ and LN$_\mathrm{dos}$ exhibit excellent reproduction performance 
with $\rho > 0.95$, and spec$_1$ shows correlation coefficients of 
at least $\rho = 0.9$ (Fig.~\ref{fig2_mu_exploration}A), which 
is substantially better than $\rho(r_N,\mu_\mathrm{ext})$. 
%the most simple\textcolor{gray}{/artificial} 
%identity model, i.e., using the input as the output, 
%$r(t) \coloneqq \mu_\mathrm{ext}(t)$. 
The small differences between the three top models 
can be better assessed from the RMS distance measure.
For large input variance $\sigma_\mathrm{ext}^2$ the two LN models perform best
(cf. Fig.~\ref{fig2_mu_exploration}A, top, and for an example,
\ref{fig2_mu_exploration}C). For weak input variance and large mean (small 
$\sigma_\mathrm{ext}$, large $\bar\mu$) the spec$_2$ model outperforms the LN models, unless the variation magnitude $\vartheta_\mu$ is very large.  
For small mean $\bar\mu$, where transient activity is interleaved with periods of quiescence,
% (cf. \ref{fig2_mu_exploration}C), 
%where small and zero spike rates occur already for moderate 
%input variation strength $\vartheta_\mu$, 
the
LN$_\mathrm{exp}$ model performs best, except for weak variations $\vartheta_\mu$, where LN$_\mathrm{dos}$ is slightly better (see Fig.~\ref{fig2_mu_exploration}A, bottom). 

Stronger differences in performance emerge when considering faster changes of the mean input
$\mu_\mathrm{ext}(t)$ (i.e., for
small $\tau^\mu_\mathrm{ou}$), see Fig.~\ref{fig2_mu_exploration}B, and 
for examples, Fig.~\ref{fig2_mu_exploration}C. 
The  spec$_1$ model again performs worst with 
    $\rho$ values even below the input/output correlation
    baseline $\rho(r_N,\mu_\mathrm{ext})$ for large mean input $\bar\mu$ 
    (cf. Fig.~\ref{fig2_mu_exploration}B, left).
    The spec$_1$ spike rate typically decays too slowly (cf. Fig.~\ref{fig2_mu_exploration}C).  
The three better performing models differ as follows: 
    for large input variance and mean (large $\sigma_\mathrm{ext}$ and  
    $\bar\mu$),  
    where the spike rate response to the input is rather fast 
    (cf. increased $\rho(r_N,\mu_\mathrm{ext})$), 
    the performance of all three models in terms of $\rho$ is very high, but  
    the RMS distance measure indicates that 
    LN$_\mathrm{exp}$ is the most accurate model (cf. Fig.~\ref{fig2_mu_exploration}B, top). 
    %-- at least for stronger input variations $\vartheta_\mu$. 
    For weak mean input LN$_\mathrm{exp}$ is once again the top model 
    while LN$_\mathrm{dos}$ and, 
    especially noticeable, spec$_2$ show a performance decline
%     \todo{@MA: interprete/explain, cf. rectify} 
    (see example in Fig.~\ref{fig2_mu_exploration}C).
    For weak input variance (Fig.~\ref{fig2_mu_exploration}A, bottom), where 
    significant (oscillatory) excursions of the spike rates in response to changes 
    in the mean 
    input %$\mu_\mathrm{ext}(t)$ 
    can be observed (see also Fig.~\ref{fig1_example}),
    we obtain the following benchmark contrast: for large mean drive $\bar\mu$ the spec$_2$ model performs best, except for large variation amplitudes $\vartheta_\mu$, 
    at which LN$_\mathrm{exp}$ is more accurate. Smaller mean input on the 
    other hand corresponds to the most sensitive regime where periods of quiescence 
    alternate with rapidly increasing and decaying spike rates. % due to 
    %switches between sub- and suprathreshold total/effective mean drive through $\mu_\mathrm{ext}$ and $\langle w \rangle$. The 
    The LN$_\mathrm{exp}$ model shows the most robust and accurate spike rate 
    reproduction in this setting, 
    while LN$_\mathrm{dos}$ and spec$_2$ each exhibit decreased 
    correlation and larger RMS distances -- spec$_2$ even for moderate 
    input variation intensities $\vartheta_\mu$. 
    The slowness approximation underlying 
    the spec$_2$ model likely induces an error due to 
  the fast external input changes in comparison with the rather slow 
  intrinsic time scale by the dominant eigenvalue, 
$\tau_\mathrm{ou}^\mu=5$~ms vs. $1/|\mathrm{Re}\{\lambda_1\}| \approx 15$~ms 
(cf. visualization of the spectrum in Sect.~\textit{Spectral~models}).
% (cf. Fig.~\ref{fig7_spectral}C).  % removed due to fig appearance in order
Note that for these weak inputs 
    the distribution of the spike rate is rather 
    asymmetric (cf. Fig.~\ref{fig2_mu_exploration}B). 
Interestingly the LN$_\mathrm{dos}$ model performs worse than LN$_\mathrm{exp}$ for large mean input variations (i.e., large $\vartheta_\mu$) in general, and only slightly better for small input variance and mean input variations that are not too large and fast. % (cf. Fig.~\ref{fig2_mu_exploration}). 
    
% Most of the inflections %in the measures vs. variation 
% %amplitudes $\vartheta_\mu$ graphs that correspond to larger decrease in performance 
% %of the models 
% in the spec$_2$ performance curves %and LN$_\mathrm{dos}$ 
% correspond to the start of \todo{@MA: rephrase start of ...} 
% significantly frequent network quiescence, i.e., where an asymmetry in the 
% spike rate distribution shows up due to the lower bound of zero (cf. Fig.~\ref{fig2_mu_exploration}A,B). \todo{@MA: This link is not really obvious in the figure (maybe omit or clarify w/o the figure (not shown))} 
%
% \textcolor{gray}{Note that removing the spike rate rectification of the 
% spec$_2$ model would yield significantly negative rates in the 
% sensitive regime for small mean input $\bar\mu$, small external noise $\sigma_\mathrm{ext}$ and fast OU dynamics, i.e., small correlation time.} \todo{@MA: The rectification part seems more appropriate in the Methods section. Suggest to omit here.}  
%
% In this configuration the slowness assumption seems to be violated due to 
% the fast external input changes in comparison with the rather slow 
% intrinsic time scale of the dominant eigenvalue, 
% $1/|\mathrm{Re}(\lambda_1)| \approx 20$~ms vs. $\tau_\mathrm{ou}^\mu=5$~ms 
% (cf. Fig.~\ref{fig7_spectral}C). \todo{@MA: remove seems perhaps, and maybe rephrase intrinsic time scale of eigenvalue} 

We would like to note that decreasing the Gaussian filter width $\sigma_t$ to smaller values, e.g., 
fractions of a millisecond, can lead to a strong performance decline for the spec$_2$ model because of its explicit dependence on first and second order time derivatives of the mean input.
% \footnote{\textcolor{gray}{Internal remark: This does not affect the LN$_\mathrm{dos}$ model (more than what can be expected from faster variations), see exploration results with $\sigma_t = 0.5, 0.1$ ms (not shown).}} % and since an unfiltered OU process realization is nowhere differentiable. 

% \todo{note on how much results depend on chosen Gauss filter width (not shown)\footnote{Fabian currently (14.7.) runs width$=0.1, 0.5, 1$ms since the last value (used so far) quite strongly acts on the OU -- see \url{1ms_gauss_smoothing.png} in artwork folder.}

%\textcolor{gray}{potentially excluded further remarks:
%\begin{itemize}
%  \item rms bounded by below due to finite size (ostojic11)
%  \item correlation for small OU standard deviations: finite size effect (decrease for small $\vartheta_\mu$ -- why exactly?)
%\end{itemize}
%}

Furthermore, we show how the adaptation parameters affect the reproduction 
performance of the different models in Fig.~\ref{fig3_adapt_timescale}. 
The adaptation time constant $\tau_w$ and spike-triggered adaptation increment $b$ are varied simultaneously (keeping their product constant) such that the average spike rate and adaptation current, and thus the spiking regime, remain comparable for all parametrizations. As expected, the accuracy of the derived models decreases for faster adaptation current dynamics, due to the adiabatic approximation that relies on sufficiently slow 
adaptation (cf. Sect.~\secmethods). Interestingly however, the performance of all reduced models (except spec$_1$) declines only slightly as the adaptation time constant decreases to the value of the membrane time constant (which means the assumption of separated time scales underlying the adiabatic approximation is clearly violated). This kind of robustness is particularly pronounced for input with 
large baseline mean $\bar \mu$ and small noise amplitude $\sigma_\mathrm{ext}$, cf. Fig.~\ref{fig3_adapt_timescale}B.

\subsection*{Performance for variations of the input variance} %\textcolor{gray}{/Nonstationary input variance}
%\todo{gray title suggests that we consider varying input variance \textit{in addition}}

For perfectly balanced excitatory and inhibitory synaptic coupling 
the contribution of presynaptic activity to the mean input $\mu_\mathrm{syn}$
is zero by definition, but the input variance 
$\sigma_\mathrm{syn}^2$ %the impact of the delayed spike rate, $r_d$, on the mean is by definition none and 
is always positively (linearly) 
affected by a presynaptic spike rate -- even for a negative synaptic efficacy $J$ (cf. Eq.~\eqref{eq_res_musyn_sigmasyn}). 
% (cf. Eq.~\eqref{eq_res_musyn_sigmasyn}). \todo{Not clearly visible in the eq.} 
To assess the performance of the derived models in this scenario, but within the 
reference setting of an uncoupled population, 
% \textcolor{gray}{of aEIF neurons},
we consider constant external mean drive $\mu_\mathrm{ext}$ and let 
the variance $\sigma_\mathrm{ext}^2(t)$ evolve according to a filtered
OU process (such as that used for the mean input $ \mu_\mathrm{ext} $ in the previous section) with parameters %\textcolor{gray}{that correspond to mean} 
$\overline{\sigma^2}$ and 
%\textcolor{gray}{and standard deviation} 
$\vartheta_{\sigma^2}$ of the stationary normal distribution $\mathcal N(\overline{\sigma^2}, \vartheta_{\sigma^2}^2)$, correlation time $\tau^{\sigma^2}_\mathrm{ou}$ and 
Gaussian filter standard deviation $\sigma_t$ as before. 

The results of two input parametrizations are shown in Fig.~\ref{fig4_sigma2_ou}.
For large input mean $\mu_\mathrm{ext}$ and rapidly varying variance $\sigma_\mathrm{ext}^2(t)$ 
the spike rate response of the aEIF population is very well reproduced by the FP model
and, to a large extent, by the spec$_2$ model 
(cf. Fig.~\ref{fig4_sigma2_ou}A). This may be attributed to the fact that the latter model 
depends on
the first two time derivatives of the input variance 
$\sigma^2_\mathrm{ext}$. %The latter is not present in the model spec$_1$ which 
The LN models cannot well reproduce the rapid spike rate 
% deviations/
excursions in this setting, and the spec$_1$ model performs worst, exhibiting time-lagged spike rate dynamics compared to $r_N(t)$ which leads to a very small value of correlation coefficient $\rho$ (below the input/output correlation 
baseline $\rho(r_N,\sigma_\mathrm{ext}^2)$).
%does not reproduce the dynamics well 
%but shows a time lagged spike rate for this example leading to 
%\textcolor{gray}{almost uncorrelation/}a very small correlation coefficient $\rho$. \todo{@MA: move small spec1 correlation 
%to fig. caption}
%\textcolor{blue}{The LN cascade based models are 
%derived for the mu filter and do not follow the spike rate dynamics well......}\todo{@JL: explain and refer to filters for example}.
For smaller mean input $\mu_\mathrm{ext}$ and 
moderately fast varying variance $\sigma^2_\mathrm{ext}(t)$ (larger correlation time $\tau_\mathrm{ou}^{\sigma^2}$)
the fluctuating 
aEIF population spike rate is again nicely reproduced by the FP model 
while the rate response of the spec$_2$ model exhibits over-sensitive behavior to changes in
the input variance, as indicated by the large RMS distance
(see Fig.~\ref{fig4_sigma2_ou}B). This effect is
even stronger for faster variations, i.e., 
smaller $\tau_\mathrm{ou}^{\sigma^2}$ 
(cf. supplementary visualization S1 Figure). The LN models 
perform better in this setting, 
%but still do not 
%capture the full spike rate amplitudes  
and the spec$_1$ model (again) performs worst in terms of correlation coefficient $ \rho $  
due to its time-lagged spike rate response.

It should be noted that the lowest possible value of the input standard deviation, i.e., $\sigma_\mathrm{ext}$ (plus a nonnegative number in case of recurrent 
input) cannot be chosen completely freely but must be large enough 
($\gtrsim 0.5$~mV/$\sqrt{\text{ms}}$) for our parametrization. 
% \todo{Check this limit. This paragraph might seem better in Methods, but there is no suitable sect. where it would fit. It is OK here.}). 
This is due to 
theoretical reasons (Fokker-Planck formalism) 
and practical reasons (numerics for Fokker-Planck solution and for calculation of the derived quantities, such as $ r_\infty $). 
%work in this regime 
%requiring discretization on the order of round-off errors making significant 
%problems
%-- especially for small mean input).
% }

\subsection*{Oscillations in a recurrent network} %\textcolor{gray}{/Network-generated spike rate oscillations}
%\todo{be positive and reclarify connection to uncoupled...}
%\todo{indicate right osc example as ''example close to Hopf bifurcation'' in figure}
%\todo{outlook E-I: for increased variance always possible to good regime..., especially for multipop with balanced input...} 

To demonstrate the applicability of the low-dimensional models for network analyses 
we consider a recurrently coupled population of aEIF neurons that 
produces self-sustained network oscillations 
by the interplay of strong excitatory feedback and 
spike-triggered adaptation or, alternatively, by delayed recurrent synaptic inhibition \cite{Augustin2013,Ladenbauer2015}. 
The former oscillation type is quite sensitive to changes in input, adaptation and especially coupling parameters for the current-based type of synaptic coupling considered here and due to lack of (synaptic) inhibition and refractoriness. For example, a small increase in coupling strength can lead to a dramatic (unphysiologic) increase in oscillation amplitude because of strong recurrent excitation. Hence we consider a difficult setting here to evaluate the reduced spike rate models -- in particular, when the network operates close to a bifurcation.

In Fig.~\ref{fig5_network_oscillations}A we present two example parametrizations from a region (in parameter space) that is characterized by stable oscillations. This means the network exhibits oscillatory spike rate dynamics for constant external input moments $\mu_\mathrm{ext}$ and $\sigma_\mathrm{ext}^2$. 
%One of these parametrizations is very close to a Hopf bifurcation, which constitutes a very difficult setting ... 
%Note that  and Due to lack of inhibition and refractoriness, ... 
%
%For thre, exact quantitative agreement is not 
%expected -- not even for the (complex) FP model.
%Nevertheless, 
The derived models reproduce the limit cycle behavior of the aEIF network surprisingly well, except for small frequency and amplitude deviations 
(FP, spec$_2$, LN$_\mathrm{dos}$, LN$_\mathrm{exp}$) and 
larger frequency mismatch (spec$_1$), see Fig.~\ref{fig5_network_oscillations}A, top.
%
%
% we show two differently 
%parametrized example networks that exhibit stable oscillatory spike rate 
%dynamics for 
%constant external input moments $\mu_\mathrm{ext}$ and $\sigma_\mathrm{ext}^2$. 
%
%
%We picked one parameter point within a region of stable oscillations and one close to a Hopf bifurcation, which constitutes a very difficult setting ...
%All derived models well reproduce the limit cycle behavior of the aEIF network, except for small frequency and amplitude deviations 
%(FP, spec$_2$, LN$_\mathrm{dos}$, LN$_\mathrm{exp}$) and 
%larger frequency mismatch (spec$_1$), see Fig.~\ref{fig5_network_oscillations}A.
%
%\todo{improve/rigorosify this sentence} 
%Nevertheless ... it looks great!!!
%
For weaker input moments and increased spike-triggered adaptation strength 
the network is closer to a Hopf bifurcation \cite{Augustin2013,Ladenbauer2015}. 
It is, therefore, not surprising that the differences in oscillation period 
and amplitude are more prominent (cf. Fig.~\ref{fig5_network_oscillations}A, bottom). 
The bifurcation point of the LN$_\mathrm{exp}$ model is 
slightly shifted, shown by the slowly damped oscillatory 
convergence to a fixed point. This suggests that the bifurcation parameter value 
of each of the derived models is not far from
the true critical parameter value of the aEIF network but can 
quantitatively differ (slightly) in a model-dependent way.

The second type of oscillation is generated by delayed synaptic inhibition\cite{Brunel2003JN}
and does not depend on the (neuronal) inhibition that is provided by an 
adaptation current. To demonstrate this independence  
the adaptation current was disabled 
(by setting the parameters $a=b=0$) for the two respective examples that are shown in 
Fig.~\ref{fig5_network_oscillations}B. Similarly as for the previous
oscillation type, the low-dimensional models (except spec$_1$) reproduce the spike rate limit 
cycle of the aEIF network surprisingly well, in particular for 
weak external input (see Fig.~\ref{fig5_network_oscillations}B, top). 
For larger external input and stronger inhibition with shorter delay 
the network operates close to a Hopf bifurcation, leading to 
larger differences in oscillation amplitude and frequency in a model-dependent way 
(Fig.~\ref{fig5_network_oscillations}B, bottom). 
Note that the intermediate (Fokker-Planck) model very well 
reproduces the inhibition-based type of oscillation which demonstrates the 
applicability of the underlying mean-field approximation.
We would also like to note that enabling the adaptation current dynamics  
(only) leads to decreased average spike rates but does not affect the 
reproduction accuracy.

%What to conclude... only two examples, but previous comprehensive evaluations provide a more thorough ... because ...
We would like to emphasize that the previous comprehensive evaluations for an uncoupled population provide a deeper insight on the reproduction performance -- also for a recurrent network -- than the four examples shown here, as explained in the Sect.~\textit{Performance for variations of the mean input}.
For example, the (improved) reproduction performance for increased input variance in the uncoupled setting (cf. Fig.~\ref{fig2_mu_exploration}) informs about the reproduction performance for networks of excitatory and inhibitory neurons that are roughly balanced, i.e., where the overall input mean is rather small compared to the input standard deviation.

% flatex input end: [section/results_exploration.tex]

% flatex input: [section/results_implementation.tex]
\subsection*{Implementation and computational complexity}

We have developed efficient implementations of the derived models
using the Python programming language and 
by employing the library Numba for low-level machine acceleration \cite{Lam2015}.
These include:
(i) the numerical integration of the Fokker-Planck model using an accurate 
finite volume scheme with implicit time discretization (cf. Sect.~\secmethods{}), 
(ii) the parallelized precalculation of the quantities required by 
the low-dimensional spike rate models and 
(iii) the time integration of the latter models, 
as well as example scripts demonstrating (i)--(iii). 
The code is available as open source software under a free license at GitHub: \href{\githuburl}{\githuburl}
% \todofinal{fill repository}

% \todo{scaling both for no of neurons and populations for the different model classes}
% 
% \textcolor{blue}{we have it!!!! all !!! available free open github. Note that Python code for all models presented here is made available: 
% \href{\githuburl}{\githuburl}
% }.
% 
% \textcolor{blue}{We have developed 
% an accurate numerical method based on the Scharfetter-Gummel 
% finite volume scheme which provides an optimum way to discretize 
% drift-diffusion equations.}

With regards to computational cost, summarizing the results of several aEIF network parametrizations,
the duration to generate population activity time series for the 
low-dimensional spike rate models is usually several orders of magnitude smaller 
%reduced by at least 2-3 orders of magnitude 
compared to numerical simulation of the original aEIF network 
and a few orders of magnitude smaller in comparison to the numerical solution of the FP model. 
%integration of the finite volume discretized 
%Fokker-Planck model. 
For example, considering a population of $50,000$ coupled neurons with 2\% connection probability, a single simulation run of 5~s 
% (i.e., one noise realization) 
and the same integration time step across the models, %and implementations, 
the computation times amounted to $1.1-3.6$~s for 
the low-dimensional models (with order -- fast to slow -- LN$_\mathrm{exp}$, spec$_1$, LN$_\mathrm{dos}$, spec$_2$), about $100$~s %97sec.
for the FP model and roughly %$ \sim $
$ 1500$~s %1450s w/o build synapses, 1702s total
for the aEIF network simulation on a dual-core laptop computer. 
The time difference to the network simulation substantially increases with the
numbers of neurons and connections, and with spiking activity within the 
network due to the extensive propagation of synaptic events. 
%In comparison to the FP model the speedup amounts to XX orders of magnitude (time difference) for one population. 
Note that the speedup becomes even more pronounced 
with increasing number of populations, 
where the runtimes of the FP model and the aEIF network simulation 
scale linearly and the low-dimensional models show a sublinear runtime increase 
due to vectorization of the state variables 
representing the different populations. 

The derived low-dimensional (ODE) spike rate models are very efficient to 
integrate given that the required input-dependent parameters are available as 
precalulated look-up quantities. 
For the grids used in this contribution, the precomputation time was 40 min. for the cascade (LN$_\mathrm{exp}$, LN$_\mathrm{dos}$) 
models and 120 min. for the spectral (spec$_1$, spec$_2$) models, both on a hexa-core desktop computer.  
The longer calculation time for the spectral models was due to the finer internal grid for the mean input (see \supplement).
% Precomputation for a reasonably large grid of input values 
% $(\mu,\sigma)$ can take significant time. For the grids
% % \footnote{\textcolor{gray}{The grid for the spectral models needed to be denser because ... (correct?).}} 
% used in this contribution it amounted to 40~min.
% % \footnote{\textcolor{gray}{about 2 hours on a dual-core laptop computer}} 
% for the cascade (LN$_\mathrm{exp}$, LN$_\mathrm{dos}$) models, 
% and 120~min. for the spectral (spec$_1$, spec$_2$) models due to 
% the finer internal grid for the mean input (see \supplement), 
% both on a hexa-core desktop computer. 
% \todo{@MA/FB: run on hexacore ``desktop'' risha once for comparison}

Note that while the time integration of the spec$_2$ model is on the same order as 
for the other low-dimensional models its implementation complexity is larger 
because of the many quantities it depends on, cf. Eqs.~\eqref{eq_meth_beta2}--\eqref{eq_meth_betac}. 

% \todo{ensure finally that all parameters/symbols are specified in the methods/params Sect.}
% flatex input end: [section/results_implementation.tex]

% flatex input end: [section/results.tex]

% flatex input: [section/discussion.tex]
\section*{Discussion}

%\todo{briefly repeat extensions?}
%\todo{clarify robustness of LN models}
%\todo{clarify recommendations}
%\todo{relate open loop (findings) to recurrency again?}
%\todo{add impossible heterog.}

In this contribution we have developed four low-dimensional models that 
approximate the spike rate dynamics of coupled aEIF neurons 
and retain all parameters of the underlying model neurons. 
These simple spike rate models were derived in two different ways from a 
Fokker-Planck PDE that describes the evolving membrane voltage distribution 
in the mean-field limit of large networks, and is complemented by an ODE for the population-averaged slow adaptation current.
% which carries over to the 
% low-dimensional spike rate descriptions where it composes one of the 
% few dynamic variables. } 
Two of the reduced spike rate models (spec$_1$ and spec$_2$) were obtained by a 
truncated spectral decomposition of the Fokker-Planck operator 
assuming vanishingly slow (for spec$_1$) or moderately slow (for spec$_2$) changes of the input moments. 
%, extending \cite{Mattia2002,Schaffer2013,Mattia2016Arxiv} 
%for an adaptation current, 
%time-varying external input and 
%delayed coupling (cf. Sect.~\secmethods). 
The other two reduced models (LN$_\mathrm{exp}$ and LN$_\mathrm{dos}$) are described by a cascade of 
linear filters (one for the input mean and another for its standard deviation) and a nonlinearity which were 
 %\tj{not really analytical -- better numerically or semi-analytically?}{} 
%analytically determined 
derived from the Fokker-Planck equation, and subsequently the filters were semi-analytically approximated.
% in different ways.  
% \textcolor{gray}{
%(using exponentials or damped oscillating functions, respectively).
% }.
Our approaches build upon \cite{Mattia2002,Schaffer2013,Mattia2016Arxiv} as well as \cite{Ostojic2011PLOS}, and extend those methods for 
adaptive nonlinear integrate-and-fire neurons that are 
% an adaptation current, 
% a membrane voltage nonlinearity and 
sparsely coupled with distributed delays (cf. Sect.~\secmethods).

We have compared the different spike rate representations for a 
range of biologically plausible input statistics and found that 
three of the reduced models (spec$_2$, LN$_\mathrm{exp}$ and LN$_\mathrm{dos}$) 
accurately reproduce the spiking activity of the underlying 
aEIF population while one model (spec$_1$) shows the least accuracy. 
Among the best models, the simplest (LN$_\mathrm{exp}$) was the most 
robust %\todo{maybe we should explain what we mean by robust: it did not show exaggerated/wild deflections (due to oversensitivity) as sometimes shown by the other 2 models} 
and (somewhat surprisingly) overall outperformed spec$_2$ and LN$_\mathrm{dos}$ -- especially in the 
sensitive regime of rapidly changing sub- and suprathreshold mean drive 
and in general for rapid and strong input variations. 
The LN$_\mathrm{exp}$ model did not exhibit exaggerated deflections in that regime as compared to the other two models.
This result is likely due to the importance of the quantitatively 
correct decay time of the filter for the mean input in the LN$_\mathrm{exp}$ model, while the violations of the slowness 
assumptions for the spec$_2$ and LN$_\mathrm{dos}$ models seem more harmful in this regime.
In the strongly mean-driven regime, however, the best performing model was 
spec$_2$ for variations both in the mean drive (as long as those variations are not too strong and fast) and for variations of the input variance. 
% \textcolor{gray}{The latter may be understood by considering the importance 
% of damped oscillations in this regime which are supported through the 
% imaginary part of the eigenvalues for the spectral models and through the damped oscillatory mean input filter of the 
% LN$_\mathrm{dos}$ model.} \todo{Why do spec1 and LNdos do not perform clearly better than LNexp then? }
% The LN$_\mathrm{dos}$ model performed worse than LN$_\mathrm{exp}$ for large mean input variations (i.e., large $\vartheta_\mu$) in general, and only slighly better for small input variance and mean input variations that are not too large and fast (cf. Fig.~\ref{fig2_mu_exploration}). 

We have also demonstrated that the low-dimensional models well reproduce the dynamics 
of recurrently coupled aEIF populations in terms of asynchronous states (see  Fig.~\ref{fig1_example}) 
and spike rate oscillations (cf. Fig.~\ref{fig5_network_oscillations}), %\textcolor{gray}{that are caused by an excitation-adaptation loop}, 
where %for the latter 
mild deviations at critical (bifurcation) parameter values are expected due to the 
approximative nature of the model reduction. 
%Note that most insight about the reproduction performance ...

The computational demands of the 
low-dimensional models are very modest in comparison to the aEIF network and 
also to the integration of the Fokker-Planck PDE, for which we have developed a novel finite volume discretization scheme. 
% The implementation complexity, 
% however, is higher for the spectral models, especially spec$_2$ requires 
% a large number of quantities to be precalculated and looked up in each 
% timestep. 
We would like to emphasize that any change of a parameter value 
for input, coupling or adaptation current does not require renewed precomputations. 
To facilitate the application of the presented models we have made available 
implementations 
that precompute all required quantities and numerically integrate 
the derived low-dimensional spike 
rate models as well as the  %\textcolor{gray}{full/complete/original} 
Fokker-Planck equation, together with example (Python) scripts, 
as open source software. 
%\todo{do not forget to add a link to the (final) repository before submission.}

%\tm{complete this blue part }
%\textcolor{blue}{
%erweiterungen..
%...
% for which we have developed 
%a novel finite volume scheme
%...
%spectral solver nice -- allows application of mattia2002 for general (linear and nonlinear) if neurons
%... 
%}

%APPLICATIONS: stability analyses and multi-pop. networks
Since the derived models are formulated in terms of simple ODEs,
they allow to conveniently perform linear stability analyses, e.g., 
based on the eigenvalues of the Jacobian matrix of the respective vector field. 
%This shows the potential of the derived models 
% for characterizing network states, 
In this way network states can be rapidly characterized by 
quantifying the bifurcation structure of the population dynamics --
including regions of the parameter space where multiple fixed points and/or 
limit cycle attractors co-exist. For a
characterization of stable network states 
by numerical continuation 
and an assessment of their controllability through neuromodulators
using the LN$_\mathrm{exp}$ model 
see \cite{Ladenbauer2015}~ch.~4.2 and \cite{Ladenbauer2016}. 
%\todo{briefly summarize the results of the bifurcation study in one sentence from the diss? JL: unfortunately no time...}
%
%\todo{Also mention here as application (again): simulations of (large) multi-pop. networks (to obtain pop.-averaged activity time series) }
Furthermore, the low-dimensional models are well suited to be employed in
large neuronal networks of multiple populations for efficient simulations of population-averaged activity time series.
Overall, the LN$_\mathrm{exp}$ model seems a good candidate for that purpose considering its accuracy and robustness, as well as its computational and implementational simplicity.

%\textcolor{gray}{FP spectral solver: ``advancing the approach in \cite{Ostojic2011JN,Schaffer2013} '' -- explain the advance better}

% \subsection*{Applications}
% \begin{itemize}
%  \item study osc .... describe results of JLdiss chap4, similar to frontiers but using...
% \item I oscillations: const delay or biexp (outline how to do this) \& E-I oscillations [Brunel1999,Augustin2013] 
% \end{itemize}

%\tm{CURRENT TODO: 1) order bulletpoints, 2) add citations and complete keyword list for each discussed (other) paper, 3) make bulletpoints for the extensions etc, 5) make sentences from the latter two}

\subsection*{Extensions}

\paragraph{Heterogeneity} We considered a homogeneous population of neurons in the sense that the parameter values across model neurons are identical except those for 
% coupling
synaptic input. Thereby we assume that neurons with similar dynamical properties can be grouped into populations \cite{Harris2015}. %maybe also [Lefort2009Neuron]. 
Heterogeneity is incorporated by distributed synaptic delays, by 
sparse random coupling,
and by fluctuating %\textcolor{gray}{/noisy} 
external inputs for each neuron. The (reduced) population models further allow for heterogeneous synaptic strengths that are sampled from a Gaussian distribution and can be included in a straightforward way \cite{Gigante2007PRL,Augustin2013} (see also Sect.~\secmethods). Distributed values for other parameters (of the isolated model neurons within the same population) are currently not supported. 
%homogeneity in the model parameters (per pop), but: 
%grouping of similar neurons possible which yields multi-population setup;
%%analogy to gerstner talk in prague15
%already included: heterog. delays and weights and input noise 
%%for noisy (const) input current, cf roxin quenched noise paper?

\paragraph{Multiple populations} The presented mean-field network model can be easily adjusted for multiple populations. In this case we obtain a low-dimensional ODE 
for each population and the overall synaptic moments for population $ k $ become
\begin{equation} \label{eq_disc_musyn_sigmasyn}
\mu_{\mathrm{syn},k} = \mu_{\mathrm{ext},k}(t) + \sum_l J_{kl} K_{kl} r_{d,kl}(t), \qquad 
\sigma^2_{\mathrm{syn},k} = \sigma_{\mathrm{ext},k}^2(t) +  \sum_l J_{kl}^2 K_{kl} r_{d,kl}(t),
\end{equation}  
where $ J_{kl} $ is the synaptic strength for the $K_{kl}$ neurons from population $ l $ targeting neurons from population $ k $ and $ r_{d,kl} $ is the delayed spike rate of population $ l $ affecting population $ k $ (cf. Eq.~\eqref{eq_res_musyn_sigmasyn}). For each pair of coupled populations we may consider identical or distributed delays (using distributions from the exponential family) as well as identical or distributed synaptic strengths (sampled from a Gaussian distribution).
%\textcolor{gray}{For example, differentially parametrized two-pop. network of E,I (Frontiers, JL Diss ch.4.2).}

\paragraph{Synaptic coupling} % \textcolor{gray}{and connectivity}} 
Here we described synaptic interaction by delayed (delta) current pulses with delays sampled from an exponential distribution. This description leads to a fluctuating overall synaptic input current with white noise characteristics. Interestingly, for the mean-field dynamics this setting is very similar to considering exponentially decaying synaptic currents with a decay constant that matches that of the delay distribution, although the overall synaptic input current is a colored noise process in that case, see \cite{Biggio2017} and, for an intuitive explanation \cite{Roxin2011PD}. %for effective delay (intuition); 

A conductance-based model of synaptic coupling can also be considered in principle \cite{Augustin2013,Richardson2004}, which results in a multiplicative noise process for the overall synaptic input. This, however, would in general impede the beneficial concept of precalculated ``look-up'' quantities that are unaffected by the input and coupling parameters.
%even though effective ways of including c-b syn. models exist, leading to ...

It should be noted that most current- or conductance-based models of synaptic coupling (including the one considered here) can produce unphysiologically large amounts of synaptic current in case of high presynaptic activity, unless the coupling parameters are carefully tuned.   
%Potential problem, especially for multi-population networks: unbounded input for large presynaptic activity due to simple ``standard'' synaptic coupling models (decaying, and also conductance-based).
This problem can be solved, for example, by considering a (more realistic) model of synaptic coupling based on \cite{Destexhe1994}, from which activity-dependent coupling terms can be derived for the mean-field and reduced population models \cite{Ladenbauer2015} ch.~4.2. Using that description ensures robust simulation of population activity time series without having to fine-tune the coupling parameter values, which is particularly useful for multi-population network models. In this contribution though we used for simplicity a basic synaptic coupling model that has frequently been applied in the mean-field literature.
%\textcolor{gray}{Repeat comment on sparse connectivity: topology can be arbitrary but vanishing noise correlations are required (achieved, e.g., by less sparse connections and more heterogeneity).}

\paragraph{Input noise process}
The Gaussian stochastic process driving the individual neurons could 
also be substituted by colored noise, which would lead to a Fokker-Planck model with increased dimensionality \cite{Schwalger2015JCN}. However, this would require more complex and computationally expensive numerical schemes not only to solve that model but also for the different dimension reduction approaches. 
% For example, it would be required to calculate the eigenmodes of a 1+k-dimensional continuous Fokker-Planck operator, where k is the number of additional variables

\paragraph{Slow adaptation} To derive low-dimensional models of population activity we approximated the adaptation current by its population average, justified by its slow dynamics compared to the other time scales of the system. This approximation is equivalent to a first order moment closure method \cite{Nicola2015}. In case of a faster adaptation time scale the approximation may be improved by considering second and higher order moments \cite{Hertaeg2014,Nicola2015}.

\paragraph{Population size} The mean-field models %of population activity 
presented here can well reproduce the dynamics of population-averaged state variables 
(that is, spike rate, mean membrane voltage, and mean adaptation current) for large populations ($N \to \infty$ in the derivation). Fluctuations of those average variables 
due to the finite size of neuronal populations, however, are not captured. %(see, e.g., Figs.~\ref{LC_figFP} and \ref{LC_fig7}). 
Hence, it would be interesting to extend the mean-field models so as to reproduce these (so-called) finite size effects, for example, by incorporating an appropriate stochastic process \cite{Mattia2002} 
or using concepts from \cite{Schwalger2017}. % ,Dumont2017}.

\paragraph{Cascade approach}
For uncoupled EIF populations (without an adaptation current) and constant input standard deviation it has been shown that the LN cascade approximation performs well for physiological ranges of %parameter regions 
amplitude and time scale for mean input variations \cite{Ostojic2011PLOS}.
Our results for the cascade models are consistent with \cite{Ostojic2011PLOS}, but the performance is substantially improved for the sensitive low (baseline) input regime 
(LN$_\mathrm{exp}$ and LN$_\mathrm{dos}$, also in absence of adaptation), and damped oscillatory behavior (including over- and undershoots) is accounted for by the LN$_\mathrm{dos}$ model.

To achieve these improvements we semi-analytically fit the linear filters derived from the Fokker-Planck equation using exponential and damped oscillator functions considering a range of input frequencies. The approximation can be further improved by using more complex 
functions, such as a damped oscillator with two time scales. That, however, can lead to less robustness (i.e., undesired model behavior) for rapid and strong changes of the input moments %\textcolor{gray}{due to adaptation of the filter parameters to the input} 
(cf. Sect.~\secmethods). % \textit{Cascade models} in 

LN cascade models are frequently applied in neuroscience to describe population activity, and the model components are often determined from %, e.g., reverse correlation methods to determine linear filters from ...
electrophysiological recordings using established techniques. The methodology presented here contributes to establishing quantitative links between networks of spiking neurons, a mesoscopic description of population activity and recordings at the population level. %and also single cell level \cite{Brette2005}

\paragraph{Spectral approach}
Here we provide a new numerical solver for the eigenvalue problem of the Fokker-Planck operator and its adjoint. This allows to compute the full spectrum together with associated eigenfunctions 
and is applicable to nonlinear integrate-and-fire models, extending \cite{Mattia2002,Ostojic2011JN,Schaffer2013}.

Using that solver the spec$_2$ model, which is based on two eigenvalues, 
can be further improved by interpolating its coefficients, 
Eqs.~\eqref{eq_meth_beta2}--\eqref{eq_meth_betac}, 
around the double eigenvalues at the spectrum's real-to-complex 
transition.
% that are present at those 
% parameters for which the spectrum changes from real to complex. 
This interpolation would effectively smooth the quantities 
-- e.g., preventing the jumps and kinks that are present in 
the visualization of Sect.~\textit{Spectral models} --
% (e.g., the jumps and kinks in Fig.~\ref{fig7_spectral}C)  % removed due to fig appearance in order
and is expected to increase the spike rate reproduction 
accuracy (particularly for weak mean input) beyond 
what was reported in this contribution.

The spec$_2$ model can also be
extended to yield a third order ODE with everywhere 
smooth coefficients by considering an additional eigenvalue (cf. Sect.~\textit{\secspecprops}).

Moreover, the spec$_2$ model, and more generally the whole spectral 
decomposition approach, can be extended to account 
for a refractory period in the presence of time-varying total input moments, 
e.g., by building upon previous attempts \cite{Brunel2000,Mattia2002,Apfaltrer2006}. 
 
Furthermore, it could be beneficial to explicitly quantify 
the approximation error due to the slowness assumption 
that underlies the spec$_2$ model 
by 
integration of the (original)
spectral representation of the Fokker-Planck model. %, 
% Eqs.~\eqref{eq_meth_spec_a_full},\eqref{eq_meth_spec_rate_full}, . 

Both reduced spectral models allow for a refined description 
of the mean adaptation current dynamics, cf. Eq.~\eqref{eq_results_w_mean}, by
replacing the mean membrane voltage $\langle V \rangle$ with its 
steady-state value $\langle V \rangle_\infty$, using that 
the membrane voltage distribution is available through the
eigenfunctions of the Fokker-Planck operator.
% 
% and thus available.
% the 
% mean membrane voltage $\langle V \rangle$ due to its 
% 
% 
% the membrane voltage distribution 
% 
% 
% Optionally, since the membrane voltage distribution $p(V,t)$ is available 
% through the eigenfunction expansion of Eq.~\eqref{eq_meth_spec_p_expand_phi}, it can 
% be used to calculate the mean $\langle V \rangle$. This allows
% to replace the mean adaptation current dynamics, Eq.~\eqref{eq_meth_w_mean_p_inf}, 
% that depends on the steady-state value $\langle V \rangle_\infty$,
% by Eq.~\eqref{eq_meth_w_mean_p}. 
% This possibility carries over to the 
% reduced models of the following Sects. and might be beneficial 
% for smaller values for the adaptation time constants 
% $\tau_w$ than the one considered here. 

The numerical eigenvalue solver can be extended in a straightforward way to yield quantities 
that are required by the original spectral representation of the Fokker-Planck model
% , Eq.~\eqref{eq_meth_spec_a_full}, 
and by the corresponding stochastic equation for
% adjusted for 
finite population size \cite{Mattia2002}.

\subsection*{Alternative derived models}

In addition to the work we build upon 
\cite{Mattia2002,Schaffer2013,Mattia2016Arxiv,Ostojic2011PLOS} (cf. Sect.~\secmethods) there are a few other approaches to derive spike rate models from populations of spiking neurons.
% \underline{first closely related approaches:}
% \begin{itemize}
%   \item ensure mattia2002, schaffer13 and ostojic11plos are included also in discussion
%  \item spec1 without adaptation and instantaneous coupling slower in recurrency also 
%   shown in schaffer et al fig...
%   \item knight1996,2000,2000?
% \end{itemize}
% \underline{then intermediate:}
Some methods also result in an ODE system, taking into account (slow) neuronal adaptation \cite{Gigante2007PRL,Nesse2008,Nicola2015,Zerlaut2017,Buchin2010} 
or disregarding it \cite{Montbrio2015}. 
The settings differ from the work presented here in 
that (i) the intrinsic neuronal dynamics are adiabatically neglected 
\cite{Nesse2008,Nicola2015,Zerlaut2017,Gigante2007PRL}, %or (only) spike-triggered adaptation 
%\cite{Gigante2007PRL}. 
(ii) only uncoupled populations
\cite{Buchin2010} or all-to-all connected networks \cite{Nesse2008,Nicola2015,Montbrio2015} are assumed 
in contrast to sparse connectivity, and 
(iii) (fixed) heterogeneous instead of 
fluctuating input is considered \cite{Montbrio2015}. 
Notably, these previous methods yield rather qualitative agreements 
with the underlying spiking neuron population activity except for \cite{Montbrio2015} 
where an excellent quantitative reproduction for (non-adaptive) quadratic integrate-and-fire oscillators with quenched input randomness is reported. 
%\textcolor{gray}{(though not for white noise drive)}. 

% \underline{then more distant approaches with adaptation:}
% \begin{itemize} 
% \item Gigante2007PRL (sparse coupling, first order rate eq. (no damped osc.), quasi-stationary assumption: qualitative prediction, not quantitative; no subthreshold adaptation; qualitative analysis of bistable fp and limit cycles)
%  \item nicola2013jcompneurosci (is already above -- mean adapt and synaptic with steady state firing rate)
% \item nesse2008 (all-to-all, neuronal dynamics approx. by its steady-state value and only synaptic and neuronal variables remain, qualitative agreement not quantiative)
% \item Buchin2010 (uncoupled, rate-dependent M and AHP channels, fair agreement)
% \item \textcolor{red}{other refs from mendeley add here...}
% \end{itemize}
% 
% 
% \underline{then more distant approaches without adaptation:}
% \begin{itemize}
% \item montbrio2015 (report excellent agreements for 2d ODE for all-to-all coupled QIF neurons, quenched input noise; quite good agreement also for white noise input)
% \item \textcolor{red}{other refs from mendeley add here...}
% \end{itemize}

Other approaches yield mesoscopic representations of population activity  
%derived from neuronal network dynamics  
in terms of model classes that are substantially
less efficient to simulate and more complicated to analyze than
low-dimensional ODEs \cite{Zhang2015,Gerstner2000,Iyer2013,Augustin2013,Nicola2015,Naud2012,Schwalger2017}. 
The spike rate dynamics in these models has been described (i) by a rather 
complex ODE system that depends on a stochastic jump process 
derived for integrate-and-fire neurons without adaptation \cite{Zhang2015}, 
(ii) by PDEs for recurrently connected aEIF \cite{Augustin2013} or Izhikevich \cite{Nicola2015} neurons, (iii) by an integro-PDE with 
displacement for non-adaptive neurons \cite{Iyer2013} or (iv) 
by integral equations that represent the (mean) activity of 
coupled phenomenological spiking neurons without \cite{Gerstner2000} 
and with adaptation \cite{Naud2012,Schwalger2017}.
% for which an extension was developed to 
%that have also been extended for neuronal adaptation \cite{Naud2012}. 

% Furthermore, a PDE representation for all-to-all connected aEIF networks
% have been derived based on a moment closure approximation \cite{Nicola2013Front} 
% which 

% 
% \begin{itemize}
% 
% \item  integral or PDE together: more expensive...
% 
% \item Gerstner integral eq. approach cf. gerstner00nc and naud plos, Pozzorini2015?, Tilo in prep for FS effects) -- quantitative accurate approach, phenomenological neuron description, integral eq instead of ODE
% 
% \item Zhang2015 (4-dimensional ODE that includes a jump process as well and its integration is rather complex)
% 
% 
%  \item nicola2013frontiers (see jldiss/heterog.  input not noise process) -- combine maybe with hertaeg... as both use moment closure for adaptation! -- ode form
% %  \item nicola2014[SIAM J. APPLIED DYNAMICAL SYSTEMS]: birfucation analysis of the 2d system above
%  \item nicola2015[SIAM J. APPL. MATH.] (as frontiers but with white noise input) -- ode form
% 
% 
% \end{itemize}

Furthermore, the stationary condition of a noise-driven population of
adaptive EIF neurons \cite{Hertaeg2014,Richardson2009,Rosenbaum2016Front} and the first order spike rate response to weak input modulations \cite{Richardson2009,Rosenbaum2016Front} have been analyzed 
using the Fokker-Planck equation. Ref. \cite{Hertaeg2014} also considered a refined approximation of the (purely spike-triggered) adaptation current including higher order moments.

It may be interesting for future studies to explore ways to extend the presented methods and relax some of the underlying assumptions, in particular, considering (i) the diffusion approximation (via shot noise input, e.g., \cite{Nykamp2001,Richardson2010}), (ii) the Poisson assumption (e.g., using the concept from \cite{Renart2007} in combination with results from \cite{Ladenbauer2014}) and (iii) (noise) correlations (see, e.g., \cite{Rosenbaum2016Nat}).

% flatex input end: [section/discussion.tex]

% flatex input: [section/methods.tex]
\section*{\secmethods}

% \todo{make all equations more beautiful here, in the results and in the appendix (e.g. use declarmathoperator instead of text/mathrm, smaller spacings for fvm indices etc.}

\label{sec_methods}

Here we present all models in detail -- the aEIF network (\textit{ground truth}), the mean-field FP system (\textit{intermediate model}) and the low-dimensional models: spec$_1$, spec$_2$, LN$_{\mathrm{exp}}$, LN$_{\mathrm{dos}}$ -- including step-by-step derivations %\textcolor{gray}{from the \textit{ground truth} model} 
and essential information on the respective numerical solution methods. An implementation of these models using Python is made available at GitHub: \href{\githuburl}{\githuburl}
%(cf. Sect. Methods). 
% provided at \cite{codeDOI}.\todo{add github DOI citation here}

% flatex input: [section/methods_network.tex]
\subsection*{Network model} %\textcolor{gray}{/Single neuron dynamics/Ground truth}

We consider a large (homogeneous) population of $N$ synaptically coupled %adaptive exponential 
%integrate-and-fire (aEIF) 
aEIF model neurons \cite{Brette2005}. 
Specifically, for each neuron ($i=1, \dots ,N$), 
the dynamics of the membrane voltage $V_i$ is described by
\begin{equation}
C \frac{dV_i}{dt} = I_{\mathrm{L}}(V_i) + I_{\mathrm{exp}}(V_i) - w_i
+ I_{\mathrm{syn},i}(t), 
\label{eq_meth_V_single}
\end{equation}
where the capacitive current through the membrane with capacitance $C$ equals the 
sum of three ionic currents %$I_{\mathrm{ion}}$ 
and the synaptic current $I_{\mathrm{syn},i}$. The ionic currents consist of 
% old version further below
%$I_{\mathrm{ion}}(V_i) \coloneqq I_{\mathrm{L}}(V_i) + I_{\mathrm{exp}}(V_i) - w_i$. 
a linear leak current 
$I_{\mathrm{L}}(V_i) = -g_{\mathrm{L}} (V_i-E_{\mathrm{L}})$ 
with conductance $g_{\mathrm{L}}$ and reversal potential $E_{\mathrm{L}}$, 
a nonlinear term %an exponential term 
%\textcolor{gray}{/current} 
$I_{\mathrm{exp}}(V_i) = g_{\mathrm{L}} \, \Delta_{\mathrm{T}} \, 
%e^{\tfrac{V-V_{\mathrm{T}}}{\Delta_{\mathrm{T}}}} - w.
% \exp \left( \frac{V_i-V_{\mathrm{T}}}{\Delta_{\mathrm{T}}} \right)
\exp \left( (V_i-V_{\mathrm{T}})/\Delta_{\mathrm{T}} \right)$ 
that approximates the rapidly increasing $\mathrm{Na}^+$ 
current at spike initiation with threshold slope factor $\Delta_{\mathrm{T}}$ 
and effective threshold voltage $V_{\mathrm{T}}$,
and the adaptation current $w_i$ which reflects a slowly 
deactivating $\mathrm{K}^+$ current.
The adaptation current evolves according to
\begin{equation}
\tau_{w} \frac{dw_i}{dt} = a(V_i-E_w) -w_i,
\label{eq_meth_w_single}
\end{equation}
% 
%on a timescale given by ...
with adaptation time constant $\tau_{w}$. Its strength depends on the subthreshold membrane voltage via conductance $a$. $E_w$ denotes its reversal potential.
When $V_i$ increases beyond $V_\mathrm{T}$, it diverges to infinity %$\infty$ 
in finite time due to the exponentially increasing current $I_{\mathrm{exp}}(V_i)$, 
which defines a spike. 
In practice, however, \index{neuronal spike} the spike is said to occur when $V_i$ 
reaches a given value $V_{\mathrm{s}}$ -- the spike voltage. % \geq V_{\mathrm{T}}$. 
%In real neurons threshold crossing of the membrane voltage elicits a positive 
% feedback cycle where more and more voltage-sensitive $\mathrm{Na}^+$-channels open ...
The downswing of the spike is not explicitly modeled; instead, when 
$V_i \geq V_{\mathrm{s}}$, the membrane voltage $V_i$ is instantaneously reset to a 
lower value $V_{\mathrm{r}}$.
%\todo{@JL: if required extend this in a 
%short way to spike shape (implies the last sentence of this paragraph, too)}}
At the same time, the adaptation current $w_i$ is incremented by a value of parameter
$b$, which implements suprathreshold (spike-dependent) activation of the adaptation current. 

Immediately after the reset, $V_i$ and $w_i$ are %[possibly] 
clamped (i.e., remain constant) for a short refractory period $T_{\mathrm{ref}}$, and 
subsequently governed again by Eqs.~\eqref{eq_meth_V_single} and \eqref{eq_meth_w_single}.
At the end of the \secmethods{} section we describe how (optionally) a spike shape can be included in the aEIF model, together with the associated small changes for the models derived from it. 

%\textcolor{orange}{new synaptic description (check):}
To complete the network model 
% (Eq.~\eqref{eq_meth_V_single}) 
the 
% total 
synaptic current in Eq.~\eqref{eq_meth_V_single} needs to be specified: for each cell it is given 
by the sum of recurrent and external input, 
$I_{\mathrm{syn},i} = I_{\mathrm{rec},i}(t) + I_{\mathrm{ext},i}(t)$.
%The total synaptic current $I_{\mathrm{syn}}$ 
% It consists %is made up of 
% \textcolor{gray}{the sum of}
% \textcolor{blue}{
% [new recurrency]
Recurrent synaptic input is received from $K$ other neurons of the network, 
that are connected in a sparse ($K \ll N$) and uniformly random way, 
and is modeled by 
\begin{equation}
I_{\mathrm{rec},i} = C \sum_{j} J_{ij} \sum_{t_j} \delta(t-t_j-d_{ij}),
\end{equation}
where  $\delta$ denotes the Dirac delta function.
Every spike by one of the $K$ presynaptic neurons with 
indices $j$ and spike times $t_j$ causes a 
postsynaptic membrane voltage jump 
% weighted by  
% the coupling strength $J_{ij}$
of size $J_{ij}$. The coupling strength is positive (negative) for excitation (inhibition) and of small magnitude. Here it is chosen to be constant, i.e., $J_{ij} = J$. 
% with amplitude given by the coupling strength 
% $J_{ij}$
%\footnote{\textcolor{gray}{$J$ or 
%$J_{ij}$ seem a bit misleading since $j$ is the postsynaptic index, so $G_{ij}$ would be an alternative. \\Also: one could absorb 
%the factor $C$ into the coupling strength (such that the membrane potential jump is only weighted and not given 
%by $J_{ij}$ -- othen coupling matrix in unit current instead of voltage.}}
% (positive/negative for excitation/inhibition) 
Each of these 
%\textcolor{gray}{spike-triggered}
membrane voltage deflections occur after a time delay $d_{ij}$ 
that 
takes into account (axonal and dendritic) spike propagation times
and
is sampled (independently) 
from a probability distribution $p_d$. 
In this work we use exponentially distributed delays, 
i.e., $p_d(\tau) =  \exp(-\tau/\tau_d) / \tau_d$ 
(for $\tau \ge 0$) with mean delay $\tau_d$.
% }
% \textcolor{gray}{[old recurrency]
% Recurrent synaptic input is received from other neurons of the network, 
% $I_{\mathrm{rec},i} = C \sum_{j} J_{ij} \sum_m \delta(t-t_j^m-d_{ij})$,
% i.e., every spike by one of $K$ presynaptic neurons 
% with 
% indices $\{j\}$ and spike times $\{t_j^m\}$
%  causes a 
% postsynaptic membrane voltage jump 
% % weighted by  
% % the coupling strength $J_{ij}$
% of \textcolor{gray}{small} size $J_{ij}$ -- the coupling strength (positive/negative for excitation/inhibition) --
% % with amplitude given by the coupling strength 
% % $J_{ij}$
% %\footnote{\textcolor{gray}{$J$ or 
% %$J_{ij}$ seem a bit misleading since $j$ is the postsynaptic index, so $G_{ij}$ would be an alternative. \\Also: one could absorb 
% %the factor $C$ into the coupling strength (such that the membrane potential jump is only weighted and not given 
% %by $J_{ij}$ -- othen coupling matrix in unit current instead of voltage.}}
% % (positive/negative for excitation/inhibition) 
% after a 
% time delay $d_{ij}$ has passed, 
% accounting for the (axonal and dendritic) spike propagation times.} \todo{We consider random and sparse connectivity, i.e., $ K \ll N $.}

The second type of synaptic input is a fluctuating current generated from network-external neurons,  
% $I_{\mathrm{syn},i}(t) = I_{\mathrm{rec},i}(t; \{t_j^k\}) 
% + I_{\mathrm{ext},i}(t)$. 
% (if a network is considered) 
\begin{equation}
I_{\mathrm{ext},i} = C [ \mu_{\mathrm{ext}}(t) + \sigma_{\mathrm{ext}}(t) \xi_{\mathrm{ext},i}(t)],
\end{equation} 
% is modeled by 
with time-varying moments $\mu_{\mathrm{ext}}$ and $\sigma_{\mathrm{ext}}^2$, 
and unit
Gaussian white noise process $\xi_{\mathrm{ext},i}$. 
The latter is 
% fluctuates independently 
uncorrelated %\textcolor{gray}{/independent}
with that of other neurons $ j \ne i $, %$\xi_j$ \textcolor{gray}{(i.e., for $i\ne j$) 
i.e., $\big \langle \xi_{\mathrm{ext},i}(t) \xi_{\mathrm{ext},j}(t+\tau) \big \rangle =  \delta(\tau) \delta_{ij}$, where 
$\langle \cdot \rangle $ denotes expectation 
(w.r.t. the joint ensemble of noise realizations at 
times $t$ and $t+\tau$) and $\delta_{ij}$ is the Kronecker delta.
This external current, for example, accurately approximates the input generated from a large number of  
%produced by large numbers of 
%\textcolor{gray}{(network-external)} 
independent 
Poisson neurons that produce instantaneous postsynaptic potentials of small magnitude, cf. \cite{Ladenbauer2014}.

The spike rate $r_N$ of the network %\textcolor{gray}{with size $N$} 
is defined as the population-averaged number of 
emitted spikes per time interval $[t,t+\Delta T]$,
\begin{equation}
r_N(t) = \frac{1}{N} \sum_{i=1}^{N} \frac{1}{\Delta T}
    \int_{t}^{t+\Delta T} \sum_{t_i} \delta(s-t_i) \, ds,
\end{equation}
where the interval size $\Delta T$ is practically chosen small 
enough to 
capture the dynamical structure 
and large enough to yield a comparably smooth 
time evolution for a finite network, i.e., $N\!<\!\infty$.

%An example \textcolor{gray}{network} is illustrated 
%through time series of 
%single neuron and population quantities in Fig.~\ref{fig1_intro}A-X.
 
% including
% single neuron trajectories and population histograms for the state variables 
% as well as the spike rate $r_N^{\Delta t}(t)$
% , 
% an aEIF population neglecting recurrent coupling, 
% i.e., $I_{\mathrm{rec},i} \equiv 0$ and thus $I_{\mathrm{syn},i} \equiv I_{\mathrm{ext},i}$
% with time-varying external moments \textcolor{gray}{(denoted as 
% $\mu_{\mathrm{syn}}(t)=\mu_{\mathrm{ext}}(t)$ and 
% $\sigma_{\mathrm{syn}}(t) = \sigma_{\mathrm{ext}}(t)$)} 
% is shown in Fig.~\ref{fig1_intro}A-X. 

We chose values for the 
% \textcolor{gray}{aEIF} 
neuron model parameters to describe cortical pyramidal cells, which exhibit ``regular spiking'' behavior and spike frequency adaptation \cite{Badel2008JN,Destexhe2009,Wang2003}. For the complete 
parameter specification see Table~\ref{table1_paramsall}.

All network simulations were performed using the Python software 
BRIAN2\cite{Stimberg2014,Goodman2009} with C++ code 
generation enabled for efficiency. 
The aEIF model Eqs.~\eqref{eq_meth_V_single} and \eqref{eq_meth_w_single} 
were discretized using the Euler-Maruyama method with equidistant time step 
$\Delta t$ and initialized with $w_i(0) = 0$ and $V_i(0)$ that is (independently) sampled from a Gaussian initial distribution $p_0(V)$ with mean $V_\mathrm{r}\!-\!\delta V$ and standard deviation $\delta V/2$ where $\delta V = V_\mathrm{T} - V_\mathrm{r}$. Note that the models derived 
in the following Sects. do not depend on this particular 
initial density shape but allow for an 
arbitrary (density) function $p_0$. 

%\todo{revisit actual simulation model python code if further details are important to mention here, e.g., initialization, randomness etc (same for alle other models)}

\begin{table}[!ht]
% \begin{adjustwidth}{-2.25in}{0in} % Comment out/remove adjustwidth environment if table fits in text column.
\caption{
{\bf Parameter values used throughout the study.}}
\begin{tabular}{|l|c|c|}
\hline
Name & Symbol & Value \\ \hline \hline
\multicolumn{3}{|c|}{\bf Network model} \\ \hline 
Number of neurons  & $N$ & 50,000 \\ \hline
Membrane capacitance  & $C$ & 200~pF \\ \hline
Leak conductance & $g_\mathrm{L}$ & 10~nS \\ \hline
Leak reversal potential & $E_\mathrm{L}$ & $ -65 $~mV \\ \hline
Threshold slope factor & $\Delta_\mathrm{T}$ & 1.5~mV \\ \hline
Threshold voltage & $V_\mathrm{T}$ & $ -50 $~mV \\ \hline
Spike voltage & $V_\mathrm{s}$ & $ -40 $~mV \\ \hline
Reset voltage & $V_\mathrm{r}$ & $ -70 $~mV \\ \hline
Subthreshold adaptation conductance$^1$ & $a$ & 4~nS \\ \hline
Spike-triggered adaptation increment$^1$ & $b$ & 40~pA \\ \hline
Adaptation reversal potential & $E_w$ & $ -80 $~mV \\ \hline
Adaptation time constant & $\tau_w$ & 200~ms \\ \hline
Refractory period$^2$ & $T_\mathrm{ref}$ & 0~ms \\ \hline
Gaussian filter width for external input & $\sigma_t$ & 1~ms \\ \hline
Discretization time step & $\Delta t$ & 0.05~ms \\ \hline
Spike rate estimation bin width & $\Delta T$ & 1~ms \\ \hline
\hline
\multicolumn{3}{|c|}{\bf Fokker-Planck model} \\ \hline
Membrane voltage lower bound & $V_\mathrm{lb}$ & $ -200 $~mV \\ \hline
Finite-volume membrane voltage spacing & $\Delta V$ & 0.028~mV \\ \hline
Discretization time step & $\Delta t$ & 0.05~ms \\ \hline
\hline
\multicolumn{3}{|c|}{\bf Low-dimensional models} \\ \hline
Discretization time step & $\Delta t$ & 0.01~ms \\ \hline
Membrane voltage spacing$^3$ & $\Delta V$ & 0.01~mV \\ \hline
Spacing of mean input$^3$ & $\Delta \mu$ & 0.025~mV/ms \\ \hline
Spacing of input standard deviation$^3$ & $\Delta \sigma$ & 0.1~mV/$\sqrt{\mathrm{ms}}$ \\ \hline
% \hline
% \multicolumn{3}{|c|}{\bf Spectral solver} \\ \hline
% Numerical param1 (e.g. eps)$^3$ & $yyy$ & XX~unit \\ \hline
% Numerical param2 (e.g. tol)$^3$ & $yyy$ & XX~unit \\ \hline
% Numerical param3 (e.g. no of eigenvalues)$^3$ & $yyy$ & XX~unit \\
% Numerical param4 (e.g.) or better in appendix?
% \hline
\end{tabular}
\begin{flushleft} 
$^1$If not specified otherwise. 
$^2$A nonzero refractory period is not supported by the spec$_2$ model. %For a full model set in the systematic comparison of Sect. Results (a nonzero refractory period is not supported by spec$_2$). %\todo{@MA: please rephrase}
$^3$Parameters for precalculation of model quantities (before simulation). 
\\
The values of coupling parameters ($ K $, $ J $, $ \tau_d $) are specified in the captions of Figs.~\ref{fig1_example} and \ref{fig5_network_oscillations}, the values of parameters for the external input, $\mu_\mathrm{ext}$ or ($\bar\mu$, $\tau^\mu_\mathrm{ou}$, $\vartheta_\mu$), and $\sigma_\mathrm{ext}$ or ($\overline{\sigma^2}$, $\tau^{\sigma^2}_\mathrm{ou}$, $\vartheta_{\sigma^2}$), are provided in each figure (caption).
\end{flushleft}
\label{table1_paramsall}
% \end{adjustwidth}
\end{table}
%\todo{@Fabian: check and complete all param. values (except precalc.) in Table 1}
% \todo{could add ranges for input and coupling parameters (+9) ?}
%\todo{@MA: fill spec numerics params and check for others}
% flatex input end: [section/methods_network.tex]

% flatex input: [section/methods_fokkerplanck_meanfield.tex]
\subsection*{\secmeanfield}

\paragraph{Adiabatic approximation}

%\todo{similarly to mean-field assumptions put slowness already in network model and ``use'' that here?}
The time scales of (slow) $\mathrm{K}^+$ channel kinetics which are effectively described by the adaptation current 
$w_i$, cf. Eq.~\eqref{eq_meth_w_single}, are typically much larger than the faster 
membrane voltage dynamics modeled 
by Eq.~\eqref{eq_meth_V_single}, i.e., $\tau_w \gg  C/g_\mathrm{L}$  
%\todo{remove $\tau_m$ or is it used somewhere?}
\cite{Brown1980,Sanchez-Vives2000,Sanchez-Vives2000a,Stocker2004}. %,LaCamera2006}.
This observation justifies to replace the individual adaptation current $w_i$ in 
Eq.~\eqref{eq_meth_V_single} by 
its average across the population, $\langle w \rangle_N = 1/N \sum_{i=1}^N w_i(t)$,
%\textcolor{gray}{[alternative text is available in JL thesis chapter 4.2 Sect. 
%\textit{Adiabatic approximation}]}
in order to reduce computational demands and enable further analysis. 
% \todo{Suggest one of the following: have finite $ N $ here together with explanation how to calculate  $ \langle V \rangle_N $ (taken from next sect.) and have limit $ N\to\infty $ entirely in the next sect. (clearer, but doubles the eq. for mean adapt. current) -- or -- do not indicate subscript $ N $ at all (to have same eq. for finite and infinite $ N $; sloppy but saves space). I prefer the former.}
%In the limit 
%of large networks the population-averaged adaptation current 
%$\langle w \rangle = \lim_{N\to\infty} \langle w \rangle_N$ is then governed by 
The mean adaptation current is then governed by
\cite{Augustin2013,Ladenbauer2014,Gigante2007PRL,Brunel2003PRE}
%\begin{equation}
%\frac{d{\langle w \rangle}}{dt} = \frac{a 
%\left(\langle V \rangle -E_w \vphantom{^2} \right) - 
%  \langle w \rangle}{\tau_{w}} + b\, r(t), \label{eq_meth_w_mean_p}
%\end{equation}
\begin{equation}
\frac{d{\langle w \rangle}_N}{dt} = \frac{a 
\left(\langle V \rangle_N -E_w \vphantom{^2} \right) - 
  \langle w \rangle_N}{\tau_{w}} + b\, r_N(t), \label{eq_meth_w_mean_finite}
\end{equation}
%\todo{discuss dependence on spike rate ``bin'' size $\Delta T$ in terms of the formulation in the previous Sect.}
where $\langle V \rangle_N$ denotes the time-varying population average of the 
membrane voltage 
of non-refractory neurons. %\textcolor{gray}{and $r_N$ is assumed to be calculated with 
%sufficiently small $\Delta T$}.
%\footnote{\textcolor{gray}{maybe correct the 
%$\langle V \rangle$ and $\langle w \rangle$ sum formulaes.}}, 
%$\langle V \rangle = \lim_{N\to\infty} 1/N \sum_{i=1}^N V_i(t)$, and 
%$r(t) = \lim_{N\to\infty} r_N(t)$ is the spike rate \textcolor{gray}{of the network}. 

The dynamics of the population-averaged adaptation 
current reflecting 
the non-refractory proportion of neurons are well captured by 
Eq.~\eqref{eq_meth_w_mean_finite} %considering the refractory or non-refractory 
% proportion of trials, or both, % which also only reflects the 
% proportion of trials $P(t)$
as long as $T_{\mathrm{ref}}$ is small compared to $\tau_w$. In this
(physiologically plausible) case $\langle w \rangle_N$ from Eq.~\eqref{eq_meth_w_mean_finite} 
can be considered equal 
to the average adaptation current over the refractory proportion of neurons
\cite{Ladenbauer2014,Augustin2013}. %\todo{add further refs}

\paragraph{Mean field limit}

For large networks ($N\to\infty$) 
the recurrent input
can be approximated by a mean part with additive fluctuations, 
$I_\mathrm{rec,i}/C \approx J K r_d(t) + J \sqrt{K r_d(t)} \xi_{\mathrm{rec},i}(t)$
with delayed spike rate 
\begin{equation} \label{eq_fp_rate_convolve_delay}
r_d = r \ast p_d,
\end{equation}
i.e., 
the spike rate convolved with the delay distribution, 
and unit white Gaussian noise process $\xi_{\mathrm{rec},i}$ 
that is uncorrelated to that of any other neuron 
\cite{Augustin2013,Brunel2000,Mattia2002,Gigante2007PRL}.

The step is valid under the assumptions of
(i) sufficiently many incoming synaptic connections ($K\gg 1$)
with small enough weights $|J_{ij}|$ 
in comparison with $V_\mathrm{T}-V_\mathrm{r}$ and sufficient presynaptic activity
(diffusion approximation)
%\todo{check: replaced (arbitrary high) $V_s$ by $V_T$}
%\todo{ok that we do not consider sufficiently high spike rate here?}, 
(ii) that neuronal spike trains can be %well 
approximated by independent 
%\textcolor{gray}{piece-wise homogeneous/inhomogeneous} 
Poisson processes
%\textcolor{gray}{/have approximately 
%exponentially distributed inter-spike intervals} 
(Poisson assumption) 
and (iii) that the
%\footnote{\textcolor{gray}{Refine: specific
%connectivity distribution/probability is missing and delay/weight sampling could be merged herein}} such that 
correlations between the fluctuations of synaptic inputs for different neurons vanish 
% \textcolor{gray}{(conditioned on a common spike rate)} 
(mean-field limit).
%\footnote{
%\textcolor{gray}{
%possible addition: 
%% (i) complete total synaptic current cf. membrane voltage eq and results section (ii) 
%Justification why we do not have a $N+1$-dimensional FP eq.:
%In other words, under the assumptions above the network dynamics 
%is given by $2N$ Eqs.~~\eqref{eq_meth_V_single},\eqref{eq_meth_w_single} + reset. 
%which are independent conditioned on the delayed 
%spike rate $r_d$.
%}} 
% \todo{clarify what exact network topology we use. nowhere sparse random is named (except here as an example)}
The latter assumption is fulfilled by sparse and uniformly random synaptic connectivity, but also when synaptic strengths $J_{ij}$ and delays $ d_{ij} $ are independently distributed (in case of less sparse or random connections) \cite{Mattia2002}.
%topology can be arbitrary but vanishing noise correlations are required (achieved, e.g., by less sparse connections and more heterogeneity)
% Here we consider synaptic delays that are sampled independently from a continuous distribution $p_d$, and identical synaptic strengths, $J_{ij} = J$.  
% \todo{move identical strengths and distribution delay sampling already in network model?}
%

This approximation of the recurrent input allows to replace the 
overall synaptic current in Eq.~\eqref{eq_meth_V_single} by 
$I_\mathrm{syn,i}= C[\mu_\mathrm{syn}(t,r_d) + \sigma_\mathrm{syn}(t,r_d) \xi_i(t)]$ 
with overall %\textcolor{gray}{(sum of external and recurrent)} 
synaptic moments 
\begin{equation} \label{eq_meth_musyn_sigmasyn}
\mu_\mathrm{syn} = \mu_\mathrm{ext}(t) + J K r_d(t), \qquad 
\sigma^2_\mathrm{syn} = \sigma_\mathrm{ext}^2(t) +  J^2 K r_d(t),
\end{equation}  
and (overall) unit Gaussian white noise $\xi_i$ that is uncorrelated to that of any 
other neuron. Here we have used that external $I_{\mathrm{ext},i}$ and recurrent synaptic current $I_{\mathrm{rec},i}$ are independent from each other.

The resulting mean-field dynamics of the membrane voltage is given by 
\begin{equation}
\frac{dV_i}{dt} = \frac{I_{\mathrm{L}}(V_i) + I_{\mathrm{exp}}(V_i) - \langle w \rangle}{C}
+ \mu_\mathrm{syn}(t,r_d) + \sigma_\mathrm{syn}(t,r_d) \xi_i(t), 
\label{eq_meth_V_meanfield}
\end{equation}
and corresponds to
a McKean-Vlasov type of equation
with distributed delays \cite{Touboul2012} 
and discontinuity
due to the reset mechanism 
\cite{Delarue2015}
that complements the dynamics of $V_i$ as before.
The
population-averaged adaptation current $\langle w \rangle = \lim_{N\to\infty} \langle w \rangle_N$ 
is governed by
\begin{equation}
\frac{d{\langle w \rangle}}{dt} = \frac{a 
\left(\langle V \rangle -E_w \vphantom{^2} \right) - 
  \langle w \rangle}{\tau_{w}} + b\, r(t), \label{eq_meth_w_mean_p}
\end{equation} %\todo{add: where $ \langle V \rangle $ is calculated as
%$\langle V \rangle = \int_{-\infty}^{V_{\mathrm{s}}} v p(v,t) dv / \int_{-\infty}^{V_{\mathrm{s}}} p(v,t) dv$ and ..., instead of note below?}
with mean membrane voltage (of non-refractory neurons), 
$\langle V \rangle = \lim_{N\to\infty} \langle V \rangle_N$, 
and spike rate $r = \lim_{N\to\infty, \Delta T\to 0} r_N(t)$.  

Remarks: Instead of exponentially distributed synaptic delays 
we may also consider other continuous densities $p_d$, 
identical delays, $p_d(\tau) = \delta(\tau-d)$ with $d>0$, or no delays at all, $p_d(\tau) = \delta(\tau)$. Instead of identical synaptic strengths one may also consider strengths $J_{ij}$ that are drawn independently 
from a normal distribution with mean $J_m$ and variance $J_v$ instead, in which case the overall synaptic moments become
$\mu_\mathrm{syn} = \mu_\mathrm{ext}(t) + J_m K r_d(t)$ and 
$\sigma^2_\mathrm{syn} = \sigma_\mathrm{ext}^2(t) + (J_m^2 + J_v) K r_d(t)$, cf. \cite{Augustin2013,Gigante2007PRL}.

\paragraph{Continuity equation}

In the membrane voltage evolution, Eq.~\eqref{eq_meth_V_meanfield}, 
individual neurons are exchangeable as they are described 
by the same stochastic equations and are coupled 
to each other exclusively through the (delayed) spike rate via the 
overall synaptic moments $\mu_\mathrm{syn}$ and $\sigma_\mathrm{syn}^2$.
%\todo{should we go from population to trials (of a prototypical neuron) here to make 
%it clearer and thereby explaining the reduction of $N+1$ to $1+1$ FP variable count?}
% allowing the interpretation 
% of a prototypyical neuron that represents the network in the sense that 
% different sample paths of the eq above 
% fokker-planck eq.: prototypical neuron $V$! instead of $N+1$-dim. FP
% the neurons all follow same dynamics except for different realizations
Therefore, the adiabatic and mean-field approximations allow us to represent the collective 
%\textcolor{gray}{neuronal} 
dynamics of a large network 
% ($N\to\infty$), 
by a (1+1-dimensional)
Fokker-Planck equation 
\cite{Augustin2013,Brunel2000,Mattia2002,Gigante2007PRL},
\begin{equation} \label{eq_meth_fp_pde}
\frac{\partial}{\partial t} p(V,t) + \frac{\partial}{\partial V} q_p(V,t) = 0 
\qquad \text{for} \quad V \in (-\infty, V_\mathrm{s}], \; t>0, %[V_\mathrm{lb}, \mathrm{s}], \; t>t_0,
\end{equation} 
% \todo{start with $V_\mathrm{lb}$ right from the beginning omitting $-\infty$? -- note that $V_\mathrm{s}$ could also be avoided in the FP model and be replaced by positive $\infty$}
which describes the evolution of the probability density $p(V,t)$ to find a neuron in 
state $V$ at time $t$ (in continuity form). 
The probability flux is given by 
\begin{equation} \label{eq_meth_fp_flux}
q_p(V,t) = \left( \frac{I_{\mathrm{L}}(V) + I_{\mathrm{exp}}(V)}{C} 
+ \mu_\mathrm{tot}(t) \right) p(V,t) - \frac{\sigma_\mathrm{tot}^2(t)}{2} \frac{\partial}{\partial V} p(V,t),
% q_p(V,t) = \left( \frac{I_{\mathrm{L}}(V) + I_{\mathrm{exp}}(V) - \langle w \rangle}{C} 
% + \mu_\mathrm{syn}(t,r_d) \right) p(V,t) - \frac{\sigma_\mathrm{syn}^2(t,r_d)}{2} \frac{\partial}{\partial V} p(V,t).
\end{equation}
with \textit{total} input mean and standard deviation,
\begin{align} \label{eq_total_input_mean}
\mu_\mathrm{tot}(t) &= \mu_\mathrm{syn}(\mu_\mathrm{ext}(t),r_d(t)) - \langle w \rangle(t)/C \\
\sigma_\mathrm{tot}(t) &= \sigma_\mathrm{syn}(\sigma_\mathrm{ext}(t),r_d(t)). \label{eq_total_input_std}
\end{align} 
Note that the mean adaptation current (simply) 
subtracts from the synaptic mean
% $ \mu_\mathrm{syn} -\langle w \rangle/C $
in the drift term, cf.~\eqref{eq_meth_V_meanfield}.

The mean adaptation current evolves according to 
Eq.~\eqref{eq_meth_w_mean_p}
with time-dependent mean membrane voltage (of the non-refractory 
neurons)
% , $ \langle V \rangle$,} 
\begin{equation} \label{eq_mean_V_fp}
\langle V \rangle = \frac{\int_{-\infty}^{V_{\mathrm{s}}} v p(v,t) dv}{
\int_{-\infty}^{V_{\mathrm{s}}} p(v,t) dv}.
\end{equation}
The spike rate $r$ is obtained by the probability 
flux through $V_s$, 
\begin{equation} \label{eq_meth_fp_rate}
r(t)=q_p(V_s,t).
\end{equation}
%
%Additionally we assume that synaptic delays 
%$d_{ij}$ are 
%either constant ($d_{ij} \equiv d \ge 0$) or 
%sampled independently from 
%a continuous distribution 
%$p_d$\footnote{\textcolor{gray}{merge const delays as delta distribution,too? and add explicitly the case of no delays, i.e., 
%$p_d(\tau) = \delta(\tau)$?.
%The total synaptic moments 
%are given by 
%\begin{equation} \label{eq_meth_musyn_sigmasyn}
%\mu_\mathrm{syn} = \mu_\mathrm{ext}(t) + J K r_d(t), \qquad 
%\sigma^2_\mathrm{syn} = \sigma_\mathrm{ext}^2(t) +  J^2 K r_d(t)
%\end{equation} 
%where 
%the delayed spike rate  $r_d$
%equals the spike rate $r(t)$ convolved with the delay distribution $p_d$, i.e., 
%$r_d = r \ast p_d$. In this model the spike rate is obtained by the probability 
%flux at $V_s$, that is, 
%\begin{equation} \label{eq_meth_fp_rate}
%r(t)=q_p(V_s,t).
%\end{equation}
To account for 
the reset condition of the aEIF neuron dynamics and ensuring that probability mass is conserved, Eq.~\eqref{eq_meth_fp_pde} is complemented 
by the reinjection condition,
\begin{align} \label{eq_meth_fp_reinjection}
q_p(V_{\mathrm{r}}^+,t) - q_p(V_{\mathrm{r}}^-,t) 
= q_p(V_{\mathrm{s}}, t-T_{\mathrm{ref}}),
\end{align}
where $q_p(V_\mathrm{r}^{+}) \coloneqq \lim_{V \searrow  V_{\mathrm{r}}} q_p(V)$ and $q_p(V_\mathrm{r}^{-}) \coloneqq \lim_{V \nearrow V_{\mathrm{r}}} q_p(V)$,
an absorbing boundary at $V_s$,
\begin{equation} \label{eq_meth_fp_absorbing}
p(V_{\mathrm{s}},t) = 0,
\end{equation} and 
a natural (reflecting) boundary condition, 
\begin{equation} \label{eq_meth_fp_reflecting}
    \lim_{V \to -\infty} q_p(V,t) = 0. % alt: lb instead of infty
\end{equation} 
Together with the initial membrane voltage distribution $p(V,0) = p_0(V)$ 
and mean adaptation current $\langle w \rangle(0) = 0$
the 
Fokker-Planck mean-field model is now completely specified. % \textcolor{gray}{for any 
%suitable choice of delay distribution $p_d$}.
%\textcolor{gray}{For a visualization of the ...
%see also Fig.~XYZ\footnote{\textcolor{gray}{This figure could 
%contain the typical boundary condition 
%visualization which would be easire if we took $V_\mathrm{lb}$ instead of $-\infty$}}}.

Note that $p(V,t)$ only reflects the proportion of neurons 
which are not refractory at time $t$, given by 
$P(t) = \int_{-\infty}^{V_{\mathrm{s}}} p(v,t) dv \ = 1 - \int_{t-T_\mathrm{ref}}^t r(s) ds$ ($<1$ for 
$T_{\mathrm{ref}} > 0$ and $r(t)>0$). The total probability density 
that the membrane voltage is $V$ at time $t$ is given by 
$p(V,t) + p_{\mathrm{ref}}(V,t)$ with refractory density 
$p_{\mathrm{ref}}(V,t) = (1-P(t)) \, \delta (V-V_{\mathrm{r}})$.
%Since $p(V,t)$ does not integrate to unity in general, the mean
%membrane voltage %with respect to $p(V,t)$ 
%in Eq.~\eqref{eq_meth_w_mean_p} is calculated as
%$\langle V \rangle = \int_{-\infty}^{V_{\mathrm{s}}} v p(v,t) dv / P(t)$.
%
%The dynamics of the population-averaged\textcolor{gray}{/mean} adaptation 
%current $\langle w \rangle (t)$ reflecting 
%the non-refractory proportion of trials is well captured by 
%eq.~\eqref{eq_meth_w_mean_p} %considering the refractory or non-refractory 
%% proportion of trials, or both, % which also only reflects the 
%% proportion of trials $P(t)$
%as long as $T_{\mathrm{ref}}$ is small compared to $\tau_w$. In this
%(physiologically plausible) case $\langle w \rangle(t)$ can be considered equal 
%to the average adaptation current over the refractory proportion of trials.
%[add refs including frontiers and jneurophys]
At the end of the \secmethods{} section we describe how an (optional) spike shape extension for the aEIF model changes the calculation of $p_{\mathrm{ref}}$ and $\langle V \rangle$.

In practice we consider a finite reflecting lower barrier $V_\mathrm{lb}$ instead of 
negative infinite for the numerical solution (next section)
and for the low-dimensional approximations of the Fokker-Planck PDE
(cf. sections below). 
% for the support of the membrane voltage density 
$V_\mathrm{lb}$ is chosen sufficiently small in order to not distort the free diffusion of the 
membrane voltage for values below the reset, 
i.e., $V_\mathrm{lb} \ll V_r$. The density $p(V,t)$ is then supported on $[V_\mathrm{lb},V_s]$ for 
each time $t$, and in all expressions above 
$V\to -\infty$ is replaced by $V_\mathrm{lb}$.
%\todo{should we start with $V_\mathrm{lb}$ right from the beginning omitting $-\infty$?}

% flatex input end: [section/methods_fokkerplanck_meanfield.tex]

% flatex input: [section/methods_fokkerplanck_fvm.tex]
\paragraph{Finite volume discretization}

In this work we focus on low-dimensional approximations of the FP model. To obtain a reference for the reduced models it is, however, valuable to solve the (full) FP system, Eqs.~\eqref{eq_meth_fp_pde}--\eqref{eq_meth_fp_reflecting}. 
Here we outline an accurate and robust method of solution that exploits 
the linear form of the FP model in contrast to 
previously described numerical schemes 
% to solve the time-dependent 
% Fokker-Planck eq. for integrate-and-fire populations 
% have been described 
\cite{Marpeau2009,Caceres2010}
% . However, these 
% schemes 
which both require rather small time steps due to the steeply 
increasing exponential current $I_\mathrm{exp}$ in the flux $q_p$ close 
to the spike voltage $V_\mathrm{s}$. 
We first discretize the (finite) 
domain $[V_\mathrm{lb}, V_\mathrm{s}]$ 
into $N_V$ equidistant grid cells $[V_{m-\frac{1}{2}}, V_{m+\frac{1}{2}}]$ 
with centers $V_m$ ($m=1,\dots,N_V$) that satisfy $V_1 < V_2 < \dots < V_{N_V}$, 
where $V_{\frac{1}{2}} = V_\mathrm{lb}$ and 
$V_{N_V+\frac{1}{2}} = V_\mathrm{s}$ 
are the outmost cell borders.
Within each cell the numerical 
approximation of $p(V, t)$ is assumed to be constant 
and corresponds to the average value denoted by $p(V_m, t)$. 
% and corresponds
% to the average value denoted by $p_m(t) = \int_{V_{m-\frac{1}{2}}}^{V_{m+\frac{1}{2}}}p(V, t) dV$. 
Integrating Eq.~\eqref{eq_meth_fp_pde} combined with Eq.~\eqref{eq_meth_fp_reinjection}
over the volume of cell $m$, and applying the divergence theorem, yields
\begin{align}
% &\int\limits_{\mathcal{C}_i}\frac{\partial p(V,t)}{\partial t}\mathrm{d}V + \int\limits_{\mathcal{C}_i}\frac{\partial q_p(V,t)}{\partial V}\mathrm{d}V=0 \notag\\
% \iff \hspace{0.5cm}&
\frac{\partial}{\partial t}  p(V_m,t)
=\frac{q_p(V_{m-\frac{1}{2}},t)-q_p(V_{m+\frac{1}{2}},t)}{\Delta V} 
+ \delta_{mm_\mathrm{r}}\frac{1}{\Delta V}q_p(V_{N_V+\frac{1}{2}}, t-T_\mathrm{ref})
,  \label{eq_meth_impl_semidiscrecte_fp}
\end{align}
where $\Delta V$ 
% \textcolor{gray}{= V_{m+1}-V_m = 
% V_{m+\frac{1}{2}}-V_{m-\frac{1}{2}}}$ 
is the grid spacing and $m_\mathrm{r}$ corresponds to the index of the 
cell that contains the reset voltage $V_\mathrm{r}$.
% \textcolor{gray}{that is assumed to match a cell center}
% , i.e., $V_{m_\mathrm{r}} = V_\mathrm{r}$
%\textcolor{gray}{and $\delta_{nm}$ denotes the Kronecker delta.}
To solve Eq.~\eqref{eq_meth_impl_semidiscrecte_fp} forward in time 
the fluxes at the borders of each cell 
% and \textcolor{gray}{right of} the 
% $N_V$-th cell \textcolor{gray}{(for the flux reinjection at the reset)}
need to be approximated.
% \todo{the reset cell further needs the ``rate'' flux} 
Since the Fokker-Planck PDE belongs to the class of 
drift-diffusion equations this can be accurately achieved by the
first order Scharfetter-Gummel
flux \cite{Scharfetter1969,Gosse2013},
% that involves the locally exact stationary solution,
\begin{align}\label{eq_meth_impl_scharfetter_gummel_flux}
q_p(V_{m+\frac{1}{2}},t) = v_{m+\frac{1}{2}}\,\frac{p(V_m, t)-p(V_{m+1},t)\,\exp \left(-v_{m+\frac{1}{2}}\Delta V/D \right) }{1-\exp \left(-v_{m+\frac{1}{2}}\Delta V/D \right)}\, , 
\end{align}
where $v_{m+\frac{1}{2}}(t)
% =v(V_{m+\frac{1}{2}}) 
= \left[I_{\mathrm{L}}(V_{m+\frac{1}{2}}) + I_{\mathrm{exp}}(V_{m+\frac{1}{2}}) 
% -\langle w \rangle 
\right]/C 
+ \mu_\mathrm{tot}(t,r_d(t),\langle w \rangle(t))$ 
% - \langle w \rangle]/C 
% + \mu_\mathrm{syn}(t,r_d)$ 
and $D(t)=\frac{1}{2}\sigma_\mathrm{tot}^2(t,r_d(t))$
% \todo{if total moments are not defined in fp sect. anymore remember to change this part again}
denote the drift and diffusion coefficients, respectively (cf. Eq.~\eqref{eq_meth_fp_flux}).
% that involves the locally exact stationary solution,
This exponentially fitted scheme \cite{Gosse2013} 
% \todo{rephrase: no (is usual definition)} 
is globally first order convergent \cite{Farrell1991}
and yields for large drifts, $|v_{m+\frac{1}{2}}|\Delta V\gg D$, 
the upwind flux, sharing its stability properties. For vanishing drifts, on the other hand, the centered difference method is recovered \cite{Gosse2013}, 
leading to more accurate solutions than the upwind scheme 
in regimes of strong diffusion.

For the time discretization we rewrite Eq.~\eqref{eq_meth_impl_semidiscrecte_fp} (with Eq.~\eqref{eq_meth_impl_scharfetter_gummel_flux})
in vectorized form and approximate the involved time derivative as
first order backward difference to ensure numerical stability. 
This yields in each time step of length $\Delta t$
% \textcolor{gray}{($t_n \to t_{n+1}$)} 
a linear system for the values 
$\mathbf{p}^{n+1}$
of the (discretized) probability density at
%membrane voltage values at
% the $N_V$ cell centers, 
$t_{n+1}$, 
% $\mathbf{p}^{n+1}$, 
given the values 
$\mathbf{p}^{n}$
at the previous time step $t_n$, 
and the spike rate at the time $t_{n+1-n_\mathrm{ref}}$ 
for which the refractory period has just passed, 
% \todo{possible shortcut: end here (just before the eq.), saying that the lin. system is tridiag. and solved using \texttt{banded\_solve}; rest to Appendix; but this way is fine, too}
% \todo{@FB double check}
\begin{align} \label{eq_meth_impl_fvm_timeint}
% \frac{\partial p(V_i, t)}{\partial t}  \approx \frac{\underline{p}(t_{n+1})-\underline{p}(t_{n})}{\Delta_t} &=  \mathbf{G}\underline{p}(t_{n+1})\, \\
(\mathbf{I}-\frac{\Delta t}{\Delta V} \, \mathbf{G}^n)\mathbf p^{n+1}
=\mathbf p^n + \boldsymbol g^{n+1-n_\mathrm{ref}},
\end{align}
with vector elements $p^{n}_m = p(V_m, t_{n})$, 
$m=1,\dots,N_V$, and 
$g^{n+1-n_\mathrm{ref}}_m = \delta_{mm_\mathrm{r}} 
\frac{\Delta t}{\Delta V} r(t_{n+1-n_\mathrm{ref}})$. The refractory period in 
time steps is given by
$n_\mathrm{ref} = \lceil T_\mathrm{ref}/\Delta t \rceil$, where the 
brackets denote the ceiling function, and $\mathbf{I}$ is the identity matrix. 
% \todo{and the matrix $ \mathbf{G}^n $ contains ... (see Appendix)}.
This linear equation can be efficiently solved 
with runtime complexity $\mathcal{O}(N_V)$ due to the 
tridiagonal structure of $\mathbf{G}^n \in\mathbb{R}^{N_V\times N_V}$
% the system matrix 
% $\mathbf{I}-\Delta t/\Delta V \, \mathbf{G}^n \in\mathbb{R}^{N_V\times N_V}$, 
which contains the discretization of the membrane voltage (cf.
Eqs.~\eqref{eq_meth_impl_semidiscrecte_fp}, \eqref{eq_meth_impl_scharfetter_gummel_flux}), 
including the absorbing and reflecting boundary conditions 
(Eqs.~\eqref{eq_meth_fp_absorbing},\eqref{eq_meth_fp_reflecting}).
% Eqs.~\eqref{eq_meth_fp_absorbing},\eqref{eq_meth_fp_reflecting}
For details we refer to \supplement.

The spike rate, Eq.~\eqref{eq_meth_fp_rate}, in this representation is obtained by evaluating the 
Scharfetter-Gummel flux, 
Eq.~\eqref{eq_meth_impl_scharfetter_gummel_flux}, 
at the spike voltage $V_\mathrm{s}$, taking into account 
the absorbing boundary condition, Eq.~\eqref{eq_meth_fp_absorbing}, 
and introducing an auxiliary ghost cell \cite{LeVeque2002}, with center $V_{N_V+1}$,
% \coloneqq V_\mathrm{s}+\Delta V/2$
which yields
\begin{equation} \label{eq_meth_impl_fvmrate}
r(t_{n+1}) = q_p(V_{N_V+\frac{1}{2}},t_{n+1}) = v_{N_V+\frac{1}{2}}\,\frac{1+
\exp(-v_{N_V+\frac{1}{2}}\Delta V/D)}{1-\exp(-v_{N_V+\frac{1}{2}}\Delta V/D)}\,p_{N_V}^{n+1} ,
\end{equation} 
where the drift and diffusion coefficients, $v_{N_V+\frac{1}{2}}$ and $D$, are evaluated at 
$t_{n}$. The mean membrane voltage (of non-refractory neurons), 
Eq.~\eqref{eq_mean_V_fp},
used for the dynamics of the mean adaptation current, Eq.~\eqref{eq_meth_w_mean_p}, 
is calculated by $\langle V \rangle(t_n) = \sum_{m=1}^{N_V} V_m p^{n}_m / 
\sum_{m=1}^{N_V} p^n_m $.

Practically, we use the initialization $p^{0}_m = p_0(V_m)$ and 
solve in each time step the linear system, Eq.~\eqref{eq_meth_impl_fvm_timeint}, 
using the function \texttt{banded\_solve} from the Python library SciPy \cite{Oliphant2007}. %, exploiting the tridiagonal structure of the matrix $\mathbf{G}^n$. 
Note that (for a recurrent network or time-varying external input) the tridiagonal matrix $\mathbf{G}^n$ has to be constructed 
in each time step $t_n$, which can be time consuming -- especially 
for small $\Delta V$ and/or small $\Delta t$. Therefore, we employ low-level
virtual machine acceleration for this task through the Python package Numba \cite{Lam2015} which yields an efficient implementation.

Remark: for a vanishing refractory period $T_\mathrm{ref} = 0$ the matrix 
$\mathbf{G}^n$ would lose its tridiagonal structure 
% \textcolor{gray}{in the last element of the $m_\mathrm{r}$-th row} 
due to the instantaneous reinjection, 
cf. Eq.~\eqref{eq_meth_impl_fvmrate}. 
In this case we enforce a minimal refractory period of one time step, 
$T_\mathrm{ref}=\Delta t$, 
which is an excellent approximation if the time step is chosen sufficiently small 
and the spike rate does not exceed biologically plausible values. 

% flatex input end: [section/methods_fokkerplanck_fvm.tex]

% flatex input: [section/methods_reduction.tex]
\subsection*{Low-dimensional approximations}

%\todo{This section should be improved, @JL do you might try? / cf. also beginning of Sect. Results}
%\todo{$ \mu, \sigma $ as placeholder, explain look-up at given time point in forward integration, e.g., using $ \mu_{\mathrm{tot}} $ etc.}

%REDUCE THIS SECTION: IMPORTANT ARE THE 3. AND LAST PARAGRAPH PLUS COMMONNESS THAT REDUCED MODELS TRANSFORM INPUT IN TERMS OF MOMENTS MU, SIGMA (WHICH INCLUDE REC. COUPLING) TO SPIKE RATE OUTPUT IN AN ODE SYSTEM. PRECALC CONCEPT ALREADY MENTIONED IN RESULTS AND SHOULD BE REPEATED IN MORE COMPACT WAY. THEN SPECIFY IN SECT. BELOW WHICH ARE THE PRECALC QUANTITIES PER MODEL.

%\todo{Consistency w.r.t. $ \mu, \sigma $ here and in Results (eff, tot, syn, etc.)}
In the following sections we present two approaches of how simple spike rate models can be derived from the  Fokker-Planck mean-field model described in the previous section,
cf. Eqs.~\eqref{eq_fp_rate_convolve_delay},\eqref{eq_meth_musyn_sigmasyn},\eqref{eq_meth_w_mean_p}--\eqref{eq_meth_fp_reflecting}.
%In order to derive simple spike rate models 
%\textcolor{gray}{in terms of a low-dimensional 
%ordinary differential equation} from the 
%\textcolor{gray}{(full)} mean-field model, 
%cf. Eqs.~\eqref{eq_meth_w_mean_p}-\eqref{eq_meth_fp_reflecting} 
%\textcolor{gray}{(at least)} two 
%approaches are available.
%
% \textcolor{gray}{One approach uses the spectral decomposition of the Fokker-Planck 
% operator $\LL$ \cite{Mattia2002}, the other approach is based on a cascade of linear input filters and a nonlinearity,
% which can be extracted from 
% % linear rate response and steady-state solution of 
% the Fokker-Planck equation \cite{Ostojic2011PLOS}.}
% \todo{This is basically a copy from Results and can be omitted, however, one could add the references here (which we left out in results) -- we would like to allow readers start with methods first and go to results afterwards as well}
%
%These derived models describe the spike rate dynamics by low-dimensional ODEs and do not express 
%the evolution of the entire membrane voltage distribution 
%\textcolor{gray}{in constrast to the Fokker-Planck model where 
%$p(V,t)$ is available at each time point $t$}. 

%\footnote{
%\textcolor{gray}{Here we should show in a 2d plot for $r_\infty$ and $\langle V \rangle_\infty$ in the 
%$\mu$, $\sigma$ plane how the overall/effective moments in the example from Fig1 (or 
%optimally Fig3 oscillations) evolve -- maybe only one example model (e.g. LNexp).}}.

The derived models %(which are outlined in the following Section) 
are described by low-dimensional ordinary differential equations (ODEs) 
%(for the spike rate $r(t)$\textcolor{gray}{, the population-averaged adaptation current
%$\langle w\rangle(t)$, and the delayed spike rate $r_d(t)$, and 
%an auxiliary state variable depending on the particular approach}) 
which depend on a number of quantities defined in the plane of (generic) input mean and standard deviation $(\mu, \sigma)$. 
%\footnote{\textcolor{gray}{The different text order of first 
%completely describe the low-dim. models (the next two subsections) 
%and then the commonality (this subsection) has one weakness: 
%the second spectral 
%model depends on the mean adaptation in a more complex way than the other 
%two and the $\langle V \rangle_\infty$-dependence of the adaptation has to be introduced 
%before. Therefore I chose the reversed order.}}.
%All derived models %(i.e., of both reduction classes) 
%require as quantities \todo{could give only these two as examples} 
To explain this concept more clearly we consider, as an example, the steady-state spike rate, which is a quantity required by all reduced models.
%
% \textcolor{blue}{[new version]
The steady-state spike rate as a function of 
$ \mu $ and $ \sigma $, 
%\todo{discuss other form: 
%$\lim_{t \to \infty}r(t; \mu_\mathrm{tot}\!\leftarrow\!\mu, \sigma_\mathrm{tot}\!\leftarrow\!\sigma)$ (also for the steady state mean membrane voltage)}
 \begin{equation} \label{eq_ss_spikerate}
 r_\infty(\mu,\sigma) := \lim_{t \to \infty}r(t; \mu_\mathrm{tot}\!=\!\mu, \sigma_\mathrm{tot}\!=\!\sigma),
 \end{equation}
denotes the stationary value of Eq.~\eqref{eq_meth_fp_rate} under 
replacement of the (time-varying) total moments $\mu_\mathrm{tot}$ and
$\sigma_\mathrm{tot}^2$ in the probability flux $q_p$, Eq.~\eqref{eq_meth_fp_flux}, 
by (constants) $\mu$ and $\sigma^2$, respectively. 
Thus the steady-state spike rate $r_\infty$ %\textcolor{gray}{as defined above} 
effectively corresponds 
to that of an uncoupled EIF population whose membrane voltage is governed by 
$dV_i/dt = [I_{\mathrm{L}}(V_i) + 
I_{\mathrm{exp}}(V_i)]/C + \mu + \sigma \xi_i(t)$ plus reset condition, i.e., adaptation and 
synaptic current dynamics are detached. 
For a visualization of $ r_\infty(\mu,\sigma) $ see Fig.~\ref{fig6_lowdim_intro}.
% \textcolor{gray}{Note, that for the actual steady-state spike rate of a particular 
% network of aEIF neurons
% (that converges to a fixed point), 
% $\lim_{t\to\infty} r(t)$, the 
% parametrization $ \mu $ and $ \sigma $ in Eq.~\eqref{eq_ss_spikerate} 
% cannot be chosen arbitrarily. Instead 
% it is determined self-consistently, i.e., $\mu=\lim_{t\to\infty}\mu_\mathrm{tot}(t)$ and $\sigma=\lim_{t\to\infty}\sigma_\mathrm{tot}(t)$, where 
% the stationary values for mean adaptation $\langle w \rangle$ and 
% delayed spike rate $r_d$ are taken into account 
% (cf. Eqs.~\eqref{eq_total_input_mean}, \eqref{eq_total_input_std}).}

When simulating the reduced models these quantities %\textcolor{gray}{($r_\infty$ and others)} 
need to be evaluated for each discrete time point $t$ at a 
certain value of $(\mu, \sigma)$ which depends on the overall synaptic moments $\mu_\mathrm{syn}(t)$, $\sigma_\mathrm{syn}^2(t)$ and on the mean adaptation current $\langle w \rangle(t)$ in a model-specific way (as described in the following Sects.). 
%\textcolor{gray}{For example, the spectral based models require these evaluations at the total input mean and standard deviation, $ \mu = \mu_\mathrm{tot}(t) $ and 
%$ \sigma = \sigma_\mathrm{tot}(t) $, respectively (e.g., $r_\infty(\mu_\mathrm{tot}(t), \sigma_\mathrm{tot}(t))$).} 
An example trajectory of $ r_\infty $ in the $(\mu,\sigma)$ space 
for a network showing stable spike rate oscillations is shown in Fig.~\ref{fig5_network_oscillations}. 

Importantly, %the $(\mu,\sigma)$-dependent 
these quantities depend on the parameters of %\textcolor{gray}{(recurrent and external)} 
synaptic input ($ J $, $ K $, $ \tau_d $, $ \mu_\mathrm{ext} $, $ \sigma_\mathrm{ext} $) and adaptation current ($ a $, $ b $, $ \tau_w $, $ E_w $) only through their arguments $(\mu, \sigma)$. Therefore, for given parameter values of the %\textcolor{gray}{(non-adaptive)} 
EIF model ($C$, $g_\mathrm{L}$, $E_\mathrm{L}$, 
$\Delta_\mathrm{T}$, $V_\mathrm{T}$, $V_\mathrm{r}$, $T_\mathrm{ref}$) 
% \textcolor{gray}{and numerics ($V_\mathrm{lb}$, 
% $V_\mathrm{s}$)} 
we precalculate those quantities on a (reasonably 
large and sufficiently dense)
grid of $\mu$ and $\sigma$ values, and access them during time integration
by interpolating the quantity values stored in a table. This greatly reduces the computational complexity and %\textcolor{gray}{thus} 
enables rapid numerical simulations.
 
%independent of the parameters for external input ($\mu_{\mathrm{ext}}(t)$,  
%$\sigma_{\mathrm{ext}}(t)$), synaptic coupling ($K$, $J$, $p_d$) 
%and adaptation current ($a$, $b$, $\tau_w$). 
%Therefore, for given 
%neuronal membrane properties
%($C$, $g_\mathrm{L}$, $E_\mathrm{L}$, 
%$\Delta_\mathrm{T}$, $V_\mathrm{T}$, $V_\mathrm{r}$, $T_\mathrm{ref}$) 
%\textcolor{gray}{and numerical ones ($V_\mathrm{lb}$, 
%$V_\mathrm{s}$)} the quantities can be precalculated in a (reasonably 
%large and sufficiently high resolved)
%$(\mu,\sigma)$-rectangle and during simulation then be accessed 
%by interpolating the values stored in a table for arbitrary 
%explorations of input, adaptation and coupling parameter space 
%-- e.g. to assess stability of asynchronous or oscillatory 
%spike rate dynamics through linear analysis.

The derived low-dimensional models describe the spike rate dynamics %\textcolor{gray}{but/}
and generally do not express 
the evolution of the entire membrane voltage distribution. 
%\textcolor{gray}{in constrast to the Fokker-Planck model where 
%$p(V,t)$ is available at each time point $t$}.  
%
% [NEW VERSION]
Therefore, the mean adaptation dynamics, 
which depends on 
the density 
$p(V,t)$ (via $\langle V \rangle$, cf. Eq.~\eqref{eq_meth_w_mean_p}) is adjusted through approximating 
the mean membrane voltage $\langle V \rangle$ 
by
the expectation over the 
% \textcolor{gray}{(non-refractory)} 
steady-state 
distribution,
\begin{equation}
\langle V \rangle_\infty = \frac{\int_{-\infty}^{V_{\mathrm{s}}} v p_\infty(v) dv}
{\int_{-\infty}^{V_{\mathrm{s}}} p_\infty(v) dv}, \label{eq_meth_V_inf}
\end{equation}
which is valid for sufficiently slow adaptation current dynamics \cite{Ladenbauer2014,Brunel2003PRE}.
The steady-state distribution is defined as
$p_\infty(V) = \lim_{t\to\infty}p(V,t; \mu_\mathrm{tot}\!=\!\mu, \sigma_\mathrm{tot}\!=\!\sigma)$, 
representing the stationary membrane voltages of 
%\textcolor{gray}{the non-refractory neurons in} 
an uncoupled EIF population 
for generic input mean $\mu$ and standard deviation $\sigma$. %\textcolor{gray}{(cf. 
%definition of steady-state spike rate $r_\infty$ above)}. 
%\textcolor{gray}{which can be conveniently precalculated and} 
The mean adaptation current in all reduced models is thus governed by
\begin{equation}
\frac{d{\langle w \rangle}}{dt} = \frac{a 
\left(\langle V \rangle_\infty -E_w \vphantom{^2} \right) - 
  \langle w \rangle}{\tau_{w}} + b\, r(t), \label{eq_meth_w_mean_p_inf}
\end{equation}
%\textcolor{gray}{(same as Eq.~\eqref{eq_results_w_mean})}, 
where the evaluation of quantity $\langle V \rangle_\infty$ in terms 
of particular values for $\mu$ and $\sigma$ at a given time $t$ 
is model-specific (cf. following Sects.).
Note again that the calculation of $ \langle V \rangle_\infty $ slightly changes when considering an (optional) spike shape extension for the aEIF model, as described at the end of the \secmethods{} section.

The %\textcolor{gray}{partial differential} 
Fokker-Planck model 
does not restrict the form of the 
delay distribution $p_d$, except that the convolution 
with the spike rate $r$, Eq.~\eqref{eq_fp_rate_convolve_delay}, 
has to be well defined. Here, however, we aim at 
specifying the complete network dynamics %\textcolor{gray}{(i.e., of $r$, $\langle w \rangle$ and $r_d$)} 
in terms of a low-dimensional ODE system. 
%\textcolor{gray}{instead of partial or integro-differential equations}. 
%of ordinary differential equations 
Exploiting the exponential form of the delay 
distribution $p_d$ we obtain a simple ordinary differential equation for the delayed spike rate, 
\begin{equation} \label{eq_meth_delay}
 \frac{d r_d}{dt} = \frac{r - r_d}{\tau_d},
\end{equation}
which is equivalent to the convolution $r_d=r \ast p_d$.

Note that more generally any delay distribution 
from the exponential 
%\todo{removed (not existent): (or gamma)}
family allows to represent the 
delayed spike rate $r_d$ by an equivalent ODE instead of 
a convolution integral \cite{MacDonald1978}. 
Identical delays, $r_d(t) = r(t-d)$, are also possible 
but lead to delay differential equations.
Naturally, in case of no delays, we simply have $ r_d(t) = r(t) $.

To simulate the reduced models standard explicit time discretization schemes can be 
applied -- directly to the first order equations of the LN$_{\mathrm{exp}}$ model, and for the other models 
(LN$_{\mathrm{dos}}$, spec$_1$, spec$_2$) -- to the respective equivalent (real) first order systems. 
We would like to note that when using the explicit Euler method to integrate any of the latter three 
low-dimensional models a sufficiently small integration time step $\Delta t$ is required to prevent
oscillatory artifacts.
%
%the two effective timescales that all models (except \textcolor{blue}{LNexp}) contain (e.g., 
%that of $r$ and $s$ in Eq.~\eqref{eq_meth_spec1_2dsys} for the simple spectral model)
%can have very different amplitudes
%leading to oscillatory artefacts 
%as the explicit Euler method is not A-stable. 
Although the explicit Euler method works well for the 
parameter values used in this contribution, 
we have additionally implemented the method of Heun, i.e., the 
explicit trapezoidal rule, which is second order accurate. 
\subsection*{Spectral models}

\paragraph{Eigendecomposition of the Fokker-Planck operator}

Following and extending \cite{Mattia2002} 
we can specify the Fokker-Planck operator $\LL$ 
\begin{equation} \label{eq_meth_spec_operator}
\LL(\mu,\sigma) [p] =  -\frac{\partial}{\partial V} \left [ \left( \frac{I_{\mathrm{L}}(V) + 
I_{\mathrm{exp}}(V) }{C} 
+ \mu \right) p \right ] 
+ \frac{\sigma^2}{2} \frac{\partial^2 p}{\partial V^2},
\end{equation}
for an uncoupled EIF population receiving
(constant) input $(\mu, \sigma)$, cf. Sect.~\textit{Low-dimensional approximations}.
This operator allows to rearrange the 
% (full) 
FP dynamics of the recurrent aEIF network, Eq.~\eqref{eq_meth_fp_pde}, 
as 
\begin{equation} \label{eq_meth_spec_fullfp}
\frac{\partial p}{\partial t} = \LL(\mu_\mathrm{tot}(t),\sigma_\mathrm{tot}(t)) [p]
\end{equation}
which depends on 
the (time-varying) total input moments $\mu_{\mathrm{tot}}(t,r_d,\langle w \rangle)$ and $\sigma_{\mathrm{tot}}^2(t,r_d)$, cf. 
Eq.~\eqref{eq_total_input_mean},\eqref{eq_total_input_std},
% \begin{equation} \label{eq_meth_spec_input}
% \mu_{\mathrm{tot}} \coloneqq \mu_\mathrm{syn}(t,r_d) - \langle w \rangle/C, 
% \qquad\quad \sigma_{\mathrm{tot}}^2 \coloneqq \sigma_\mathrm{syn}^2(t,r_d)
% \end{equation} 
in the drift and diffusion coefficients, respectively. 
% \textcolor{gray}{where $\partial_x$ denotes the partial derivative w.r.t. $x$}. 

For each value of $(\mu,\sigma)$ the operator $\LL$ possesses
an infinite, discrete set of eigenvalues $ \lambda_n $ in the left complex half-plane including 
zero \cite{Mattia2002}, i.e., 
$\mathrm{Re}\{\lambda_n\} \le 0$, 
and associated  eigenfunctions $ \phi_n(V)$ ($n =0,1,2,\dots$) 
%$\{[V_\mathrm{lb},V_\mathrm{s}] \ni V \mapsto \phi_n(V)\}$ 
satisfying
\begin{equation} \label{eq_meth_spec_eigeneq_sparse}
\LL [\phi_n] = \lambda_n \phi_n. 
\end{equation}
Furthermore, the boundary conditions, 
Eq.~\eqref{eq_meth_fp_reinjection}--\eqref{eq_meth_fp_reflecting} 
have to be fulfilled for each eigenfunction $\phi_n$
separately, i.e., the absorbing boundary at the spike voltage,
% \begin{equation} \label{eq_meth_spec_phi_reinjcond}
% \lim_{V \downarrow V_{\mathrm{r}}} q_{\phi_n}(V) - 
% \lim_{V \uparrow V_{\mathrm{r}}} q_{\phi_n}(V) 
% = q_{\phi_n}(V_{\mathrm{s}}) 
% \end{equation}
\begin{equation} \label{eq_meth_spec_phi_absorb}
\phi_n(V_\mathrm{s}) = 0,
\end{equation} 
and the reflecting barrier at the (finite) lower bound voltage,
\begin{equation} \label{eq_meth_spec_qlb_reflect}
q_{\phi_n}(V_\mathrm{lb}) = 0, 
\end{equation}
must hold. 
The eigenflux is given by 
% Moreover, 
% the eigenflux $q_{\phi_n}$, i.e., 
% Eq.~\eqref{eq_meth_fp_flux} with eigenfunction $ \phi_n $ instead of 
% density $ p $, through the spike voltage has to be reinjected into 
% the reset voltage,
% Moreover, 
% the eigenflux 
% % $q_{\phi_n}$ 
\begin{equation} \label{eq_meth_eigenflux_new}
q_{\phi_n}(V) = \left( \frac{I_{\mathrm{L}}(V) + I_{\mathrm{exp}}(V)}{C} 
+ \mu \right) \phi_n(V) - \frac{\sigma^2}{2} \frac{\partial}{\partial V} \phi_n(V),
% q_p(V,t) = \left( \frac{I_{\mathrm{L}}(V) + I_{\mathrm{exp}}(V) - \langle w \rangle}{C} 
% + \mu_\mathrm{syn}(t,r_d) \right) p(V,t) - \frac{\sigma_\mathrm{syn}^2(t,r_d)}{2} \frac{\partial}{\partial V} p(V,t).
\end{equation}
i.e., the flux
$q_p$ of Eq.~\eqref{eq_meth_fp_flux} with eigenfunction $\phi_n$ 
and (constant) generic input moments ($\mu,\sigma^2$) 
instead of density $p$ and (time-varying) total input moments, respectively.
Moreover, 
the eigenflux $q_{\phi_n}$ has to be reinjected 
% (cf. Eq.~\eqref{eq_meth_fp_flux}) 
into the reset voltage, cf. Eq.~\eqref{eq_meth_fp_reinjection},
% 
% Eq.~\eqref{eq_meth_fp_flux} with eigenfunction $ \phi_n $ instead of 
% density $ p $, through the spike voltage has to be reinjected into 
% the reset voltage,
% 
\begin{equation} \label{eq_meth_spec_phi_reinjcond}
q_{\phi_n}(V_\mathrm{r}^+) - 
q_{\phi_n}(V_\mathrm{r}^-) 
= q_{\phi_n}(V_{\mathrm{s}}), 
\end{equation}
where we have neglected the refractory 
period, i.e., $T_\mathrm{ref} = 0$.
% $q_{\phi_n}(V_\mathrm{r}^{\pm})$ denotes the limit 
% $V \downarrow V_\mathrm{r}$ from above ($+$) or $V \uparrow V_\mathrm{r}$ from
% below ($-$),
Note that incorporating a refractory period $T_\mathrm{ref} > 0$ 
is straightforward only for the simplified case 
of vanishing total input moment variations, 
$\dot \mu_\mathrm{tot} \approx \dispdot {\sigma^2_\mathrm{tot}}\approx 0$,
which is described in the following section
and is not captured here in general. 
% and lastly the reflecting barrier at small voltage values,
% \begin{equation}
% \lim_{V\to-\infty} q_{\phi_n}(V) = 0, 
% \end{equation}

The spectrum of $\LL$ is shown in Fig.~\ref{fig7_spectral}A and 
further discussed in Sect.~\textit{\secspecprops}.

Defining the non-conjugated\cite{Risken1996} inner product 
% \textcolor{gray}{of two functions $\phi, \psi: [V_\mathrm{lb},V_\mathrm{s}] \to \CC$,}
% $\langle \psi, \phi \rangle = \int_{-\infty}^{V_\mathrm{s}} \psi(v) \phi(v) dv$ 
$\langle \psi, \phi \rangle = \int_{V_\mathrm{lb}}^{V_\mathrm{s}} \psi(v) \phi(v) dv$ 
yields the corresponding adjoint operator $\LL^*(\mu,\sigma)$ given by \cite{Mattia2002}
\begin{equation} \label{eq_meth_spec_adjoint_operator}
\LL^* = \left( \frac{I_{\mathrm{L}}(V) + I_{\mathrm{exp}}(V)}{C} 
+ \mu
% _{\mathrm{tot}}\left(t,r_d,\langle w \rangle\right) 
 \right) \frac{\partial}{\partial V}
+ \frac{\sigma^2
% _{\mathrm{tot}}^2(t,r_d)
}{2} \frac{\partial^2}{\partial V^2},
% \LL^* = \left( \frac{I_{\mathrm{L}}(V) + I_{\mathrm{exp}}(V) - \langle w \rangle}{C} 
% + \mu_\mathrm{syn}(t) \right) \frac{\partial}{\partial V}
% + \frac{\sigma_\mathrm{syn}^2(t)}{2} \frac{\partial^2}{\partial V^2}
% (V,t) = \left( \frac{I_{\mathrm{L}}(V) + I_{\mathrm{exp}}(V) - \langle w \rangle}{C} 
% + \mu_\mathrm{syn}(t) \right) p(V,t) - \frac{\sigma_\mathrm{syn}^2(t)}{2} \frac{\partial}{\partial V} p(V,t).
\end{equation}
which satisfies 
$\langle \psi, \LL \phi \rangle = \langle \LL^* \psi,  \phi \rangle$ for any 
complex-valued functions $\psi$ and $\phi$ that are sufficiently smooth on
$[V_\mathrm{lb},V_\mathrm{s}]$.
$\LL^*$ has the same set of
eigenvalues $\lambda_n$ as $\LL$  
but distinct associated eigenfunctions $\psi_n(V)$, 
% \textcolor{gray}{(supported on $(-\infty,V_\mathrm{s}]$)}, 
% \textcolor{gray}{(supported on $[V_\mathrm{lb},V_\mathrm{s}]$)}, 
i.e., 
\begin{equation} \label{eq_meth_spec_eigeneq_adjoint_sparse}
\LL^*[ \psi_n] = \lambda_n \psi_n,
\end{equation} 
which have to satisfy three boundary conditions, 
\begin{equation} \label{eq_meth_spec_psi_bc1}
\psi_n(V_\mathrm{s}) = \psi_n(V_\mathrm{r}),
\end{equation}
\begin{equation} \label{eq_meth_spec_dpsi_lb_bc2}
% \lim_{V\to-\infty }\frac{\partial \psi_n}{\partial V}(V) = 0,
\frac{\partial \psi_n}{\partial V}(V_\mathrm{lb}) = 0,
\end{equation}
\begin{equation} \label{eq_meth_spec_adj_bc_cont}
\frac{\partial \psi_n}{\partial V}(V_\mathrm{r}^+) = 
\frac{\partial \psi_n}{\partial V}(V_\mathrm{r}^-)
% \lim_{V\downarrow V_\mathrm{r}}\frac{\partial \psi_n}{\partial V}(V) = 
% \lim_{V\uparrow V_\mathrm{r}}\frac{\partial \psi_n}{\partial V}(V)
\end{equation}
that are determined by 
integrating $\langle \psi_n, \LL[\phi_n] \rangle$ 
by parts and equalling with $\langle \LL^* [\psi_n],  \phi_n \rangle$ 
using the conditions of $\LL$, Eqs.~\eqref{eq_meth_spec_phi_absorb}--\eqref{eq_meth_spec_phi_reinjcond}.
Note that the last condition, Eq.~\eqref{eq_meth_spec_adj_bc_cont}, ensures a continuous derivative of 
$\psi_n$ at $V_\mathrm{r}$ in contrast to the eigenfunctions 
$\phi_n$ of $\LL$, that have a kink at the reset due to reinjection condition, Eq.~\eqref{eq_meth_spec_phi_reinjcond}, as shown in Fig.~\ref{fig7_spectral}B. 

The eigenfunctions of $\LL$ and $\LL^*$ 
are pairwise orthogonal and in the following 
(without loss of generality) assumed to be 
scaled according to the biorthonormality 
condition, 
\begin{equation} \label{eq_meth_spec_biorth}
\langle \psi_n, \phi_m \rangle = \delta_{nm}.
\end{equation}

The membrane voltage probability density can now be expanded 
onto the (moving) eigenbasis of $\LL$ \cite{Mattia2002,Knight1996},
\begin{equation} \label{eq_meth_spec_p_expand_phi}
p(V,t) = \sum_{n=0}^\infty \alpha_n(t) \phi_n(V), 
\end{equation}
where each eigenfunction $\phi_n$ depends on time 
via the total input moments $\mu=\mu_\mathrm{tot}(t,r_d,\langle w \rangle)$,  $\sigma^2=\sigma^2_\mathrm{tot}(t,r_d)$, and 
the projection coefficients are given by 
$\alpha_n = \langle \psi_n, p \rangle$ and 
particularly $\alpha_0 = 1$ \cite{Mattia2002}.
Deriving $\alpha_n$ with respect to time (for $n=1,2,\dots$) and using 
the expansion, Eq.~\eqref{eq_meth_spec_p_expand_phi}, the Fokker-Planck Eq.~\eqref{eq_meth_spec_fullfp} as well as the definition 
of the adjoint operator yields 
an infinite-dimensional equation for the
complex-valued projection coefficients 
$\boldsymbol{\alpha}(t) = (\alpha_1(t), \, \alpha_2(t), \, \dots )^T$,
%, $ \alpha_n \in \mathbb C $
\begin{equation} \label{eq_meth_spec_a_full}
\dot{\boldsymbol{\alpha}} = \left ( \mathbf \Lambda + 
  \mathbf C_\mu \, \dot \mu_{\mathrm{tot}} + \mathbf C_{\sigma^2} \, 
  \dispdot{\sigma^2_{\mathrm{tot}}} \right ) \boldsymbol\alpha + \mathbf c_\mu \, \dot \mu_{\mathrm{tot}} 
  + \mathbf c_{\sigma^2} \, \dispdot{\sigma^2_{\mathrm{tot}}}.
\end{equation}
This dynamics is initialized 
by $\alpha_n(0) = \langle \psi_n , p_0 \rangle$, 
and is complemented by (i) an expression for the spike rate, 
\begin{equation} \label{eq_meth_spec_rate_full}
r(t) = r_\infty + \mathbf{f} \cdot \boldsymbol{\alpha},
\end{equation}
that is obtained from Eq.~\eqref{eq_meth_fp_rate} (using Eq.~\eqref{eq_meth_spec_p_expand_phi}), 
and (ii) by the mean adaptation and 
delayed spike rate dynamics, Eqs.~\eqref{eq_meth_w_mean_p_inf},\eqref{eq_meth_delay}.
% \textcolor{gray}{that yield $p_0(V) = \sum_{n=0}^\infty \alpha_n(t_0) \phi_n(V)$}.
The dots in Eqs.~\eqref{eq_meth_spec_a_full} and \eqref{eq_meth_spec_rate_full} 
denote 
time derivatives (e.g.,, $\dispdot{\sigma^2_{\mathrm{tot}}} = d \sigma^2_{\mathrm{tot}} / dt$) and a non-conjugated scalar 
product of complex vectors (i.e., $\mathbf{f} \cdot \boldsymbol{\alpha} = \sum_{n=1}^\infty f_n \alpha_n$), respectively. 
The matrix $ \mathbf{\Lambda} = \mathrm{diag}(\lambda_1, \lambda_2, \dots)$ 
contains the eigenvalues of $\LL$, 
the matrices $ \mathbf C_\mu $ and $ \mathbf C_{\sigma^2} $ 
have elements $(\mathbf C_x)_{n,m} = \langle \partial_x \psi_n, \, \phi_m \rangle$ 
for 
$x \in \{ \mu, \sigma^2 \}$ and 
$n,m \in \mathbb N$ 
with partial derivative $\partial_x = \partial / \partial x$, 
the vectors $ \mathbf c_\mu $ and $ \mathbf c_{\sigma^2} $ consist 
of components 
$c^x_n = \langle \partial_x \psi_n, \, \phi_0 \rangle$. 
%The quantities which all depend on time via $(\mu, \sigma^2)$ are 
%the eigenvalues $\mathbf{\Lambda} = \mathrm{diag}(\lambda_1, \lambda_2, \dots)$, 
%inner products between the eigenfunctions of $\LL^*$ and $\LL$, 
%$(\mathbf C_x)_{n,m} = \langle \partial_x \psi_n, \, \phi_m \rangle$, 
%$(\mathbf c_x)_{n} = \langle \partial_x \psi_n, \, \phi_0 \rangle$ 
%(with $\partial_x \coloneqq \partial / \partial x$ for 
%$x \in \{ \mu, \sigma^2 \}$ and 
%$n,m \in \mathbb N$),
%$n,m\ge 1$, 
The steady-state spike rate is given by
% \todo{@MA: add $r(t) \propto p'(V_\mathrm{s})$ in previous full FP sect.}
$r_\infty = q_{\phi_0}(V_\mathrm{s})$, i.e., the flux of 
% = 
% -\frac{\sigma^2}{2} \partial_V \phi_0(V_\mathrm{s})$
%   -\frac{\sigma^2}{2} \frac{\partial\phi_0}{\partial V}(V_\mathrm{s})$ 
the eigenfunction $\phi_0 = p_\infty$ 
%   of 
% $\LL$ 
that represents the stationary membrane voltage distribution 
$p_\infty$ and corresponds to
the (stationary) eigenvalue $\lambda_0=0$ \cite{Mattia2002}. 
The vector $ \mathbf{f} $ contains the (nonstationary)
eigenfluxes evaluated at the spike voltage,
% spike voltage fluxes \todo{rephrase sp. volt. fluxes more clearly perhaps} of $\LL$'s eigenfunctions, 
$f_n = q_{\phi_n}(V_\mathrm{s})$.
% \textcolor{gray}{= 
% -\partial_V \phi_n(V_\mathrm{s}) \sigma^2/2$ 
% -\frac{\sigma^2}{2} \, \partial_V \phi_n(V_\mathrm{s})}$
% -\frac{\sigma^2}{2} \, \frac{\partial\phi_n}{\partial V}(V_\mathrm{s})$ 
% \textcolor{gray}{using the absorbing boundary, Eq.~\eqref{eq_meth_spec_phi_absorb}}. 

Note that the quantities ($ \mathbf{\Lambda}, \mathbf{C}_\mu, \mathbf{C}_{\sigma^2}, \mathbf c_\mu, \mathbf c_{\sigma^2}, r_\infty, \mathbf{f} $, $\langle V \rangle_\infty$) 
all depend on time in Eqs.~\eqref{eq_meth_spec_a_full},\eqref{eq_meth_spec_rate_full},\eqref{eq_meth_w_mean_p_inf} via the total input moments 
$(\mu_{\mathrm{tot}}, \sigma_{\mathrm{tot}}^2)$. Particularly, 
at time $t$ the (biorthonormal) solution of the eigenvalue problems for $\LL$ and its adjoint $\LL^*$, 
Eqs.~\eqref{eq_meth_spec_eigeneq_sparse}--\eqref{eq_meth_spec_phi_reinjcond} and \eqref{eq_meth_spec_eigeneq_adjoint_sparse}--\eqref{eq_meth_spec_adj_bc_cont} 
with $n\in\mathbb N_0$, is required for $\mu=\mu_{\mathrm{tot}}(t, r_d, \langle w \rangle)$, $\sigma=\sigma_{\mathrm{tot}}(t, r_d)$. 
Because $\LL$ is a real operator its spectrum contains only real eigenvalues 
and/or complex conjugated pairs depending on $(\mu,\sigma)$. This property carries over to the 
eigenfunctions and therefore also to the components of 
all quantities above implying
for example that the scalar product $\mathbf f \cdot \boldsymbol \alpha$ is always real-valued.

Although the spectral representation, 
Eqs.~\eqref{eq_meth_spec_a_full},\eqref{eq_meth_spec_rate_full}, 
is fully equivalent to the original 
(partial differential) 
Fokker-Planck equation without refractory period
and while it contains only time derivatives 
it is still an infinite-dimensional
and 
furthermore generally an implicit model. 
% for the spike rate. 
% \todo{@MA: refine this passage, 
% as for exp. distr. delays we have an explicit model $\implies$ 
% use this for low-dim. ERE; HOWEVER what about diffusive vs. regular mode couple transition? discuss this only!} 
% \textcolor{gray}{not an 
% ordinary differential equation of the form 
% $\dot{\boldsymbol{\alpha}} = \mathbf{g}(\boldsymbol{\alpha}, r_d)$ 
% but rather implicitly described $\dot{\boldsymbol{\alpha}} =
% \mathbf{h}(\boldsymbol{\alpha}, \dot{\boldsymbol{\alpha}}, r_d)$} 
% due to the dependence on $\dot\mu_{\mathrm{tot}}$ 
% and 
% $\dot{\sigma^2}_{\mathrm{tot}}$. 
% on 
% $r$ \textcolor{gray}{(via $\dot\mu_\mathrm{syn}$ and $\dispdot{\sigma^2_\mathrm{syn}}$ 
% respectively)}. 
Therefore, to derive an explicit low-dimensional 
ordinary differential model 
for the spike rate $r(t)$, 
it is not sufficient to truncate the 
expansion in Eq.~\eqref{eq_meth_spec_p_expand_phi} after, e.g., two terms but 
additional assumptions have to be considered.
% Here we describe two different 
% means of how to come up with such a 
% % low-dimensional 
% model based on the spectral 
% representation of the Fokker-Planck system, Eqs.~\eqref{eq_meth_spec_a_full} and 
% \eqref{eq_meth_spec_rate_full}. 

% 
% approximated\textcolor{gray}{/recovered}
% % under this approximation 
% by solving 
% Eq.~\eqref{eq_meth_spec_rate_full} and its first order time derivative 
% for $f_1 \alpha_1$ and $f_2 \alpha_2$, and 
% plug the thereby obtained projection coefficients
% % $\alpha_1$ and $\alpha_2$, 
% into the expansion 
% Eq.~\eqref{eq_meth_spec_p_expand_phi} which yields $p(V,t) = p_\infty(V) + 
% \alpha_1 \phi_1(V) + \alpha_2 \phi_2(V)$. This allows 
% to replace the second adaptation current approximation, Eq.~\eqref{eq_meth_w_mean_p_inf} by the first one, 
% Eq.~\eqref{eq_meth_w_mean_p} 
% \textcolor{gray}{(replacing $\langle V \rangle_\infty$ 
% with the average over $p$)} which might be beneficial for faster
% adaptation dynamics.
% which carries over to the two spectral models cf next sections...

\paragraph{Basic model: one eigenvalue, negligible input variations}

The first and simpler, derived spectral model 
is based on \cite{Schaffer2013} and requires the strong assumption of 
vanishing changes of the total input moments, 
$\dot \mu_\mathrm{tot} \approx 0$, $\dispdot{\sigma^2_\mathrm{tot}} \approx 0$. Under this approximation the 
projection coefficient dynamics, Eq.~\eqref{eq_meth_spec_a_full}, simplifies to 
\begin{equation} \label{eq_meth_spec_a_exp}
\dot \alpha_n = \lambda_n \alpha_n
\end{equation}
for $ n \in \mathbb N $.
% Justified by the ordered spectrum (cf. properties below) 
Considering 
only the dominant (nonzero) eigenvalue
\begin{equation} \label{eq_meth_lambda1}
% \lambda_1 = \mathrm{argmin}_{\lambda_n \ne 0} |\mathrm{Re}\{\lambda_n\}|, 
\lambda_1 = \mathrm{argmin} \{ |\mathrm{Re}\{\lambda_n\}| : \lambda_n \ne 0 \}, 
\end{equation} 
i.e., the
``slowest mode'', 
% \footnote{\textcolor{gray}{cf. spectrum ... due to the lower bound we cannot just 
% say $\lambda_1 = \lambda_1$ but this depends on $\mu, \sigma$ -- note the 
% romanian numbers for the extracted eigenvalues}} 
we obtain from Eqs.~\eqref{eq_meth_spec_rate_full} and \eqref{eq_meth_spec_a_exp} 
that
% \eqref{eq_meth_spec_a_full} and 
$r(t) = r_\infty + m_1 \, \mathrm{Re}\{f_1 \alpha_1\}$
% \todo{is the scalar prod. of 2 complex numbers always equal to the real part of their product?}
% \begin{equation}
% \dot \alpha_1 = \lambda_1 \alpha_1
% \qquad \text{and} \qquad 
% m_1 = 
% \begin{cases}
% 1 \ \mathrm{for} \ \lambda_1 \in \mathbb R \\
% 2 \ \mathrm{for} \ \lambda_1 \in \mathbb C \setminus \mathbb R
% \end{cases} 
% \end{equation}
with $m_1 = 1$ if $\lambda_1 \in \mathbb R$ 
and $m_1 = 2$ if $\lambda_1 \in \mathbb C \setminus \mathbb R$.
Here we have included for a complex eigenvalue $\lambda_1$ also
its complex conjugate $\overline \lambda_1$ that has
% is an eigenvalue (cf. previous Sect.) 
the projection coefficient $\overline {\alpha_1}(t)$. They 
jointly yield a zero imaginary part in the scalar product of
Eq.~\eqref{eq_meth_spec_rate_full}. 
% since $\LL$ is a real operator.
Defining $\tilde r(t) = r_\infty + m_1 f_1 \alpha_1$ yields the complex first 
order equation for the spike rate \cite{Schaffer2013}, 
\begin{equation} \label{eq_meth_spec1}
\dot{\tilde r} = \lambda_1 (\tilde r - r_\infty), \qquad r(t) = \mathrm{Re}\{\tilde r\}.
\end{equation}

While this derivation is based on neglecting changes of the total input moments,
% in the final spike rate model 
% Eq.~\eqref{eq_meth_spec1} 
their 
% actual 
time-variation 
% of the moments is 
% (artificially) 
is effectively reintroduced 
in the two quantities (dominant eigenvalue and steady-state rate),
i.e., 
$\lambda_1(\mu_{\mathrm{tot}},\sigma_{\mathrm{tot}})$ and
$r_\infty(\mu_{\mathrm{tot}},\sigma_{\mathrm{tot}})$
with $\mu_{\mathrm{tot}}(t)$ and $\sigma_{\mathrm{tot}}(t)$
% % \textcolor{gray}{and $\langle V \rangle_\infty(\mu,\sigma)$ (the latter for the adaptation current)} 
according to 
% $\mu_\mathrm{tot}(t,r_d,\langle w \rangle)$, 
Eqs.~\eqref{eq_total_input_mean},\eqref{eq_total_input_std}.
Therefore, the spike rate evolution, 
Eq.~\eqref{eq_meth_spec1}, is complemented with the dynamics of 
mean adaptation current $\langle w \rangle$ and delayed spike rate $r_d$, 
Eqs.~\eqref{eq_meth_w_mean_p_inf},\eqref{eq_meth_delay},
where the former involves the third
$(\mu_\mathrm{tot},\sigma_\mathrm{tot})$-dependent
quantity $\langle V \rangle_\infty$. See Figs.~\ref{fig6_lowdim_intro},\ref{fig7_spectral}C for the involved quantities depending on (generic) input $\mu$, $\sigma$.

% by
% $\mu_{\mathrm{tot}} = \mu_\mathrm{syn}(t) - 
%   \langle w \rangle(t)/C$ and $\sigma_{\mathrm{tot}}=\sigma_\mathrm{syn}(t)$. \todo{explicate this a bit more in words perhaps; also probably use ref. to tot. moments defined above}
We call this first derived low-dimensional spike rate model, i.e., Eqs.~\eqref{eq_meth_spec1},
\eqref{eq_meth_w_mean_p_inf},\eqref{eq_meth_delay}, spec$_1$.
It is very simple in comparison with the full 
Fokker-Planck system in the spectral representation, 
Eqs.~\eqref{eq_meth_spec_a_full}--\eqref{eq_meth_spec_rate_full},
\eqref{eq_meth_w_mean_p_inf},\eqref{eq_meth_delay}, in the sense that it does not depend on 
``nonstationary'' quantities of $\LL$ or its adjoint $\LL^*$ except for 
the dominant eigenvalue $\lambda_1$.
%   
% This model does depend on the ``stationary'' quantities 
% steady-state spike rate $r_\infty$ and mean membrane voltage
% $\langle V\rangle_\infty$ but\footnote{todo: combine with quantities above to include 
% there also mean V for the adaptation?} not on ``nonstationary'' quantities 
% of $\LL$ or its adjoint $\LL^*$ except for the dominant eigenvalue $\lambda_1$.

% \textcolor{gray}{Remark 1: 
% $\mathrm{Re}(\lambda_1) < 0$ (see below) and therefore 
% for constant input indeed the model converges 
% to the steady-state rate as expected.}

Note that under the assumption of vanishing input moment variations 
the dynamics of the expansion coefficients $\alpha_n$, Eq.~\eqref{eq_meth_spec_a_full}, 
is simply exponentially decaying in time, cf. 
Eq.~\eqref{eq_meth_spec_a_exp}. This allows to incorporate a refractory period $T_\mathrm{ref}>0$ 
into the spectral decomposition framework 
by inserting the eigenbasis
expansion of the membrane voltage distribution, Eq.~\eqref{eq_meth_spec_p_expand_phi},
into the reinjection condition of the Fokker-Planck model, 
Eq.~\eqref{eq_meth_fp_reinjection}, 
using $\alpha_n(t) = \alpha_n(0)\exp(\mathbf \lambda_n t)$
and the absorbing boundary, Eq.~\eqref{eq_meth_spec_phi_absorb}.
This generalizes the reinjection condition for the eigenfunctions 
$\phi_n$ of $\LL$ from Eq.~\eqref{eq_meth_spec_phi_reinjcond} to
\begin{equation} \label{eq_meth_spec_phi_reinjcond_ref}
q_{\phi_n}(V_\mathrm{r}^+) - 
q_{\phi_n}(V_\mathrm{r}^-) 
= q_{\phi_n}(V_{\mathrm{s}})  \exp(-\lambda_n T_\mathrm{ref}),
% \lim_{V \downarrow V_{\mathrm{r}}} q_{\phi_n}(V) - 
% \lim_{V \uparrow V_{\mathrm{r}}} q_{\phi_n}(V) 
% = q_{\phi_n}(V_{\mathrm{s}})  \exp(-\lambda_n T_\mathrm{ref})
\end{equation} 
which was applied in \cite{Schaffer2013},
and 
the corresponding boundary condition
for the adjoint operator's eigenfunctions $\psi_n$
from Eq.~\eqref{eq_meth_spec_psi_bc1} to 
\begin{equation} \label{eq_meth_spec_psi_bc1_ref}
\psi_n(V_\mathrm{s})  = \psi_n(V_\mathrm{r}) \exp(-\lambda_n T_\mathrm{ref}).
\end{equation}

\paragraph{Advanced model: two eigenvalues, slow input moment variations}

Another possibility to derive a low-dimensional spike rate model is based on 
ongoing work of Maurizio Mattia. 
He has recognized that under the weaker 
(compared to the basic spec$_1$ model above)
assumption of small but not vanishing 
input moment changes a real-valued second order ordinary differential 
equation for the spike rate $r(t)$  
can be consistently derived \cite{Mattia2016Arxiv}. Here we 
extend this approach to account for neuronal adaptation, 
time-varying external input moments 
and delay distributions.

The steps in a nutshell (for the detailed derivation 
see \supplement) 
are (i) taking the derivative of Eq.~\eqref{eq_meth_spec_a_full} (once) and of
\eqref{eq_meth_spec_rate_full} (twice) w.r.t. time, 
(ii) considering only the first two dominant eigenvalues $\lambda_1$ and 
$\lambda_2$, i.e., neglecting all (faster) eigenmodes that correspond to 
eigenvalues with larger absolute real part (``modal approximation''), 
(iii) assuming slowly changing input moments, i.e., small 
$\dot{\mu}_\mathrm{tot}$ and $\dispdot{\sigma^2_\mathrm{tot}}$ that
allow to consider projections coefficients of that order, i.e.,
$\alpha_{n} = \OO(\dot{\mu}_\mathrm{tot})$ and 
$\alpha_{n}=\OO( \dispdot{\sigma^2_\mathrm{tot}} )$ for $n=1,2$, 
and therefore to consider only linear occurences of $\dot{\mu}_\mathrm{tot}$, 
$\dispdot{\sigma^2_\mathrm{tot}}$, $\alpha_1$, $\alpha_2$, and 
neglect terms of second and higher order. 
The slowness 
approximation implies 
that neither the external moments $\mu_\mathrm{ext}(t)$, 
$\sigma^2_\mathrm{ext}(t)$ nor the 
(delayed) spike rate $r_d(t)$ nor the population-averaged adaptation current
$\langle w \rangle(t)$ should change very fast (cf. Eqs.~\eqref{eq_total_input_mean},\eqref{eq_total_input_std},\eqref{eq_meth_musyn_sigmasyn}). 
Note that the dynamics of $\langle w \rangle$ 
is assumed to be slow already, cf. Sect.~\textit{\secmeanfield}.
% For biologically plausible timescales (cf. Results) 
% of the external and recurrent input this approximation is quite good and the 
% (mean) adaptation current dynamics is assumed to be slow 
% already in Sect.~\secmeanfield.
% which are proportional to the factors 
% $\dot{\mu}\dot{\sigma^2}$, $\dot{\mu}^2$, 
% $(\dot{\sigma^2})^2$, $\dot{\mu}\boldsymbol{\alpha}$, 
% $\dot{\sigma^2}\boldsymbol{\alpha}$, 
% $\ddot{\mu}\boldsymbol{\alpha}$, $\ddot{\sigma^2}\boldsymbol{\alpha}$, 
% $\dot{\mu}\dot{\boldsymbol{\alpha}}$, $\dot{\sigma^2}\dot{\boldsymbol{\alpha}}$}.

Under these approximations we obtain for the spike rate dynamics
the following real second order ODE,
\begin{equation} \label{eq_meth_spec2}
\beta_2 \, \ddot r + \beta_1 \, \dot r + \beta_0 \, r 
= r_\infty - r - \beta_c, 
% \left 
% ( \langle w \rangle, \,  r_d, \dot \mu_\mathrm{ext}, \, \ddot \mu_\mathrm{ext}, \, 
% \dispdot{\sigma_\mathrm{ext}^2}, \, \dispddot {\sigma_\mathrm{ext}^2} \right ),
\end{equation}
that is complemented with the mean 
adaptation current and delayed spike rate dynamics, 
Eqs.~\eqref{eq_meth_w_mean_p_inf},\eqref{eq_meth_delay}, 
and we call the model spec$_2$.
The coefficients 
% \footnote{\textcolor{blue}{these coeffs. are derived
% for synaptic coupling with delays that are independently drawn from an exponential 
% distribution with with decay constant $\tau_d$ (for zero or constant delays see \supplement). }}  
\begin{align} \label{eq_meth_beta2} 
\beta_2 = &\, \, D,  \\
\beta_1 =  &\,-T+DM\frac{b}{C}-\frac{R}{\tau_d},  \label{eq_meth_beta1}  \\
\beta_0 = &-DM\frac{b}{\tau_wC}-\frac{b}{C}H_\mu
  +\frac{1}{\tau_d}\left( KJH_\mu
  +KJ^2H_{\sigma^2} \right)
  +\frac{R}{\tau_d^2}, \label{eq_meth_beta0} \\
\beta_{\text{c}} = &\,
\,r_d \left(-\frac{1}{\tau_d}\left( KJH_\mu
  +\,KJ^2H_{\sigma^2}\right) 
  -\frac{R}{\tau_d^2} \right) \nonumber \\ 
&-\left(\ddot{\mu}_{\text{ext}}+\frac{a\left(\langle V \rangle_\infty-E_w\right) -\langle w \rangle}{\tau_w^2 C}
  \right)DM-\dispddot{\sigma^2_{\text{ext}}}DS \label{eq_meth_betac} \\
&+\left(\dot{\mu}_{\text{ext}}-\frac{a\left(\langle V \rangle_\infty-E_w\right)-\langle w \rangle}{\tau_w C}\right) H_\mu
  +\dispdot{\sigma^2_{\text{ext}}}H_{\sigma^2}, \nonumber
\end{align}
% $\beta_2 = D$, $\beta_1 = -T+DMb/C-R/\tau_d$, 
% $\beta_0 = -DMb/(\tau_wC)-H_\mu b/C
%   +\frac{1}{\tau_d}\left( KJH_\mu
%   +KJ^2H_{\sigma^2} \right)
%   +R/\tau_d^2$ 
%   and $\beta_c$ 
% $\beta_c \left ( \langle w \rangle, \,  r_d, \dot \mu_\mathrm{ext}, \, \ddot \mu_\mathrm{ext}, \, 
% \dispdot{\sigma_\mathrm{ext}^2}, \, \dispddot {\sigma_\mathrm{ext}^2} \right )$ 
depend on the 
% (combined) spectral  
(lumped) quantities 
$D = 1/\lambda_1 \cdot 1/\lambda_2$, 
$T = 1/\lambda_1 + 1/\lambda_2$, 
$M = \partial_\mu r_\infty+\mathbf{f}\cdot\mathbf{c}_\mu$, 
$S = \partial_{\sigma^2} r_\infty+\mathbf{f}\cdot\mathbf{c}_{\sigma^2}$,
$R = DMKJ + DSKJ^2$, 
$H_\mu = TM+DMa \partial_\mu\langle V\rangle_\infty/(\tau_wC)-DF_\mu$, 
$H_{\sigma^2} = TS+DMa \partial_{\sigma^2}\langle V \rangle_\infty/(\tau_wC) -DF_{\sigma^2}$, 
$F_\mu = \mathbf{f}\cdot\mathbf{\Lambda}\,\mathbf{c}_{\mu}$ 
and 
$F_{\sigma^2} = \mathbf{f}\cdot\mathbf{\Lambda}\,\mathbf{c}_{\sigma^2}$.
Here the diagonal eigenvalue matrix $\mathbf\Lambda = \mathrm{diag}(\lambda_1, \lambda_2)$ 
and the vectors 
$\mathbf f = (f_1, f_2)^T$, 
$\mathbf c_{\mu} = (c^{\mu}_1, c^{\mu}_2)^T$, 
$\mathbf c_{\sigma^2} = (c^{\sigma^2}_1, c^{\sigma^2}_2)^T$
% representing real or complex quantities 
are two-dimensional in contrast to infinite as in the original dynamics, 
Eqs.~\eqref{eq_meth_spec_a_full} and \eqref{eq_meth_spec_rate_full}. 
These individual quantities, that also include $\langle V \rangle_\infty$ 
(cf. Eq.~\eqref{eq_meth_w_mean_p_inf}), $r_\infty$ and derivatives of both w.r.t. generic mean $\mu$ and variance $\sigma^2$, 
% \textcolor{gray}{$D$, $T$, $M$, $R$, $H_\mu$, 
% $H_{\sigma^2}$, $F_\mu$ and $F_{\sigma^2}$} 
are evaluated at the total input moments 
$\mu=\mu_{\mathrm{tot}}(t,r_d,\langle w \rangle)$, $\sigma^2 = \sigma_{\mathrm{tot}}^2(t,r_d)$.
A relevant subset of individual and lumped quantities 
% \textcolor{gray}{that are functions of generic input $(\mu,\sigma$)} 
is shown in Figs.~\ref{fig6_lowdim_intro},\ref{fig7_spectral}C.

The four coefficients, Eqs.~\eqref{eq_meth_beta2}--\eqref{eq_meth_betac}, are real-valued because we 
define $\lambda_2$ as the second dominant 
eigenvalue conditioned that $(\lambda_1, \lambda_2)$ compose 
either a real or a complex conjugated eigenvalue pair, i.e., 
\begin{equation} \label{eq_meth_lambda2}
 \lambda_2 = \mathrm{argmin}\{|\mathrm{Re}\{\lambda_n\}| \, : \, \lambda_n \ne \lambda_1 \ \ \text{s.t.} \ \ \lambda_n \ne 0, \ \lambda_1 + \lambda_n \in \RR \},
% \lambda_2 = \mathrm{argmin}_{\lambda_n \ne \lambda_1 \ \text{s.t.} \ \lambda_n \ne 0, \ \lambda_1 + \lambda_n \in \RR}|\mathrm{Re}(\lambda_n)|,
\end{equation} 
% \todo{check for ambiguity of argmin (which is the set?)}
where $ \lambda_1 $ is obtained as for the (basic) spec$_1$ model, cf. Eq.~\eqref{eq_meth_lambda1}.
This condition ensures that all related complex quantities 
occur in vectors of two complex conjugate components 
(for example, $f_1 = \overline{ f_2}$) implying that the
scalar products above are real-valued 
(e.g., $\mathbf f \cdot \mathbf c_\mu \in \RR$) and therefore all (nine) lumped quantities, too. Note that this specific
definition of $\lambda_2$ is required only for integrate-and-fire neuron models 
that have a lower bound different from the reset, i.e., $V_\mathrm{lb} < V_\mathrm{r}$, as discussed in the following section.
% \textcolor{gray}{Otherwise for a real-complex eigenvalue mixture the coefficients, 
% Eqs.~\eqref{eq_meth_beta2}--\eqref{eq_meth_betac}, would take complex values.}

% \todo{add Note on precalculation procedure: individual quantities can be precalced, lumped ones as dep. on recurrency, adaptation and input have to be built upon the precalced...}

% \todo{clarify adaptation dep. and at total moments (cf. spec1)}

The coefficients of the spec$_2$ model require -- in addition to eigenvalues $ \lambda_n $ and 
steady-state rate $r_\infty$ (cf. basic 
 spec$_1$ model, Eq.~\eqref{eq_meth_spec1}) -- quantities that involve the first and second 
eigenfunctions of the Fokker-Planck operator $\mathcal L$ and its adjoint $\mathcal L^*$. 
% a real-complex mixture which would lead to complex \textcolor{gray}{(instead of real)} 
% coefficients, Eqs.~\eqref{eq_meth_beta2}--\eqref{eq_meth_betac}, i.e., 
Additionally, $\beta_1$, $\beta_0$ and $\beta_c$ contain the parameters 
of membrane voltage ($C$) and mean adaptation current dynamics 
($a$, $b$, $\tau_w$, $E_w$) as well as of the recurrent coupling 
($K$, $J$, $\tau_d$) and, importantly, explicit dependencies 
on the population-averaged adaptation current 
$\langle w \rangle$ and the delayed spike rate $r_d$ (that is in addition to implicitly via $\mu_\mathrm{tot}$ and $\sigma_\mathrm{tot}$). Furthermore
$\beta_c$ depends on the first two time derivatives of the external input moments 
$\mu_\mathrm{ext}(t)$ and $\sigma^2_\mathrm{ext}(t)$. 
% \todo{polish this sentence later and optionally move passage from appendix (before Sect. ``Second order ordinary differential equation'' about the param dep. lumped quants. (runtime) based onprecalculation}
This explicit occurence of neuronal and coupling parameters, state variables and input moment derivatives in the 
coefficients is not expressed in the 
basic spectral model (spec$_1$), 
Eq.~\eqref{eq_meth_spec1}. 
% \textcolor{gray}{nor in the cascade based models, Eqs.~\eqref{...}}\todo{@MA: check whether this holds true 
% also for the final dosc model}.
A consequence 
% of the dependence on the 
% first and second order time derivatives 
% of the external input moments $\mu_\mathrm{ext}(t)$ and 
% $\sigma^2_\mathrm{ext}(t)$ 
is that for the (advanced) model spec$_2$ 
the external moments 
have to be provided twice differentiable or in case of non-smooth time series 
to be filtered (e.g., see Sect.~\textit{Performance for variations of the mean input}). 
% \textcolor{gray}{which can be interpreted as 
% part of the model specification}.

The particular coefficients, 
Eqs.~\eqref{eq_meth_beta2}--\eqref{eq_meth_betac}, 
are specific for the choice of an exponential delay 
distribution, as indicated by the occurrence of the 
mean delay $\tau_d$ in $\beta_1$, $\beta_0$ and 
$\beta_c$. Other choices such as identical delays or no delays
are described in the supplementary material \supplement.
%
%where for example coefficients for instantaneous coupling\textcolor{gray}{, 
%i.e., zero delay,} are given.

Note that the original (infinite-dimensional) spectral dynamics, Eq.~\eqref{eq_meth_spec_a_full},\eqref{eq_meth_spec_rate_full}, 
assumes for the refractory period a value of $T_\mathrm{ref} = 0$ 
which carries over to the same choice for the spec$_2$ model, whereas $T_\mathrm{ref} > 0$ is not supported (yet). 

The spike rate is by definition nonnegative, however, 
% \textcolor{gray}{integrating} 
the model  spec$_2$, Eq.~\eqref{eq_meth_spec2},
can yield negative rates $r(t)$, 
especially for small total mean input $\mu_\mathrm{tot}$ and fast (external) input changes, e.g., when 
$\dot \mu_\mathrm{ext}$ is large. We explicitly avoid that behaviour by setting both 
the spike rate of this model, $r$, and its derivative, $\dot r$, to zero
whenever $r(t) < 0$ 
% \textcolor{gray}{which guarantees a nonnegative and continuous spike rate}, 
and continue the integration of the differential equation afterwards.

\paragraph{\secspecprops}

In the previous two sections we have developed two 
spike rate models based on approximations
of the Fokker-Planck system's spectral 
representation, Eqs.~\eqref{eq_meth_spec_a_full},\eqref{eq_meth_spec_rate_full} under different slowness assumptions. 
In the derivation of the (simple) model spec$_1$, cf. 
Eq.~\eqref{eq_meth_spec1}, temporal variations of the total input moments 
are completely neglected while the (advanced) model spec$_2$, cf. Eq.~\eqref{eq_meth_spec2}, 
incorporates (slow) changes 
of the total input moments through linear 
terms 
(proportional to $\dot \mu_{\mathrm{tot}}$ or $\dispdot{\sigma^2_{\mathrm{tot}}}$) 
and neglects (faster) quadratic and higher order ones.
% \textcolor{gray}{i.e., $\dot \mu^2 \approx 0$, \
% $(\dispdot{\sigma^2})^2 \!\! \approx 0$, \ $\dot \mu \dispdot{\sigma^2} \approx 0$.}
% \textcolor{gray}{Furthermore, $\beta_1$, $\beta_0$, $\beta_c$ depend 
%on the adaptation current parameters $a$, $b$ and $\tau_w$, i.e., the 
%population-averaged adaptation current $\langle w \rangle$ is considered in the second 
%spectral model beyond subtracting from the input mean 
%(see also the explicit dependence on $\langle w \rangle$ in 
%% the coefficient of $\beta_c$ in 
%Eq.~\eqref{eq_results_spectral2}).}
% 
% In the next Sect. we describe two different 
% means of how to come up with a  
% low-dimensional 
% model based on the spectral 
% representation of the Fokker-Planck system, Eqs.~\eqref{eq_meth_spec_a_full} and 
%  \eqref{eq_meth_spec_rate_full}. 

Both 
% derivations are based on the 
slowness approximations
% of slowly changing 
% total input moments 
imply that the eigenvalue matrix $\boldsymbol\Lambda$ is the dominant 
term in the homogeneous part of Eq.~\eqref{eq_meth_spec_a_full}. 
Therefore the eigenvalues approximately correspond to decay time constants 
$1/|\mathrm{Re}\{\lambda_n\}|$ in their real part and in case of complex eigenvalues they 
contribute 
damped oscillatory components with frequency 
$|\mathrm{Im}\{\lambda_n\}|/(2\pi)$ through their imaginary part to the dynamics. 
% in the next Section, Eqs~\eqref{eq_meth_spec1},\eqref{eq_meth_spec2}, 
% are 
% depend on the spectrum of the 
% Fokker-Planck operator $\LL$.
% and
% (for the advanced model, Eq.~\eqref{eq_meth_spec2})
% its adjoint $\LL^*$. 
How the spectrum of the Fokker-Planck operator $\LL$ depends on the 
input moments therefore gives insights into the behavior 
% and limitations 
of the derived spike rate models 
spec$_1$ and spec$_2$, 
and of the FP model with slow adaptation in general. 

Here we summarize the main properties of the eigenvalues $\lambda_n(\mu,\sigma)$ 
for the 
(uncoupled, nonadaptive) 
EIF neuron model as a function of 
generic input mean $\mu$ and standard deviation $\sigma$ 
(cf. Sect.~\textit{Low-dimensional approximations}) that are shown in 
Fig.~\ref{fig7_spectral}A. 
Note that at time $t$ the total input
% \textcolor{gray}{, $\mu=\mu_\mathrm{tot}(t,r_d(t),\langle w \rangle(t))$ 
% and $\sigma=\sigma_\mathrm{tot}(t,r_d(t))$,} 
determines the particularly effective 
eigenvalue, $\lambda_n(\mu_\mathrm{tot}(t),\sigma_\mathrm{tot}(t))$, in dependence of 
external input moments, (delayed) spike rate and adaptation current, cf. 
Eqs.~\eqref{eq_total_input_mean},\eqref{eq_total_input_std},\eqref{eq_meth_musyn_sigmasyn}. 
% \todo{EIF def. before I guess; Important: extend this sentence because it is hard for the reader to understand the link here to (uncoupled, nonadaptive) EIF model in response to noise, justifying the next page of observations} 
We thereby extend the findings of \cite{Mattia2002} concerning the 
perfect integrate-and-fire neuron 
with reflecting lower barrier at the reset voltage (PIFb), i.e., 
$dV_i/dt = \mu + \sigma \xi_i(t)$ with $V_\mathrm{lb} = V_\mathrm{r}$. 
% and $I_\mathrm{L}=I_\mathrm{exp}=0$ in Eq.~\eqref{eq_meth_V_single}. 
% -- 
% to the EIF neuron that involves a leak current $I_\mathrm{L}(V)$, a nonlinearity $I_\mathrm{exp}(V)$ and no lower bound (for the numerics the latter is chosen 
% far below the reset voltage, $V_\mathrm{lb} \ll V_\mathrm{r}$).
% that are shown in Fig.~\ref{fig7_spectral}
% \todo{@MA: extension into discussion}
% (extending the findings of \cite{Mattia2002} concerning the perfect 
% integrate-and-fire neuron with lower bound at the reset (PIFb)).
% summarized 
% as follows (extending the results of [ref mattia02] 
% \textcolor{gray}{for the perfect integrate-and-fire 
% neuron with lower bound at the reset}):
% \todo{below we should have subscript tot or (with brief explanation) mu, sigma as placeholder}
\begin{enumerate} 
 \item The eigenvalue $\lambda_0=0$ exists for all (generic) input moments $\mu$,$\sigma^2$ with stationary membrane voltage 
  distribution as corresponding eigenfunction, 
  $\phi_0 = p_\infty$ (see Fig.~\ref{fig7_spectral}A,B), the respective adjoint eigenfunction is constant, $\psi_0 = 1$, cf. \cite{Mattia2002}.
  The other, nonstationary eigenvalues $\lambda_n$ with $n \ge 1$ have 
  negative real parts for all $(\mu,\sigma)$ which yields stability of 
  the stationary distribution for constant total input moments.
  For sufficiently small mean input $\mu_\mathrm{min}$ the eigenvalues $\lambda_n(\mu_\mathrm{min}, \sigma)$ are real-valued for all $n$ and $\sigma$, 
    i.e., no (damped) oscillatory spike rate transient 
%     \textcolor{gray}{for an uncoupled population} 
    in that regime are present (consistent with \cite{Mattia2002}), cf. Fig.~\ref{fig7_spectral}A.
 \item Two classes of modes can be 
%  \todo{merge point 4 and 5?}
 distinguished: eigenvalues of the first kind occur in couples which 
 are real-valued at  $\mu_\mathrm{min}$ but merge for increasing mean input 
 $\mu$ 
%    \textcolor{gray}{(for constant $\sigma$ and $n$)} 
   to become a complex conjugated pair 
   of decreasing 
    absolute real part and almost linearly increasing imaginary part (see 
    Fig.~\ref{fig7_spectral}A). 
    Thus they correspond in this situation to damped oscillatory 
    dynamics with increased frequency $|\mathrm{Im}\{\lambda_n\}|/(2\pi)$ 
    and decay time constant $1/|\mathrm{Re}\{\lambda_n\}|$ 
    for stronger mean input. Note that there is no single 
    critical parameter for the real-to-complex transition, instead that depends 
    on the input mean $\mu$ and standard deviation $\sigma$ as well as on 
    the eigenvalue index $n$ in contrast to the PIFb neuron
    where $\mu=0$ (alone) induces the transition \cite{Mattia2002}. 
    We call this type of eigenvalue 
    \textit{regular} because it is observed also 
    for very simple integrate-and-fire models (such as PIFb). 
%      and also with lower bound=reset 
\\
 The second type of eigenvalue is real for the whole input parameter space of $(\mu,\sigma)$.
 The corresponding decay time constant is large if noise 
  dominates while 
  for stronger mean input 
  the respective dynamics is negligibly fast. 
  Note that by setting the lower bound $V_\mathrm{lb}$ equal to the reset voltage,
  this eigenvalue class is completely removed for the EIFb model, 
  which is the EIF membrane voltage description with $V_\mathrm{lb} = V_\mathrm{r}$ 
  (not used here). This explains why 
  this new type of eigenvalue has not been described
  in \cite{Mattia2002}, and
 suggests a relationship 
%  of such an eigenmode 
  to the diffusion of the membrane voltage for 
  hyperpolarized neuronal states. The latter correspondence is further supported 
  from the significant values of the respective 
  eigenfunction $\phi_n$ below 
%   \textcolor{gray}{the reinjection-induced kink at} 
  the reset voltage $V_\mathrm{r}$ 
  in contrast 
  to the eigenfunctions of regular eigenvalues (cf. Fig.~\ref{fig7_spectral}B). 
  Therefore we label this second type of eigenvalue \textit{diffusive}.
 \item \label{item_specprops_modeclass_switching} The input noise intensity 
 $\sigma$ controls the spectrum's mixture of the two eigenvalue classes as follows:
    weak noise favors regular modes, i.e., the dominant two eigenvalues 
    are pairs of real (for smaller mean input $\mu$) or complex conjugated eigenvalues (for larger $\mu$) 
    while the diffusive modes are irrelevantly fast in this regime 
    (cf. Fig.~\ref{fig7_spectral}A, left). 
    Increased input fluctuations, i.e., a larger $\sigma$, on the other hand leads to 
    a spectrum with dominant (``slowest'' eigenvalue $\lambda_1$) of the diffusive 
    kind for small mean input $\mu$
%     \footnote{this behaviour in the noise-dominated regime is consistent with 
%     the idea that the purely real eigenvalues describe the 
%     fluctuations at membrane voltages smaller than $V_\mathrm{r}$} 
    while for larger $\mu$ the dominant
    mode is from the regular class  
    (being real or complex depending on 
    $\mu$, $\sigma$, $n$), see Fig.~\ref{fig7_spectral}A, right. 
%     cf. Fig.~\ref{fig7_spectral}A.
%     \footnote{
% %     \textcolor{gray}{refer to this property in the following (discontinuity of imag. part of lambda1 
%     but also real part of lambda2}}. 
    Furthermore an increased noise strength $\sigma$
    leads to a smaller decay time constant $1/|\mathrm{Re}\{\lambda_n\}|$ for 
    regular modes, i.e., their respective contribution to the 
    spike rate dynamics is faster in the fluctuation-dominated regime than in 
    the drift-dominated one, whereas for diffusive modes the opposite holds true. 
%     (indicating that larger noise 
%     is more consistent with allows 
%     forslowness assumptions as exploited in the spectral1 model w.r.t. to input changes)
%   \item \textcolor{blue}{[discuss the eigenfunctions and the discontinuity in $\phi$ but not $\psi$ maybe too?] cf. Fig.}
\end{enumerate}

The specific definition of the second dominant eigenvalue $\lambda_2$, 
Eq.~\eqref{eq_meth_lambda2}, is necessary to ensure real coefficients 
in the  spec$_2$ model, Eqs.~\eqref{eq_meth_beta2}--\eqref{eq_meth_betac},
for regions in $(\mu,\sigma)$-space where the first dominant eigenvalue $\lambda_1$ 
is of the 
(real) diffusive type and the second is part of a complex conjugate couple (for an example see Fig.~\ref{fig7_spectral}A, right).

Extracting the required dominant eigenvalues $\lambda_1$ (for both spectral 
models: spec$_1$, spec$_2$) and $\lambda_2$ (for  spec$_2$ model only) according to 
Eqs.~\eqref{eq_meth_lambda1},\eqref{eq_meth_lambda2} in the 
$(\mu,\sigma)$-plane 
leads to points of instantaneous changes 
in the imaginary part 
($\lambda_1$ and $\lambda_2$) and the real part 
(only $\lambda_2$) due to transitions from a dominant diffusive 
mode for small mean input $\mu$ to a regular eigenvalue (or pair) becoming dominant 
for larger $\mu$ (see Fig.~\ref{fig7_spectral}A,C and the last property above). 
These discontinuities (which lie on a one-dimensional curve $\mu(\sigma)$) could be avoided by either restricting to $V_\mathrm{lb}=V_\mathrm{r}$ 
(no diffusive modes) or by deriving 
a third spectral model based on three eigenvalues: 
the dominant regular pair together with the 
dominant diffusive eigenvalue. Making the latter extension is straightforward
by using the same steps and slowness approximation as for the model 
spec$_2$ and would yield a 3rd order ODE for the spike rate $r(t)$ 
with smooth 
% $\mu_\mathrm{tot}(t),\sigma_\mathrm{tot}(t)$-
coefficients and is expected to show increased reproduction accuracy (especially for small mean input) compared to the model spec$_2$.

% Additional quantities \todo{additional to what? lambdas are shown in C, too} that are required by the advanced spectral model 
% and depend only on the total input $\mu$ and $\sigma$
% \textcolor{gray}{(i.e., not on adaptation current and coupling parameters} 
% have always real values and are shown in Fig.~\ref{fig7_spectral}C.
% %

The properties above imply for the low-dimensional models 
spec$_1$ and spec$_2$, that both enable damped oscillatory 
spike rates for mean-dominated input (for large $\mu$) since then
the first two dominant eigenvalues ($\lambda_1$ and $\lambda_2$) have nonzero imaginary parts and are complex conjugates of each other 
(see Fig.~\ref{fig7_spectral}C). Furthermore in this case the effective time constant, 
$1/|\mathrm{Re}\{\lambda_{1}\}| = 1/|\mathrm{Re}\{\lambda_{2}\}|$, is large (especially for small input 
standard deviation $\sigma$).
% , i.e., the 
% spike rate $r$ responds to input changes. 
For noise-dominated input, i.e., when $\sigma$ (and not $\mu$) 
is large, on the other hand, the corresponding spike rate dynamics 
is fast and does not contain an oscillatory component.

\paragraph{Numerical solver}

Here we present 
% \textcolor{gray}{for the first time} 
a numerical solution method of the 
% \textcolor{gray}{(input) parameter-dependent}
Fokker-Planck boundary eigenvalue problem 
(BEVP) 
for the operator $\LL(\mu,\sigma)$, Eqs.~\eqref{eq_meth_spec_eigeneq_sparse}--\eqref{eq_meth_spec_phi_reinjcond}, 
and its adjoint $\LL^*(\mu,\sigma)$,  Eqs.~\eqref{eq_meth_spec_eigeneq_adjoint_sparse}--\eqref{eq_meth_spec_adj_bc_cont}.
% \todo{indicate eqs., and then hint to the aspects for which the approach is general (nonlin. IF etc.)}
The solution of the two BEVPs in terms of eigenvalues $\lambda_n$ and associated (biorthonormal) eigenfunctions ($\phi_n(V)$, $\psi_n(V)$) as well as quantities that are 
derived from those and are required by the models spec$_1$ and spec$_2$, 
is obtained 
for a rectangle of the input mean $\mu$ and standard deviation $\sigma$ 
(see Sect.~\textit{Low-dimensional Approximations}). 
Note that the numerical method is not restricted to the EIF neuron model 
and supports other integrate-and-fire models as well 
(e.g., perfect, leaky or quadratic). 

The eigenequation $\LL(\mu,\sigma) [\phi] = \lambda \phi$,
i.e., Eq.~\eqref{eq_meth_spec_eigeneq_sparse} 
(omitting the eigenvalue index $n$), 
represents a second order ODE (cf. Eq.~\eqref{eq_meth_spec_operator}), that is equivalent to the following first order system for the eigenflux $q_\phi(V)$ 
and the eigenfunction $\phi(V)$,
% 
% where we 
% omitting eigenindex
% 
% 
% 
% 
% \todo{@MA: remove redundancy/make consistent with model-common part three SEcts. before: e.g., specify discrete grid instead; note also that $ \mu_{min} $ was used already above} % $(\mu,\sigma) \in $
% $\mathcal R = [\mu_\mathrm{min}, \mu_\mathrm{max}] \times [\sigma_\mathrm{min}, \sigma_\mathrm{max}]$
% by numerically solving the 
% % \textcolor{gray}{(second order)} 
% eigenequation 
% \eqref{eq_meth_spec_eigeneq_sparse} using Eq.~\eqref{eq_meth_spec_operator} \todo{maybe comment on placeholder / ``open-loop controlled'' mu, sigma here for clarity}
% \textcolor{gray}{and omitting the eigenvalue index $n$}, i.e.,
% 
% \begin{equation} \label{eq_meth_spec_eigeneq_dense}
% \LL_{\mu,\sigma}[\phi] = 
%  - \frac{\partial}{\partial V} q_\phi = 
% - \frac{\partial}{\partial V} \left [ 
% \left ( \frac{I_{\mathrm{L}}(V) + I_\mathrm{exp}(V)}{C} + 
% \mu \right ) \phi
% - \frac{\sigma^2}{2} \frac{\partial \phi}{\partial V}
% \right ]
% =
% \lambda \phi,
% \end{equation}

% We reformulate the complex-valued second order Eq.~\eqref{eq_meth_spec_eigeneq_dense} 
% to an equivalent first order system
% \todo{@MA: ensure that fvm matrix if that is 
% defined is named differentially than the matrix below} for 
% $q_\phi(V)$ and $\phi(V)$, %which gives
\begin{align} 
% in components:
% \label{eq_meth_spec_eigeneq_sys1}
% % -q_{\phi}'(V)
% & -\frac{\partial q_{\phi}}{\partial V} = \lambda \phi \\
% & -\frac{\partial \phi}{\partial V} = \frac{q_\phi - (g(V) + \mu)\phi}{\sigma^2/2} 
% \label{eq_meth_spec_eigeneq_sys2}
%
% matrix notation:
\label{eq_meth_spec_eigeneq_sys}
% -q_{\phi}'(V)
& -\frac{d}{dV} 
% \mathbf{
  \begin{pmatrix}
q_{\phi}
\\
\phi
\end{pmatrix} = 
% \begin{pmatrix}
% \lambda \phi
% \\
% \dfrac{q_\phi - (g(V) + \mu)\phi}{\sigma^2/2} 
% \end{pmatrix}
% = 
\underbrace{
\begin{pmatrix}
0 & \lambda
\\
\frac{2}{\sigma^2}  & - 2\frac{g(V) + \mu}{\sigma^2}
\end{pmatrix} }_{ = \mathbf{A}}
\begin{pmatrix}
q_{\phi}
\\
\phi
\end{pmatrix}
\end{align}
with coefficient matrix $\mathbf{A}(V)$ that has a nonlinear 
component through $g(V) = [I_\mathrm L(V) + I_\mathrm{exp}(V)]/C$ 
that contains leak and exponential (membrane) currents. 
Here it was used 
the form of the eigenflux $q_\phi$ (cf. Eq.~\eqref{eq_meth_eigenflux_new}), and 
that 
% \textcolor{red}{TODO: make better with prob. flux}
% $q_\phi = \left[ \left (I_{\mathrm{L}} + I_{\mathrm{exp}}\right)/C 
% + \mu \right] \phi - \sigma^2/2 \;
% % \frac{\partial}{\partial V} 
% \partial_V \phi$ and 
$\LL(\mu,\sigma) [\phi] = -\partial_V q_\phi$ for generic input moments 
$\mu$ and $\sigma^2$ (cf. Eq.~\eqref{eq_meth_spec_operator}).
% Sect.~\textit{Eigendecomposition of the Fokker-Planck operator}).

% ...for each $(\mu,\sigma) \in \mathcal R$ such that the boundary conditions, 
% Eqs.~\eqref{eq_meth_spec_phi_reinjcond}--\eqref{eq_meth_spec_qlb_reflect}, hold, as follows. 

Basically a direct discretization, e.g., by a finite difference approximation,  
of the membrane voltage derivatives in the system~\eqref{eq_meth_spec_eigeneq_sys} 
can be applied. In combination with the boundary conds., Eqs.~\eqref{eq_meth_spec_phi_absorb}--\eqref{eq_meth_spec_phi_reinjcond}, 
this leads to a (sparse) matrix eigenvalue problem that allows for
application of standard (Arnoldi iteration based)
numerical solvers.
However, the convergence properties of this approach are very poor 
in the sense that extremely small voltage steps $\Delta V$ have to be chosen for 
the finite differences. Thus, the technique is inefficient as huge 
systems appear or even inaccurate due to amplified round-off errors by 
ill-conditioned matrices. 
% As an alternative

Here we propose an alternative  
solution procedure which is based on a reformulation 
of the system~\eqref{eq_meth_spec_eigeneq_sys}
with corresponding boundary 
conditions, Eqs.~\eqref{eq_meth_spec_phi_absorb}--\eqref{eq_meth_spec_phi_reinjcond}, 
as a 
% nonlinear 
complex-valued algebraic root finding problem 
\begin{equation} \label{eq_meth_spec_root}
\lambda \mapsto q_\phi(V_\mathrm{lb}; \lambda) \overset{!}{=} 0
\end{equation} 
whose solutions are the eigenvalues $\lambda_n$.
To evaluate the nonlinear function (the left hand side of this equation) for 
an arbitrary $\lambda \in \CC$ (not necessary an eigenvalue) 
Eq.~\eqref{eq_meth_spec_eigeneq_sys}
is integrated backward starting from 
the spike voltage $V_\mathrm{s}$ (initializing one component
to satisfy the absorbing boundary cond., 
Eq.~\eqref{eq_meth_spec_phi_absorb}, i.e., 
$\phi(V_\mathrm{s})=0$, and another component 
that can be chosen arbitrarily, 
$q_\phi(V_\mathrm s) \in \CC \setminus \{0\}$, 
due to the linearity of the problem)  via the 
reset $V_\mathrm{r}$ (where the reinjection cond., Eq.~\eqref{eq_meth_spec_phi_reinjcond}, is 
enforced, i.e.,
$q_{\phi}(V_\mathrm{r}^-) 
= q_{\phi}(V_\mathrm{r}^+) - q_{\phi}(V_{\mathrm{s}})$, 
which induces a discontinuity in $q_\phi$ at the reset voltage) 
finally to the lower bound voltage $V_\mathrm{lb}$. 
There, a nonzero value of $q_\phi(V_\mathrm{lb}) \in \CC$
indicates that $\lambda$ is not an eigenvalue since in this case $\phi(V)$ 
violates the reflecting boundary condition, Eq.~\eqref{eq_meth_spec_qlb_reflect}. 
$q_\phi(V_\mathrm{lb}) = 0$ on the other hand shows
that $\lambda$ is an eigenvalue with 
corresponding eigenfunction $\phi(V)$ and eigenflux $q_\phi(V)$. 
Note that when considering a nonzero refractory period the 
generalized version of the 
reinjection cond., Eq.~\eqref{eq_meth_spec_phi_reinjcond_ref}, i.e., 
$q_{\phi}(V_\mathrm{r}^-) 
= q_{\phi}(V_\mathrm{r}^+) - q_{\phi}(V_{\mathrm{s}})
e^{-\lambda T_\mathrm{ref}}
$, is enforced instead of the expression above. The latter 
is only valid for the  spec$_1$ model, Eq.~\eqref{eq_meth_spec1} (see 
Sect.~\textit{Basic model: one eigenvalue, negligible input variations}) 
and makes the Fokker-Planck eigenvalue problem nonlinear 
due to the exponentiation of $\lambda$ in Eq.~\eqref{eq_meth_spec_phi_reinjcond_ref}.

% ($T_\mathrm{ref}=0$ both 
% be chosen which recovers also the boundary 
% cond.~\eqref{eq_meth_spec_phi_reinjcond}.

The (complex-valued) root finding problem, Eq.~\eqref{eq_meth_spec_root} can be solved 
numerically to yield a target eigenvalue $\lambda_n$ 
using an iterative procedure (for example a
variant of Newton's 
method) given that a sufficiently close initial approximation 
$\tilde \lambda_n \in \CC$ is available. 
In our Python implementation we apply Powell's hybrid 
method as implemented in MINPACK wrapped through 
SciPy \cite{Oliphant2007} 
% with finite difference step size $\Delta \lambda$ and 
% relative convergence tolerance $\varepsilon$ 
to the equivalent
real system of the two variables 
$\mathrm{Re}\{\lambda\}$ and $\mathrm{Im}\{\lambda\}$.
% for the real and imaginary 
% part of $q_\phi(V_\mathrm{lb}; \lambda)$ as a function of the 
% real and imaginary part of $\lambda$ respectively}. 

Appropriate initial approximations $\tilde \lambda_n$ 
can be achieved (i) by exploiting that for sufficiently small 
generic mean input $\mu_\mathrm{min}$ all eigenvalues 
% \textcolor{gray}{($\lambda_1, \lambda_2, \dots $)} 
have zero imaginary part 
(see Sect.~\textit{\secspecprops}). In that case the eigenvalues 
are given by the roots $\{\lambda_0, \lambda_1, \dots \}$ of
$q(V_\mathrm{lb}; \lambda)$ -- the one-dimensional function 
of the 
real-valued eigenvalue candidate $\lambda \in (-\infty,0]$ -- 
which are obtained, for example, by (dense) evaluation of that function in a sufficiently large (sub)interval below zero. 
Furthermore (ii) all eigenvalues 
depend continuously on the (input) parameters $\mu$ and $\sigma$ 
(cf. Fig.~\ref{fig7_spectral}A), i.e., for a 
small step in the respective parameter space the solution $\lambda_n$ 
of Eq.~\eqref{eq_meth_spec_root} 
for the last parametrization is a very good 
initial value $\tilde \lambda_n$ for the current parametrization. 
With this initialization the solution of Eq.~\eqref{eq_meth_spec_root} is typically found in a few steps of the chosen Newton-like method (see \supplement{} for details). 

To efficiently and accurately evaluate $q_\phi(V_\mathrm{lb}; \lambda)$ 
in each iteration (of the root finding algorithm), we perform an exponential (backward) integration 
of the system~\eqref{eq_meth_spec_eigeneq_sys}. The resulting scheme 
is based on truncating the Magnus expansion of the exact solution 
after one term \cite{Hochbruck2010}. Here
matrix exponential function evaluations of the form 
$\exp[\mathbf A(V)\Delta V]$ occur that are 
calculated using an analytic expression 
\cite{Bernstein1993}. 
Note that ODE solvers that either 
do not exploit the linear structure of 
Eq.~\eqref{eq_meth_spec_eigeneq_sys} at all 
(such as the explicit Euler method but also 
higher order Runge-Kutta methods) 
or utilize linearity only in one variable (e.g., \cite{Richardson2007})
have poor convergence behaviour and thus
require very small step sizes $\Delta V$ 
due to the strong nonlinearity $g(V)$ 
% \textcolor{gray}{in the 
% coefficient matrix.}
% $\mathbf{A}$ 
as a consequence of the large value of the 
current $I_\mathrm{exp}$ close to $V_\mathrm{s}$ and 
significantly large absolute 
eigenvalues $|\lambda_n|$, respectively. 
% Note that a semi-exponential integration (cf. \cite{Richardson2007}) where 
% the $q_\phi$ dynamics is integrated using the explicit Euler method and (only) $\phi$ is calculated using a (one-dimensional) exponential 
% scheme does require extremely small membrane voltage steps 
% $\Delta V$ to accurately determining the eigenfunctions 
% and is thus not efficient 
% enough for the large input rectangle $\mathcal R$.
% 
% 
% A major factor for efficiency and accuracy of the solver 
% is the particular numerical way in which the backward integration of the 
% system~\eqref{eq_meth_spec_eigeneq_sys}
% and corresponding boundary 
% conditions, Eqs.~\eqref{eq_meth_spec_phi_absorb}--\eqref{eq_meth_spec_phi_reinjcond},
% is performed. General purpose (nonlinear) numerical ODE solver such as Euler's method
% or (step-size controlled) Runge-Kutta methods require very accurate
% 
% since this 
% corresponds to one function evaluation of $\lambda \mapsto q(V_\mathrm{lb}; \lambda)$
% and determines therefore the cost as well as a bound on the 
% numerical accuracy of a single Newton-like step. 
% 
% The major computational effort is made by (accurately) 
% evaluating $q_\phi(V_\mathrm{lb}; \lambda^{(\gamma)})$, 
% i.e., backward integrating of Eq.~\eqref{eq_meth_spec_eigeneq_sys}
% with corresponding boundary 
% conditions, Eqs.~\eqref{eq_meth_spec_phi_absorb}--\eqref{eq_meth_spec_phi_reinjcond},
% in iteration $\gamma$. 

% exponential backward integration we refer to the 
% \supplement. 

For more information on the numerical solver 
we refer to \supplement, where 
details regarding the adjoint operator, the exponential integration, the initial eigenvalue determination  and the (parallelized) 
treatment of the input parameters are included. 
% Particularly how (i) the initial values $\tilde \lambda_n$ are determined, (ii) the exponential integration is 
% performed and (iii) the derived quantities are practically computed.

% 
% \todo{@MA: double check that all parameters here and in appendix are specified in the implementation section}
% flatex input end: [section/methods_spectral.tex]
 
% flatex input: [section/methods_cascade.tex]
\subsection*{Cascade models} %Linear-nonlinear cascade models
% \tj{check: overall syn. current, $ \approx $ vs $ = $}
%\vspace{0.3cm}
%{\large \textbf{Adaptive linear-nonlinear cascade}} 
%\phantomsection  \addcontentsline{toc}{subsubsection}{Adaptive linear-nonlinear cascade} 
%\vspace{-0.1cm}

Linear-Nonlinear (LN) cascade models of neuronal activity are often applied in neuroscience, %\cite{Chichilnisky2001,Pillow2008}, 
because they are simple and efficient, and the model components can be estimated using established experimental procedures
\cite{Chichilnisky2001,Pillow2008,Ostojic2011PLOS}.
%For example, the linear filter components can be obtained (experimentally with a real neuron or computationally using a model neuron) by applying a reverse correlation method \cite[see, e.g.,][chapter 2.2]{Dayan2001}.
Here we use the LN cascade as an ansatz to develop a low-dimensional model and we determine its components from the underlying Fokker-Planck model. %\todo{improve}
%
%It should be mentioned that the LN approach enables to compare (and calibrate) the components of the (population) spike rate model to experimentally obtained estimates using established experimental protocols \cite{Chichilnisky2001,Ostojic2011PLOS} \textcolor{gray}{in addition to the relation to the underlying spiking aEIF neuron presented here}. %\todo{Gray part elsewhere, e.g. Intro/Discussion}
%%whose parameters can be fitted using electrophysiological recordings and standard (step and ramp) stimulation protocols \cite{Brette2005}
%For example, estimates for the filters $D_\mu$ and $D_\sigma$ can be obtained using a reverse correlation method \cite[see, e.g.,][chapter 2.2]{Dayan2001}, by applying an input current $I(t) = \mu(t) + \sigma(t) \xi(t)$ to the neuron (prototype of a population) with white-noise temporal statistics for $ \xi $ and small correlation times for the independent processes $ \mu $ and $ \sigma $, and computing the spike triggered averages of $ \mu $ and $ \sigma $. 
%\textcolor{gray}{The LN models presented here further relate to existing cascade models frequently used to describe neuronal activity in the literature [REFS].}
%
This section builds upon \cite{Ostojic2011PLOS} and extends that approach for recurrently coupled aEIF neurons; specifically, by taking into account an adaptation current and variations of the input variance. Furthermore, we consider %\textcolor{gray}{/propose} 
an improved approximation of the derived linear filters and include an (optional) explicit description of the spike shape, cf. \cite{Ladenbauer2015} (ch 4.2). 

%Reduction: LN cascade Ansatz (opening new window to experimenters, since L,N components can be estimated via established procedures), comment on alternative spectral approach with pros and cons (aLN: minimal model, extra relation to wealth of existing models / experimental data, less ``direct'') \\
%EXP vs. DOSC fitting to obtain ODE representation (for convenient solution and analysis)
%
%PDE nice, but not optimal considering multi-population networks / stability analyses...
%We next reduce the \textit{intermediate} Fokker-Planck PDE-ODE model system %from the previous section 
%to a low-dimensional ODE system that can be numerically solved much faster (using standard ODE solvers) and conveniently analyzed. %using standard methods and powerful ...
%The reduction approach we present is based on a Linear-Nonlinear (LN) cascade, in which the population ...
The cascade models considered here produce spike rate output by applying to the time-varying mean $\mu_{\mathrm{syn}}$ and standard deviation $\sigma_{\mathrm{syn}}$ of the (overall) synaptic input, cf. Eq.~\eqref{eq_meth_musyn_sigmasyn}, separately %successively 
a linear temporal filter, $D_\mu$ and $D_\sigma$, %respectively, 
followed by a common nonlinear function $F$. That is,
%\textcolor{gray}{, for each population\tj{suggest to remove}},\tj{Recurrent coupling could already be included here, but this way it is easier. (MA: I agree, I like this Sect.)} 
%
\begin{align} \label{LN_cascade_eq}
r(t) &= F \left(\mu_{\mathrm{f}}, \sigma_{\mathrm{f}}, \langle w \rangle \vphantom{^2}\right), \\
\mu_{\mathrm{f}}(t) &= D_\mu \ast \mu_{\mathrm{syn}}(t), \label{LN_cascade_eq2} \\
\sigma_{\mathrm{f}}(t) &= D_\sigma \ast \sigma_{\mathrm{syn}}(t), \label{LN_cascade_eq3} %\\
%\textcolor{gray}{\frac{d{\langle w \rangle}}{dt}} &\textcolor{gray}{= H\left(\mu_{\mathrm{f}}, \sigma_{\mathrm{f}}, \langle w \rangle \vphantom{^2}\right),} \label{LN_cascade_eq4}
%\langle V \rangle (t) &= H \left(\mu_{\mathrm{f}}, \sigma_{\mathrm{f}}, \langle w \rangle \vphantom{^2}\right), \label{LN_cascade_eq1} 
\end{align}
%
%\tj{the Ansatz for $H$ is not clear to me, the original mean adapt. dynamics 
%does depend on $\langle V \rangle$ and $r$... I suggest to remove $H$ here and 
%use the approximation of the adaptation current of the common Sect. Low-Dimension Approx. as 
%foundation}
%\tj{$\mu_f$ and $\sigma_f$ should maybe be given names here?}
where $\mu_{\mathrm{f}}$ and $\sigma_{\mathrm{f}}$ denote the filtered mean and filtered standard deviation of the input, respectively. 
$ D_\mu \ast \mu_{\mathrm{syn}}(t) = \int_0^\infty D_\mu(\tau) \mu_{\mathrm{syn}}(t-\tau) d\tau $ is the convolution between $ D_\mu $ and $ \mu_{\mathrm{syn}} $.
The filters $ D_\mu $, $ D_\sigma $ are adaptive 
%\tj{is adaptive the 
%right word here/I thought this is only adequate when the time-dependent input not the constant baseline is considered}{} 
in the sense that they depend on the mean adaptation current $ \langle w \rangle $ and on the (arbitrary) baseline input %\tj{maybe add: \textit{arbitrary} baseline}{} 
in terms of baseline mean $ \mu_{\mathrm{syn}}^0 $ and standard deviation $ \sigma_{\mathrm{syn}}^0 $. %which may change (slowly) over time. 
For improved readability these dependencies are not explicitly indicated in Eqs.~\eqref{LN_cascade_eq2}, \eqref{LN_cascade_eq3}.
%\tj{Alternative:  $\mu_{\mathrm{f}}(t) = [D_\mu (\mu_{\mathrm{syn}}^0,\sigma_{\mathrm{syn}}^0,\langle w \rangle) \ast \mu_{\mathrm{syn}}](t)$,  $ \sigma_{\mathrm{f}}(t) = [D_\sigma (\mu_{\mathrm{syn}}^0,\sigma_{\mathrm{syn}}^0,\langle w \rangle) \ast \sigma_{\mathrm{syn}}](t)$. MA: suggest to stick with the version that is currentl used in the text -- not the expanded version} 
Note, that the nonlinearity $ F $ also depends on $ \langle w \rangle $, which is governed by Eq.~\eqref{eq_meth_w_mean_p}.
%Recall, that for the (mean-field) network the input moments $ \mu_{\mathrm{syn}}(t, r_d), \sigma_{\mathrm{syn}}(t, r_d) $ depend on the delayed spike rate (cf ...). 
Since the mean adaptation current depends on the mean membrane voltage $ \langle V \rangle $ we also consider a nonlinear mapping $ H $ for that population output quantity, %\todo{...to close the ansatz}{} %cf. Eq.~\eqref{LN_cascade_eq1}. 
\begin{equation} \label{LN_cascade_eq1}
\langle V \rangle (t) = H \left(\mu_{\mathrm{f}}, \sigma_{\mathrm{f}}, \langle w \rangle \vphantom{^2}\right).
\end{equation}
For the derivation below it is instructive to first consider an uncoupled population, i.e., the input moments do not depend on $ r_d $ for now. %(cf. Eq.~\eqref{eq_meth_musyn_sigmasyn}\tj{omit eq. reference here if it is already 
%cf.'ed above}). %\todo{Eqs. for $\mu_{\mathrm{syn}}$, $\sigma_{\mathrm{syn}}$ from common part above here}). 
In particular, the input statistics are described by $\mu_{\mathrm{syn}}(t) = \mu_{\mathrm{syn}}^0 + \mu_{\mathrm{syn}}^1(t)$ and $\sigma_{\mathrm{syn}}(t) = \sigma_{\mathrm{syn}}^0 + \sigma_{\mathrm{syn}}^1(t)$.
In the following, we derive the components $ F $, $ D_\mu $ and $ D_\sigma $ 
from the Fokker-Planck model 
%in the linear limit of vanishing amplitude 
for small amplitude variations $\mu_{\mathrm{syn}}^1$, $\sigma_{\mathrm{syn}}^1$ %around their baseline values $\mu_0$, $\sigma_0$, 
and for a slowly varying adaptation current (as already assumed). We then approximate the derived linear filter components using suitable functions such that the convolutions can be expressed in terms of simple ODEs. Finally, we account for time-varying baseline input ($\mu_{\mathrm{syn}}^0(t)$, $\sigma_{\mathrm{syn}}^0(t)$) and for recurrent coupling in the resulting low-dimensional spike rate models.

%which depends on 
%the overall input moments
%\begin{equation} \label{eq_meth_spec_input}
%\mu \coloneqq \mu_\mathrm{syn}(t,r_d) - \langle w \rangle/C, 
%\qquad\quad \sigma^2 \coloneqq \sigma_\mathrm{syn}^2(t,r_d)
%\end{equation} 
%MENTION SOMEWHERE THAT THE FILTERS SHOULD ADAPT TO CHANGING ``BASELINE'' INPUT AND DEPEND ALSO ON THE MEAN ADAPT. CURRENT. 

%Here, we determine/extract the components of the LN model from the underlying Fokker-Planck model. [ELSEWHERE: similarly as in \cite{Ostojic2011PLOS}.]
%...use analytical results %and efficiently obtainable numerical results
%for the Fokker-Planck model described above to determine the components of the LN model, similarly as in \cite{Ostojic2011PLOS}. %and we compare these results with those from the reverse correlation method -> FOR FILTERS ONLY (W/O ADAPTATION).
%The nonlinear function $F$ is derived in the adiabatic limit of very slow variations of $\mu_{\mathrm{syn}}$, 
%$\sigma_{\mathrm{syn}}$, and accounts for neuronal adaptation.  
%
%We extend the results obtained for (uncoupled) EIF neurons in \cite{Ostojic2011PLOS} to networks of aEIF neurons, and derive two low-dimensional model variants %of a low-dimensional spike rate model, 
%by approximating the linear filter components in suitable ways such that the convolutions can be expressed in terms of simple ODEs. %also somewhat different compared to \cite{Ostojic2011PLOSCB}
%%and we compare the different methods/spike rate models in terms of reproduction accuracy.  

\paragraph{Deriving the components of the cascade} 

%To determine the components of the cascade 
We first expand $ F $ in %$F(x, y, z)$ in
Eq.~\eqref{LN_cascade_eq} around the baseline %\tj{the term steady-state seems 
%in the uncoupled and not time-dependent input a bit odd} 
$\mu_{\mathrm{f}}=\mu_{\mathrm{syn}}^0$,  $\sigma_{\mathrm{f}}=\sigma_{\mathrm{syn}}^0$, $\langle w \rangle=\langle w \rangle_0$ to linear order, 
assuming that the amplitudes of $\mu_{\mathrm{syn}}^1 $ and 
$\sigma_{\mathrm{syn}}^1 $ are small, and the mean adaptation current varies %(very) 
slowly compared to the input moments, %$\mu_{\mathrm{syn}}^1(t), \sigma_{\mathrm{syn}}^1(t) $, 
to obtain
the approximation for Eq.~\eqref{LN_cascade_eq},
%\footnote{\textcolor{gray}{Due to slow adaptation ($\langle w \rangle(t) = \langle w \rangle_0 + \langle w \rangle_1(t)$ with vanishing $ \langle w \rangle_1 $) we can neglect the expansion term in the direction of $\langle w \rangle$ and replace $\langle w \rangle_0=\langle w \rangle$ in the remaining terms of the approximation.}}
\begin{align} \label{LN_derivation_eq1}
r(t) \approx F(\mu_{\mathrm{syn}}^0, \sigma_{\mathrm{syn}}^0, \langle w \rangle) &+ D_\mu \ast \mu_{\mathrm{syn}}^1(t) \,\frac{\partial}{\partial \mu} F(\mu_{\mathrm{syn}}^0, \sigma_{\mathrm{syn}}^0, \langle w \rangle) \\
&+ D_\sigma \ast \sigma_{\mathrm{syn}}^1(t) \,\frac{\partial}{\partial \sigma} F(\mu_{\mathrm{syn}}^0, \sigma_{\mathrm{syn}}^0, \langle w \rangle). \nonumber
%\\
% &= r_\infty (\mu_0 - \langle w \rangle/C, \sigma_0) + D_\mu \ast \mu_{\mathrm{syn}}(t) \frac{\partial}{\partial \mu} 
%r_\infty (\mu_0 - \langle w \rangle/C, \sigma_0) + D_\sigma \ast \sigma_{\mathrm{syn}}(t) \frac{\partial}{\partial \sigma} 
%r_\infty (\mu_0 - \langle w \rangle/C, \sigma_0), \label{L_derivation_eq1}
\end{align}
Due to slow adaptation ($\langle w \rangle(t) = \langle w \rangle_0 + \langle w \rangle_1(t)$ with vanishing $ \langle w \rangle_1 $) we have neglected the expansion term in the direction of $\langle w \rangle$ and replaced $\langle w \rangle_0=\langle w \rangle$ in the approximation above. Note also that, without loss of generality, we have assumed normalized filters, $ \int_{0}^{\infty} D_\mu (\tau) d\tau  = \int_{0}^{\infty} D_\sigma (\tau) d\tau = 1 $.
%DEPENDENCE ON ADAPT. NOT INDICATED EXPLICITLY ABOVE
%(where $F(\mu_0, \sigma_0)$ %= r_0$ 
%gives the baseline spike rate %generated by %a synaptic current with constant mean and variance determined by 
%%the synaptic moments $\mu_0$ and $\sigma_0$.
%which can change slowly over time due to adaptation).
%%Note that $r_0$ may vary slowly over time due to adaptation, accounted for by the function $F$. %accounts for the slowly varying adaptation current (cf. assumption A1) as shown below. 
%In the same linear limit and for a slowly varying adaptation current %(cf. assumption A1) 
Under the same assumptions the output from the Fokker-Planck model (Eqs.~\eqref{eq_meth_w_mean_p}--\eqref{eq_meth_fp_reflecting}) can be approximated as %\footnote{\textcolor{blue}{Maybe replace $\mu_{\mathrm{tot}}^0 = \mu_0 - \langle w \rangle/C$, $\sigma_{\mathrm{tot}}^0 = \sigma_0$ everywhere below.}}
%[Ostojic 2011 p.2-3]
\begin{align} \label{LN_derivation_eq2}
r(t) &\approx r_\infty \left( \mu_{\mathrm{tot}}^0, \sigma_{\mathrm{tot}}^0 \right) + R_\mu \ast \mu_{\mathrm{syn}}^1(t) + R_\sigma \ast \sigma_{\mathrm{syn}}^1(t), \\
\frac{d{\langle w \rangle}}{dt} &\approx \frac{a 
\left(\langle V \rangle_\infty -E_w \vphantom{^2} \right) - 
  \langle w \rangle}{\tau_{w}} + b\, r(t), \label{aLN_mean_w_eq} \\
\mu_{\mathrm{tot}}^0(t) &= \mu_{\mathrm{syn}}^0 - \langle w \rangle/C, \qquad  \sigma_{\mathrm{tot}}^0 = \sigma_{\mathrm{syn}}^0,  \label{bl_musig_tot_eq} 
\end{align}
%where $\langle w \rangle$ is given by Eq.~\eqref{aLN_mean_w_eq}. 
%\textcolor{gray}{(Eq.~\eqref{aLN_mean_w_eq} same as Eqs.~\eqref{eq_results_w_mean}, \eqref{eq_meth_w_mean_p_inf})}
where $r_\infty$ and $\langle V \rangle_\infty$ are the steady-state spike rate and mean membrane voltage of a population of EIF neurons in response to an input of \textit{total} mean $\mu_{\mathrm{tot}}^0 $ %= \mu_0 - \langle w \rangle/C$ 
plus Gaussian white noise with standard deviation $ \sigma_{\mathrm{tot}}^0 $. %$\sigma_{\mathrm{tot}}^0 = \sigma_0$. 
In particular, $\langle V \rangle_\infty$ reflects the mean over all nonrefractory neurons, 
%$\langle V \rangle_\infty = \int_{-\infty}^{V_{\mathrm{s}}} v p_\infty(v) dv / \int_{-\infty}^{V_{\mathrm{s}}} p_\infty(v) dv$.
%\begin{equation} \label{LN_V_mean_inf}
%\textcolor{gray}{\langle V \rangle_\infty = \frac{\int_{-\infty}^{V_{\mathrm{s}}} v p_\infty(v) dv}{\int_{-\infty}^{V_{\mathrm{s}}} p_\infty(v) dv}}
%\end{equation} 
cf. Eq.~\eqref{eq_meth_V_inf}. %(in Sect.~\textit{Low-Dimensional Approximations}).
$R_\mu$ and $R_\sigma$ are the so-called linear rate response functions of %an EIF 
the population %(a population of) EIF neurons to 
for weak modulations of the input mean and standard deviation around 
%$\mu = \mu_0 - \langle w \rangle/C$ and $\sigma = \sigma_0$, 
$ \mu_{\mathrm{tot}}^0 $ and $ \sigma_{\mathrm{tot}}^0 $, respectively 
\cite{Fourcaud-Trocme2003,Richardson2007,Ostojic2011PLOS}. %Brunel2001? 
%
%with effective mean $\mu_{\mathrm{syn}}(t) - \langle w \rangle/C$ and standard deviation $\sigma_{\mathrm{syn}}(t)$\footnote{
%Note that the mean adaptation current subtracts from the input mean but does not affect the input standard deviation, justified by the slow adaptation time scale (cf. adiabatic approximation above).
%
%ACTUALLY R SHOULD BE FOR AEIF BUT WE USE THE TRICK HERE THAT WE CONSIDER EIF WITH MEAN INPUT MUSYN-MEANW (CF. ADIAB. APPROX) 
%-- THEREFORE, WOULDN'T IT BE CLEANER TO CONSIDER AN ANSATZ WITH MUEFF IN EQ 4.67?
%NOTE THAT THE SLOWLY VARYING AV. ADAPTATION CURRENT (WITH TIMESCALE >> FILTER TIMESCALE) IS SUBTRACTED FROM THE MEAN INPUT AS WE USE R CALCULATED FROM EIF MODEL
%-------------
%Note that because the adaptation time scale is large %(much larger that those of the linear rate response functions) 
%these four functions can be calculated for EIF neurons subject to Gaussian white noise input, 
%where the mean adaptation current appears as a slowly varying parameter which subtracts from the synaptic mean input but does not affect the input standard deviation.
%-------------
%This method can be used here since the adaptation component of the aEIF neuron model can be handled by the nonlinearity $F$ in the LN model as shown above.
%From Eqs.~\eqref{L_derivation_eq1} and \eqref{L_derivation_eq2} we obtain
Comparing Eqs.~\eqref{LN_derivation_eq1} and \eqref{LN_derivation_eq2} we obtain for the nonlinearity $F$ and the linear filters $D_\mu$, $D_\sigma$,
%\begin{equation} \label{L_components_eq}
%D_\mu(t) = \frac{R_\mu(t)}{\frac{\partial}{\partial \mu} F(\mu_0, \sigma_0)}, 
%\qquad 
%D_\sigma(t) = \frac{R_\sigma(t)}{\frac{\partial}{\partial \sigma} F(\mu_0, \sigma_0)}.
%\end{equation}
\begin{align} \label{LN_components_eq}
F(\mu,\sigma,\langle w \rangle) &= r_\infty \left( \mu - \langle w \rangle/C, \sigma \right), \\
D_\mu(t) &= \frac{R_\mu(t)}{\frac{\partial}{\partial \mu} r_\infty \left( \mu_{\mathrm{tot}}^0, \sigma_{\mathrm{tot}}^0 \right)}, 
\label{LN_components_eq2} \\
D_\sigma(t) &= \frac{R_\sigma(t)}{\frac{\partial}{\partial \sigma} r_\infty \left( \mu_{\mathrm{tot}}^0, \sigma_{\mathrm{tot}}^0 \right)}. \label{LN_components_eq3}
\end{align}
Furthermore, the function $ H $ in Eq.~\eqref{LN_cascade_eq1} is given by 
\begin{equation} \label{LN_components_eq4}
H(\mu,\sigma,\langle w \rangle) = \langle V \rangle_\infty \left( \mu - \langle w \rangle/C, \sigma \right),
\end{equation}
ensuring that the mean membrane voltage corresponds with the instantaneous spike rate estimate of the model (in every time step).
%where $\langle w \rangle$ evolves according to Eq.~\eqref{aLN_mean_w_eq}, \textcolor{gray}{i.e., the function $ H $ in Eq.~\eqref{LN_cascade_eq4} is given by the right hand side of Eq.~\eqref{aLN_mean_w_eq}. Note, that $ H $ depends on $ \mu_{\mathrm{f}} $, $ \sigma_{\mathrm{f}} $ because the spike rate $ r $ is a function of these variables (cf. Eqs.~\eqref{LN_cascade_eq}, \eqref{LN_components_eq})}. 
%WHY DO WE ACTUALLY NEED THE FIRST LIMIT ABOVE, $F$ CAN BE DETERMINED HERE AS WELL! MAYBE BETTER START WITH THIS LIMIT AND THEN EXPLAIN THAT IT ALSO WORKS EXACTLY IN THE OTHER LIMIT.
Note that $R_\mu$ and $R_\sigma$ depend on $ \mu_{\mathrm{tot}}^0 $ and $ \sigma_{\mathrm{tot}}^0 $ (which is again not explicitly indicated for improved readability). %(and thus on $\langle w \rangle$).

Fortunately, the quantities 
$r_\infty$, $\langle V \rangle_\infty$, $R_\mu$, and $R_\sigma$ %are all evaluated as a function of input mean $\mu_0 - \langle w \rangle/C, \sigma_0$ %with $\langle w \rangle$ given by Eq.~\eqref{aLN_mean_w_eq}, 
can be calculated from the Fokker-Planck equation 
using an efficient numerical method %that has been developed based on the Fokker-Planck equation 
\cite{Richardson2007}. In particular, for $r_\infty$ and $\langle V \rangle_\infty$ we need to solve a linear boundary value problem (BVP), and $R_\mu(t)$, $R_\sigma(t)$ are calculated in the Fourier domain, where we need to solve two linear BVPs to obtain $\hat{R}_\mu(f)$, $\hat{R}_\sigma(f)$ for each frequency $ f $. It is worth noting that the refractory period is included in a straightforward way and does not increase the complexity of the BVPs to be solved, see \cite{Richardson2007,Ladenbauer2015} (and our provided code). %\todo{refer to github here, again?}

\paragraph{Approximating the filter components}

To express the LN model with adaptation, Eqs.~\eqref{LN_cascade_eq}--\eqref{LN_cascade_eq1}, \eqref{eq_meth_w_mean_p} and \eqref{LN_components_eq}--\eqref{LN_components_eq4}, in terms of a low-dimensional ODE system %which is easier to solve and analyze, 
we next  
approximate the linear filters 
$D_\mu$ and $D_\sigma$ using suitable functions. %and subsequently take into account that the filter components adjust to changing baseline input parameters $\mu_0$ and $\sigma_0$.
The shapes of the true filters (proportional to $R_\mu$ and $R_\sigma$, cf. Eqs.~\eqref{LN_components_eq2}, \eqref{LN_components_eq3}) for different input parameter values ($ \mu = \mu_{\mathrm{tot}}^0 $, $ \sigma = \sigma_{\mathrm{tot}}^0 $) are shown in Fig.~\ref{fig8_cascade}A,B.
%\todo[color=green]{double check units in fig caption for $R_\mu$ and $R_\sigma$, e.g., by comparison with Eq.~\eqref{LN_derivation_eq2} -- and also for the fourier transformed versions $\hat R$}
%\todo[color=green]{the color gray in the fig 7 A seems to be too dark to distinguish it from the damped oscillatory filter sometimes (and different grays seem to be used -- make consistent also with fig part B)}

We first consider the linear filter $D_\mu$ (Eq.~\eqref{LN_components_eq2}) and apply the approximation 
%$D_\mu(t;\mu_{\mathrm{tot}}^0,\sigma_{\mathrm{tot}}^0) \approx A_\mu(\mu_{\mathrm{tot}}^0,\sigma_{\mathrm{tot}}^0) \exp \left(-t/\tau_{\mu}(\mu_{\mathrm{tot}}^0,\sigma_{\mathrm{tot}}^0)\right)$, 
\begin{equation} \label{mu_filter_expappx_eq}
D_\mu(t) \approx A_\mu \exp \left(-t/\tau_{\mu}\right),
\end{equation}
motivated by the exponential decay exhibited by $R_\mu$, particularly for large input variance compared to its mean. Note that $ D_\mu $ depends on $ \mu_{\mathrm{tot}}^0 $, $ \sigma_{\mathrm{tot}}^0 $ and therefore the scaling parameter $A_\mu$ and the time constant $\tau_{\mu}$ both depend on $ \mu_{\mathrm{tot}}^0 $, $ \sigma_{\mathrm{tot}}^0 $, which is not explicitly indicated.
$A_\mu$ and $\tau_{\mu}$ may be determined analytically using asymptotic results for the Fourier transform $\hat{R}_\mu$ of the linear rate response %of EIF neurons 
for vanishing and very large frequencies, respectively \cite{Ostojic2011PLOS, Fourcaud-Trocme2003},
\begin{equation}
\lim_{f \to 0} \hat{R}_\mu(f) = \frac{\partial}{\partial \mu} 
r_\infty \left( \mu_{\mathrm{tot}}^0, \sigma_{\mathrm{tot}}^0 \right), \qquad
\lim_{f \to \infty} \hat{R}_\mu(f) = \frac{r_\infty \left( \mu_{\mathrm{tot}}^0, \sigma_{\mathrm{tot}}^0 \right)}{i 2 \pi f \Delta_{\mathrm{T}}}.
\end{equation}
%Specifically, for vanishing frequency we use
%$\lim_{f \to 0} \hat{R}_\mu(f) = \frac{\partial}{\partial \mu} 
%r_\infty (\mu_0 - \langle w \rangle/C, \sigma_0)$, and for large frequency we have $\lim_{f \to \infty} \hat{R}_\mu(f) = r_0 / (i 2 \pi f \Delta_{\mathrm{T}} \tau_{\!\mathrm{m}})$.  
To guarantee that these asymptotics are matched by the Fourier transform 
$A_\mu \tau_{\mu} / \left(1 + i 2 \pi f \tau_{\mu}\right)$ of the exponential, taking into account the scaling factor in Eq.~\eqref{LN_components_eq2}, we obtain
\begin{equation}
A_\mu = \frac{1}{\tau_{\mu}},  \qquad
\tau_{\mu} = \frac{\Delta_{\mathrm{T}}}{r_\infty \left( \mu_{\mathrm{tot}}^0 , \sigma_{\mathrm{tot}}^0 \right)}
\frac{\partial}{\partial \mu} 
r_\infty \left( \mu_{\mathrm{tot}}^0 , \sigma_{\mathrm{tot}}^0 \right).
\end{equation}
Note that matching the zero frequency limit in the Fourier domain is equivalent to the natural requirement that the time integral $\int_0^\infty D_\mu (\tau) d\tau$ of the linear filter is reproduced exactly, that is, the approximation is normalized appropriately.
An advantage of this approximation is that it is no longer required to calculate the linear rate response function $R_\mu$ explicitly. On the other hand, as only the limit  
$f \to \infty$ is used for fitting in addition to the normalization constraint, the approximation of the linear filter can be poor for a range of intermediate frequencies (in the Fourier domain), %which leads to a poor approximation in the time domain, 
in particular for small input mean and standard deviation (see Fig.~\ref{fig8_cascade}A here, and Fig.~4B,C in \cite{Ostojic2011PLOS}). 
%$\tau_{\mu}$ values which are much too large 
To improve the approximation for intermediate frequencies we use the same normalization condition, which fixes the parameter $A_\mu = 1/\tau_{\mu}$, and we determine $\tau_{\mu}$ by a least-squares fit of $\hat{D}_\mu$ over the range of frequencies $f \in [0, 1]$~kHz.
In both cases, using the approximation \eqref{mu_filter_expappx_eq} %we obtain a minimal LN cascade based spike rate model, for which 
the filtered mean input $\mu_{\mathrm{f}}(t) = D_\mu \ast \mu_{\mathrm{syn}}(t)$ can be equivalently obtained by solving the simple scalar ODE, 
\begin{equation} \label{aLNexp_filter_ode}
\frac{d \mu_{\mathrm{f}}}{dt} = \frac{\mu_{\mathrm{syn}} - \mu_{\mathrm{f}}}{\tau_{\mu}}.
\end{equation}
Recall that $ \tau_{\mu} $ depends on $ \mu_{\mathrm{tot}}^0 $, $ \sigma_{\mathrm{tot}}^0 $.
This exponentially decaying filter is part of the LN$_{\mathrm{exp}}$ cascade model variant.

A shortcoming of the approximation \eqref{mu_filter_expappx_eq}, \eqref{aLNexp_filter_ode} above is that it cannot reproduce damped oscillations exhibited by the true linear filter, %$D_\mu$, %(with corresponding peaks at finite frequencies shown by its Fourier transform), 
in particular, for large input mean and small variance (see Fig.~\ref{fig8_cascade}A).
%These oscillations cannot be reproduced by a single exponential function.
Therefore, we introduce an alternative approximation using a damped oscillator function, % to approximate the full linear filter, 
\begin{equation} \label{mu_filter_dosappx_eq}
D_\mu(t) \approx B_\mu \exp (-t/\tau) \cos (\omega t). 
\end{equation}
Note that here $ B_\mu $, $ \tau $ and $ \omega $ depend on $ \mu_{\mathrm{tot}}^0 $, $ \sigma_{\mathrm{tot}}^0 $, which is not explicitly indicated. 
We fix the scaling parameter $B_\mu$ by (again) requiring that the approximation is normalized to reproduce the time integral of the true linear filter $ D_\mu $, %(as above), 
which yields
$B_\mu = \left(1 + (\tau^2 \omega^2) \right) /\tau$.
The remaining two parameters $\tau$ and $\omega$ are determined such that the dominant oscillation frequency is reproduced. Specifically, %$\tilde{\tau}$ and $\tilde{f}$ are adjusted 
the approximation should match %the values of the full linear filter 
$\hat{D}_\mu$ at the frequencies 
$f_{\mathrm{R}} = \mathrm{argmax}_f \,\left| \mathrm{Re}\left\{\hat{D}_\mu(f) \right\}\right|$ and 
$f_{\mathrm{I}} = \mathrm{argmax}_f \,\left| \mathrm{Im}\left\{\hat{D}_\mu(f) \right\}\right|$ in the Fourier domain as close as possible. 
We would like to note that using the method of least-squares over a range of frequencies instead can generate approximated filters which decay to zero instantly, particularly for large input mean and small variance (not shown). For such inputs a damped oscillator with a single decay time constant is too simple to fit the complete, rather complex linear filter shape. %and a delta like filter shape is the best least sq. solution there.
With \eqref{mu_filter_dosappx_eq} the filtered mean input can be obtained by solving the %complex 
second order ODE
%\begin{equation} \label{aLNdosc_filter_ode}
%\frac{d \tilde{\mu}_{\mathrm{f}}}{dt} = \left( (2\pi \tilde{f})^2 \tilde{\tau} + 
%\frac{1}{\tilde{\tau}} \right) \mu_{\mathrm{syn}} + \left( i 2\pi \tilde{f} - \frac{1}{\tilde{\tau}} \right) \tilde{\mu}_{\mathrm{f}}, 
%\qquad \mu_{\mathrm{f}}(t) = \mathrm{Re} \left( \tilde{\mu}_{\mathrm{f}}(t) \right),
%\end{equation}
\begin{equation} \label{aLNdosc_filter_ode}
\ddot \mu_{\mathrm{f}} + \frac{2}{\tau} \dot \mu_{\mathrm{f}} + \left( \frac{2}{\tau^2} + \omega^2 \right) \mu_{\mathrm{f}}
= \frac{1 + \tau^2 \omega^2}{\tau} \left( \frac{\mu_{\mathrm{syn}}}{\tau} + \dot \mu_{\mathrm{syn}} \right).
\end{equation}
Using this damped oscillator filter gives rise to the LN$_{\mathrm{dos}}$ cascade model variant. 
%[MENTION RELATIONSHIP TO DISS. VERSION: complex ODE representation ]

Considering the linear filter $D_\sigma$ (Eq.~\eqref{LN_components_eq3}) we use the approximation $D_\sigma(t) \approx A_\sigma \exp \left(-t/\tau_{\sigma}\right)$, with $A_\sigma = 1/\tau_{\sigma}$ and $\tau_{\sigma}$ determined by a least-squares fit of $\hat{D}_\sigma$ for frequencies $f \in [0, 1]$~kHz, %(similarly as above), 
as long as $\frac{\partial}{\partial \sigma} 
r_\infty \left( \mu_{\mathrm{tot}}^0 , \sigma_{\mathrm{tot}}^0 \right) > 0$. When this condition is not fulfilled, which occurs for large input mean and small variance, the full linear filter cannot be properly fit by an exponential function. This may be seen by the asymptotic behavior \cite{Fourcaud-Trocme2005} %[SIMPLER EXPLANATION?]
\begin{equation} \label{sigma_asymptotics_eq}
%\lim_{f \to 0} \hat{R}_\sigma(f) = \frac{\partial}{\partial \sigma} 
%r_\infty \left( \mu_{\mathrm{tot}}^0 , \sigma_{\mathrm{tot}}^0 \right), \qquad
\lim_{f \to \infty} \hat{R}_\sigma(f) = \frac{\sigma_{\mathrm{tot}}^0 r_\infty \left( \mu_{\mathrm{tot}}^0 , \sigma_{\mathrm{tot}}^0 \right)}{i 2 \pi f  \Delta_{\mathrm{T}}^2},
\end{equation}
which implies negative $ \lim_{f \to \infty} \hat{D}_\sigma(f) $ (cf. Eq.~\eqref{LN_components_eq3})
that cannot be approximated for nonnegative $ \tau_{\sigma} $. In this case  
 we use $\tau_{\sigma} \to 0$ which effectively yields $D_\sigma(t) \approx \delta (t)$, justified by the observation that the full filter rapidly relaxes to zero (see Fig.~\ref{fig8_cascade}B).  %(for both model variants). 
%Sigma filter is approximated using a scaled delta function, since it decays rapidly over large mu0 sigma0 space.  
The linear filter application can be implemented by solving 
\begin{equation} \label{aLNexp_sigmafilter_ode}
\frac{d \sigma_{\mathrm{f}}}{dt} = \frac{\sigma_{\mathrm{syn}} - \sigma_{\mathrm{f}}}{\tau_{\sigma}},
\end{equation}
where $\sigma_{\mathrm{f}}(t) = D_\sigma \ast \sigma_{\mathrm{syn}}(t)$ is the filtered input standard deviation. This filter is used in both model variants (LN$_{\mathrm{exp}}$ and LN$_{\mathrm{dos}}$).
Note (again) that $ \tau_{\sigma} $ depends on $ \mu_{\mathrm{tot}}^0 $, $ \sigma_{\mathrm{tot}}^0 $.

%\begin{figure}[ht!]
%\centering
%\includegraphics[width=1.0\textwidth,draft]{linear_response_functions_and_fits.png}
%\caption[Linear rate response functions of EIF neurons subject to white noise input.]{\textbf{Linear rate response functions of EIF neurons subject to white noise input.}
%Linear rate response functions
%for weak modulations of the input mean around $\mu$ with constant $\sigma$: $R_\mu(t; \mu, \sigma)$ in kHz/mV (\textbf{A}, black) and for weak modulations of the input standard deviation around $\sigma$ with constant $\mu$: $R_\sigma(t; \mu, \sigma)$ in kHz/(mV$\cdot \sqrt{\mathrm{ms}}$) (\textbf{B}, black). These functions are calculated in the Fourier domain for a range of modulation frequencies [$\hat{R}_\mu(f; \mu, \sigma)$ in 1/mV and $ \hat{R}_\sigma(f; \mu, \sigma)$ in 1/(mV$\cdot\sqrt{\mathrm{ms}}$)] 
%%and the absolute value of their Fourier transforms 
%%$\big| \hat{R}_\mu(f; \mu, \sigma) \big|$ and $\big| \hat{R}_\sigma(f; \mu, \sigma) \big|$ 
%(insets, black; absolute values are shown), which are fit using an exponential function in a semi-analytical way exploiting asymptotic results for $\hat{R}_\mu$ (\textbf{A}, blue dashed), and numerically (\textbf{A} and \textbf{B}, light blue). 
%In addition, $\hat{R}_\mu$ is fit using a damped oscillator function (\textbf{A}, red). The details of the fitting procedures are described in the text. 
%}
%\label{LC_fig4}
%\end{figure}

\paragraph{Extension for changing input baseline and recurrent coupling}

%to large variations of the input parameters
%Finally... Extending the cascade model with these components to larger variations of the synaptic moments $\mu_{\mathrm{syn}}$ and $\sigma_{\mathrm{syn}}$, we next 
In the derivation above we considered that the synaptic input mean and standard deviation, $\mu_{\mathrm{syn}}(t)$ and $\sigma_{\mathrm{syn}}(t)$, vary around $\mu_{\mathrm{syn}}^0$ and $\sigma_{\mathrm{syn}}^0$ with small magnitudes.
To extend the LN cascade model(s) to inputs that show large deviations %of $\mu_{\mathrm{syn}}$ and $\sigma_{\mathrm{syn}}$ 
from their baseline values %$\mu_0$ and $\sigma_0$, 
%we evaluate the estimate $\langle V \rangle_\infty$ of the mean membrane voltage in Eq.~\eqref{aLN_mean_w_eq} at $\mu_{\mathrm{eff}}(t) = \mu_{\mathrm{f}} - \langle w \rangle/C$ and 
%$\sigma_{\mathrm{eff}}(t) = \sigma_\mathrm{f}$
%%$\left(\mu_{\mathrm{f}} - \langle w \rangle/C, \sigma_{\mathrm{f}}\right)$, 
%%$\mu_{\mathrm{f}} - \langle w \rangle/C$ and $\sigma_{\mathrm{f}}$ 
%(instead of $\mu_{\mathrm{tot}}^0 $, $\sigma_{\mathrm{tot}}^0$) 
%in every timestep, to correspond with the instantaneous spike rate estimate of the model.
%In addition, 
we let the linear filters %adjust to the time-varying synaptic moments % (in the following way):
adjust to a changing input baseline in the following way: using the exponentially decaying mean input filter (LN$_{\mathrm{exp}}$ model) the filter parameters $\tau_{\mu}$ and $\tau_{\sigma}$ are evaluated at %$\mu_{\mathrm{eff}}(t)$, $\sigma_{\mathrm{eff}}(t)$ 
\begin{equation} \label{eq_eff_input_mom}
\mu_\mathrm{eff}(t) = \mu_\mathrm{f} - \langle w \rangle(t)/C , \qquad
\sigma_\mathrm{eff}(t) = \sigma_\mathrm{f}
\end{equation} 
in every time step, i.e., these parameters adapt to the \textit{effective} input moments. %[suggestion in \cite{Ostojic2011PLOS}] 
Using the damped oscillating mean input filter (LN$_{\mathrm{dos}}$ model), on the other hand, the filter parameters $\tau$, $ \omega $ and $\tau_{\sigma}$ are evaluated at $\mu_{\mathrm{tot}}(t)$, % = \mu_{\mathrm{syn}} - \langle w \rangle/C$, 
$\sigma_{\mathrm{tot}}(t)$ %= \sigma_\mathrm{syn} $
given by Eqs.~\eqref{eq_total_input_mean}, \eqref{eq_total_input_std}, i.e., these parameters adapt directly to the \textit{total} input moments, assuming that these moments do not fluctuate too vigorously. 
Note that because of these adjustments we do not need to consider a (particular) input baseline. 
Two remarks are in place: (i) the parameters of the damped oscillator cannot be adapted to a changing input baseline using the \textit{effective} input mean (with $\mu_{\mathrm{f}}$ given by Eq.~\eqref{aLNdosc_filter_ode}), because this can lead to stable oscillations (for an uncoupled population) and thus decreased reproduction performance (not shown); 
(ii)
for input moments that change very rapidly the reproduction performance of the LN$_{\mathrm{dos}}$ model variant may be improved by alternatively evaluating the parameters $\tau$, $ \omega $ and $\tau_{\sigma}$ at $ \mu_a(t)= \mu_{\mathrm{f}}^a - \langle w \rangle/C$, $\sigma_{\mathrm{eff}}(t)$, with $ \mu_{\mathrm{f}}^a $ governed by Eq.~\eqref{aLNexp_filter_ode} (combining LN$_{\mathrm{exp}}$ and LN$_{\mathrm{dos}}$), cf. \cite{Ladenbauer2015} (ch.~4.2).

%MENTION THE SMALL DIFFERENCE TO DISS. AND THAT THE DISS. VERSION MAY WORK BETTER FOR VERY RAPIDLY CHANGING INPUTS AND WITH REFR. PERIOD.

Finally, recurrent coupling within the population is included (in both model variants) by the dependence of the synaptic input moments on the delayed spike rate, $\mu_{\mathrm{syn}}(t, r_d)$, $\sigma_{\mathrm{syn}}^2(t, r_d)$, with $ r_d $ given by Eq.~\eqref{eq_meth_delay}. For the LN$_{\mathrm{dos}}$ model we can then replace $ \dot \mu_{\mathrm{syn}} $ in Eq.~\eqref{aLNdosc_filter_ode} by 
\begin{equation} \label{musyndot_rec_eq}
\dot \mu_{\mathrm{syn}} = \dot \mu_{\mathrm{ext}} + J K \dot r_d = \dot \mu_{\mathrm{ext}} + J K \frac{r-r_d}{\tau_d}.
\end{equation}
In case of identical (constant) propagation delays within the population this term would be 
$ \dot \mu_{\mathrm{syn}} = \dot \mu_{\mathrm{ext}} + J K \dot r(t-d) $ and in case of recurrent coupling without delays we would have 
\begin{equation} \label{musyndot_rec_nodelay_eq}
\dot \mu_{\mathrm{syn}} = \dot \mu_{\mathrm{ext}} + J K \left[ \frac{\partial \, r_\infty}{\partial \mu} \left( 
\dot \mu_{\mathrm{f}} - \frac{a 
\left(\langle V \rangle_\infty -E_w \vphantom{^2} \right) - 
  \langle w \rangle}{C \tau_{w}} + \frac{b\, r}{C} \right) + 
\frac{\partial \, r_\infty}{\partial \sigma} \frac{\sigma_{\mathrm{syn}} - \sigma_{\mathrm{f}}}{\tau_\sigma} \right].
\end{equation}

To summarize both LN cascade models, the population spike rate and mean membrane voltage are described by Eqs.~\eqref{LN_cascade_eq} and \eqref{LN_cascade_eq1}, using Eqs.~\eqref{LN_components_eq}, \eqref{LN_components_eq4} and \eqref{eq_meth_w_mean_p}, respectively. 
For the LN$_{\mathrm{exp}}$ model input filtering is governed by Eqs.~\eqref{aLNexp_filter_ode} and \eqref{aLNexp_sigmafilter_ode}, where the filter parameters are evaluated at $\mu_{\mathrm{eff}}(t)$, $\sigma_{\mathrm{eff}}(t)$ (Eq.~\eqref{eq_eff_input_mom}).
For the LN$_{\mathrm{dos}}$ model input filtering is described by Eqs.~\eqref{aLNdosc_filter_ode} and \eqref{aLNexp_sigmafilter_ode}, where the filter parameters are evaluated at $\mu_{\mathrm{tot}}(t)$, $\sigma_{\mathrm{tot}}(t)$ given by Eqs.~\eqref{eq_total_input_mean}, \eqref{eq_total_input_std}.
The system for the recurrent network under consideration is closed by Eqs.~\eqref{eq_meth_musyn_sigmasyn} and \eqref{eq_meth_delay} which relate population spike rate output and overall synaptic input moments. 
Note that both LN systems here are fully equivalent to the respective ones specified in the Sect.~\textit{Model reduction}.

\paragraph{Further remarks}

In order to efficiently simulate the derived cascade rate models it is highly recommended to precalculate the quantities which are needed in each time step on a $ (\mu,\sigma) $-rectangle (see Sect.~\textit{Low-Dimensional Approximations} above). For the LN$_{\mathrm{exp}}$ model these quantities are the filter time constants $\tau_{\mu}$, $\tau_{\sigma}$, and for the LN$_{\mathrm{dos}}$ model we need the quantities $\tau$, $\omega$ and $\tau_{\sigma}$ (all displayed in Fig.~\ref{fig8_cascade}C). Both models additionally require the steady-state quantities $r_\infty$ and $\langle V \rangle_\infty$ shown in Fig.~\ref{fig6_lowdim_intro}. An efficient implementation to obtain these quantities using Python with the package Numba for low-level virtual machine acceleration is available. 
% at GITHUB LINK. \todo{refer to github here} %The typical duration for these precalculations on a physiologically meaningful and sufficiently dense $ (\mu,\sigma) $-grid (using the provided code) amounts to about 45 min. on a hexa-core desktop computer. 
Recall, that changing any parameter value of the external input, the recurrent synaptic input or the adaptation current does not require renewed precomputations.    
%Note that %$r_\infty$, $\langle V \rangle_\infty$, $\tau_{\!\mu}$, $\tilde{f}$, $\tilde{\tau}$, $\tau_{\!\sigma}$ 
%the quantities $\langle V \rangle_\infty$, $\tau_{\mu}$, $\omega$, $\tau$, and $\tau_{\sigma}$ depend on the strengths of synaptic input and adaptation current only via the effective or total input moments. To reduce the computational complexity when numerically solving the LN model forward in time one can conveniently pre-compute these functions, together with $r_\infty$, for a range of values for effective input mean and input standard deviation (see Fig.~\ref{...}) and use look-up tables during time integration. Changing any parameter value of the external input, the recurrent synaptic input or the adaptation current does not require renewed pre-computations, but a change of any of the following (neuron model) parameters does: $C$, $g_{\mathrm{L}}$, $E_{\mathrm{L}}$, $\Delta_{\mathrm{T}}$, $V_{\mathrm{T}}$, $V_{\mathrm{r}}$, $V_{\mathrm{s}}$, $T_{\mathrm{ref}}$.
%\todo{Maybe remark on similarity between some spec and casc quantities}

If desired, it is also possible to obtain initial values for the variables of the cascade models (LN$_\mathrm{exp}$ and LN$_\mathrm{dos}$ variants) that correspond to a given initial distribution of membrane voltage and adaptation current values $\lbrace V_i(0)\rbrace$, $\lbrace w_i(0) \rbrace$ of a population of $N$ aEIF neurons. We can calculate $\langle w \rangle(0) = 1/N \sum_i w_i(0)$ and determine 
$\mu_{\mathrm{f}}(0)$, $\sigma_{\mathrm{f}}(0)$ by requiring that the initial membrane voltage distribution of the respective LN model $p_\infty \left(V; \,\mu_{\mathrm{f}}(0) - \langle w \rangle(0)/C, \sigma_{\mathrm{f}}(0)\right)$ matches the %empirical 
initial voltage distribution from the aEIF population as close as possible (e.g., using the Kolmogorov–Smirnov statistic). For the LN$_\mathrm{dos}$ model we additionally set $\dot \mu_{\mathrm{f}}(0) = 0$. 
%The initial values for the filter parameters are obtained at  $ \tau $, $ \omega $, $ \tau_\sigma $ at $ \mu=\mu_{\mathrm{f}}(0) - \langle w \rangle(0)/C $, $ \sigma = \sigma_{\mathrm{f}}(0) $.
 %... obtain the imaginary part of $\tilde{\mu}_{\mathrm{f}}(0)$ as $\mathrm{Im} \left( \tilde{\mu}_{\mathrm{f}}(0) \right) = 2\pi \tilde{f} \tilde{\tau}\mu_{\mathrm{f}}(0)$,  
%$\tilde{\mu}_{\mathrm{f}}(0) = \mu_{\mathrm{f}}(0)\left(1 + i 2\pi \tilde{f} \tilde{\tau}\right)$ 
%using the steady-state solution of Eq.~\eqref{aLNdosc_filter_ode}. 
This means we assume vanishing changes in the input history which underlies the initial membrane voltage distribution and filter parameters in the LN models (i.e., $\dot \mu_{\mathrm{syn}} \approx 0$, $\dot \sigma_{\mathrm{syn}} \approx 0$ for a sufficiently long time interval prior to $ t=0 $).

The components of the LN model are derived in the limit of small amplitude variations of $\mu_{\mathrm{syn}}$ and $\sigma_{\mathrm{syn}}$. However, the approximation also provides an exact description of the population dynamics 
for very slow variations of $\mu_{\mathrm{syn}}$ and $\sigma_{\mathrm{syn}}$, where the spike rate, mean membrane voltage and adaptation current are well approximated by their steady-state values in each time step. % cf. \cite{Ostojic2011PLOS}.
Here we approximated the derived linear filters using exponential and damped oscillator functions. 
We would like to note that, for a given baseline input ($ \mu_{\mathrm{tot}}^0 $, $ \sigma_{\mathrm{tot}}^0 $) the filter application using the latter function (Eq.~\eqref{mu_filter_dosappx_eq}) can be equivalently described by a complex-valued ODE \cite{Ladenbauer2015} (ch.~4.2). %, which, however, can lead to decreased reproduction performance when the baseline input % 
%changes very rapidly.
Furthermore, the true linear filters $ D_\mu $ and $ D_\sigma $ can be substantially better approximated by a damped oscillator function with two time scales (i.e., two exponentials) each. In these three cases, however, the ODE representation for the filter application can lead to 
decreased reproduction performance when the baseline input %($ \mu_{\mathrm{tot}}^0 $, $ \sigma_{\mathrm{tot}}^0 $) 
changes very rapidly (due to increased sensitivity to variations of the filter parameters).

\subsection*{Spike shape extension (optional)}

%How to use a spike shape and not clamp w:
In this contribution the membrane voltage spike shape has been neglected (typical for IF type neuron models) by clamping $V_i$ and $w_i$ during the refractory period, justified by the observation that it is rather stereotyped and its duration is very brief. Furthermore, the spike shape is believed to contain little information compared to the time at which the spike occurs. Nevertheless, it can be incorporated in the aEIF model in a straightforward way using the following reset condition, as suggested in \cite{Richardson2009}:
%Since the membrane voltage spike shape, which is usually disregarded in IF type models, can affect the adaptation current dynamics we here introduce a small model extension. 
%Instead of applying the reset condition $V_i:= V_{\mathrm{r}}$, $w_i := w_i + b$ whenever $V_i \geq V_{\mathrm{s}}$ and subsequently clamping $V_i$, $w_i$ at those values for the refractory period $T_{\mathrm{ref}}$, 
When $V_i$ reaches the spike voltage $V_{\mathrm{s}}$ from below we let $V_i$ decrease linearly from $V_{\mathrm{s}}$ to $V_{\mathrm{r}}$ during the refractory period and increment the adaptation current $w_i \leftarrow w_i + b$ at the onset of that period. %, after which the dynamics of $w_i$ is governed again by . 
That is, $V_i$ and $w_i$ are not clamped during the refractory period, instead, $V_i$ has a fixed time course and $w_i$ is incremented by $b$ and then governed again by Eq.~\eqref{eq_meth_w_single}.   
This modification implies %for the adiabatic approximation presented above 
that the average membrane voltage in Eq.~\eqref{eq_meth_w_mean_p} needs to be calculated over all neurons (and not only the nonrefractory ones), that is, $\langle V \rangle$ is calculated with respect to $p + p_{\mathrm{ref}}$, where
$p_{\mathrm{ref}}(V,t) = \int_0^{T_{\mathrm{ref}}} r(t-s) \delta \left(V-V_{\mathrm{sp}} (s)\right) ds$ with spike trajectory 
$V_{\mathrm{sp}}(t) = V_{\mathrm{s}} + (V_{\mathrm{r}}-V_{\mathrm{s}}) t/T_{\mathrm{ref}}$, cf. \cite{Richardson2009}. 
The same applies to the 
steady-state mean membrane potential in Eqs.~\eqref{eq_results_w_mean}, \eqref{eq_meth_w_mean_p_inf} and \eqref{aLN_mean_w_eq}, i.e., 
$\langle V \rangle_\infty$ is then given by 
\begin{equation} \label{eq_mean_v_spikeshape}
  \langle V \rangle_\infty = 
  \int_{-\infty}^{V_{\mathrm{s}}} v p_\infty(v) dv + 
  \left( 1 - \int_{-\infty}^{V_{\mathrm{s}}} p_\infty(v) dv \right)
  \frac{V_{\mathrm{r}}+V_{\mathrm{s}}}{2},
\end{equation} 
instead of Eq.~\eqref{eq_meth_V_inf}. % \textcolor{gray}{(and \eqref{LN_V_mean_inf})}.
Notably, the accuracy of the adiabatic approximation (Eq.~\eqref{eq_meth_w_single}) does not depend on the refractory period $T_{\mathrm{ref}}$ in this case.
%\footnote{Why we do not use this in the first place:
%1) Having a spike shape is nice, but the linear dependence of the adaptation current on the subthreshold voltage via $a$ is not ``supposed'' to be extended to the suprathreshold part, because for this contribution we have the spike triggered increments $b$. Therefore, clamping both, $V$ and $w$, during the refractory period seems most reasonable for the aEIF model (from the neuroscientist perspective). Mathematically the spike shape extension would be the ``cleaner'' way, considering the adiabatic approximation. 
%2) Consistency with previous chapters.}
That type of spike shape can therefore be considered in the FP model and the low-dimensional models in a straightforward way without significant additional computational demand. Note, however, that for the spec$_2$ model a nonzero refractory period 
is not supported
%can lead to %considerable decrease in 
% decreased reproduction performance 
(see above).   
For an evaluation of the spike shape extension in terms of reproduction accuracy of the LN models see \cite{Ladenbauer2015} (Fig.~4.15).
% flatex input end: [section/methods.tex]

% flatex input: [section/acknowledgements.tex]
\section*{Acknowledgements}

% the following was not allowed and we had to remove it in the submission process
We thank Maurizio Mattia for several fruitful discussions, 
sharing his unpublished manuscript on the advanced 
spectral model and the permission to apply it to our system. 
Additionally we thank Srdjan Ostojic for providing his insights and code related 
to the eigenvalue problem of the Fokker-Planck operator. 
We would finally like to acknowledge that Douglas J. Sterling wrote
the initial version of the software framework which was used to integrate and 
compare the different models. 
% We futhermore Günter Bärwolff for suggesting the Scharfetter-Gummel finite volume method.
% We thank Max Mustermann for helpful comments on the manuscript.

This work was supported by German Research Foundation (Collaborative Research
Center no. 910). 
% flatex input end: [section/acknowledgements.tex]

\paragraph{Supporting Information}

% \todo{add code URL into plos first page left column, maybe also here as supporting url }

\paragraph{\supplement. Supplementary methods.}
A) Numerical integration of the time-dependent Fokker-Planck equation.
B) Derivation of the model spec$_2$ based on the Fokker-Planck operator.
C) Numerical solver for the nonlinear Fokker-Planck eigenvalue problem.
% \todofinal{sync names with actual pdf}

\paragraph{S1 Figure. Fast changes of the input variance.}

%\nolinenumbers

% references (florian recommended to replace month and url in the final bbl file)
%*flatex input: [modelcomparison.bbl]

% flatex input end: [modelcomparison.bbl]

\newpage

% flatex input: [section/figures.tex]
% TIP: The following table includes only the minimum and maximum dimensions 
% allowed. To align your figure with the text column of the PDF version of the 
% article, make it no wider than 5.2 inches (13.2 cm). 
% see below Dimensions on:
%http://journals.plos.org/ploscompbiol/s/figures#dimensions

%Resolution and quality
% Submit TIFF at 300-600 ppi (no greater) at the desired dimensions. For EPS, 
% zoom in to view the quality of the figure.

% DO NOT INCLUDE GRAPHICS IN YOUR MANUSCRIPT
% - Figures should be uploaded separately from your manuscript file. 
% - Figures generated using LaTeX should be extracted and removed from the PDF before submission. 
% - Figures containing multiple panels/subfigures must be combined into one image file before submission.
% For figure citations, please use "Fig." instead of "Figure".
% See http://www.plosone.org/static/figureGuidelines for PLOS figure guidelines.

% => FINALLY: FIGURES/Captions (not the images themselves though) SHOULD BE 
% INCLUDED/CITED AT THE POSITION WHERE THEY FIRST OCCUR (from PLOS template)

\section*{Figures}

% the following is NOT true (see template pdf file)
% \todo{Note that we have to change the beginning of the caption (currently Figure) into Fig (without period).}

\newcommand{\maxfigwidth}{19.05cm}

%ht!
\begin{figure*}[ht]
\begin{adjustwidth}{-2.25in}{0in}
\includegraphics[width=\maxfigwidth]{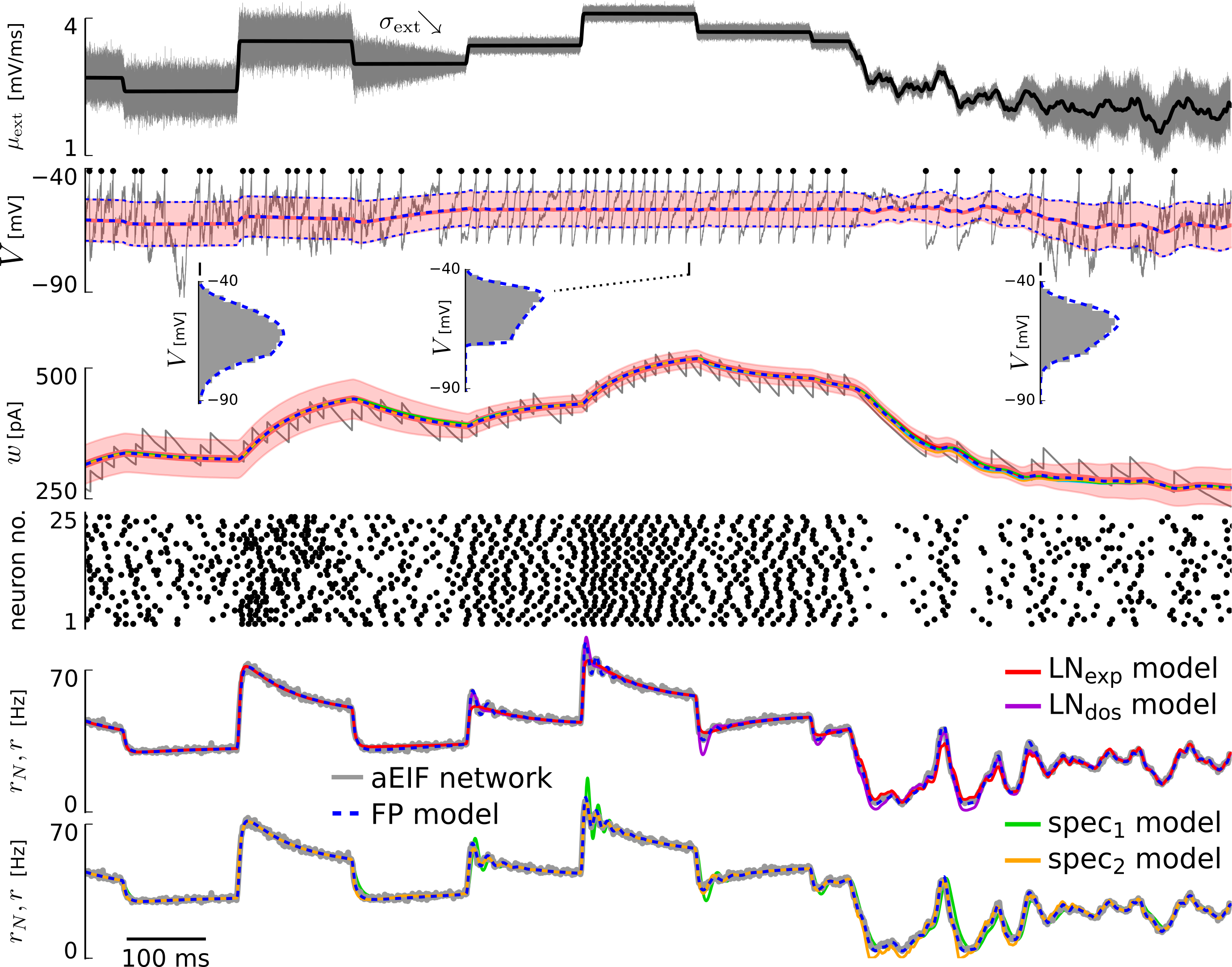}
\end{adjustwidth}
\end{figure*}
\begin{figure*}
\caption{\textbf{Example of aEIF network response and output of derived models for varying input.} 
From top to bottom: Mean
input $\mu_\mathrm{ext}$ (black) %which is first step-like
%and then turns into a smoothed Ornstein-Uhlenbeck signal (black) 
together with input standard deviation $\sigma_\mathrm{ext}$ (gray, visualized 
for one neuron by sampling the respective white noise process $\xi_\mathrm{ext,i}$).
%which changes from large (4 $\mathrm{mV}/\sqrt{\mathrm{ms}}$) to small
%(1 $\mathrm{mV}/\sqrt{\mathrm{ms}}$) (gray). 
$ 2^\mathrm{nd}$ row: Membrane voltage $ V $ of one neuron (gray, with spike times highlighted by black dots) and
membrane voltage statistics from 
the excitatory coupled aEIF population of 50,000 neurons (red) and from the FP model (blue dashed): mean $ \pm $ standard deviation
over time, as well as voltage histograms (gray) and probability densities 
$p(V,t)$ (blue dashed) at three indicated time points.
%the same for the Fokker-Planck (FP) model in blue (dotted lines).
%Below: Voltage histograms of network simulations (gray) and 
%probability densities of the FP model (dotted blue)). 
%Thrid row: 
$ 3^\mathrm{rd}$ row: Adaptation current $w$ of one neuron (gray) and
mean adaptation currents of all models $ \pm $ standard deviation for the aEIF network (shaded area). 
Note that differences in the mean adaptation currents of the different models are hardly recognizable. 
$ 4^\mathrm{th}$ row: Spike times of a subset 
of 25 neurons randomly chosen from the network. 
Below: Spike rate $ r $ of the LN cascade based models (LN$_\mathrm{exp}$, LN$_\mathrm{dos}$) and the 
spectral models (spec$_1$, spec$_2$) in
comparison to the FP model and the aEIF network ($r_N$). 
The values of the coupling parameters were 
$J=$0.05~mV,  
$K=$100,
$\tau_d=$3~ms.
}
\label{fig1_example}
\end{figure*}
% \todo{adapt symbol of the network spike rate + caption $r_N$}
% \todo{indicate coupling param. values ($ K,J,\tau_d $)}

\begin{figure*}[ht]
\begin{adjustwidth}{-2.25in}{0in}
\includegraphics[width=\maxfigwidth]{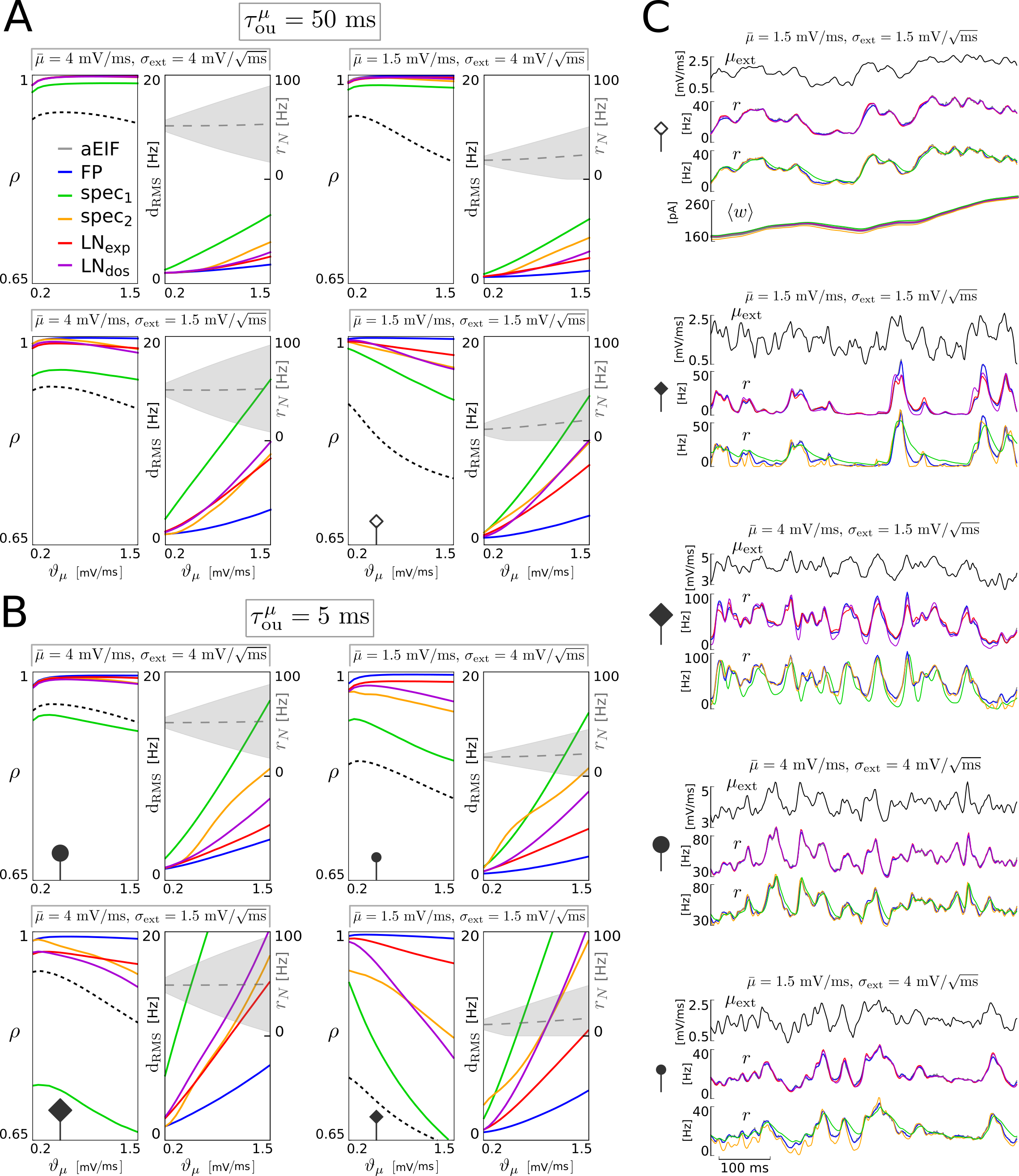}
\end{adjustwidth}
\end{figure*}
\begin{figure*}
\caption{\textbf{Reproduction accuracy of the reduced models for variations of the mean input.}
Pearson correlation coefficient ($\rho$) and root mean square
distance ($\mathrm{d}_{\mathrm{RMS}}$) between the spike rate time series $r_N(t)$ of the aEIF population and $r(t)$ 
of each derived model (FP, spec$_1$, spec$_2$, LN$_\mathrm{exp}$, LN$_\mathrm{dos}$) 
for different strengths of baseline mean input $\bar\mu$, input standard deviation $\sigma_{\mathrm{ext}}$, 
mean input variation $\vartheta_\mu$ and for a large value of time constant $\tau^\mu_{\mathrm{ou}}$
(\textbf{A}, moderately fast variations) as well as a smaller value (\textbf{B}, rapid variations). 
%Each point in parameter space, is specified by the 
%parameters $\bar\mu$, $\tau^\mu_{\mathrm{ou}}$ and $\vartheta_\mu$ of
%the stochastic process of Eq.~\eqref{eq_results_ou_mu} and standard deviation 
%of the external input $\sigma_{\mathrm{ext}}$. 
%
% the full Fokker-Planck
%system and the reduced spike rate models. To exclude errors due to 
%initialization mismatches both measures were computed after a 
%transient period of one second. Instead of averaging the response
%over a number of instances of OU processes with the same parameter set 
%($\bar\mu$, $\tau^\mu_{\mathrm{ou}}$,$\theta_\mu$) we simulated 
%for 60 seconds, which by far (orders of magnitude) exceeds the
%longest (internal and external) timescales of the system.
%\textbf{A}: Exploration for $\tau^\mu_{\mathrm{ou}}=50$ms. 
%For Four different parameter combinations of $\bar\mu$ (large/small) 
%and $\sigma_\mathrm{ext}$ (large/small) we vary $\vartheta_\mu$ 
%between 0.2 $\mathrm{mV}/\sqrt{\mathrm{ms}}$ and 
%1.5 $\mathrm{mV}/\sqrt{\mathrm{ms}}$ and compute the corresponding
%$\rho$ (left) $\mathrm{d}_{\mathrm{RMS}}$ (right) for the 
%set of reduced models with respect to the population rate which 
%was averaged over 1 ms time bins. 
The input-output correlation (between $ \mu_{\mathrm{ext}} $ and $ r_N $) is included as a reference (black dashed lines), and  
mean $ \pm $ standard deviation of the population spike rate $ r_N $ are indicated
(gray dashed lines, shaded areas).
For each parametrization, activity time series of 60~s duration were generated (from 50,000 aEIF neurons and each derived model),
from which the first second was omitted (each) to exclude transients, since the initial conditions of the models were not matched.
%of network simulations, the FP model and the reduced models were not
%matched. Instead a transient period of at least $1000$~ms was allowed to elapse before
%any measures were computed.
% \textcolor{gray}{Spike rate bin size for the aEIF population was 1~ms.}
Representative time series examples are shown on the right side with parameter values indicated (\textbf C), where empty and filled symbols correspond to large and 
small correlation time $\tau^\mu_{\mathrm{ou}}$, respectively, relating the examples 
to the panels \textbf A and \textbf B. The adaptation current traces were 
excluded in all but the first example to allow for a larger number of parameter points.
}
\label{fig2_mu_exploration}
\end{figure*}

\begin{figure*}[ht]
\begin{adjustwidth}{-2.25in}{0in}
\includegraphics[width=\maxfigwidth]{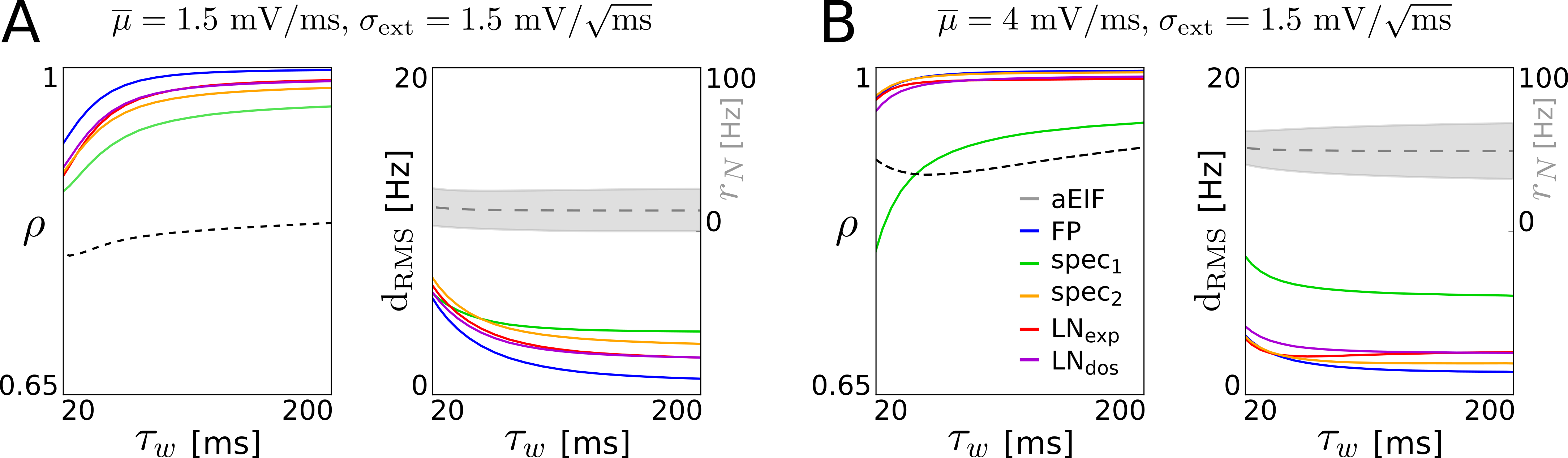}
\end{adjustwidth}
\end{figure*}
\begin{figure*}
\caption{\textbf{Effect of adaptation current timescale on reproduction accuracy.}
Performance measures and population spike rate statistics 
(cf. Fig.~\ref{fig2_mu_exploration}A,B) as a function 
of the adaptation time constant $\tau_w$, that takes values between 
20~ms (equal to the membrane time constant) and 200~ms (used throughout the rest of the study). The spike-triggered adaptation increment $b$ 
was co-varied (antiproportional to $\tau_w$) such that the product $\tau_w \, b = $~8~pAs is fixed for all shown parametrizations.
The input mean $\mu_\mathrm{ext}(t)$ fluctuates with
timescale $\tau^\mu_{\mathrm{ou}} =$~50~ms and strength $\vartheta_\mu=$~0.54~mV/ms 
(same value as for examples in Fig.~\ref{fig2_mu_exploration}C) 
around a smaller (\textbf A) and a larger (\textbf B) baseline mean $\bar \mu$, while the input deviation $\sigma_\mathrm{ext}$ is constant. 
Note that the rightmost parametrization of \textbf A corresponds to 
Fig.~\ref{fig2_mu_exploration}C (top example) and is contained in 
Fig.~\ref{fig2_mu_exploration}A (bottom right) while that of \textbf B 
is shown in Fig.~\ref{fig2_mu_exploration}A (bottom left).
}
\label{fig3_adapt_timescale}
\end{figure*}

\begin{figure*}[ht]
\begin{adjustwidth}{-2.25in}{0in}
\includegraphics[width=\maxfigwidth]{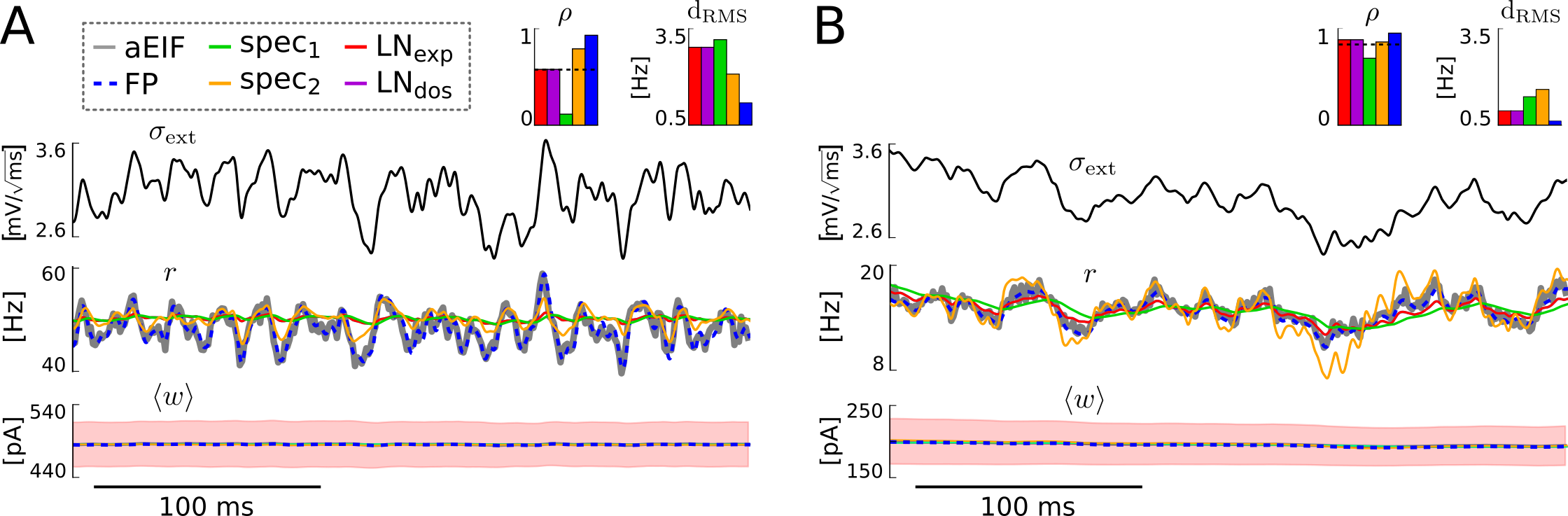}
\end{adjustwidth}
\end{figure*}
\begin{figure*}
\caption{\textbf{Performance for variations of the input variance.}
Time series of population spike rate and mean adaptation current from the different models in response to varying $\sigma^2_{\mathrm{ext}}$ for large mean input and rapid variations, $\mu_{\mathrm{ext}}=$ 4~$\mathrm{mV}/\mathrm{ms}$, 
 $\tau^{\sigma^2}_\mathrm{ou}=$ 5~ms (\textbf{A}) and for small mean input and moderately fast variations, $\mu_{\mathrm{ext}}=$ 1.5~$\mathrm{mV}/\mathrm{ms}$, 
  $\tau^{\sigma^2}_\mathrm{ou}=$ 50~ms (\textbf{B}). The values for the remaining input parameters were 
  $\bar\sigma^2_{\mathrm{ext}}=$ 9~$\mathrm{mV}^2/\mathrm{ms}$, $\vartheta_{\sigma^2}$ = 2~$\mathrm{mV}^2/\mathrm{ms}$.
  For the aEIF population $ \langle w \rangle \pm $ standard deviation are visualized (red shaded areas). Note that the mean adaptation time series of all models 
  as well as the spike rates of the cascade based models are on top of each other. 
%traces of model the output quantities in response to a varying
%$\sigma^2_{\mathrm{ext}}$ and constant $\mu_{\mathrm{ext}}$ input.
%\textbf{A}: The input mean is constant ($\mu_{\mathrm{ext}}=$1.5 $\mathrm{mV}/\mathrm{ms}$)
%whereas $\sigma^2_\mathrm{ext}$ is an OU process 
%($\bar\sigma^2_{\mathrm{ext}}=$ 9 $\mathrm{mV}^2/\mathrm{ms}$, $\tau^{\sigma^2}_\mathrm{ou}=$ 50 ms,
% $\vartheta_{\sigma^2}$ = 2 $\mathrm{mV}^2/\mathrm{ms}$). 
The indicated Pearson correlation coefficients (with dashed input-output correlation) and root mean
square distances were calculated from simulated spike rate time series of 60~s duration from which the first
second was excluded, as the initial conditions of the models were not matched.  
In \textbf A the correlation $\rho$ (but not the distance 
$\mathrm d_\mathrm{RMS}$) between the spike rates 
of the model spec$_1$ and the aEIF population) is strongly decreased 
due to a small time lag between the 
two time series which is difficult to see in the figure. 
%For computing the measures ($\rho$, $\mathrm{d}_{\mathrm{RMS}}$)
%we simulted for 60 seconds and excluded a transient period of 1 s
%(see results: colored bar chart). An excerpt of 300 ms of the varying $\sigma^2_{\mathrm{ext}}$ input (top)
%is displayed together with the rate response (middle) and the adaptation traces for the aEIF population, the FP model and the reduced spike rate models (bottom). The red shaded area illustrates the standard deviation of the network adaptation.
%\textbf{B}: Similar setting for different parameters: 
%$\bar\sigma^2_{\mathrm{ext}}=$ 9 $\mathrm{mV}^2/\mathrm{ms}$,,
%$\vartheta_{\sigma^2}$ = 2 $\mathrm{mV}^2/\mathrm{ms}$.
}
\label{fig4_sigma2_ou}
\end{figure*}

% \todo{$r_\infty$ instead of $r$ in heat map 2x (also A)}
\begin{figure*}[ht]
\begin{adjustwidth}{-2.25in}{0in}
\includegraphics[width=\maxfigwidth]{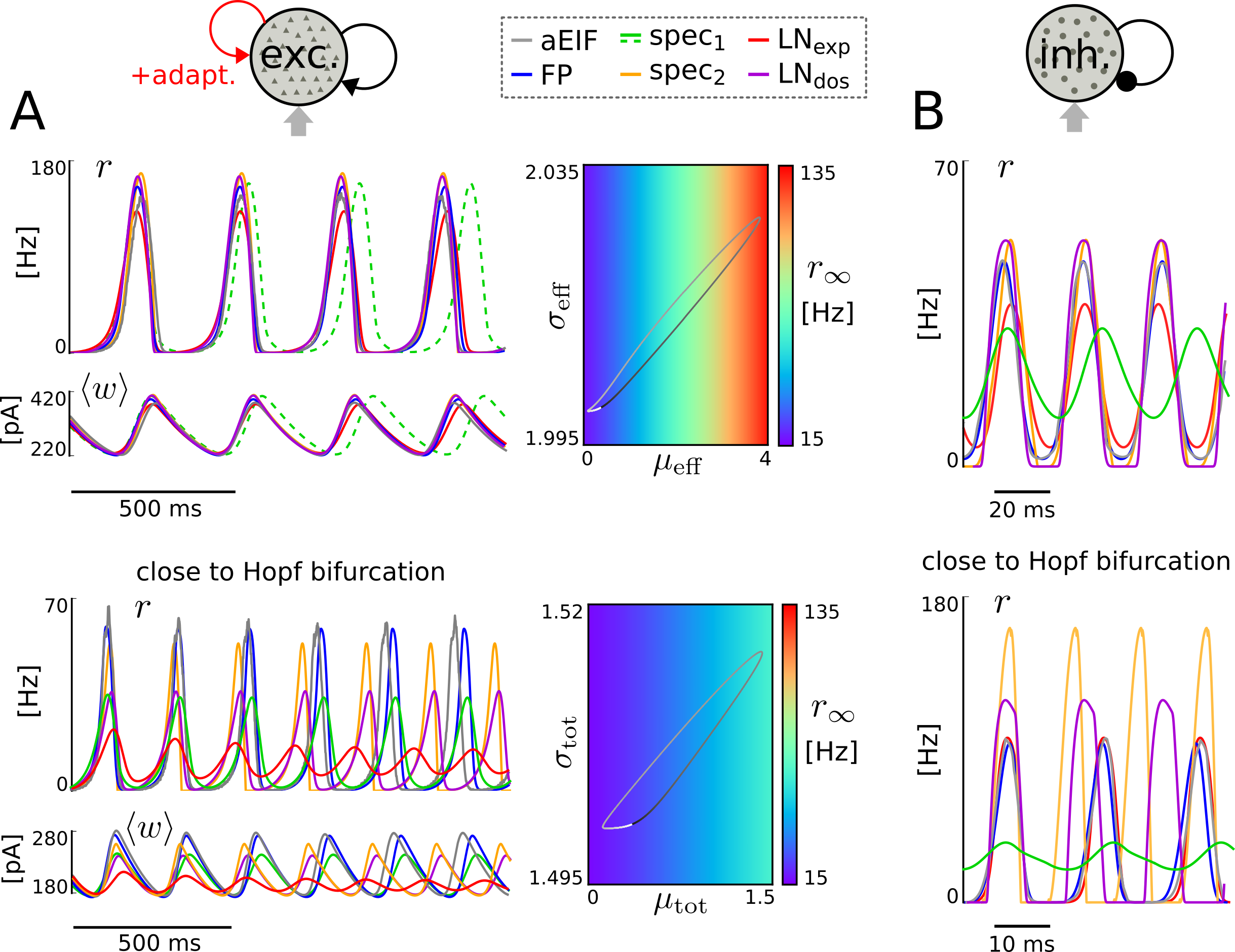}
\end{adjustwidth}
\end{figure*}
\begin{figure*}
\caption{\textbf{Network-generated oscillations.}
Oscillatory population spike rate and mean adaptation current %(bottom)
of 50,000 excitatory coupled aEIF neurons and each of the derived models (for constant 
external input moments) generated by the interplay of
recurrent excitation/adaptation current (\textbf A) and by delayed recurrent inhibition (\textbf B).
In addition, the limit cycle of the LN$_{\mathrm{exp}}$ model is shown in terms of the (quantity) steady-state spike rate $r_\infty$ as a function of 
effective input moments $\mu_\mathrm{eff}$, $\sigma_\mathrm{eff}^2 $ (\textbf{A}, top) and for the $\mathrm{spec}_2$ model 
in dependence of the total input moments $\mu_\mathrm{tot}$, $\sigma_\mathrm{tot}^2$ (\textbf{A}, bottom). 
The phase of the cycle is visualized by grayscale color code (increasing phase from black to white). 
The values for the input, adaptation and coupling parameters were 
$\mu_\mathrm{ext}=$~1.5~$\mathrm{mV}/\mathrm{ms}$, 
$\sigma_\mathrm{ext}=$~2~$\mathrm{mV}/\sqrt{\mathrm{ms}}$, $a=$~3~nS, $b=$~30~pA (\textbf{A}, top), 
$\mu_\mathrm{ext}=$~1.275~$\mathrm{mV}/\mathrm{ms}$,
$\sigma_\mathrm{ext}=$~1.5~$\mathrm{mV}/\sqrt{\mathrm{ms}}$, $a=$~3~nS, $b=$~60~pA (\textbf{A}, bottom),
$K=$~1000, $J=$~0.03~mV, $ \tau_d = 3$~ms 
(\textbf A, both). In \textbf B adaptation was removed ($a=b=0$) and delays were identical $d_{ij} = d$; input and coupling parameter values were 
$\mu_\mathrm{ext}=$~1.5~mV/ms, $\sigma_\mathrm{ext}=$~1.5~mV/$\sqrt{\mathrm{ms}}$, $K=$~1000, $J=-0.0357$~mV, $d=$~10~ms (top) and $\mu_\mathrm{ext}=$~3~mV/ms, $\sigma_\mathrm{ext}=$~2~mV/$\sqrt{\mathrm{ms}}$, $K=$~1000, $J=-0.087$~mV, $d=$~5~ms (bottom).
%
%Additionally we show the limit cycles in terms of spike rate as a 
%function of () for the
%LN$_{\mathrm{exp}}$ model, and, in dependence of 
%($\mu_\mathrm{tot},\sigma_\mathrm{tot} $) for the $\mathrm{spec}_2$ model. The color code illustrates
%the amplitude of the spike rate and the gray 
%scale of the line illustrates time during the period of one 
%oscillation/limit cycle. \textbf{A}: Constant input 
%scenario with $\mu_\mathrm{ext}$= 1.5 $\mathrm{mV}/\mathrm{ms}$ and
%$\sigma_\mathrm{ext}$ = 2 $\mathrm{mV}/\sqrt{\mathrm{ms}}$.
%The values of the coupling parameters were $K=$ 1000, $J=$ 0.03 mV, $ \tau_d =$
%and the adaptation strengths were $a$ = 3 nS and $b$ = 30 pA.
%\textbf{B}: Similar setting as in \textbf{A} but for different input
%parameters and adaptation strengths: 
%$K$ = 1000, $J$ = 0.03 mV, $a$ = 3 nS, $b$ = 60 pA.
}
\label{fig5_network_oscillations}
\end{figure*}

% 
% \begin{figure*}[ht]
% \begin{adjustwidth}{-2.25in}{0in}
% \includegraphics[width=0.9\textwidth]{figures/Fig5.png}
% \end{adjustwidth}
% \end{figure*}
 \begin{figure}[ht]
 \includegraphics[width=0.8\textwidth]{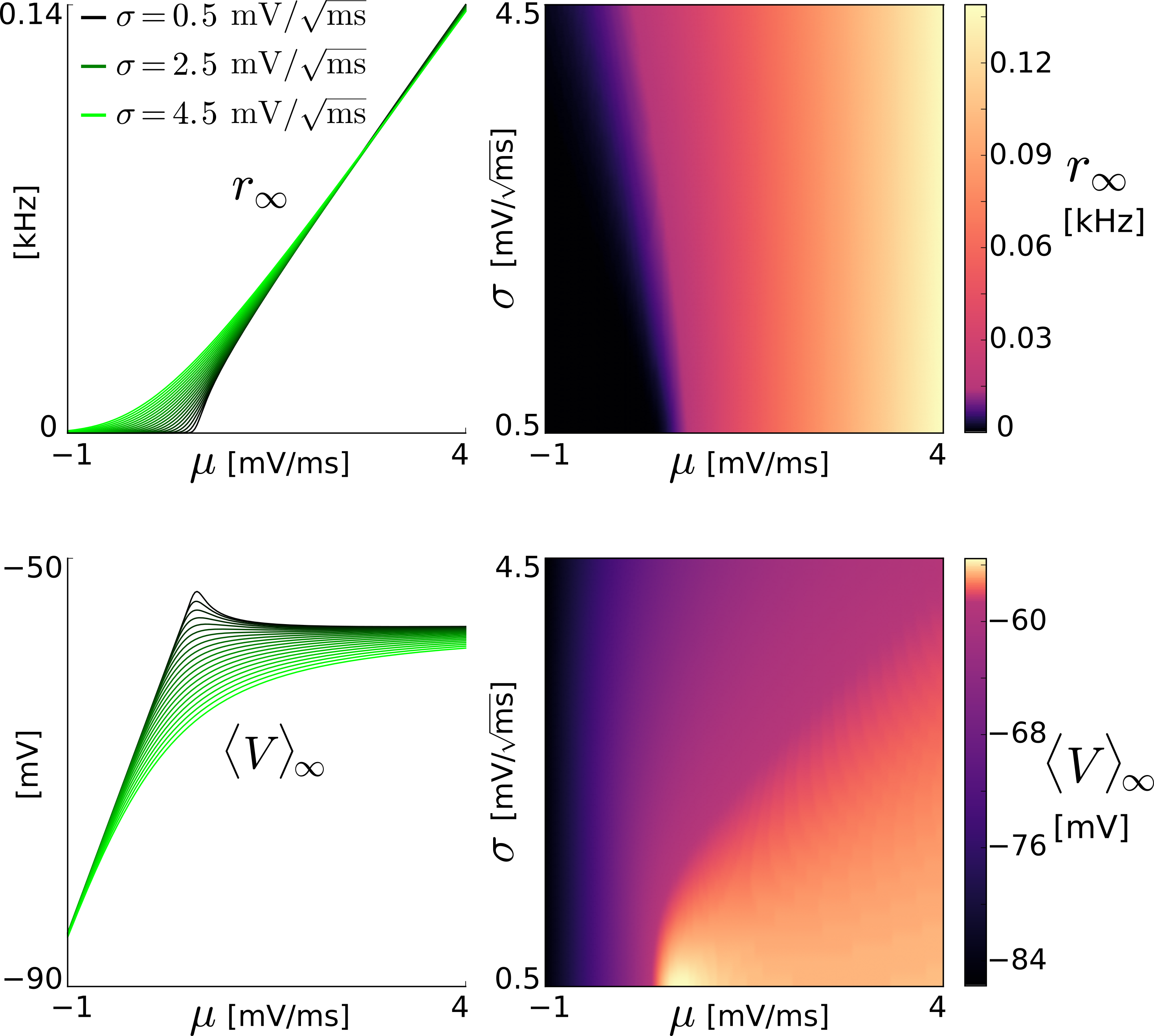}
 \caption{\textbf{Steady-state spike rate and mean membrane voltage for a population of EIF neurons.}
 $ r_\infty $ and $ \langle V \rangle_\infty $ for an uncoupled 
 population of EIF neurons (aEIF with $ a=b=0 $) as a function of 
 (generic) input mean $ \mu $ and standard deviation $ \sigma $, calculated from the (steady-state) Fokker-Planck equation, 
 shown in two different representations (left and right, each).
 }
 \label{fig6_lowdim_intro}
 \end{figure}

\begin{figure*}[ht]
% \begin{center}
\begin{adjustwidth}{-2.25in}{0in}
\includegraphics[width=\maxfigwidth]{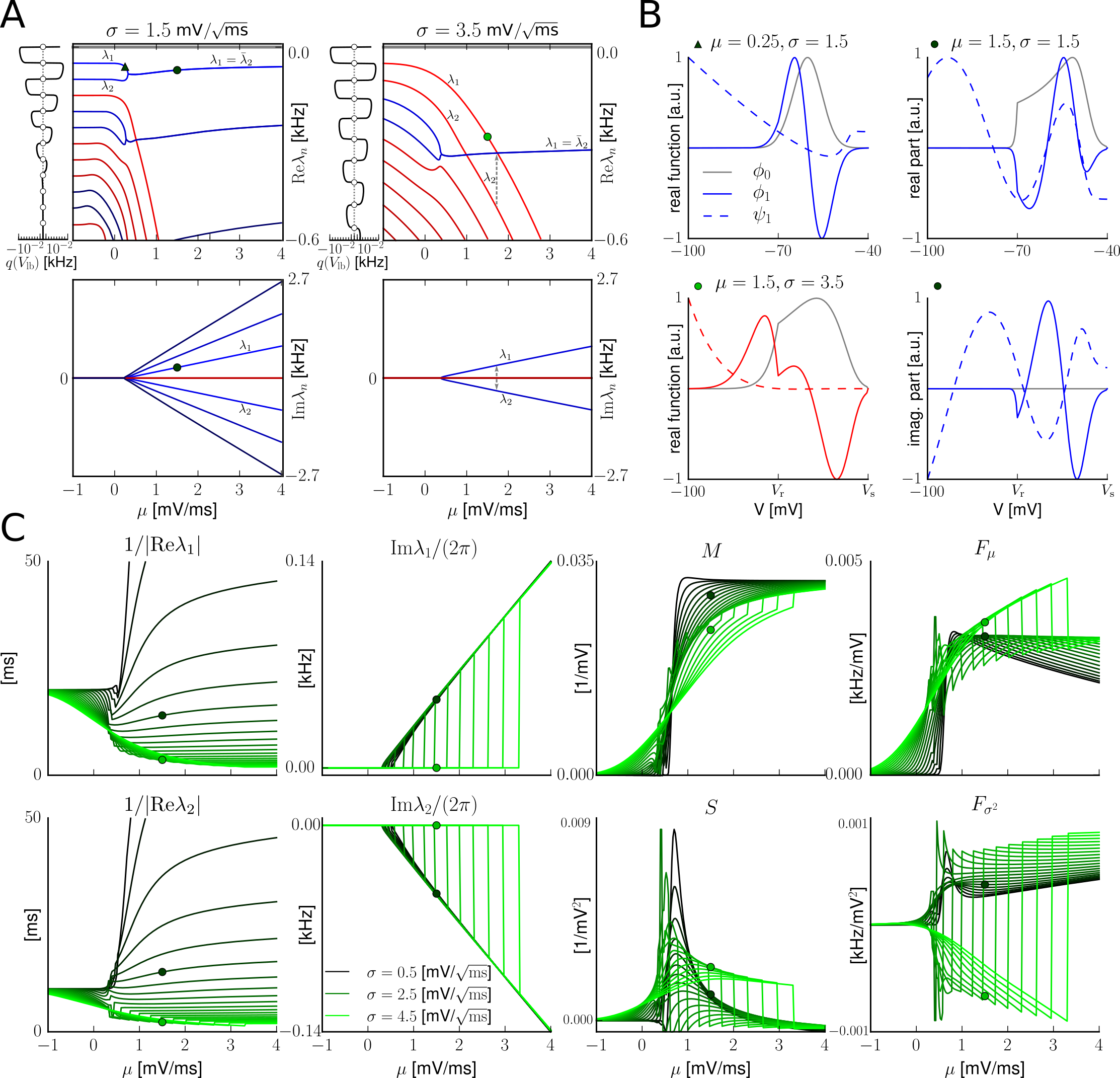}
\end{adjustwidth}
% \end{center}
\end{figure*}
\begin{figure*}
\caption{\textbf{Spectrum of the Fokker-Planck operator $\LL$ and related quantities.} 
\textbf{A}: regular eigenvalues of $\LL$ (blue) and diffusive ones (red) with real and 
imaginary part (top and bottom, respectively) as a function of the 
mean input $\mu$ for small noise intensity $\sigma$ (left) and larger input
fluctuations (right). 
The first two dominant eigenvalues $\lambda_1$, $\lambda_2$ are indicated together with
discontinuities in the real ($\lambda_2$) and imaginary part ($\lambda_1$ and 
$\lambda_2$), respectively.
The stationary eigenvalue $\lambda_0=0$ is shown in gray. 
Note that the value of the mean input $\mu$ at which the eigenvalue 
$\lambda_n$ changes from real to complex values depends on the noise amplitude $\sigma$ and the eigenvalue index $n$ which is difficult to see in the figure.
The narrow winding curves attached to the left side of the respective 
spectra represent the lower bound flux $q(V_\mathrm{lb})$ 
for $\mu_\mathrm{min} = -1.5$~mV/ms as a function of (real) eigenvalue candidate 
$\lambda$. The flux axis 
has a logarithmic scale between the large ticks 
(absolute values between $10^{-10}$ and $10^{-2}$~kHz) 
and is linear around the dashed zero value.
The open circles denote the eigenvalues, i.e., those $\lambda$ that satisfy $q(V_\mathrm{lb})=0$. Note that $q(V_\mathrm{lb})$ ranges over several orders 
of magnitude. 
\textbf{B}: stationary eigenfunction $\phi_0 = p_\infty$ (gray) 
and nonstationary eigenfunctions $\phi_1$ of $\LL$ and $\psi_1$ of $\LL^*$ 
corresponding to the first dominant eigenvalue 
$\lambda_1$ for three different input parameter values 
indicated by the triangle and circles in \textbf{A} (same units 
of $\mu$ and $\sigma$ as therein). 
The eigenfunctions are biorthonormalized, but (only) for visualization 
truncated at $V=-100$~mV and furthermore individually scaled 
to absolutely range within the unit interval of arbitrary units [a.u.]. 
% The omitted $[\mu]=\mathrm{mV/ms}$ and $[\sigma]=\mathrm{mV/}\sqrt{\mathrm{ms}}$.
\textbf{C}: first and second dominant eigenvalues $\lambda_1$, $\lambda_2$ with real and imaginary part (that are also indicated in \textbf{A}), as well as additional (real-valued) quantities of the model 
spec$_2$ ($M$, $S$, $F_\mu$, $F_{\sigma^2}$) as a function of 
input mean $\mu$ and noise strength $\sigma$ in steps of 0.2~mV/$\sqrt{\mathrm{ms}}$ 
from small values (black) to larger ones (green). 
The dots indicate identical parameter values to the spectra of \textbf{A} 
and the eigenfunctions of \textbf{B} 
(darker: $\sigma=1.5$~mV/$\sqrt{\mathrm{ms}}$, brighter green: 
$\sigma=3.5$~mV/$\sqrt{\mathrm{ms}}$).
}
\label{fig7_spectral}
\end{figure*}

\begin{figure*}[ht]
% \begin{center}
\begin{adjustwidth}{-2.25in}{0in}
\includegraphics[width=\maxfigwidth]{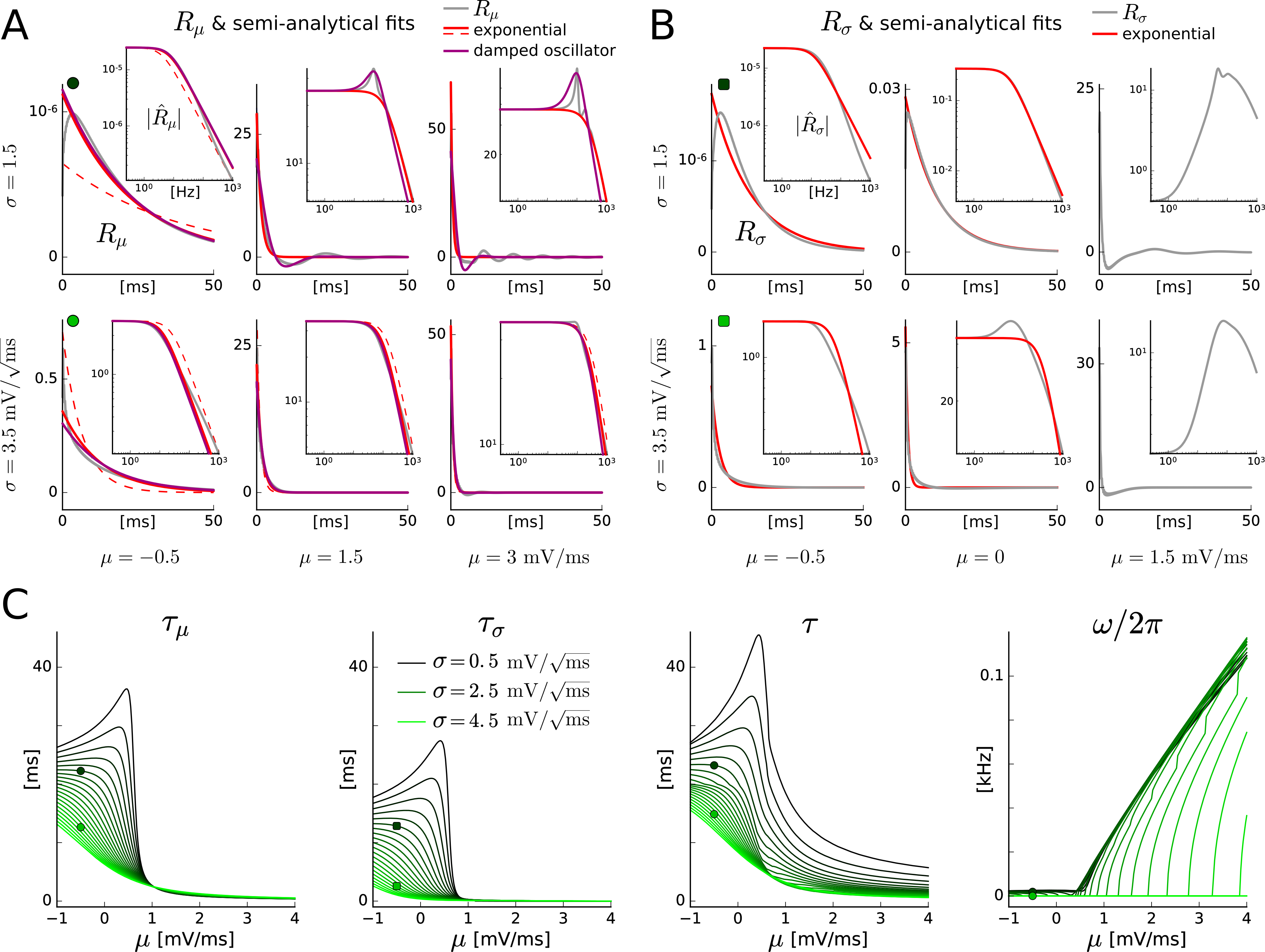}
\end{adjustwidth}
% \end{center}
\end{figure*}
\begin{figure*}
\caption{\textbf{Linear rate response functions and quantities for the cascade models.} 
Linear rate response functions of EIF neurons subject to white noise input
for modulations of the input mean around $\mu$ with constant $\sigma$: $R_\mu(t; \mu, \sigma)$ in kHz/V (\textbf{A}, gray) and for modulations of the input standard deviation around $\sigma$ with constant $\mu$: $R_\sigma(t; \mu, \sigma)$ in kHz/(V$\cdot \sqrt{\mathrm{ms}}$) (\textbf{B}, gray). 
These functions are calculated in the Fourier domain for a range of modulation frequencies [$\hat{R}_\mu(f; \mu, \sigma)$ in 1/V and $ \hat{R}_\sigma(f; \mu, \sigma)$ in 1/(V$\cdot\sqrt{\mathrm{ms}}$)] 
%and the absolute value of their Fourier transforms 
%$\big| \hat{R}_\mu(f; \mu, \sigma) \big|$ and $\big| \hat{R}_\sigma(f; \mu, \sigma) \big|$ 
(insets, gray; absolute values are shown), 
and 
fit using an exponential function exploiting asymptotic results for $\hat{R}_\mu$ (\textbf{A}, red dashed), as well as considering a range of frequencies (\textbf{A} and \textbf{B}, red solid). 
In addition, $\hat{R}_\mu$ is fit using a damped oscillator function (\textbf{A}, violet). The details of the fitting procedures are described in the text.
\textbf{C}: quantities ($\tau_{\mu}$, $\tau_{\sigma}$, $\tau$ and $\omega$) from the linear filter approximations (cf. \textbf{A}, \textbf{B}),  required for the LN$_\mathrm{exp}$ and LN$_\mathrm{dos}$ model variants (Eqs.~\eqref{aLNexp_filter_ode}, \eqref{aLNdosc_filter_ode} and \eqref{aLNexp_sigmafilter_ode}), as a function of $\mu$ and $\sigma$. 
}
\label{fig8_cascade}
\end{figure*}

% strangely, without the two lines below the last caption is not displayed
\begin{figure*}[ht]
\end{figure*}
% flatex input end: [section/figures.tex]

%\section*{References}
% Either type in your references using
% \begin{thebibliography}{}
% Text
% \end{thebibliography}
%
% OR
%
% Compile your BiBTeX database using our plos2015.bst
% style file and paste the contents of your .bbl file
% here.
% 
% \begin{thebibliography}{10}
% \bibitem{bib1}
% Devaraju P, Gulati R, Antony PT, Mithun CB, Negi VS. Susceptibility to SLE in South Indian Tamils may be influenced by genetic selection pressure on TLR2 and TLR9 genes. Mol Immunol. 2014 Nov 22. pii: S0161-5890(14)00313-7. doi: 10.1016/j.molimm.2014.11.005
% 
% \bibitem{bib2}
% Huynen MMTE, Martens P, Hilderlink HBM. The health impacts of globalisation: a conceptual framework. Global Health. 2005;1: 14. Available: http://www.globalizationandhealth.com/content/1/1/14.
% 
% \end{thebibliography}

% \vfill
% \input{section/figures}
\clearpage
\newpage

\section*{\supplement: Supplementary Methods}

\subsection*{Numerical integration of the time-dependent Fokker-Planck equation}

\paragraph{Numerical approach}
 
We begin with three equations of the main text:
Eq.~\eqref{eq_meth_impl_semidiscrecte_fp},
\begin{align}
% &\int\limits_{\mathcal{C}_i}\frac{\partial p(V,t)}{\partial t}\mathrm{d}V + \int\limits_{\mathcal{C}_i}\frac{\partial q_p(V,t)}{\partial V}\mathrm{d}V=0 \notag\\
% \iff \hspace{0.5cm}&
\frac{\partial}{\partial t}  p(V_m,t)
 =\frac{q_p(V_{m-\frac{1}{2}},t)-q_p(V_{m+\frac{1}{2}},t)}{\Delta V} 
+ \delta_{mm_\mathrm{r}}\frac{1}{\Delta V}q_p(V_{N_V+\frac{1}{2}}, t-T_\mathrm{ref})
,  \label{eq_app_fpsemidisc}
\end{align}
describes the membrane voltage discretization 
of the Fokker-Planck Eq.~\eqref{eq_meth_fp_pde} and includes the 
reinjection condition, Eq.~\eqref{eq_meth_fp_reinjection},
for each grid cell $[V_{m-\frac{1}{2}}, V_{m+\frac{1}{2}}]$ 
with center $V_m$ 
($m=1,\dots,N_V$) 
and cell spacing $\Delta V$.
Secondly, Eq.~\eqref{eq_meth_impl_scharfetter_gummel_flux}, 
\begin{align}
q_p(V_{m+\frac{1}{2}},t) = v_{m+\frac{1}{2}}\,\frac{p(V_m, t)-p(V_{m+1},t)\,\exp(-v_{m+\frac{1}{2}}\Delta V/D) }{1-\exp(-v_{m+\frac{1}{2}}\Delta V/D)},
\label{eq_app_scharfetter_gummel_flux}
\end{align}
represents the Scharfetter-Gummel flux approximation with 
drift and diffusion coefficients, $v_{m+\frac{1}{2}}$ and $D$, respectively.
Lastly, Eq.~\eqref{eq_meth_impl_fvm_timeint},
\begin{align} \label{eq_app_fvm_timeint}
% \frac{\partial p(V_i, t)}{\partial t}  \approx \frac{\underline{p}(t_{n+1})-\underline{p}(t_{n})}{\Delta t} &=  \mathbf{G}\underline{p}(t_{n+1})\, \\
(\mathbf{I}-\frac{\Delta t}{\Delta V} \, \mathbf{G}^n)\mathbf p^{n+1}
=\mathbf p^n + \boldsymbol g^{n+1-n_\mathrm{ref}},
\end{align}
is the linear system after time discretization 
that is solved for $p^{n+1}_m = p(V_m, t_{n+1})$ 
in each timestep $t_n\to t_{n+1}$, where $g^{n+1-n_\mathrm{ref}}_m = \delta_{mm_\mathrm{r}} 
\frac{\Delta t}{\Delta V} r(t_{n+1-n_\mathrm{ref}})$ 
contains the flux reinjection. The coefficients of the
system matrix $\mathbf{G}^n$ are obtained as follows. 

\paragraph{Boundary conditions}

The
absorbing boundary condition, $p(V_{\mathrm{s}},t) = 0$ (cf. Eq.~\eqref{eq_meth_fp_absorbing}), at cell border
$V_{N_V+\frac{1}{2}}=V_\mathrm{s}$, is discretized through 
linear interpolation between the last cell $N_V$ and a 
ghost cell that is introduced 
% \textcolor{gray}{with left border $V_{N_V+\frac{1}{2}}$} and
with center $V_{N_V+1} = V_\mathrm{s}+\Delta V/2$, yielding for the ghost value
\begin{equation} \label{eq_app_ghost_absorbing}
p(V_{N_V + 1}, t) = -p(V_{N_V}, t).
\end{equation}
The reflecting boundary cond. of Eq.~\eqref{eq_meth_fp_reflecting}, $q_p(V_\mathrm{lb},t) = 0$, at cell border 
$V_\frac{1}{2}=V_\mathrm{lb}$ is discretized 
by inserting a second ghost cell with center 
$V_0 = V_\mathrm{lb}-\Delta V/2$ and 
by setting the flux in that cell, i.e., Eq.~\eqref{eq_app_scharfetter_gummel_flux} for $m=0$, to zero, which gives 
for the ghost value
\begin{equation} \label{eq_app_ghost_reflecting}
p(V_0, t) = p(V_1, t) \exp(-v_\frac{1}{2} \Delta V / D).
\end{equation}

Furthermore, the spike rate, $r(t)=q_p(V_s,t)$, in this representation is given by evaluating the 
Scharfetter-Gummel flux, 
Eq.~\eqref{eq_app_scharfetter_gummel_flux}, 
at $V_{N_V+\frac{1}{2}}$, using
the ghost value from the discretized absorbing boundary, Eq.~\eqref{eq_app_ghost_absorbing}, which yields 
\begin{equation} \label{eq_app_fprate}
r(t) = q_p(V_{N_V+\frac{1}{2}},t) = v_{N_V+\frac{1}{2}}\,\frac{1+
\exp(-v_{N_V+\frac{1}{2}}\Delta V/D)}{1-\exp(-v_{N_V+\frac{1}{2}}\Delta V/D)}\,p(V_{N_V}, t) ,
\end{equation} 

\paragraph{Semi-implicit time discretization}

Inserting
the Scharfetter-Gummel flux representation, Eq.~\eqref{eq_app_scharfetter_gummel_flux}, into 
Eq.~\eqref{eq_app_fpsemidisc}, using Eq.~\eqref{eq_app_fprate} (at $t-T_\mathrm{ref}$) and approximating the time derivative 
with first order backward differences, 
results in 
\begin{align} \label{eq_app_fvm_fullint}
\begin{aligned}
\frac{p^{n + 1}_m-p^n_m}{\Delta t}\quad 
& = \quad \frac{v_{m - \frac{1}{2}}}{\Delta V}\,\left(\frac{p^{n + 1}_{m-1}-p^{n + 1}_m\,\exp(-v_{m - \frac{1}{2}}\Delta V/D)}
{1-\exp(-v_{m - \frac{1}{2}}\Delta V/D)}\right)\\
& - \quad \frac{v_{m + \frac{1}{2}}}{\Delta V}\,\left(\frac{p^{n + 1}_m-p^{n + 1}_{m+1}\,\exp(-v_{m + \frac{1}{2}}\Delta V/D)}
{1-\exp(-v_{m + \frac{1}{2}}\Delta V/D)}\right)\\
&  + \quad \delta_{mm_\mathrm{r}} 
\frac{1}{\Delta V} r(t_{n+1-n_\mathrm{ref}}),
\end{aligned}
\end{align}
where the drift and diffusion coefficients, $v_{m\pm\frac{1}{2}}$ and $D$, 
respectively, are evaluated at $t_{n}$ here and in the following, 
which corresponds more precisely to a semi-implicit time discretization. 
% \textcolor{gray}{ to ensure a linear and tridiagonal 
% system structure per timestep}
% to allow for instantaneous coupling ($d_{ij}\equiv 0$) without losing linearity of the discretized problem 
% nor the tridiagonal structure of $\mathbf{G^n}$.

Collecting terms for inner grid cells, $m=2,\dots,N_V-1$, gives the following elements of the tridiagonal matrix $\mathbf{G}^n$ from Eq.~\eqref{eq_app_fvm_timeint}:
\begin{align}
\mathbf{G}^n_{m,m}&=\frac{v_{m + \frac{1}{2}}}{\exp(-v_{m + \frac{1}{2}}\Delta V/D)-1}+\frac{v_{m - \frac{1}{2}} \exp(-v_{m - \frac{1}{2}}\Delta V/D)}{\exp(-v_{m - \frac{1}{2}}\Delta V/D)-1}, \\
\mathbf{G}^n_{m,m-1}&=\frac{v_{m - \frac{1}{2}}}{1-\exp(-v_{m - \frac{1}{2}}\Delta V/D)},\\
\mathbf{G}^n_{m,m+1}&= \frac{v_{m + \frac{1}{2}} \exp(-v_{m + \frac{1}{2}}\Delta V/D)}{1-\exp(-v_{m + \frac{1}{2}}\Delta V/D)} .
\end{align}

The remaining nonzero elements, i.e., those in the first and last row of $\mathbf{G}^n$,
are obtained by using the ghost cell values from the discretized boundary conditions. 
Inserting Eq.~\eqref{eq_app_ghost_reflecting} into Eq.~\eqref{eq_app_fvm_fullint} 
with $m=1$ yields for the reflecting boundary,
\begin{align}
\mathbf{G}^n_{1,1}&=\frac{v_{\frac{3}{2}}}{\exp(-v_{\frac{3}{2}}\Delta V/D)-1},\\
\mathbf{G}^n_{1,2}&=\frac{v_{\frac{3}{2}}\exp(-v_{\frac{3}{2}}\Delta V/D)}{1-\exp(-v_{\frac{3}{2}}\Delta V/D)}.
\end{align}
Note that these coefficients 
for the first row of 
$\mathbf{G}^n$ are alternatively also obtained by setting the term $q_p(V_{\frac{1}{2}},t)$ 
to zero in Eq.~\eqref{eq_app_fpsemidisc} for $m=1$ which allows to 
skip the introduction of the auxiliary ghost cell for the reflecting boundary.

For the absorbing boundary we insert Eq.~\eqref{eq_app_ghost_absorbing} into Eq.~\eqref{eq_app_fvm_fullint} with $m=N_V$, resulting in
\begin{align}
\mathbf{G}^n_{N_V, N_V-1}&=\frac{v_{N_V-\frac{1}{2}}}{1-\exp(-v_{N_V-\frac{1}{2}}\Delta V/D)}\\
\mathbf{G}^n_{N_V, N_V}&=v_{N_V - \frac{1}{2}}\,\frac{\exp(-v_{N_V - \frac{1}{2}}\Delta V/D)}{\exp(-v_{N_V - \frac{1}{2}}\Delta V/D)-1} \\
& +v_{N_V + \frac{1}{2}}\,\frac{\exp(-v_{N_V + \frac{1}{2}}\Delta V/D)+1}{\exp(-v_{N_V + \frac{1}{2}}\Delta V/D)-1}, \notag
\end{align}
which completes the specification of the tridiagonal matrix $\mathbf{G}^n$ and 
thus the system, Eq.~\eqref{eq_app_fvm_timeint}, that is, of Eq.~\eqref{eq_meth_impl_fvm_timeint}.

\bigskip

\subsection*{Derivation of the model spec$_2$ based on the Fokker-Planck operator}
%alt title: Advanced spike rate model based on the Fokker-Planck operator

% The following approach is based on \cite{Mattia2016Arxiv} that is extended here to include neuronal adaptation, time-varying external input moments and different delay distributions.

\paragraph{Base model}
% \todo{Replace $\mathbf a$ and $a_n$ by $\boldsymbol \alpha$ and $\alpha_n$ 
% respectively (to remove confusion with subthreshold adaptation parameter $a$): 
% here and in main text, too!}
We start with the Fokker-Planck mean-field model in spectral representation, 
Eqs.~\eqref{eq_meth_spec_a_full}, \eqref{eq_meth_spec_rate_full} and 
\eqref{eq_meth_w_mean_p_inf}, 
% of the main text 
\begin{align}
\dot{\boldsymbol{\alpha}} & = \left(\mathbf{\Lambda} + \mathbf{C}_\mu\dot{\mu} + 
\mathbf{C}_{\sigma^2}\dot{\sigma^2}\right)\boldsymbol{\alpha} + 
\mathbf{c}_\mu\dot{\mu}+\mathbf{c}_{\sigma^2}\dot{\sigma^2}
\label{eq_app_a_full}
\\
r(t) & = r_\infty + \mathbf{f} \cdot \boldsymbol{\alpha}
\label{eq_app_rate_full}
\\
\dot{\braket{w}} &  = \frac{a 
\left(\langle V \rangle_\infty -E_w \vphantom{^2} \right) - 
  \langle w \rangle}{\tau_{w}} + b\, r,
\end{align}
for the (infinitely many) projection coefficients 
$(\alpha_1, \alpha_2, \dots)$, the population-averaged spike rate $r$ and 
adaptation current $\langle w \rangle$, respectively.
They depend on the quantities 
$\mathbf{\Lambda}$, $\mathbf{C}_x$, $\mathbf{c}_x$ 
(for $x \in \{ \mu,\sigma^2 \}$), $r_\infty$, $\mathbf f$ and 
$\langle V \rangle_\infty$, 
which all are evaluated at the total input moments
\begin{align} \label{eq_app_tot_mean}
\mu(t) &= \overbrace{\mu_{\text{ext}}(t)
+KJr_d(t)}^{=\mu_\mathrm{syn(t)}}
-\frac{\Braket{w}\!(t)}{C},\\
% \sigma &= \sqrt{\sigma^2_{\text{ext}}(t)
% +\sigma^2_{\text{rec}}(r_d(t))}
\sigma^2(t) &= \sigma^2_{\text{ext}}(t)
+KJ^2 \,r_d(t) =\sigma^2_\mathrm{syn}(t) \label{eq_app_tot_var}
\end{align}
omitting subscripts $_\mathrm{tot}$ here and in the following (cf. Eqs.~\eqref{eq_meth_musyn_sigmasyn},\eqref{eq_total_input_mean},\eqref{eq_total_input_std}).
Additionally, Eq.~\eqref{eq_app_a_full}, contains the (first order) time derivative 
of the total input moments, $\dot \mu$, $\dot {\sigma^2}$.

For increased generality here we do not restrict the form of the 
delay distribution $p_d$ 
(i.e., the delayed spike rate is given by $r_d = r \ast p_d$, 
cf. Eq.~\eqref{eq_fp_rate_convolve_delay})
but show specific examples that include 
exponentially distributed, identical and no delays further below.
% Therefore the delayed spike rate is initially given by 
% the convolution ,
% , but is finally replaced 
% delay distribution specific expressions.
% The total input moments are given by 
% the eigenvalues $\mathbf{\Lambda} =\text{diag}\left(\lambda_1, \lambda_2,\dots\right)$ 
% and further spectral quantities
% $\left(\mathbf{c}_x\right)_n = \Braket{\partial_x\psi_n|\phi_0}$ 
% and $\left(\mathbf{C}_x\right)_{n,m} = \Braket{\partial_x\psi_n|\phi_m}$ 
% for $x=\mu,\sigma^2$. 
% The spectral quantities are in turn dependent on the total input moments 
% $\mu$ and $\sigma^2$ which are given by
% with \textit{delayed} spike rate $r_d$ due to recurrent coupling given by
% \begin{align}
% \begin{cases}
% r_d \equiv r                           &\quad\text{(for synapses without delays)}\\
% \dot{r}_d=(r-r_d)/\tau_d    &\quad\text{(for exponentially distributed delays)}
% \end{cases}
% \end{align}

Deriving Eqs.~\eqref{eq_app_a_full} once and \eqref{eq_app_rate_full} twice 
w.r.t. time 
gives
\begin{align}
\begin{aligned}
% \dot{\boldsymbol{\alpha}} =&
% \left(\mathbf{\Lambda} 
% + \mathbf{C}_\mu\dot{\mu} 
% + \mathbf{C}_{\sigma^2}\dot{\sigma^2}\right)\boldsymbol{\alpha} 
% + \mathbf{c}_\mu\dot{\mu}+\mathbf{c}_{\sigma^2}\dot{\sigma^2}\\
\ddot{\boldsymbol{\alpha}} =& 
\Big(\partial_\mu\mathbf{\Lambda}\dot{\mu}
+\partial_{\sigma^2}\mathbf{\Lambda}\dot{\sigma^2}
+\left[\partial_\mu\mathbf{C}_\mu\dot{\mu}
+\partial_{\sigma^2}\mathbf{C}_\mu\dot{\sigma^2}\right]\dot{\mu}
+\mathbf{C}_\mu\ddot{\mu} \label{eq_app_asecondderiv_full} \\
& 
+\left[\partial_\mu\mathbf{C}_{\sigma^2}\dot{\mu}
+\partial_{\sigma^2}
\mathbf{C}_{\sigma^2}\dot{\sigma^2}\right] \dot{\sigma^2}
+\mathbf{C}_{\sigma^2}\ddot{\sigma^2}\Big)\boldsymbol{\alpha}\\
+&\left(\mathbf{\Lambda}
+ \mathbf{C}_\mu\dot{\mu}
+\mathbf{C}_{\sigma^2}\dot{\sigma^2}\right)\dot{\boldsymbol{\alpha}}
+\partial_\mu\mathbf{c}_\mu(\dot{\mu})^2
+\mathbf{c}_\mu\ddot{\mu}
+\partial_{\sigma^2}\mathbf{c}_\mu\dot{\sigma^2}\dot{\mu}\\
+&\,\partial_\mu \mathbf{c}_{\sigma^2}\dot{\mu}\dot{\sigma^2}
+\partial_{\sigma^2}\mathbf{c}_{\sigma^2}(\dot{\sigma^2})^2
+\mathbf{c}_{\sigma^2}\ddot{\sigma^2},
\end{aligned}
\end{align}
\begin{align}
% r =& \, 
% r_\infty
% +\mathbf{f}\cdot\boldsymbol{\alpha}\\
\dot{r} =& \, 
\partial_{\mu}r_\infty\dot{\mu} 
+ \partial_{\sigma^2}r_\infty\dot{\sigma^2} 
+ \left(\partial_\mu \mathbf{f}\, \dot{\mu}+\partial_{\sigma^2} \mathbf{f}\, 
\dot{\sigma^2} \right)\cdot\boldsymbol{\alpha}
+\mathbf{f}\cdot\dot{\boldsymbol{\alpha}},\\ \nonumber \\
\ddot{r} =&  
\left(\partial_{\mu\mu} r_\infty\dot{\mu} 
+ \partial_{\sigma^2\mu} r_\infty\dot{\sigma^2} \right)\dot{\mu} 
+\partial_\mu r_\infty \ddot{\mu} \nonumber \\
+& \left(\partial_{\mu\sigma^2} r_\infty\dot{\mu} 
+ \partial_{\sigma^2\sigma^2} r_\infty\dot{\sigma^2} \right)
\dot{\sigma^2}
+\partial_{\sigma^2}r_\infty\ddot{\sigma^2}  \nonumber \\
+&\, \Big(\left[ \partial_{\mu\mu}\mathbf{f}\,\dot{\mu} 
+ \partial_{\sigma^2\mu}\mathbf{f}\,\dot{\sigma^2} \right]\dot{\mu}
+\partial_\mu\mathbf{f}\, \ddot{\mu} \label{eq_app_ratesecondderiv_full} \\
& + \left[ \partial_{\mu\sigma^2}\mathbf{f}\,\dot{\mu} 
+ \partial_{\sigma^2\sigma^2}\mathbf{f}\,\dot{\sigma^2} \right]\dot{\sigma^2}
+\partial_{\sigma^2}\mathbf{f}\, \ddot{\sigma^2}\Big)\cdot\boldsymbol{\alpha} 
\nonumber \\
+&\, 2 \left(\partial_\mu\mathbf{f}\,\dot{\mu}
+\partial_{\sigma^2}\mathbf{f}\,\dot{\sigma^2}\right)\cdot\dot{\boldsymbol{\alpha}}
% + \partial_\mu \mathbf{f}\,\dot{\mu}\cdot\dot{\boldsymbol{\alpha}} 
% + \partial_{\sigma^2} \mathbf{f}\,\dot{\sigma^2}\cdot\dot{\boldsymbol{\alpha}} 
+\mathbf{f}\cdot\ddot{\boldsymbol{\alpha}} \nonumber.
\end{align}

\paragraph{Slowness and modal approximations}
Assuming slowly changing total input moments, i.e., small 
input variations $\dot{\mu}$ and $\dot{\sigma^2}$
allows to consider projections coefficients of that order:
$\alpha_n = \OO(\dot{\mu})$ and 
$\alpha_n=\OO( \dot{\sigma^2})$. We can therefore neglect all
higher order terms and in particular those
which are proportional to the following factors: 
$\dot{\mu}\dot{\sigma^2}$, $(\dot{\mu})^2$, 
$(\dot{\sigma^2})^2$, $\dot{\mu}\boldsymbol{\alpha}$, 
$\dot{\sigma^2}\boldsymbol{\alpha}$, 
$\ddot{\mu}\boldsymbol{\alpha}$, $\ddot{\sigma^2}\boldsymbol{\alpha}$, 
$\dot{\mu}\dot{\boldsymbol{\alpha}}$, $\dot{\sigma^2}\dot{\boldsymbol{\alpha}}$.\\
With this approximation Eqs.~\eqref{eq_app_a_full},\eqref{eq_app_asecondderiv_full}--\eqref{eq_app_ratesecondderiv_full} become
\begin{align}
\dot{\boldsymbol{\alpha}} =&\,\mathbf{\Lambda}\, \boldsymbol{\alpha} + 
  \mathbf{c}_\mu\dot{\mu}+\mathbf{c}_{\sigma^2}\dot{\sigma^2}, \\
\ddot{\boldsymbol{\alpha}} =&\, 
\underbrace{\mathbf{\Lambda}^2\, 
  \boldsymbol{\alpha}+\mathbf{\Lambda}\,\mathbf{c}_\mu\dot{\mu}
  +\mathbf{\Lambda}\,\mathbf{c}_{\sigma^2}\dot{\sigma^2} }_
  {=\mathbf{\Lambda} \dot{\boldsymbol{\alpha}}}+
\mathbf{c}_\mu\ddot{\mu}+\mathbf{c}_{\sigma^2}\ddot{\sigma^2}, \\
\dot{r} =& \, 
\partial_{\mu}r_\infty\dot{\mu}
+\partial_{\sigma^2}r_\infty\dot{\sigma^2}
+\underbrace
{
\mathbf{f}\cdot\mathbf{\Lambda}\,\boldsymbol{\alpha}
+\mathbf{f}\cdot\mathbf{c}_\mu\dot{\mu}
+\mathbf{f}\cdot\mathbf{c}_{\sigma^2}\dot{\sigma^2}
}
_
{=\mathbf{f}\cdot\dot{\boldsymbol{\alpha}}}, \label{eq_app_linsys1}
\\
\ddot{r} =&\,
\partial_{\mu}r_\infty\ddot{\mu}
+\partial_{\sigma^2}r_\infty\ddot{\sigma^2}  \label{eq_app_linsys2} \\ \nonumber
& +\underbrace
{
\mathbf{f}\cdot\mathbf{\Lambda}^2\,\boldsymbol{\alpha}
+\mathbf{f}\cdot\mathbf{c}_\mu\ddot{\mu}
+\mathbf{f}\cdot\mathbf{c}_{\sigma^2}\ddot{\sigma^2}
+\mathbf{f}\cdot\mathbf{\Lambda}\,\mathbf{c}_\mu\dot{\mu}
+\mathbf{f}\cdot\mathbf{\Lambda}\,\mathbf{c}_{\sigma^2}\dot{\sigma^2}
}
_{=\mathbf{f}\cdot\ddot{\boldsymbol{\alpha}}}.
\end{align}
% The former two Equations have been inserted into the latter two.
% Combining these four Equations gives 
% \begin{align}
% \mathbf{f}\cdot\mathbf{\Lambda}\,\boldsymbol{\alpha} 
% =& \; 
% \dot{r} - \, 
% \partial_{\mu}r_\infty\dot{\mu}
% -\partial_{\sigma^2}r_\infty\dot{\sigma^2}
% -\mathbf{f}\cdot\mathbf{c}_\mu\dot{\mu}
% -\mathbf{f}\cdot\mathbf{c}_{\sigma^2}\dot{\sigma^2},
% \label{eq_app_linsys1}
% \\
% \mathbf{f}\cdot\mathbf{\Lambda}^2\,\boldsymbol{\alpha}
% =& \; 
% \ddot{r} -\,
% \partial_{\mu}r_\infty\ddot{\mu}
% -\partial_{\sigma^2}r_\infty\ddot{\sigma^2}
% -\mathbf{f}\cdot\mathbf{c}_\mu\ddot{\mu}
% -\mathbf{f}\cdot\mathbf{c}_{\sigma^2}\ddot{\sigma^2} \\
% & -\mathbf{f}\cdot\mathbf{\Lambda}\,\mathbf{c}_\mu\dot{\mu}
% -\mathbf{f}\cdot\mathbf{\Lambda}\,\mathbf{c}_{\sigma^2}\dot{\sigma^2}. \label{eq_app_linsys2}
% \end{align}
% Justified by the ordered spectrum (cf. properties in the main text) 

We now consider only the first two dominant eigenvalues $\lambda_1$ and 
$\lambda_2$ (cf. Eqs.~\eqref{eq_meth_lambda1},\eqref{eq_meth_lambda2}) 
and neglect all (faster) eigenmodes corresponding to 
eigenvalues with larger absolute real part (``modal approximation'').
Therefore, we take into account only the first two components of the 
(originally infinite-dimensional) variables
$\boldsymbol{\alpha} = (\alpha_1, \alpha_2)^T$ and quantities
$\mathbf\Lambda = \mathrm{diag}(\lambda_1, \lambda_2)$, $\mathbf{f}=(f_1,f_2)^T$, 
$\mathbf c_{\mu} = (c^{\mu}_1, c^{\mu}_2)^T$, 
$\mathbf c_{\sigma^2} = (c^{\sigma^2}_1, c^{\sigma^2}_2)^T$. Note that 
here and in the following the bold symbols denote the two-dimensional 
vectors and (diagonal) matrix, respectively. 

Due to the modal approximation 
Eqs.~\eqref{eq_app_linsys1},\eqref{eq_app_linsys2} form 
a linear, two-dimensional algebraic system with 
%two 
unknowns $f_1 \alpha_1$ and $f_2 \alpha_2$.
Solving this problem and inserting the solution 
into Eq.~\eqref{eq_app_rate_full}
using $\mathbf f \cdot \boldsymbol \alpha = f_1 \alpha_1 + f_2 \alpha_2$ 
yields
% \textcolor{gray}{
% \begin{align*}
% r_\infty-r=
% -\left(\frac{1}{\lambda_1} 
% +\frac{1}{\lambda_2}\right)\tilde{b}_1 
% +\frac{1}{\lambda_1\lambda_2}\tilde{b}_2\hspace{10cm}
% \end{align*}
% with
% \begin{align*}
% \tilde{b}_1&=
% \dot{r} 
% -\partial_{\mu}r_\infty\dot{\mu}
% -\partial_{\sigma^2}r_\infty\dot{\sigma^2}
% -\mathbf{f}\cdot\mathbf{c}_\mu\dot{\mu}
% -\mathbf{f}\cdot\mathbf{c}_{\sigma^2}\dot{\sigma^2}\hspace{10cm}\\
% \tilde{b}_2&=
% \ddot{r} 
% -\partial_{\mu}r_\infty\ddot{\mu}
% -\partial_{\sigma^2}r_\infty\ddot{\sigma^2}
% -\mathbf{f}\cdot\mathbf{c}_\mu\ddot{\mu}
% -\mathbf{f}\cdot\mathbf{c}_{\sigma^2}\ddot{\sigma^2}
% -\mathbf{f}\cdot\mathbf{\Lambda}\,\mathbf{c}_\mu\dot{\mu}
% -\mathbf{f}\cdot\mathbf{\Lambda}\,\mathbf{c}_{\sigma^2}\dot{\sigma^2}
% \end{align*}
% }
\begin{align} \label{eq_app_rate_nonexpanded}
\begin{aligned}
r_\infty-r=&
-\left(\frac{1}{\lambda_1} 
+\frac{1}{\lambda_2}\right)\left[\dot{r}
-\partial_{\mu}r_\infty\dot{\mu}
-\partial_{\sigma^2}r_\infty\dot{\sigma^2}
-\mathbf{f}\cdot\mathbf{c}_\mu\dot{\mu}
-\mathbf{f}\cdot\mathbf{c}_{\sigma^2}\dot{\sigma^2}\right]\\
+&\frac{1}{\lambda_1\lambda_2}\Big[\ddot{r} 
-\partial_{\mu}r_\infty\ddot{\mu}
-\partial_{\sigma^2}r_\infty\ddot{\sigma^2} 
-\mathbf{f}\cdot\mathbf{c}_\mu\ddot{\mu}
-\mathbf{f}\cdot\mathbf{c}_{\sigma^2}\ddot{\sigma^2}\\
& \qquad -\mathbf{f}\cdot\mathbf{\Lambda}\,\mathbf{c}_\mu\dot{\mu}
-\mathbf{f}\cdot\mathbf{\Lambda}\mathbf{c}_{\sigma^2}\dot{\sigma^2}\Big].
\end{aligned}
\end{align}

This Equation represents the complete spike rate dynamics under the 
modal and slowness approximations. It involves the first two time derivatives 
of the total input moments, 
% $\dot \mu$, $\ddot \mu$, $\dot{\sigma^2}$, $\ddot{\sigma^2}$
% $\mu$ and $\sigma^2$,
\begin{align} \label{eq_app_mudot}
\dot{\mu} =&\, \dot{\mu}_{\text{ext}}(t) 
+ KJ\dot{r}_d(t) 
- \frac{\dot{\braket{w}}}{C} \\
=&\, \dot{\mu}_{\text{ext}}(t) 
+ KJ\dot{r}_d(t) - \frac{a\left(\Braket{V}_\infty-E_w\right)-
\Braket{w}}{\tau_w C}-\frac{br}{C}, \nonumber\\
\ddot{\mu} =&\, \ddot{\mu}_{\text{ext}}(t) 
+ KJ\ddot{r}_d(t) 
- \frac{\ddot{\Braket{w}}}{C},\\
=&\, \ddot{\mu}_{\text{ext}}(t) 
+ KJ\ddot{r}_d(t)
-\frac{a}{\tau_wC}\left( \partial_\mu\Braket{V}_\infty\dot{\mu}
+\partial_{\sigma^2}\Braket{V}_\infty\dot{\sigma^2}\right) \nonumber\\
& +\frac{\dot{\Braket{w}}}{\tau_wC}
-\frac{b\dot{r}}{C}, \nonumber\\ 
=&\, \ddot{\mu}_{\text{ext}}(t) 
+ KJ\ddot{r}_d(t)
-\frac{a}{\tau_wC}\left( \partial_\mu\Braket{V}_\infty\dot{\mu}
+\partial_{\sigma^2}\Braket{V}_\infty\dot{\sigma^2}\right)\\
&+\,\frac{a\left(\Braket{V}_\infty-E_w\right)-
\Braket{w}}{\tau_w^2 C}+\frac{br}{\tau_wC}-\frac{b\dot{r}}{C}, \nonumber\\              
\dot{\sigma^2} =&\,\dispdot{\sigma^2_\mathrm{ext}}(t)+KJ^2\dot{r}_d(t),\\
\ddot{\sigma^2} =&\,\dispddot{\sigma^2_\mathrm{ext}}(t)+KJ^2\ddot{r}_d(t).
\label{eq_app_varddot}
\end{align}

\paragraph{Compactification}

Expanding the terms of Eqs.~\eqref{eq_app_mudot}--\eqref{eq_app_varddot} in Eq.~\eqref{eq_app_rate_nonexpanded} yields
\begin{align}  \label{eq_app_rate_fullyexpanded}
\begin{aligned}
r_\infty-r
=&-T\dot{r}
+D\ddot{r}\\
&-D\left(\partial_\mu r_\infty 
+\mathbf{f}\cdot\mathbf{c}_\mu\right)
\Bigg(
\ddot{\mu}_{\text{ext}}
+\frac{a\left(\Braket{V}_\infty-E_w\right)
-\Braket{w}}{\tau_w^2 C} \\
& \qquad \qquad \qquad \qquad \qquad \; 
+ KJ\ddot{r}_d + \frac{br}{\tau_wC}-\frac{b\dot{r}}{C}
\Bigg)\\
&-D\big(\partial_{\sigma^2}r_\infty
+\mathbf{f}\cdot\mathbf{c}_{\sigma^2}\big)
\big(\dispddot{\sigma^2_\mathrm{ext}}
+KJ^2\ddot{r}_d\big)\\
&+\Bigg(\dot{\mu}_{\text{ext}} 
+ KJ\dot{r}_d - \frac{a\left(\Braket{V}_\infty-E_w\right)-
\Braket{w}}{\tau_w C}-\frac{br}{C}\Bigg) 
\\
& +\,\big(\dispdot{\sigma^2_\mathrm{ext}}+KJ^2\dot{r}_d\big)
\\
&\quad\cdot \Big[T\big(\partial_\mu r_\infty
+\mathbf{f}\cdot\mathbf{c}_\mu\big)+D\big(\partial_\mu r_\infty
+\mathbf{f}\cdot\mathbf{c}_\mu\big)\frac{a}{\tau_wC}\partial_\mu\Braket{V}_\infty
\\
&\qquad-D\mathbf{f}\cdot\mathbf{\Lambda}\,\mathbf{c}_{\mu}\Big] \\
&\quad\cdot\Big[T\big(\partial_{\sigma^2} r_\infty
+\mathbf{f}\cdot\mathbf{c}_{\sigma^2}\big)+D\big(\partial_\mu r_\infty
+\mathbf{f}\cdot\mathbf{c}_\mu\big)\frac{a}{\tau_wC}\partial_{\sigma^2}\Braket{V}_\infty
\\
&\qquad-D\mathbf{f}\cdot\mathbf{\Lambda}\,\mathbf{c}_{\sigma^2}\Big],
\end{aligned}
\end{align}
where we have introduced the trace $T$ and determinant $D$ of the 
inverse eigenvalue matrix $\mathbf\Lambda^{-1}$, 
\begin{align}
T             & = 1/\lambda_1 + 1/\lambda_2, \label{eq_app_lump1first} \\ %\tfrac{1}{\lambda_1} + \tfrac{1}{\lambda_2}\\
D             & = 1/\lambda_1 \cdot 1/\lambda_2,
\end{align}
for simplicity. 
With the  definitions of the additional (lumped) quantities,
\begin{align} 
& F_\mu         &=  &\quad \mathbf{f}\cdot\mathbf{\Lambda}\,\mathbf{c}_{\mu},\hspace{8cm}\\
& F_{\sigma^2}  &= &\quad \mathbf{f}\cdot\mathbf{\Lambda}\,\mathbf{c}_{\sigma^2},\\
& M             &= &\quad \partial_\mu r_\infty+\mathbf{f}\cdot\mathbf{c}_\mu, \\
& S             &= &\quad \partial_{\sigma^2} r_\infty+\mathbf{f}\cdot\mathbf{c}_{\sigma^2}, \label{eq_app_lump1last} \\
%& A             &= &\quad \frac{a\left(\Braket{V}_\infty-E_w\right)-\Braket{w}}{\tau_w C}
\label{eq_app_lump2first}
 & R & = & \quad DMKJ + DSKJ^2, \hspace{7cm}\\
 & H_\mu & =  & \quad TM+DM\frac{a}{\tau_wC}\partial_\mu\Braket{V}_\infty-DF_\mu, \\
 & H_{\sigma^2} & = & \quad TS+DM \frac{a}{\tau_wC}\partial_{\sigma^2}\Braket{V}_\infty -DF_{\sigma^2}, \label{eq_app_lump2last}
\end{align}
that are composed of the individual quantities (and of neuron and coupling parameters), 
Eq.~\eqref{eq_app_rate_fullyexpanded} can be rewritten as 
% \begin{align}
% \begin{aligned}
% r_\infty-r=&\, r\left[-DM\frac{b}{\tau_wC}-\frac{b}{C}\left(TM+DM\frac{a}{\tau_wC}\partial_\mu\Braket{V}_\infty-DF_\mu\right)\right]\\
% &+\, \dot{r}\left[-T+DM\frac{b}{C}\right]\\
% &+\, \dot{r}_d\Big[ KJ\left( TM+DM\frac{a}{\tau_wC}\partial_\mu\Braket{V}_\infty -DF_\mu \right)\\
% &\hspace{1cm}+KJ^2\left( TS+DM \frac{a}{\tau_wC}\partial_{\sigma^2}\Braket{V}_\infty -DF_{\sigma^2}\right)\Big]\\
% &+\, \ddot{r}D\\
% &-\, \ddot{r}_d\left[DMKJ+DSKJ^2 \right]\\
% &-\, DM\left(\ddot{\mu}_{\text{ext}} +\frac{a\left(\Braket{V}_\infty-E_w\right)-\Braket{w}}{\tau_w^2C} \right)\\
% &+\left( \dot{\mu}_{\text{ext}}-\frac{a\left(\Braket{V}_\infty-E_w\right)-\Braket{w}}{\tau_w C} \right)\left[ TM+DM\frac{a}{\tau_wC}\partial_\mu\Braket{V}_\infty-DF_\mu \right]\\
% &+\, \dispdot{\sigma^2_\mathrm{ext}}\left( TS +DM\frac{a}{\tau_wC}\partial_{\sigma^2}\Braket{V}_\infty-DF_{\sigma^2} \right)-DS\dispddot{\sigma^2_\mathrm{ext}}.
% \end{aligned}
% \end{align}
% By introducing the additional lumped quantities,
% \begin{align} 
% \end{align}
% it is furthermore compactified as
\begin{align} \label{eq_app_rate_compactified}
\begin{aligned}
r_\infty-r=&\, r\left(-DM\frac{b}{\tau_wC}-\frac{b}{C}H_\mu\right)\\
&+\, \dot{r}\left(-T+DM\frac{b}{C}\right)\\
&+\, \dot{r}_d\left( KJ H_\mu+KJ^2H_{\sigma^2}\right)\\
&+\, \ddot{r}D\\
&-\, \ddot{r}_d R \\
&-\, DM\left(\ddot{\mu}_{\text{ext}} +\frac{a\left(\Braket{V}_\infty-E_w\right)-\Braket{w}}{\tau_w^2C} \right)\\
&+\left( \dot{\mu}_{\text{ext}}-\frac{a\left(\Braket{V}_\infty-E_w\right)-\Braket{w}}{\tau_w C} \right)H_\mu\\
&+\, \dispdot{\sigma^2_\mathrm{ext}}H_{\sigma^2}-DS\dispddot{\sigma^2_\mathrm{ext}}.
\end{aligned}
\end{align}
% \begin{align}
% r_\infty-r=&\, r\left(-DM\frac{b}{\tau_wC}-\frac{b}{C}H_\mu\right) 
%   +\, \dot{r}\left(-T+DM\frac{b}{C}\right)\\
% &+\, \dot{r}_d\left( KJ H_\mu+KJ^2H_{\sigma^2}\right)
%   +\, \ddot{r}D
%   -\, \ddot{r}_d R \\
% &-\, DM\left(\ddot{\mu}_{\text{ext}} +\frac{a\left(\Braket{V}_\infty-E_w\right)-\Braket{w}}{\tau_w^2C} \right)\\
% &+\left( \dot{\mu}_{\text{ext}}-\frac{a\left(\Braket{V}_\infty-E_w\right)-\Braket{w}}{\tau_w C} \right)H_\mu\\
% &+\, \dispdot{\sigma^2_\mathrm{ext}}H_{\sigma^2}-DS\dispddot{\sigma^2_\mathrm{ext}} .
% \end{align}
Note that the first six lumped quantities, Eqs.~\eqref{eq_app_lump1first}--\eqref{eq_app_lump1last}, depend on the parameters of recurrent coupling and adaptation current (only) via the total
 input moments, cf. Eq.~\eqref{eq_app_tot_mean},\eqref{eq_app_tot_var}, 
while the last three, Eqs.~\eqref{eq_app_lump2first}--\eqref{eq_app_lump2last},
contain them ($K$, $J$, $a$, $\tau_w$) explicitly. Those lumped 
quantities, i.e., $R$, $H_\mu$, $H_{\sigma^2}$, can, for 
example, be evaluated during runtime for the respective 
total input moments by using precalculations of the other lumped 
quantities ($T$, $D$, $F_\mu$, $F_{\sigma^2}$, $M$, $S$) and 
of the additional individual quantities 
$\partial_{\mu}\Braket{V}_\infty$ and $\partial_{\sigma^2}\Braket{V}_\infty$, 
that all are independent of adaptation and synaptic parameters.

\paragraph{Second order ordinary differential equation}

Eq.~\eqref{eq_app_rate_compactified} leads to the final model (cf. Eq.~\eqref{eq_meth_spec2}): a real-valued 
second order equation for the spike rate $r(t)$, 
\begin{equation} \label{eq_app_spec2}
% \beta_2 \, \ddot r + \beta_1 \, \dot r + \beta_0 \, r 
% = r_\infty - r - \beta_c \left 
% ( \langle w \rangle, \,  r_d, \dot \mu_\mathrm{ext}, \, \ddot \mu_\mathrm{ext}, \, 
% \dispdot{\sigma_\mathrm{ext}^2}, \, \dispddot {\sigma_\mathrm{ext}^2} \right )
\beta_2\ddot{r}+\beta_1\dot{r}+\beta_0r = r_\infty-r - \beta_c,
\end{equation}
with coefficients 
$\beta_2$, $\beta_1$, $\beta_0$ and $\beta_c$, 
% $\beta_c \left ( \langle w \rangle, \,  r_d, \dot \mu_\mathrm{ext}, \, \ddot \mu_\mathrm{ext}, \, 
% \dispdot{\sigma_\mathrm{ext}^2}, \, \dispddot {\sigma_\mathrm{ext}^2} \right )$ 
that depend on the total input moments 
$\mu(t,r_d,\langle w \rangle)$ and $\sigma^2(t,r_d,\langle w \rangle)$ (cf. Eqs.~\eqref{eq_app_tot_mean},\eqref{eq_app_tot_var}).
% \begin{equation}
% \mu(t) = \mu_\mathrm{syn}(\mu_\mathrm{ext}(t),r_d(t)) - \langle w \rangle(t)/C, \qquad
% \sigma^2(t) = \sigma_\mathrm{syn}^2(\sigma_\mathrm{ext}^2(t),r_d(t)),
% \end{equation}
Their concrete form is determined by the delay distribution $p_d$. 
Particularly, we distinguish three cases: (i) coupling without delay, 
$p_d(\tau) = \delta(\tau)$, (ii) exponentially distributed delays,
$p_d(\tau) = \exp(-\tau/\tau_d) /\tau_d$ (for $\tau \ge 0$), and
(iii) identical delays,  $p_d(\tau) = \delta(\tau-d)$ with $d>0$.

% 
% 
% \subsection**{0) No coupling}
% Setting the coupling constants ($K$ and/or $J$) to zero and $r_d=r$ we get
% \begin{equation*}
% \beta_2\ddot{r}+\beta_1\dot{r}+\beta_0r = r_\infty-\beta_c-r\hspace{10cm}
% \end{equation*}
% with
% \begin{align*}
% \beta_0= &\, -DM\frac{b}{\tau_wC}-\frac{b}{C}\left(TM+DM\frac{a}{\tau_wC}\partial_\mu\Braket{V}_\infty-DF_\mu\right)\\
% \beta_1= &\, -T+DM\frac{b}{C}\\
% \beta_2= &\quad D\\
% \beta_{\text{c}}= &\,-\left(\ddot{\mu}_{\text{ext}}+\frac{a\left(\Braket{V}_\infty-E_w\right)-\Braket{w}}{\tau_w^2C}\right)DM-\dispddot{\sigma^2_\mathrm{ext}}DS\\
% &+\left(\dot{\mu}_{\text{ext}}-\frac{a\left(\Braket{V}_\infty-E_w\right)-\Braket{w}}{\tau_w C}\right)\left(TM+DM\frac{a}{\tau_wC}\partial_\mu\Braket{V}_\infty-DF_\mu\right)\\
% &\,+\dispdot{\sigma^2_\mathrm{ext}}\left(TS+DM\frac{a}{\tau_wC}\partial_{\sigma^2}\Braket{V}_\infty-DF_{\sigma^2}\right)\\
% \end{align*}

\paragraph{Case i) -- coupling without delays}
% In case of no delays, i.e., $r \equiv r_d$, Eq. \eqref{eq_app_rate_compactified}, becomes
% \begin{equation*}
% \beta_2\ddot{r}+\beta_1\dot{r}+\beta_0 r = r_\infty-\beta_c-r\hspace{10cm}
% \end{equation*}
% \begin{align}
% \begin{aligned}
% r_\infty-r=&\,\ddot{r}\left(D-R\right)\\
% &+\,\dot{r}\left(-T+DM\frac{b}{C}+KJ H_\mu
%   +\,KJ^2 H_{\sigma^2} \right)\\
% &+\,r\left(-DM\frac{b}{\tau_wC}-\frac{b}{C}H_\mu\right)\\
% &-\,\left(\ddot{\mu}_{\text{ext}}+\frac{A}{\tau_w}\right)DM-\dispddot{\sigma^2_\mathrm{ext}}DS \\
% &+\left(\dot{\mu}_{\text{ext}}-A\right) H_\mu 
% +\dispdot{\sigma^2_\mathrm{ext}} H_{\sigma^2}
% \end{aligned}
% \end{align}
% \textcolor{orange}{
% \begin{align*}
% r_\infty-r=&\,\ddot{r}\left(D-DMKJ-DSKJ^2\right)\\
% &+\,\dot{r}\Bigg(-T+DM\frac{b}{C}+KJ\left[TM+DM\frac{a}{\tau_wC}\partial_\mu\Braket{V}_\infty-DF_\mu\right]\\
% &\hspace{1cm}+\,KJ^2\left[TS+DM\frac{a}{\tau_wC}\partial_{\sigma^2}\Braket{V}_\infty-DF_{\sigma^2}\right]\Bigg)\\
% &+\,r\left(-DM\frac{b}{\tau_wC}-\frac{b}{C}\left[TM+DM\frac{a}{\tau_wC}\partial_\mu\Braket{V}_\infty-DF_\mu\right]\right)\\
% &-\,\left(\ddot{\mu}_{\text{ext}}+\frac{A}{\tau_w}\right)DM-\dispddot{\sigma^2_\mathrm{ext}}DS+\left(\dot{\mu}_{\text{ext}}-A\right)\left[TM+DM\frac{a}{\tau_wC}\partial_\mu\Braket{V}_\infty-DF_\mu\right]\\
% &+\dispdot{\sigma^2_\mathrm{ext}}\left[TS+DM\frac{a}{\tau_wC}\partial_{\sigma^2}\Braket{V}_\infty-DF_{\sigma^2}\right]\\
% \end{align*}
% }
For instantaneous synaptic interaction we have $r = r_d$. 
Thus, the coefficients of 
Eq.~\eqref{eq_app_spec2} are obtained by direct comparison with Eq.~\eqref{eq_app_rate_compactified}  which gives
\begin{align}
\beta_2 = &\,D-R, \\
\beta_1 =&\, -T+DM\frac{b}{C}+KJ H_\mu + KJ^2 H_{\sigma^2},\\
\beta_0 = &\, -DM\frac{b}{\tau_wC}-\frac{b}{C}H_\mu,\\
\beta_{\text{c}} = 
&\,-\left(\ddot{\mu}_{\text{ext}}+\frac{a\left(\Braket{V}_\infty-E_w\right)-\Braket{w}}{\tau_w^2 C}\right)DM-\dispddot{\sigma^2_\mathrm{ext}}DS,\\
&+\left(\dot{\mu}_{\text{ext}}-\frac{a\left(\Braket{V}_\infty-E_w\right)-\Braket{w}}{\tau_w C}\right)H_\mu 
  +\dispdot{\sigma^2_\mathrm{ext}} H_{\sigma^2}. \nonumber
\end{align}
% showing that
% % $\beta_2$, $\beta_1$ and $\beta_2$ depend (only) on ...
Note that $\beta_c$ depends explicitly on the population-averged adaptation current $\Braket{w}$ 
as well as on the first and second order time derivatives of the external moments $\mu_\mathrm{ext}$ and $\sigma_\mathrm{ext}^2$. 
% \begin{align*}
% \beta_2= &\,D-DMKJ-DSKJ^2 \\
% \beta_1= &\, -T+DM\frac{b}{C}+KJ\left(TM+DM\frac{a}{\tau_wC}\partial_\mu\Braket{V}_\infty-DF_\mu\right)\\
% &+ KJ^2\left(TS+DM\frac{a}{\tau_wC}\partial_{\sigma^2}\Braket{V}_\infty-DF_{\sigma^2}\right)\\
% \beta_0= &\, -DM\frac{b}{\tau_wC}-\frac{b}{C}\left(TM+DM\frac{a}{\tau_wC}\partial_\mu\Braket{V}_\infty-DF_\mu\right)\\
% \beta_{\text{c}}= &\,-\left(\ddot{\mu}_{\text{ext}}+\frac{a\left(\Braket{V}_\infty-E_w\right)-\Braket{w}}{\tau_w^2 C}\right)DM-\dispddot{\sigma^2_\mathrm{ext}}DS\\
% &+\left(\dot{\mu}_{\text{ext}}-\frac{a\left(\Braket{V}_\infty-E_w\right)-\Braket{w}}{\tau_w C}\right)\left(TM+DM\frac{a}{\tau_wC}\partial_\mu\Braket{V}_\infty-DF_\mu\right)\\
% &+\dispdot{\sigma^2_\mathrm{ext}}\left(TS+DM\frac{a}{\tau_wC}\partial_{\sigma^2}\Braket{V}_\infty-DF_{\sigma^2}\right)\\
% \end{align*}
\paragraph{Case ii) -- exponentially distributed delays}

Here we obtain 
the delayed rate $r_d$ by solving $\dot{r_d}=(r-r_d)/\tau_d$.
Inserting this expression together with its time derivative 
into Eq.~\eqref{eq_app_rate_compactified} results in the coefficients
\begin{align}
\beta_2 = &\, \, D, \\
\beta_1 = &\,-T+DM\frac{b}{C}-\frac{R}{\tau_d}, \\
\beta_0 = &-DM\frac{b}{\tau_wC}-\frac{b}{C}H_\mu
  +\frac{1}{\tau_d}\left( KJH_\mu
  +KJ^2H_{\sigma^2} \right)
  +\frac{R}{\tau_d^2},\\
% \beta_d = &\,-\frac{1}{\tau_d}\Bigg( KJH_\mu
%   +\,KJ^2H_{\sigma^2}\Bigg)
%   -\frac{R}{\tau_d^2} \\
\beta_{\text{c}} = &\,
\,r_d \left(-\frac{1}{\tau_d}\left( KJH_\mu
  +\,KJ^2H_{\sigma^2}\right)
  -\frac{R}{\tau_d^2} \right), \\
&-\left(\ddot{\mu}_{\text{ext}}+\frac{a\left(\Braket{V}_\infty-E_w\right)-\Braket{w}}{\tau_w^2 C}\right)DM-\dispddot{\sigma^2_\mathrm{ext}}DS \nonumber\\
&+\left(\dot{\mu}_{\text{ext}}-\frac{a\left(\Braket{V}_\infty-E_w\right)-\Braket{w}}{\tau_w C}\right) H_\mu
  +\dispdot{\sigma^2_\mathrm{ext}}H_{\sigma^2}, \nonumber
\end{align}
that correspond to those in 
Eqs.~\eqref{eq_meth_beta2}--\eqref{eq_meth_betac}.
% \begin{align}
% \beta_0= &-DM\frac{b}{\tau_wC}-\frac{b}{C}\left(TM+DM\frac{a}{\tau_wC}\partial_\mu\Braket{V}_\infty -DF_\mu \right)\\
% &+\frac{1}{\tau_d}\Bigg( KJ\left(TM+DM\frac{a}{\tau_wC}\partial_\mu\Braket{V}_\infty-DF_\mu\right)\\
% &\hspace{1cm}+KJ^2\left(TS+DM\frac{a}{\tau_wC}\partial_{\sigma^2}\Braket{V}_\infty-DF_{\sigma^2}\right) \Bigg)\\
% &-\frac{1}{\tau_d^2}\left(-DMKJ-DSKJ^2\right)\\
% \beta_1= &\,-T+DM\frac{b}{C}+\frac{1}{\tau_d}\left(-DMKJ-DSKJ^2 \right) \\
% \beta_2= &\, \, D \\
% \beta_d = &\,-\frac{1}{\tau_d}\Bigg( KJ\left(TM+DM\frac{a}{\tau_wC}\partial_\mu\Braket{V}_\infty-DF_\mu\right)\\
% &\hspace{1cm}+\,KJ^2\left(TS+DM\frac{a}{\tau_wC}\partial_{\sigma^2}\Braket{V}_\infty-DF_{\sigma^2}\right)\Bigg)\\
% &+\frac{1}{\tau_d^2}\left(-DMKJ-DSKJ^2\right) \\
% \beta_{\text{c}}= &\,-DM\left(\ddot{\mu}_{\text{ext}}+\frac{a\left(\Braket{V}_\infty-E_w\right)-\Braket{w}}{\tau_w^2 C}\right)-DS\dispddot{\sigma^2_\mathrm{ext}}\\
% &+\left(\dot{\mu}_{\text{ext}}-\frac{a\left(\Braket{V}_\infty-E_w\right)-\Braket{w}}{\tau_w C}\right)\left(TM+DM\frac{a}{\tau_wC}\partial_\mu\Braket{V}_\infty-DF_\mu\right)\\
% &+\dispdot{\sigma^2_\mathrm{ext}}\left(TS+DM\frac{a}{\tau_wC}\partial_{\sigma^2}\Braket{V}_\infty-DF_{\sigma^2}\right),
% \end{align}
% i.e., 
% % $\beta_2$, $\beta_1$ and $\beta_2$ depend (only) on ...
Here $\beta_c$ depends explicitly on the delayed spike rate $r_d$ 
(in addition to $\Braket{w}$ as well as the first and second order time derivatives of 
$\mu_\mathrm{ext}$ and $\sigma_\mathrm{ext}^2$ as in the case without delays).

\paragraph{Case iii) -- identical delays}

The delayed spike rate in this situation is given by $r_d(t) = r(t-d)$.
Inserting the first and second order time derivative of this identity
into Eq.~\eqref{eq_app_rate_compactified} yields

\begin{align}
\beta_2 = &\, \, D, \\
\beta_1 = &\,-T+DM\frac{b}{C}, \\
\beta_0 = &-DM\frac{b}{\tau_wC}-\frac{b}{C}H_\mu,\\
% \beta_d = &\,-\frac{1}{\tau_d}\Bigg( KJH_\mu
%   +\,KJ^2H_{\sigma^2}\Bigg)
%   -\frac{R}{\tau_d^2} \\
\beta_{\text{c}} = 
&-\, \ddot{r}(t-d) R 
\; +\,\dot{r}(t-d)\left( KJ H_\mu+KJ^2H_{\sigma^2}\right) \\
&-\left(\ddot{\mu}_{\text{ext}}+\frac{a\left(\Braket{V}_\infty-E_w\right)-\Braket{w}}{\tau_w^2 C}\right)DM-\dispddot{\sigma^2_\mathrm{ext}}DS \nonumber\\
&+\left(\dot{\mu}_{\text{ext}}-\frac{a\left(\Braket{V}_\infty-E_w\right)-\Braket{w}}{\tau_w C}\right) H_\mu
  +\dispdot{\sigma^2_\mathrm{ext}}H_{\sigma^2}, \nonumber
\end{align}

% \todo{@Fabian: implement this case and also the ``shifted exponentially distributed delays'' into the code: use for $\ddot r(t-d)$ and $\dot r(t-d)$ the respective value of the right hand side of the equivalent first order ODE for 
% $dr$ and $r$ before the number of timesteps corresponding to $d$ (which can be overriden in each timestep if no history is available but can also use the history this is up to you) -- for shifted exp. delays similarly (use the rhs of $\dot r_d$ and of its time derivative $\ddot r_d$ that involves $\dot r(t-d)$, cf. remark below)}

Here $\beta_c$ depends explicitly on $\ddot r(t-d)$ and $\dot r(t-d)$ 
(in addition to $\Braket{w}$ as well as the first and second order time derivatives of 
$\mu_\mathrm{ext}$ and $\sigma_\mathrm{ext}^2$ as in the case without delays). 
% Note that the total input moments in this scenario also contain 
% an (actual) delay expression since $r(t-d)$ is not a state variable at time $t$ but evaluated in the past.

\paragraph{Remarks}

\begin{itemize}

 \item Equation~\eqref{eq_app_spec2} is an ordinary differential spike rate model 
 for the cases (i) and (ii), i.e., without or exponentially distributed delays, 
 while for identical delays (case iii) delayed variables occur explicitly in $\beta_c$ and due to $r_d(t)=r(t-d)$ also 
 in any model quantity via the total input moments.
 
 \item For exponentially distributed delays with an identical shift $d$, 
 i.e., $p_d(\tau) = \exp[-(\tau-d)/\tau_d] /\tau_d$ with $\tau \ge d$, the delayed 
 spike rate $r_d$ satisfies $\dot{r_d}(t)=[r(t-d)-r_d(t)]/\tau_d$. In this situation the coefficients ($\beta_2$, $\beta_1$ and $\beta_0$) are identical to those of 
 case iii) except for $\beta_c$ which 
 is modified and depends on $r_d(t)$, $r(t-d)$, $\dot{r}(t-d)$ (and the parameter $\tau_d$), i.e., here Eq.~\eqref{eq_app_spec2} also represents a delay differential model.
 
 \item Any delay distribution $p_d$ from the exponential family 
 can be incorporated similarly as for the specific instance 
 of an exponentially distribution (cf. case ii) 
 to yield (non-delayed) coefficients of Eq.~\eqref{eq_app_spec2} 
 by using the equivalent representation of the delayed spike rate $r_d$ 
 as an ODE (system).
 
 \item The scenario of an uncoupled population is obtained
from any of the three cases by setting the number of presynaptic neurons $K$ to zero (implying $R=0$).

 \item As an alternative to derive and simulate the explicit model, Eq.~\eqref{eq_app_spec2}, 
  one can directly integrate Eq.~\eqref{eq_app_rate_nonexpanded} numerically by 
  replacing the first two time derivatives of 
  the total moments $\mu$ and $\sigma^2$ by finite (backward) differences 
  in each timestep. This approach avoids lengthy expressions and 
  might be especially useful when considering 
 multiple interacting populations irrespective of the delay distribution.
 
%  
%  \item When considering multiple interacting neuronal populations, i.e., 
%  $\mu = \mu_\mathrm{ext}(t) + \sum_\gamma K_\gamma J_\gamma r_\gamma$
\end{itemize}

% 
% 
% Another possible synaptic coupling is to use 
% constant delays $d_{ij} \equiv d > 0$ or exponentially distributed delays with 
% a constant shift, i.e., $d_{ij} = d + \tilde d_{ij}$ where $\tilde d_{ij}$ is 
% drawn from an exponential distribution. The same derivation as above can 
% be applied, but the final spike rate model 
% is not an ordinary differential eq. anymore but rather 
% involves the first two time derivatives of the spike rate $\dot r(t-d)$, $\ddot r(t-d)$, i.e., 
% Eq.~\eqref{eq_app_spec2} becomes a delay differential eq. of special form in 
% this case that can of course also be numerically integrated but 
% might not be straight-forward to analyze.
% 
% \todo{@MA: make note for the last two cases in terms of eqs for 
% const. delay case(s) -- use numerical version cf. cascade !!}
% 
% \todo{Add note that the case of multiple populations the no delay situation is
% rather complicated but numerically, the finite difference discretization of $\dot \mu$ can be taken also for this case (see const delay as reference).}

\bigskip

\subsection*{Numerical solver for the nonlinear Fokker-Planck eigenvalue problem}

% \todo{check that this section explains well enough also the case without refractory period (explicitly state this special case everywhere)}

\paragraph{Problem statement}
The (main) objective is to find the eigenvalues $\lambda_n$ of the Fokker-Planck 
operator $\LL$ which are the 
solutions of the complex-valued Eq.~\eqref{eq_meth_spec_root}, 
\begin{equation} \label{app_root}
\lambda \mapsto q_\phi(V_\mathrm{lb}; \lambda) \overset{!}{=} 0.
\end{equation} 
Each evaluation of $q_\phi(V_\mathrm{lb}; \lambda)$ 
involves a backward integration of Eq.~\eqref{eq_meth_spec_eigeneq_sys}, 
\begin{align} 
% in components:
% \label{eq_meth_spec_eigeneq_sys1}
% % -q_{\phi}'(V)
% & -\frac{\partial q_{\phi}}{\partial V} = \lambda \phi \\
% & -\frac{\partial \phi}{\partial V} = \frac{q_\phi - (g(V) + \mu)\phi}{\sigma^2/2} 
% \label{eq_meth_spec_eigeneq_sys2}
%
% matrix notation:
\label{app_eigeneq_sys}
% -q_{\phi}'(V)
& -\frac{d}{dV} 
% \mathbf{
  \begin{pmatrix}
q_{\phi}
\\
\phi
\end{pmatrix} = 
% \begin{pmatrix}
% \lambda \phi
% \\
% \dfrac{q_\phi - (g(V) + \mu)\phi}{\sigma^2/2} 
% \end{pmatrix}
% = 
\underbrace{
\begin{pmatrix}
0 & \lambda
\\
\frac{2}{\sigma^2}  & - 2\frac{g(V) + \mu}{\sigma^2}
\end{pmatrix} }_{ = \mathbf{A}}
\begin{pmatrix}
q_{\phi}
\\
\phi
\end{pmatrix},
\end{align}
which is initialized according to the absorbing boundary condition 
(cf.  Eq.~\eqref{eq_meth_spec_phi_absorb}),
\begin{equation}  \label{app_phi_absorb}
\phi(V_\mathrm{s}) = 0
\end{equation} 
and 
the arbitrary choice 
$q_\phi(V_\mathrm s) \in \CC \setminus \{0\}$ 
(due to the linearity of the problem),
and which furthermore has to take into account the 
(generalized) reinjection condition, 
Eq.~\eqref{eq_meth_spec_phi_reinjcond_ref}, 
\begin{equation} \label{app_phi_reinjcond_ref}
q_{\phi}(V_\mathrm{r}^-) 
= q_{\phi}(V_\mathrm{r}^+) - q_{\phi}(V_{\mathrm{s}})
e^{-\lambda T_\mathrm{ref}}.
\end{equation}
Note that the latter corresponds for $T_\mathrm{ref} = 0$ 
to the reinjection condition that does not include the refractory period, 
i.e., Eq.~\eqref{eq_meth_spec_phi_reinjcond}.

\paragraph{Parameter-dependent solution} \label{sec_paramspace}

The eigenvalues $\lambda_n$, the associated eigenfunctions 
$\phi_n(V)$ of $\LL$ 
and $\psi_n(V)$ of $\LL^*$ are required 
for a rectangle of input parameter values $(\mu,\sigma)$.
Using the property the eigenvalues are real-valued for sufficiently small mean input $\mu$ 
and that they furthermore 
continuously depend on both (input) parameters, mean $\mu$ and standard deviation $\sigma$, we establish the following solution algorithm: 
\begin{enumerate}
 \item Discretize the input parameter rectangle, 
 $\{ (\mu_k,\sigma_\ell) \}$, $k=1,\dots,M_\mu$, $\ell=1,\dots,M_\sigma$ 
 with small spacings $\Delta \mu$, $\Delta \sigma$.
% with $\mu_1 = \mu_\mathrm{min}$.
% equidistantly in each direction \textcolor{gray}{with 
% $\mu_1 = \mu_\mathrm{min}$, 
% $\mu_{M_\mu} = \mu_\mathrm{max}$, 
% $\sigma_1 = \sigma_\mathrm{min}$, 
% $\sigma_{M_\sigma} = \sigma_\mathrm{max}$}
% with small spacings %$\Delta \mu$, $\Delta \sigma$ 
% between the , % \in \mathcal R_\Delta$, 
 
 \item For $\mu_1 = \mu_\mathrm{min}$ evaluate
$q(V_\mathrm{lb}; \lambda)$ 
% densely 
with high resolution on a real negative interval $[\lambda_\mathrm{min},0)$ 
with sufficiently small 
$\lambda_\mathrm{min} \ll 0$ such that at least 
$N_\lambda$ eigenvalues are found. The zero-crossings of $q(V_\mathrm{lb}; \lambda)$ 
yield the eigenvalues 
$\lambda_1(\mu_1, \sigma_\ell)$, $\lambda_2(\mu_1, \sigma_\ell), \dots$, $\lambda _{N_\lambda}(\mu_1, \sigma_\ell)$, 
cf. the respectively attached axes in Fig.~\ref{fig7_spectral}A.

\item Use the computed eigenvalues $\lambda_n(\mu_1, \sigma_\ell)$ 
as initial approximations $\tilde \lambda_n$ for the 
target eigenvalues at the 
next larger mean input, $\lambda_n(\mu_2, \sigma_\ell)$ and 
iteratively solve Eq.~\eqref{app_root}
with Powell's hybrid 
method. Note this can yield a complex eigenvalue for a real initialization 
(close to the real-to-complex transition).

\item Repeat the last step by taking the eigenvalues at $\mu_{k-1}$ 
as initial approximation for $\mu_{k}$ 
where
$k=3,\dots,M_\mu$. 
\end{enumerate}

Since this procedure is independent of the sequential $\sigma_\ell$ (and $n$) 
order it can be computed in parallel for $\ell=1,\dots,M_\sigma$ 
(and $n=1,\dots,N_\lambda$). 

The nonlinear solver, Powell's hybrid method, 
approximates the Jacobian of the equivalent two-dimensional real nonlinear 
system of the complex function $q(V_\mathrm{lb}; \lambda)$ with 
finite difference step size $\Delta \lambda$ and stops iterating
when the relative convergence tolerance $\varepsilon$ is reached implying 
a solution has been found.
Since this root finding method (that we apply to solve Eq.~\eqref{app_root}) 
converges locally, the input parameter rectangle 
has to be discretized sufficiently fine, i.e, $\Delta \mu$ and $\Delta \sigma$ 
have to be small. 
Otherwise artefacts as jumps between the eigenvalue curves could occur, especially when $\lambda_n(\mu)$ is steep (cf. Fig.~\ref{fig7_spectral}A) 
and another eigenvalue is close by (in $\lambda$ space). 
% \textcolor{gray}{i.e. close to the transition of real to complex $\lambda_n$}. 
% The convergence to $\lambda_n$ is typically quite fast 
% (usually not more than 10 iterations are required)\todo{@MA: Check this}, 

% 
% The parameters.....

% We solve the root finding problem, Eq.~\eqref{eq_meth_spec_root} 
% to yield a target eigenvalue $\lambda_n$ . 
% 
% 
% 
% 
% x $\varepsilon_\mathrm{rel}$ 
% further solver params?
% 
% 
% (for example a variant of Newton's 
% method) given that a sufficiently close initial approximation 
% $\tilde \lambda_n \in \CC$ is available. 
% For example in our Python implementation we apply Powell's hybrid 
% method 
% ... finite difference step size .
% 
% $\lambda_\mathrm{min}$
% $N_\lambda$
% $\mu_\mathrm{min}$

We use for the solver parameters the values of Table~\ref{tableb1_specsolver}, 
which are suitable for the network model of this study as 
parametrized by Table~\ref{table1_paramsall} of the main text. 

\begin{table}[!ht]
% \begin{adjustwidth}{-2.25in}{0in} % Comment out/remove adjustwidth environment if table fits in text column.
\caption{
{\bf Parameter values of the spectral solver.}}
\begin{tabular}{|l|c|c|}
\hline
Name & Symbol & Value \\ \hline 
\hline
% \multicolumn{3}{|c|}{\bf Spectral solver} \\ \hline
Spacing of input rectangle, mean input & $\Delta \mu$ & 0.005~mV/ms$^1$ \\ \hline
Spacing of input rectangle, standard dev. & $\Delta \sigma$ & 0.1~mV/$\sqrt{\mathrm{ms}}$ \\ \hline
Membrane voltage discretization width & $\Delta V$ & 0.01~mV \\ \hline
Smallest mean input (real spectrum) & $\mu_\mathrm{min}$ & -1.5~mV/ms \\ \hline
Number of eigenvalues & $N_\lambda$ & 10 \\
\hline
Finite difference step size (MINPACK: EPS) & $\Delta \lambda$  & 1e-10~kHz \\ \hline
Relative convergence tolerance (MINPACK: XTOL) & $\varepsilon$ & 1e-8 \\ \hline
% Absolute convergence tolerance & $\varepsilon_\mathrm{abs}$ & 1e-4~kHz \\ \hline
Finite difference step size (quantities), mean input & $\delta \mu$ &  0.001 mV/ms \\ \hline
Finite difference step size (quantities), standard dev. & $\delta \sigma$ & 0.001~mV/$\sqrt{\mathrm{ms}}$ \\ \hline 
\end{tabular}
\begin{flushleft}
$^1$Fine spacing required (in comparison to the value of $\Delta \mu$ used for
the quantity precalculation of the cascade based models) 
due to the continuous tracking of the eigenvalues.
Note that after the calculation we downsample the spectrum to the same mean input 
resolution for comparability.
% prevent errors in sensitive regimes (close to double eigenvalue or other eigenvalue close by).
\end{flushleft}
% \begin{flushleft} 
% $^1$If not specified otherwise. 
% $^2$For a full model set in the systematic comparison of Sect. Results (a nonzero refractory period is not supported by spec$_2$). 
% $^3$Parameters for precalculation of quantities required during simulation. 
% \\
% The values of coupling parameters ($ K $, $ J $, $ \tau_d $) are specified in the captions of Figs.~\ref{fig1_example} and \ref{fig4_network_oscillations}, the values of parameters for the external input ($\bar\mu$, $\tau^\mu_\mathrm{ou}$, $\vartheta_\mu$, 
% $\overline{\sigma^2}$, $\tau^{\sigma^2}_\mathrm{ou}$, $\vartheta_{\sigma^2}$) are provided in each figure (caption).
% \end{flushleft}
\label{tableb1_specsolver}
% \end{adjustwidth}
\end{table}

\paragraph{Exponential integration} \label{sec_expint}

A major factor for efficiency and accuracy of the algorithm above 
is the particular numerical way in which the backward integration of the 
differential equation 
system~\eqref{app_eigeneq_sys} is performed since this 
corresponds to one evaluation of the nonlinear function 
$\lambda \mapsto q(V_\mathrm{lb}; \lambda)$.
% and determines therefore the cost as well as a bound on the 
% numerical accuracy of a single Newton-like step. 
% General ODE
% \todo{define this abbreviation earlier?}{} 
% solvers such as the explicit Euler method but also 
% higher order Runge-Kutta or multistep methods 
% require extremely small step sizes $\Delta V$ 
% % have poor convergence behaviour 
% due to the strong nonlinearity $g(V)$ 
% \textcolor{gray}{in the 
% coefficient matrix.}
% $\mathbf{A}$ 
% as a consequence of the exponentially increasing 
% current $I_\mathrm{exp}$ close to $V_\mathrm{s}$. 
An efficient and accurate discretization scheme is 
% found by exploiting 
% the linear form of Eq.~\eqref{eq_meth_spec_eigeneq_sys} and 
to perform exponential integration steps, i.e., 
% % \begin{equation} \label{eq_meth_spec_expint_matrix}
%   \begin{pmatrix}
% q^{m-1}
% \\
% \phi^{m-1}
% \end{pmatrix}
% % = \exp \left( \mathbf A_{m-\frac{1}{2}} \right) 
% =
% \exp
% \left(
% \begin{bmatrix}
% 0 & \lambda
% \\
% \frac{2}{\sigma^2}  & - 2\frac{g\left(V_{m-\frac{1}{2}} \right) + \mu}{\sigma^2}
% \end{bmatrix}
% \Delta V
% \right)
% \begin{pmatrix}
% q^m
% \\
% \phi^m
% \end{pmatrix}
% \end{equation}
\begin{equation} \label{app_expint}
(
q^{m-1},
\phi^{m-1})^T
=
\exp
\left[
\mathbf A \left( V_{m-\frac{1}{2}} \right) \Delta V 
\right ]
(q^{m},
\phi^{m})^T
\end{equation}
with
% \footnote{\textcolor{gray}{superscript notation because 
% subscripts are reserved for eigenvalue index}} 
$q^m = q_{\phi}(V_m)$ and $\phi^m = \phi(V_m)$ on an equidistant membrane voltage 
grid $V_m = V_\mathrm{lb}+m\Delta V$ ($m=0,\dots,N_V$ and $V_{N_V}=V_\mathrm{s}$).
This scheme involves the matrix exponential function,
$\exp(\mathbf A \Delta V) = \sum_{j=0}^\infty (\mathbf A\Delta V)^j/j!$,  
that is
inexpensively 
evaluated as an (equivalent) linear combination of 
$\mathbf A\Delta V$  and the identity matrix
% \textcolor{gray}{due to the theorem of 
% Cayley and Hamilton} 
\cite{Bernstein1993}. 
This second order convergent numerical integration scheme that 
exploits the linearity of the system, Eq.~\eqref{app_eigeneq_sys}, 
is obtained by truncating the Magnus expansion 
of the exact solution 
% \textcolor{gray}{with nonlinear coefficients} 
after one term 
% \textcolor{gray}{
and approximating the occuring integral
$\int_{V_{m-1}}^{V_{m}} \mathbf A(V) dV$ 
using the mid point rule
% }
\cite{Hochbruck2010}. 
Note that the 
matrices $\mathbf{A}(V)$ and $\mathbf{A}(\tilde V)$ 
do not commute for $V\ne\tilde V$ in general which implies that the solution of Eq.~\eqref{app_eigeneq_sys} does not have a simple exponential
representation but is rather described by an (infinite) Magnus series.
For the perfect 
integrate-and-fire model, though, the scheme, 
Eq.~\eqref{app_expint}, gives the exact solution  of Eq.~\eqref{app_eigeneq_sys}
% for arbitrary large $\Delta V$ 
% (given the reset cell is on the grid)
as the coefficient matrix $\mathbf A$ is constant due to $g(V) = 0$ in this case. 
% \todo{@MA: note on efficiency of matrix exp evaluation - caley hamilton simple expression (see paper)}
% \textcolor{gray}{Note that a semi-exponential integration (cf. \cite{RichardsonThreshIntCitation}) where 
% the $q_\phi$ dynamics is integrated using the explicit Euler method and (only) $\phi$ is calculated using a (one-dimensional) exponential 
% scheme 
% % version of Eq.~\eqref{eq_meth_spec_expint} 
% does require extremely small membrane voltage steps 
% $\Delta V$ to accurately determining the eigenfunctions 
% and is thus not efficient 
% enough for the large input rectangle $\mathcal R$.}

The integration of Eq.~\eqref{app_eigeneq_sys} is initialized 
at $V_\mathrm{s}$ with 
$\phi^{N_V} = 0$ 
to satisfy the absorbing boundary, 
Eq.~\eqref{app_phi_absorb}),
and with the arbitrary choice $q^{N_V} = 1$ 
(possible due to the linearity of both Eq.~\eqref{app_eigeneq_sys} and 
the boundary conds).
The exponential scheme, Eq.~\eqref{app_expint}, 
is then used to (backward) calculate 
$(q^m, \phi^m)$, $m={N_V-1},\dots,m_\mathrm{r}$, 
where the reset voltage is assumed to be contained in the grid, i.e.,
$V_{m_\mathrm{r}} = V_\mathrm{r}$.
At the reset voltage $V_\mathrm{r}$ the (generalized) reinjection cond., 
% $q_{\phi}(V_\mathrm{r}^-) 
% = q_{\phi}(V_\mathrm{r}^+) - q_{\phi}(V_{\mathrm{s}})
% e^{-\lambda T_\mathrm{ref}}
% $, cf. 
Eq.~\eqref{app_phi_reinjcond_ref}, 
is applied by
\begin{equation} \label{app_phi_reinjcond_discrete}
q^{m_\mathrm{r}} \leftarrow q^{m_\mathrm{r}} - 
q^{N_V} \exp(-\lambda T_\mathrm{ref}).
\end{equation}
Continuing the backward integration using the scheme of 
Eq.~\eqref{app_expint} again for 
$m=m_\mathrm{r}-1,\dots,0$ finally gives the values
$q^0$ and $\phi^0$ at $V_\mathrm{lb}$. 
Therefore, $q_0(\lambda) \overset{!}{=} 0$ corresponds to 
the root finding problem, Eq.~\eqref{app_root}, 
after (exponential) membrane voltage 
discretization,
% in semi-discrete representation 
% \textcolor{gray}{($\lambda$ has still an uncountable range $\CC$)}
and a value of $q_0=0$ indicates that $\lambda$ 
is an eigenvalue with respective 
eigenfunction $\phi(V)$ in discrete representation ($\phi^m, m=0,\dots,N_V$). 
% Combining the exponential integration, Eq.~\eqref{eq_meth_spec_expint}, 
% of system \eqref{eq_meth_spec_eigeneq_sys} with the algebraic 
% root finding problem,  Eq.~\eqref{eq_meth_spec_root}, 
% \textcolor{gray}{ changes the latter to become  $q_0(\lambda) \overset{!}{=} 0$,
% which} 
% Note that the eigen
% \textcolor{gray}{leading to 
% a jump in the flux $q$/a kink in $\phi$ at the reset}, 

\paragraph{Adjoint operator}

To calculate the eigenfunctions of the adjoint operator $\LL^*$ (cf. Eq.~\eqref{eq_meth_spec_eigeneq_adjoint_sparse}--\eqref{eq_meth_spec_adj_bc_cont}) we
assume that an eigenvalue $\lambda_n$ is given
(obtained for example using the procedure 
described in the previous two sections).
Eq.~\eqref{eq_meth_spec_eigeneq_adjoint_sparse}, i.e., 
$\LL^*[\psi_n] = \lambda_n \psi_n$, can be rewritten 
as a linear second order system for $(\psi_n, d\psi_n)^T$, 
\begin{equation} \label{app_eigeneq_adjoint_sys}
\frac{d}{dV}
\begin{pmatrix}
\psi_n
\\
d\psi_n 
\end{pmatrix}
= 
\underbrace{\begin{pmatrix}
0 & 1
\\
\frac{2 \lambda_n}{\sigma^2} 
& -2 \frac{g(V) + \mu}{\sigma^2}
\end{pmatrix}}_{ = \mathbf B}
\begin{pmatrix}
\psi_n
\\
d\psi_n  
\end{pmatrix}
\end{equation}
with $d\psi_n = \partial_V \psi_n$ and
(nonlinear) coefficient matrix $\mathbf B$. 
This system is exponentially integrated 
\textit{forwards} 
from the lower bound $V_\mathrm{lb}$ to the 
spike voltage $V_\mathrm{s}$. Specifically 
we define $\psi_n^m = \psi_n(V_m)$ and $d\psi_n^m = \partial_V \psi_n(V_m)$ 
on the same grid as 
in the previous section.
The integration is initialized
according to the boundary cond. at $V_\mathrm{lb}$
(cf. Eq.~\eqref{eq_meth_spec_dpsi_lb_bc2}), i.e., $d\psi_n^0 = 0$
together with the arbitrary choice $\psi^0_n = 1$ due to 
the linearity of the problem. 
Then we calculate the values 
$\psi_n^m$, $d\psi_n^m$ ($m=1,\dots,N_V$) using the exponential integration scheme
% \begin{equation} \label{eq_meth_spec_expint_adjoint_matrix}
% \begin{pmatrix}
% \psi^{m+1}
% \\
% d\psi^{m+1} 
% \end{pmatrix}
% = \exp
% \left(
% \begin{bmatrix}
% 0 & 1
% \\
% \frac{2 \lambda_n}{\sigma^2} 
% & -2 \frac{g\left(V_{m+\frac{1}{2}}\right) + \mu}{\sigma^2}
% \end{bmatrix}
% \Delta V
% \right)
% \begin{pmatrix}
% \psi^m
% \\
% d\psi^m
% \end{pmatrix}.
% \end{equation}

\begin{equation} \label{app_expint_adjoint}
(
\psi^{m+1}_n,
d\psi^{m+1}_n 
)^T
= \exp \left [ \mathbf B \left(V_{m+\frac{1}{2}}\right) \Delta V\right]
(\psi^m_n, 
d\psi^m_n)^T. 
\end{equation}
The (generalized) boundary cond., Eq.~\eqref{eq_meth_spec_psi_bc1_ref},
$\psi^{N_V}_n = \psi^{m_{r}}_n \exp(-\lambda_n T_\mathrm{ref})$ 
is necessarily fullfilled because $\lambda_n$ was assumed to be an eigenvalue. 
% the continuity condition, Eq.~\eqref{eq_meth_spec_adj_bc_cont}, is satisfied 
% through the integration, Eq.~\eqref{}, that does not include a 
% jump .
This implies that $\psi^m_n$ 
is the corresponding (everywhere 
continuously 
differentiable, cf. main text) eigenfunction $\psi_n(V)$ in discrete form.
% Note that the boundary conditions of the adjoint operator $\LL^*$ 
% do not impose a discontinuity in $\psi(V)$ or its derivative $d\psi(V)$
% in contrast to the operator $\LL$ that leads to eigenfunctions $\phi(V)$ with 
% jumps at the reset voltage $V_\mathrm{r}$ (see also Fig....XYZ).
Note that the generalized boundary condition above corresponds for $T_\mathrm{ref} = 0$ 
to the respective condition that does not include the refractory period, 
i.e., Eq.~\eqref{eq_meth_spec_psi_bc1}.

\paragraph{Quantities}

% To obtain 
% \textcolor{gray}{on $\mathcal R$} 
The quantities that are required by the 
spike rate models spec$_1$ (Eq.~\eqref{eq_meth_spec1}), and spec$_2$ 
(Eq.~\eqref{eq_meth_spec2}), i.e., $\lambda_1$, $\lambda_2$, $r_\infty$, $\partial_x r_\infty$, $\langle V \rangle_\infty$, $\partial_x \langle V \rangle_\infty$, $f_n$, $c^x_n$, 
for $x=\mu,\sigma^2$ and $n=1,2$,
are calculated for each 
mean $\mu_k$ and standard deviation $\sigma_\ell$ of the input rectangle as follows.

Applying the exponential integration scheme Eqs.~\eqref{app_expint},\eqref{app_phi_reinjcond_discrete} 
for the eigenvalue $\lambda_0 = 0$ gives the 
(unnormalized) eigenfunction $\hat \phi_0$ which is 
proportional to the stationary distribution $p_\infty$. 
% 
% The latter corresponds to the stationary distribution (of non-refractory neurons), $p_\infty = \phi$, after normalization, i.e., 
After normalizing $\hat\phi_0$ to yield a probability density, i.e., 
$\phi_0 = \hat \phi_0 / \int_{V_\mathrm{lb}}^{V_\mathrm{s}} \hat \phi_0(V) dV$, the 
stationary quantities, mean membrane voltage
$\langle V \rangle_\infty = \int_{V_\mathrm{lb}}^{V_\mathrm{s}} V \phi_0(V) dV$ 
and spike rate $r_\infty = q_{\phi_0}(V_\mathrm s)$, are calculated.
Practically, the latter is 
given by the (scaled) flux initialization 
of the exponential backward integration, 
$r_\infty = q^{N_V}/\int_{V_\mathrm{lb}}^{V_\mathrm{s}} \hat\phi_0(V) dV$,
% , where
% $\phi_0$ is taken before normalization. 
which is -- for a refractory period $T_\mathrm{ref} > 0$ -- denoted with 
$\tilde r_\infty$ giving after scaling the steady-state spike rate
$r_\infty = \tilde r_\infty /(1+\tilde r_\infty T_\mathrm{ref})$.
% \todo{@MA: later adapt this 
% % part to take into account clamping w or non-clamping w and spike shape vs. clamping V?}. 
The optional spike shape extension can be incorporated in this case 
% into the calculation of $\langle V \rangle_\infty$ 
using Eq.~\eqref{eq_mean_v_spikeshape} with $p_\infty = \phi_0 /(1+\tilde r_\infty T_\mathrm{ref})$.
% , where the steady-state spike rate $r_\infty$ is taken 
% before scaling.
% is the steady-state distribution of the non-refractory neurons.
%   $\langle V \rangle_\infty = 
%   \int_{-\infty}^{V_{\mathrm{s}}} v \phi_0(v) dv + 
%   \left( 1 - \int_{-\infty}^{V_{\mathrm{s}}} \phi_0(v) dv \right)
%   (V_{\mathrm{r}}+V_{\mathrm{s}})/2$
% with adapted normalization $\int_{V_\mathrm{lb}}^{V_\mathrm{s}} \phi_0(V) dV \overset{!}{=} 1$

% 
% into the 
% above calculation of $\langle V \rangle_\infty$ .

To obtain the first two dominant eigenvalues $\lambda_1$ and $\lambda_2$
we use the procedure of Sects.~\ref{sec_paramspace}, \ref{sec_expint} and calculate
a number $N_\lambda$ of (nonstationary) eigenvalues 
$\lambda_n(\mu_k,\sigma_\ell)$, $n=1,\dots,N_\lambda$ for the given input parameter 
rectangle. 
% , i.e., $N_\lambda$ 
% eigenvalues. 
These eigenvalues are sorted, for each input parameter pair 
$(\mu_k,\sigma_\ell)$ separately, such that $\lambda_1$ and $\lambda_2$ are the first and second dominant 
eigenvalue, respectively, according to Eqs.~\eqref{eq_meth_lambda1},\eqref{eq_meth_lambda2}, cf. Fig.~\ref{fig7_spectral}A,C. 
Note that the other eigenvalues, $n=3,\dots,N_\lambda$, are 
not used for the models spec$_1$ and spec$_2$. However, they 
are required within the numerical solution method to account for the points 
in input parameter space $(\mu,\sigma)$, 
where the eigenvalue class switches 
(due to $V_\mathrm{lb} \ne V_\mathrm{r}$, see main text). For example in
Fig.~\ref{fig7_spectral}A (right column, i.e., 
with large noise intensity $\sigma$) 
a diffusive mode is dominant for small mean input $\mu$ 
while for increased mean the
dominant eigenvalue (pair) is from the regular type. 
The numerical procedure described above starts 
with the dominant $N_\lambda$ (real) eigenvalues at the smallest 
mean input and then continuously tracks 
each of these eigenvalues for increasing mean input $\mu$.
Therefore, when computing only, e.g., $N_\lambda = 2$ eigenvalues 
for the previous example both would be of the diffusive kind and 
the dominant regular modes for larger $\mu$ cannot be found.

The nonstationary quantities
are based on the (already calculated) dominant eigenvalues 
$\lambda_1$ and $\lambda_2$. First the corresponding (unnormalized) eigenfunctions 
$\tilde\phi_1$ and $\tilde\phi_2$ of $\LL$  are obtained
using the exponential integration scheme, Eqs.~\eqref{app_expint},\eqref{app_phi_reinjcond_discrete} with $\lambda=\lambda_1,\lambda_2$,
as well as those of $\LL^*$ ($\psi_1$ and $\psi_2$) that are
computed via Eq.~\eqref{app_expint_adjoint}. 
The eigenfunctions of $\LL$ are then scaled according to
% $\langle \psi_n, \phi_n\rangle \overset{!}{=} 1$ 
$\phi_n = \tilde \phi_n / \langle \psi_n, \tilde \phi_n \rangle$ 
which yields (bi)orthonormal eigenfunctions, i.e., 
$\langle \psi_n, \phi_m \rangle = \delta_{nm}$, 
% , and the biorthonormality condition Eq.~\eqref{eq_meth_spec_biorth} is fulfilled for $n,m=1,2$).
and this fixes the remaining degree of freedom for 
products between quantities of $\LL$ and $\LL^*$, e.g., $c^{\sigma^2}_n f_n$. 
% compute the nonstationary quantities, i.e,
% % that are required. 
% % by the simple and 
% % advanced spike rate model
% % (Eqs.~\eqref{eq_meth_spec1} and \eqref{eq_meth_spec2}, respectively), 
% % that is. 
% % The nonstationary quantities 
% those that depend either on the eigenfunctions
% of $\LL$ (such as $f_n$) or on those of $\LL^*$ (e.g. $c^{\sigma^2}_n$). 
% The eigenfunctions 
% and therefore
% requiring the availability of $\phi_1$ and $\phi_2$ as well as $\psi_1$ and $\psi_2$. 
Note that in the  spec$_2$ model, 
Eqs.~\eqref{eq_meth_spec2}--\eqref{eq_meth_betac}, 
nonstationary quantities occur 
exclusively in such products, specifically,
% \textcolor{gray}{/products}, 
% and exclusively in
$\mathbf{f}\cdot\mathbf{c}_x = 
c^x_1 f_1 + c^x_2 f_2 $ and 
$\mathbf{f}\cdot\mathbf{\Lambda}\,\mathbf{c}_x 
= c^x_1 f_1 \lambda_1 + c^x_2 f_2 \lambda_2$ (for $x=\mu,\sigma^2$), 
and they do not enter at all the  spec$_1$ model, Eq.~\eqref{eq_meth_spec1}, 
except for the first dominant eigenvalue $\lambda_1$. 

The nonstationary quantities of $\LL$ are obtained 
% after the specific (bi)normalization
% $\phi_n \leftarrow \phi_n / \langle \psi_n, \phi_n \rangle$ by 
by $f_1 = q_{\phi_1}(V_\mathrm s)$ and $f_2=q_{\phi_2}(V_\mathrm s)$, and 
particularly (similar to $r_\infty$ above)
by ``reading off'' the respective 
(normalized) initialization values 
$f_n = q^{N_V} / \langle \psi_n, \tilde \phi_n \rangle$.
% The nonstationary quantities  for each $(\mu_k,\sigma_\ell) \in \mathcal R_\Delta$ and $n=1,2$ we are given the 
% eigenvalues $\lambda_n$ (from the precomputation above)
% and obtain $\phi_n(V)$ and $\psi_n(V)$ by using the respective 
% exponential schemes, Eqs.~\eqref{eq_meth_spec_expint}--\eqref{eq_meth_spec_phi_reinjcond_discrete} or \eqref{eq_meth_spec_expint_adjoint}. Next we normalize the eigenfunction of $\LL$ 
% % \textcolor{gray}{of $\LL$ and $\LL^*$} 
% by $\phi_n \leftarrow \phi_n / \langle \psi_n, \phi_n \rangle$ 
% yielding a biorthonormal eigenfunction set\textcolor{gray}{, i.e., 
% Eq.~\eqref{eq_meth_spec_biorth} is satisfied for $m=1,2$}. 
% This allows to obtain
% $f_1 = q_{\phi_1}(V_\mathrm s)$ and $f_2=q_{\phi_2}(V_\mathrm s)$ 
% by ``reading off'' the respective 
% normalized initialization values $f_n = q^{N_V} / \langle \psi_n, \phi_n \rangle$.

The other quantities (nonstationary of $\LL^*$ and the remaining stationary ones) involve partial derivatives 
w.r.t. $\mu$ and $\sigma^2$. They are calculated using a central 
finite difference 
approximation that is second order accurate (in the respective 
step size $\delta \mu$ or $\delta \sigma$), 
$\partial_\mu \theta
\approx [\theta_{\mu+\delta \mu,\sigma}
-\theta_{\mu-\delta \mu,\sigma}]/(2 \delta \mu)$ 
and 
$\partial_\sigma \theta 
\approx (\theta_{\mu,\sigma+\delta \sigma}
-\theta_{\mu,\sigma-\delta \sigma})/(2 \delta \sigma)$ 
for $\theta = r_\infty$, $\langle V \rangle$, $\psi_n(V)$. 
For implementation convenience we calculate $\sigma$-derivatives of 
the quantities and transform them, using the chain rule, to the 
originally required ones,
$\partial_{\sigma^2} \theta = \partial_\sigma \theta / (2\sigma)$.
For the (final) quantities $c^x_n = 
\langle \partial_x \psi_n, \, \phi_0 \rangle$ 
the stationary eigenfunction $\phi_0$ is multiplied with 
the finite difference version of $\partial_x \psi_n$. 
The latter requires for each $x=\mu,\sigma^2$ 
two (forward) integrations via Eq.~\eqref{app_expint_adjoint} (e.g., 
$\psi_n(V; \mu+\delta \mu,\sigma)$ and $\psi_n(V; \mu-\delta \mu,\sigma)$). 
% yielding 
% $(\theta_{\mu,\sigma+\delta \sigma}
% -\theta_{\mu,\sigma-\delta \sigma})/(2 \delta \sigma)$. 
% due to two different values of $x$.
% 
% Note that ...
% comparably large number $N_\lambda > 2$ diffusive ... (see remarks?)

\paragraph{Modifications}

The solutions of Eq.~\eqref{app_eigeneq_sys} or \eqref{app_eigeneq_adjoint_sys} 
can be multiplied with an arbitrary complex scalar value to yield another solution 
because the operators $\LL$ and $\LL^*$ and the corresponding boundary conds. are linear in $\phi$, 
$q$ or $\psi$, $\partial_V \psi$ respectively. Therefore the initializations 
$q^{N_V}$ or $\psi^{0}$ 
of the exponential integration schemes, Eqs.~\eqref{app_expint} or \eqref{app_expint_adjoint}, 
can be chosen arbitrarily and the (bi)normalization 
is applied a posteriori. 
In our numerical implementation 
% of the spectral solver above 
we 
specifically initialize with
$q^{N_V}=1$ only at $\mu_\mathrm{min}$ and for all other $\mu_k > \mu_\mathrm{min}$  
we start the integration with $q^{N_V}(\mu_k)=q^{N_V}(\mu_{k-1})$ where 
$q^{N_V}(\mu_{k-1})$ is taken after normalization. This 
modification 
% to the theoretically also perfectly valid unit-initialization of the flux 
allows to specify tolerance and finite 
difference parameters, 
$\varepsilon$ and $\Delta \lambda$, respectively, 
% for the nonlinear solver of Eq.~\eqref{eq_meth_spec_root}, 
that 
are appropriate for the whole input rectangle 
% $\mathcal R$ 
despite the fact that 
the magnitude of the function which is evaluated in each step of the root finding algorithm, i.e., 
$q(V_\mathrm{lb}; \lambda)$, 
depends strongly on $(\mu, \sigma, \lambda)$ (e.g., see the scales of the attached axes in Fig.~\ref{fig7_spectral}A).

% A nonzero refractory makes the eigenproblem nonlinear (in the eigenvalue) due to the reinjection condition,
%     Eq.~\eqref{eq_meth_spec_phi_reinjcond_ref} and thus requiring
%     a nonlinear method of solution as described above. 

%   \item \textcolor{gray}{Note that the amplitudes of $q(V_\mathrm{lb})$ range within several 
%     orders of magnitude depending on $(\mu, \sigma)$ and the eigenvalue index, 
%     which can be already seen from the solver's initialization, i.e., for $\mu_\mathrm{min}$ (cf. Fig.~\eqref{fig_spectral}A). To have
%     global solver tolerances $\varepsilon$ and finite difference step $\Delta \lambda$ i.e., not dependent on the particular region of $\mathcal R$ we refined the 
%     initialization above to keep $q_{N_V}=1$ only at $\mu_\mathrm{min}$ and take 
%     the $q_{N_V}$ value of the last $\mu_{k-1}$ after normalizing.}

  The numerical solver described above does not take into account the 
    fact that eigenvalues $\lambda_n$ at the transition from real to complex values 
    have multiplicity two. Therefore, at these input parameter points 
    $(\mu^*,\sigma^*)$ we calculate all 
    nonstationary quantities by nearest-neighbor interpolation to resolve 
    corresponding artefacts. Note that an even more pronounced 
    smoothing of the quantities around these points would likely be beneficial to the model performance of the model spec$_2$.
    
%    For strong input noise $\sigma$ more than the first two dominant eigenvalues 
%    can be of the diffusive type at $\mu_\mathrm{min}$ (e.g., at $\sigma_\mathrm{max}$ the first 8 are diffusive \todo{@MA: check the number}{}). For larger total 
%    mean input $\mu$, however, these diffusive modes become irrelevantly 
%    fast and (at least) the first two dominant modes are now regular. Due to the 
%    continuous tracking of the respective eigenvalues for increasing 
%    $\mu$ (see above) sufficiently many (more)
%    modes have to be calculated at $\mu_\mathrm{min}$ (e.g., for $\sigma_\mathrm{max}$ at least 
%    10 eigenvalues are required)
%    although only the first two dominant eigenvalues are finally extracted.
%    This trend is already shown in Fig.~\ref{fig6_spectral}A (right), where two 
%    eigenvalues at $\mu_\mathrm{min}$ would not be sufficient to detect 
%    the dominant eigenvalue for the larger values of the mean input $\mu$. 
  
% \todo{@MA: Explain higher order interpolation polynomial at threshold $\leftarrow$ only if the better solution explained above (taking the flux $q^{N_V}$) does not work properly}

\newpage
\section*{S1 Figure: Fast changes of the input variance}

\begin{figure*}[ht]
\begin{adjustwidth}{-2.25in}{0in}
\includegraphics[width=\maxfigwidth]{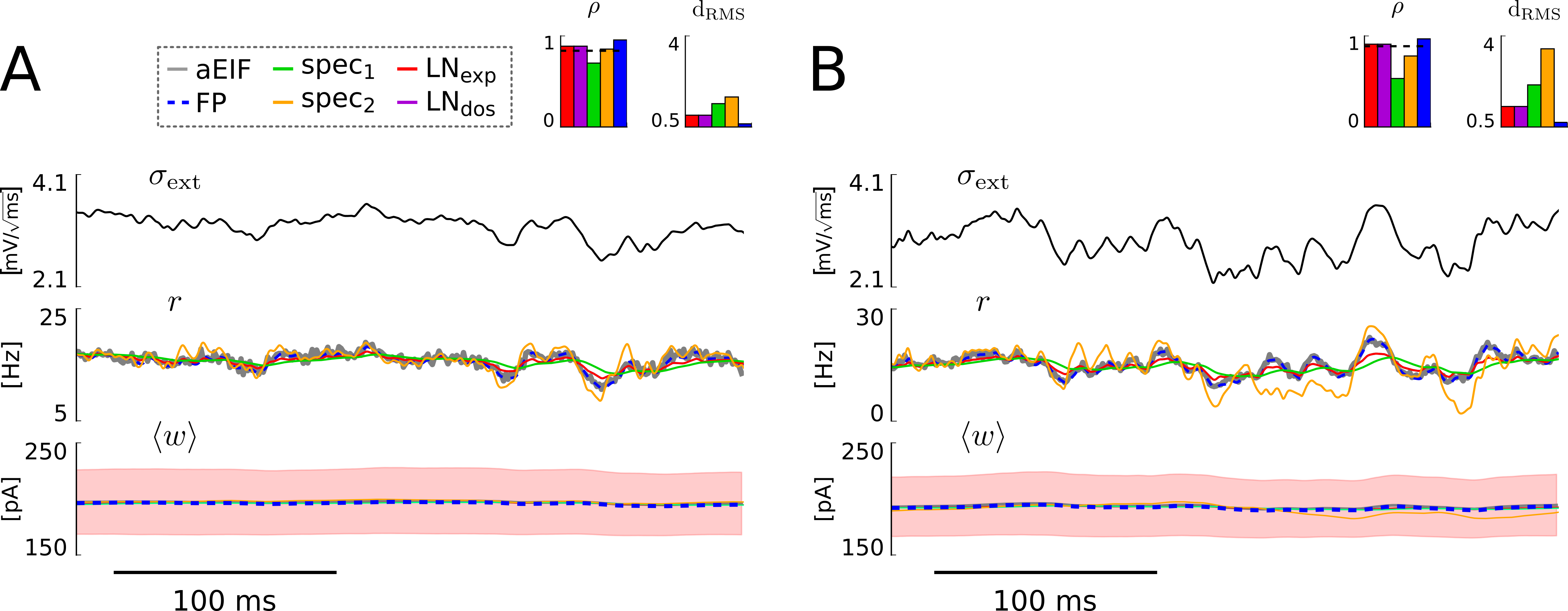}
\end{adjustwidth}
\caption*{\textbf{Fast changes of the input variance.}
Time series of population spike rate and mean adaptation current from the different models in response to weak mean 
$\mu_{\mathrm{ext}}=$ 1.5~$\mathrm{mV}/\mathrm{ms}$ and time-varying 
variance $\sigma^2_{\mathrm{ext}}$ of the input 
for moderately fast variations $\tau^{\sigma^2}_\mathrm{ou}=$ 50~ms (\textbf{A}) and 
rapid variations $\tau^{\sigma^2}_\mathrm{ou}=$ 10~ms (\textbf{B}). 
The values for the remaining parameters and the visualization style 
were as in Fig.~\ref{fig4_sigma2_ou}B of the main text which corresponds to
\textbf{A} here, except that a different realization of the OU process was used.
% To complement the results shown in Fig. 4 we provide this additional Figure to show the 
% effect of a faster varying input variance on all models but especially on the $\text{spec}_2$ model variant.
% Everything which is shown in this figure directly corresponds to the settings of Fig. 4, where (A) 
% is parametrized excactly as in Fig. 4B to see the effect of faster changes of $\sigma^2_{\mathrm{ext}}$
% shown in (\textbf{B}) where $\tau^{\sigma^2}_\mathrm{ou}$ is reduced to 10~ms. As mentioned in the 
% main text we see a strong decrease in the reproduction performance of the $\text{spec}_2$ model by 
% looking at the trace itself as well as the two provided measures the Pearson correlation
% $\rho$ which decreases and the root mean square distance ($\text{d}_\text{RMS}$) which strongly incresases.
}

\end{figure*}

\end{document}